%% file: draft.tex
\documentclass[12pt]{article}

\usepackage[margin=1in]{geometry}
\usepackage[T1]{fontenc}
\usepackage{lmodern}
\usepackage{microtype}
\usepackage{ragged2e}
\usepackage{setspace}
\usepackage{amsmath,amssymb,mathtools}
\usepackage{graphicx}
\usepackage{booktabs}
\usepackage{array}
\usepackage{tabularx}
\usepackage{threeparttable}
\usepackage{multirow}
\usepackage{float}
\usepackage{adjustbox}
\usepackage{etoolbox}
\usepackage{caption}
\usepackage{subcaption}
\usepackage{enumitem}
\usepackage{natbib}
\usepackage{xcolor}
\usepackage[colorlinks=true,allcolors=blue!55!black]{hyperref}
\usepackage[nameinlink,capitalise,noabbrev]{cleveref}
\usepackage{todonotes}
\usepackage[normalem]{ulem}

\graphicspath{{usfigs/}{./}}
\renewenvironment{abstract}%
  {\begin{center}\bfseries\abstractname\end{center}\singlespacing
   \list{}{\leftmargin=0.5in\rightmargin=0.5in\listparindent=0pt\itemindent=0pt\parsep=\medskipamount}\item\relax}%
  {\endlist}
\renewcommand{\arraystretch}{1.05}
\newcommand{\maintablesize}{\small}
\newcommand{\apptablesize}{\footnotesize}
\AtBeginEnvironment{table}{\singlespacing\maintablesize}
\AtBeginEnvironment{figure}{\singlespacing}
\newcolumntype{Y}{>{\raggedright\arraybackslash}X}
\newcommand{\notestyle}{\normalfont\small}
\newcommand{\exhibitnote}[1]{\par\noindent\begin{minipage}{\linewidth}\justifying\setlength{\parindent}{0pt}\notestyle #1\end{minipage}\par\vspace{8pt}}
\newcommand{\tablenote}[1]{\par\noindent\begin{minipage}{\linewidth}\justifying\setlength{\parindent}{0pt}\notestyle #1\end{minipage}\par\vspace{8pt}}

\newcommand{\panelheading}[1]{%
  \par\vspace{8pt}%
  {\raggedright\textit{#1}\par}%
  \vspace{4pt}%
}
\newcommand{\figlink}[2]{\par\medskip\begin{center}[Figure~\ref{#1} Around Here]\end{center}\medskip\par}
\newcommand{\figrangelink}[2]{\par\medskip\begin{center}[Figures~\ref{#1}--\ref{#2} Around Here]\end{center}\medskip\par}
\newcommand{\tablink}[2]{\par\medskip\begin{center}[Table~\ref{#1} Around Here]\end{center}\medskip\par}
\newcommand{\setappfig}[1]{%
  \setcounter{figure}{0}%
  \renewcommand{\thefigure}{#1.\arabic{figure}}%
  \renewcommand{\theHfigure}{app#1.\arabic{figure}}%
}
\newcommand{\setapptab}[1]{%
  \setcounter{table}{0}%
  \renewcommand{\thetable}{#1.\arabic{table}}%
  \renewcommand{\theHtable}{app#1.\arabic{table}}%
}

\newcommand{\checkthis}{\bgroup\markoverwith{\textcolor{yellow!35}{\rule[-0.5ex]{2pt}{2.5ex}}}\ULon}

\title{When Price Charts Rhyme: Historical Visual Retrieval\\for Stock Return Prediction\thanks{We thank Xi Dong, Lenny Kostovetsky, Lin Peng, Anastasia Zakolyukina, Daojing Zhai, and seminar participants at Baruch College, the University of South Florida, and Yeshiva University for helpful suggestions.}}
\author{
Hui Chen\thanks{MIT Sloan and NBER. Email: huichen@mit.edu.}
\and Haoyu Liu\thanks{School of Finance, Renmin University of China. Email: hyl.liu@ruc.edu.cn.}
\and Ke Wu\thanks{School of Finance, Renmin University of China. Email: ke.wu@ruc.edu.cn.}
\and Dexin Zhou\thanks{Zicklin School of Business, Baruch College. Email: dexin.zhou@baruch.cuny.edu.}
}
\date{\today}

\begin{document}
\maketitle
\thispagestyle{empty}

\begin{abstract}
We study whether visually similar price histories predict similar future returns, using a vision model to retrieve each stock's most similar charts from three decades of U.S. price histories. A representation pretrained only on natural images already retrieves neighbors whose realized returns predict the cross-section of stock returns. Aligning the representation with subsequent returns more than doubles the strategy's Sharpe ratio and outperforms a convolutional neural network trained directly on chart images. The predictability is concentrated among stocks with greater limits to arbitrage. Returns and order imbalances around portfolio formation indicate that the signal identifies temporary selling pressure that partially reverts. A sparse autoencoder traces the return-aligned signal to recurring visual states that determine which historical charts are retrieved. Our approach traces each forecast to dated historical charts, making it more transparent than direct image-to-return models and linking rule-based technical analysis with machine-learning chart reading.

\end{abstract}

\clearpage
\setcounter{page}{1}

\section{Introduction}\label{sec:introduction}

Technical analysis builds on chart reading. For example, a trader who sees a stock break out of a narrow trading range on heavy volume may recall similar charts from the past and expect the stock to behave as those earlier stocks subsequently did. The underlying premise is that visually similar price trajectories are followed by similar returns. As the saying goes, history does not repeat itself, but it often rhymes. 

A large literature evaluates this premise by translating specific chart features into trading rules or stock characteristics and testing whether they predict returns. For example, moving-average rules, trading-range breaks, and geometric formations such as head-and-shoulders carry information about subsequent returns \citep[see, e.g.,][]{brock1992,lo2000,han2013}, as do past returns, trading volume, and proximity to past price extremes \citep{debondt1985,jegadeeshtitman1993,lee2000,george2004}. However, each of these tests requires the researcher to specify the pattern before looking at the data. As a result, the evidence speaks only to particular patterns and leaves much of the information in the chart unexplored.

Testing the premise directly requires two ingredients that a chart reader supplies by judgment: a way to decide which historical charts look like the current one, and a record of what happened after those charts. Recent advances in machine learning make the first ingredient feasible at scale. \citet{jiang2023} show that convolutional neural networks (CNNs) extract return-predictive information from price and volume images, and \citet{murray2024} find that neural networks applied to historical return paths predict the cross-section of stock returns.\footnote{See also \citet{xuezhou2026}, who apply pretrained vision transformers to cryptocurrency price images.} However, these models map a price history directly into a return forecast. Historical observations affect each prediction only through the estimated parameters, so the model never explicitly compares the current chart with specific past charts, and we cannot observe the returns that followed comparable episodes. Interpretation is correspondingly difficult: \citet{jiang2023} find that standard price, risk, and liquidity characteristics jointly explain only about 10\% of the cross-sectional variation in their CNN forecasts, and a logistic approximation built from the underlying price and volume data explains less than half.

In this paper, we make this comparison explicit. We propose a historical visual retrieval framework that forecasts a stock's return from the realized returns of its most similar historical charts. For each stock and date, we render the preceding 20 trading days as a standardized chart with open-high-low-close bars, a moving-average line, and trading volume, and we use a vision encoder to map the chart into a common representation space. We then search a recursively expanding library of historical charts drawn from all stocks, industries, and time periods, retain the nearest neighbors, and average their realized subsequent five-day returns. To avoid look-ahead bias, a historical chart enters the library only after its subsequent return has been realized. In this framework, the vision encoder determines which historical charts are comparable, and the realized outcomes of those charts form the forecast.

This design brings the chart reader's premise directly to the data and combines the strengths of both literatures. As with rule-based technical signals, each forecast can be inspected: it rests on identifiable historical charts and the returns that followed them. As with machine-learning chart reading, the relevant patterns are learned from the data.

Three findings emerge. First, general visual similarity contains return information. We use DINOv2-S, a self-supervised vision transformer pretrained on a large collection of general images \citep{oquab2024dinov2}, with no stock-return supervision or finance-specific objective. Over the 2001--2025 out-of-sample period, an equal-weighted high-minus-low portfolio formed on the average outcomes of the 1,000 nearest historical charts earns 26\% per year with an annualized Sharpe ratio of 2.81. The alpha relative to the Fama--French five factors, momentum, and short-term reversal \citep{fama2015,carhart1997,jegadeesh1990,lehmann1990} is 25\%, so standard factors explain almost none of the spread. The relation remains positive across alternative pretrained encoders, neighborhood sizes, neighbor-weighting schemes, and historical libraries that draw on other stocks' charts. Visual features learned from natural images thus already organize the cross-section of price charts in an economically meaningful way.

Second, aligning the representation with subsequent returns sharply strengthens predictability. We fine-tune the encoder on 1993--1999 charts so that charts followed by closer cross-sectional ranks of subsequent five-day returns become closer neighbors in the representation space. We select the model on 2000 data and freeze its parameters before the 2001 out-of-sample period, and we refer to the resulting representation as FT-DINO. With 1,000 neighbors, the FT-DINO equal-weighted high-minus-low portfolio earns 85\% per year, with an annualized alpha of 81\% and a Sharpe ratio of 6.37, 2.3 times the pretrained benchmark. Under capped value weighting, which winsorizes market capitalization at the NYSE 80th percentile \citep{jensen2023}, the portfolio earns 28\% per year with a Sharpe ratio of 2.38, compared with 0.87 for pretrained DINO. The charts, the historical library, and the forecasting rule are unchanged. The gain therefore comes entirely from concentrating a general-purpose representation on the dimensions of visual similarity that matter for subsequent returns.

We benchmark FT-DINO against the 20-day image-based CNN of \citet{jiang2023}, which observes the same price and volume history but predicts returns directly from the image. Under equal weighting, the CNN high-minus-low portfolio earns 61\% per year with a Sharpe ratio of 4.46, compared with 85\% and 6.37 for FT-DINO, a 43\% higher Sharpe ratio. Under capped value weighting, the corresponding Sharpe ratios are 1.82 and 2.38. The two signals are not redundant. After controlling for pretrained DINO, the CNN, and standard factors, FT-DINO retains an annualized alpha of 41\% ($t$-statistic of 16.53) under equal weighting and 15\% ($t$-statistic of 8.50) under capped value weighting, about half of its raw return in both cases.\footnote{Fama--MacBeth regressions \citep{famamacbeth1973} and conditional double sorts yield the same conclusion. A linear predictor formed directly from the FT-DINO embedding performs nearly as well as retrieval, indicating that much of the gain stems from the return-aligned representation itself; retrieval nonetheless retains incremental alpha relative to it.} The two approaches also read the chart differently: the CNN concentrates on the last few trading days, whereas FT-DINO draws on the full 20-day window and relies far more on trading volume.

The predictability is concentrated where arbitrage is costly. Using a composite limits-to-arbitrage score \citep{shleifer1997}, constructed following \citet{chen2026debt} from size, Amihud illiquidity \citep{amihud2002}, dollar volume, days-to-cover \citep{hong2015dtc}, securities-lending fees \citep{davolio2002}, analyst forecast dispersion \citep{diether2002}, and turnover, we find that the annualized FT-DINO spread rises monotonically from 14\% in the lowest quintile to 119\% in the highest, more than an eightfold increase.\footnote{The spread shows a similar gradient across quintiles of the retail share of trading volume, which \citet{laarits2025} find is concentrated in hard-to-value stocks, rising from 10\% to 88\% (Appendix~\ref{app:d_retail}).} These patterns are consistent with mispricing that persists where trading frictions and non-fundamental demand make arbitrage costly and risky.

Returns and order flow around portfolio formation point to temporary price dislocations. The stocks with the highest retrieval signals are recent losers: the high-minus-low portfolio loses 5.39\% in the formation week, as order flow turns to net selling. It then earns 2.59\% over the next 20 trading days as order flow turns to net buying, and its returns over days 61 to 120 are close to zero. About 58\% of the formation-week decline thus reverses, and the reversal is not given back over six months, consistent with non-fundamental selling pressure that partially reverts \citep{coval2007}. A smaller component behaves differently. When we form signals only from the most salient retrieved neighbors, the selected stocks are recent winners bought into the formation date, and the spread earns about 0.6\% in the first month before turning negative over the next five months, consistent with salience-driven demand that later reverses \citep{bordalo2012salience,da2011}. On its own, this component predicts returns poorly, with a Sharpe ratio of 1.53, about one-quarter of that of a similarity-matched benchmark.

Ablation tests identify which visual and historical information drives the signal. Restricting retrieval to the preceding eight years or to the query stock's industry leaves the FT-DINO Sharpe ratio above 5.7, and excluding the query stock's industry leaves it close to the unrestricted 6.37. Restricting the library to the query stock's own history, by contrast, lowers it to 1.38, a 78\% decline. Cross-stock comparisons thus account for most of the retrieval signal. Within the chart, removing volume lowers the Sharpe ratio from 6.37 to 4.98, a 22\% decline, whereas removing the moving-average line has little effect and reversing the red--green color convention barely changes performance. The signal relies primarily on the geometry of the price path, and volume adds to it.

The signal remains economically meaningful under standard investability screens and transaction costs, particularly for equal-weighted portfolios. Among stocks priced at or above \$5, the equal-weighted FT-DINO portfolio earns 36\% per year with a Sharpe ratio of 3.94, and excluding the smallest 20\% of stocks yields the same return with a Sharpe ratio of 3.44. After one-way transaction costs of 10 basis points for stocks above the NYSE 80th size percentile and 20 basis points for other stocks, following \citet{jiang2023} (see also \citealp{novymarx2016}), the weekly strategy still earns 55\% per year with a Sharpe ratio of 4.12, retaining about two-thirds of its gross return. A four-week holding interval cuts turnover from 600\% to 173\% while preserving positive net performance.\footnote{The predictability also does not hinge on a particular chart length or forecast horizon. Across 20-, 40-, and 60-day chart windows, FT-DINO Sharpe ratios range from 5.86 to 6.37 for five-day returns and from 2.11 to 2.19 for 20-day returns with monthly rebalancing.}

Third, we trace the FT-DINO signal to the visual content of the charts. Each retrieval signal is tied to specific historical charts, but the reason those charts are comparable is not directly observable: FT-DINO represents each chart with a 384-dimensional dense embedding whose coordinates are not designed to encode distinct visual concepts. We therefore fit a sparse autoencoder to the FT-DINO representations using 1993--2000 data, fix it before the out-of-sample period, and decompose each dense representation into a larger set of sparsely activated visual features \citep{olson2025visionsae}. This decomposition lets us identify recurring visual structures, relate them to retrieval, and measure their importance for portfolio performance. We focus on the return-aligned representation because it generates our main results.

Our bottom-up analysis starts from the 100 sparse features with the highest return-related importance in rolling annual predictive regressions. We retain a feature when its representative real-market charts, drawn from the top 5\% of its positive activations, consistently display the same visual characteristic. The screen yields 30 visually consistent features, which we group into 11 visual states. Four states matter most for FT-DINO performance: Isolated Upward Spike, Sustained Uptrend, Late Upside Breakout, and High-Volatility Range. Removing the stocks that most strongly activate Isolated Upward Spike lowers the Sharpe ratio from 6.37 to 5.90, a 7\% decline, and removing the strongest activators of the other three states lowers it to 6.11, 6.19, and 6.24, respectively. No single state accounts for a large share of the signal. Shifting only the query representation along the selected feature directions, with the historical library held fixed, has much larger effects: moving the query away from the High-Volatility Range direction lowers the Sharpe ratio to 2.32. The sparse directions thus operate directly through the retrieval process.

The top-down analysis reverses the direction of discovery. Starting from 11 recognizable price configurations, including Up Momentum, Breakout Up, Mean Reversion, discrete winner and loser paths, MAX(1), and price-path convexity, we select real-market charts that represent each pattern and retain the sparse features that activate on them. We then validate each feature against its highest-activation real-market charts. For each configuration, we find a sparse direction that captures the pattern. These directions matter less for returns than the bottom-up features: at steering level $-10$, the average Sharpe ratio falls to 4.72, compared with 3.45 for the bottom-up features. Textbook patterns are present in the representation, but they play a smaller role in its predictions.

Finally, we use the sparse representation to organize 21 visual characteristics previously studied in asset pricing, ranging from the classical formations of \citet{lo2000} to MAX(1), overnight jumps, and price-path convexity.\footnote{The characteristics comprise eight classical formations from \citet{lo2000} (head-and-shoulders and inverse head-and-shoulders, broadening tops and bottoms, triangle tops and bottoms, and double tops and bottoms); four continuous and discrete winner and loser paths from \citet{da2014}; MAX(1) from \citet{bali2011}; negative overnight jumps from \citet{jiangzhu2017}; drop-and-rebound patterns from \citet{bremer1991}; four combinations of price direction and price-path convexity from \citet{gulenwoeppel2026}; daytime reversals from \citet{akbas2022}; and price-volume lead-lag dynamics from \citet{liao2022}.} We encode the 24.2 million out-of-sample charts from 2001--2025, cluster their sparse representations into 4,096 prototypes, and locate each characteristic in the resulting visual atlas. The atlas places researcher-defined patterns and the visual states learned from market data in a common coordinate system, so that traditional technical patterns can be read in terms of the learned representation.

Taken together, these analyses make the retrieval signal transparent at three levels. The retrieved neighbors show which historical charts support each forecast, the sparse features show which visual states make those charts comparable, and the atlas locates patterns from the technical-analysis literature among the states learned from market data. The framework thus connects research on rule-based technical signals with research on machine-learning chart reading: it keeps the flexibility of a learned representation while expressing its predictions in terms that can be compared with familiar price patterns.

Our paper contributes to four strands of literature. First, we contribute to research on return predictability from past prices and trading volume. Past returns predict future returns at different horizons, trading volume and proximity to the 52-week high carry additional information, and a technical-analysis literature studies moving-average rules, trading-range breaks, and recurring geometric formations.\footnote{See, for example, \citet{debondt1985}, \citet{jegadeeshtitman1993}, \citet{lee2000}, \citet{george2004}, and \citet{han2016} on return, volume, and moving-average signals, and \citet{brock1992}, \citet{lo2000}, and \citet{han2013} on technical analysis. \citet{sullivan1999} and \citet{bajgrowicz2012} emphasize data snooping, persistence, and transaction costs in evaluating such rules.} Each of these signals is specified before it is tested. We advance this research by shifting the object of analysis from a predefined signal to the full price chart: a learned representation determines which historical price paths are comparable, and their realized returns form the prediction. Our interpretation tools then map the learned signal back to the patterns this literature studies, so that rule-based technical indicators and machine-learned visual states can be compared within a common representation.

Second, we contribute to the use of machine learning in empirical asset pricing. Flexible models capture nonlinearities and interactions among predictors that are difficult to accommodate in conventional specifications \citep{gu2020,murray2024}. \citet{jiang2023} render price and volume histories as images and train CNNs to predict subsequent returns, and \citet{xuezhou2026} apply pretrained vision transformers to cryptocurrency price images. In these models, the historical sample shapes each forecast only through the fitted parameters. Our framework separates the representation of price histories from the forecasting rule: the encoder organizes the historical sample, whereas the forecast is the average realized return of the retrieved observations. The observations behind each prediction are therefore visible, and differences in performance can be attributed to which historical charts the representation treats as similar.

Third, we contribute to research on similarity-based judgment and forecasting. Models of memory emphasize that current cues determine which past experiences are recalled and, in turn, which outcomes shape beliefs \citep{mullainathan2002,bordalo2020}, and memory and similarity influence attention, beliefs, and trading in financial markets \citep{charles2025,jiangliu2025,charlessui2026}. Closely related forecasting methods predict the present from outcomes that followed similar historical states; \citet{chenhoberg2025}, for example, forecast aggregate market returns from historically similar news states. We extend this approach to high-dimensional, stock-level visual states, in which similarity is defined over complete price charts and the relevant histories can come from different firms, industries, and periods. Our finding that cross-stock histories carry most of the signal points to opportunities for models of similarity-based beliefs to incorporate analogies drawn across assets.

Finally, we contribute to the literature on visual financial information and the interpretation of high-dimensional models. Visual salience in price charts affects how investors weight past returns \citep{bose2022}. \citet{wang2026} use a vision model trained only on generic images to complete the hidden right-hand portion of price charts, mimicking how investors visually extrapolate a price path, and find that the implied returns predict next-day returns positively but returns over the following days negatively, consistent with a short-lived overreaction. Our retrieval signal instead predicts returns positively over the subsequent week and month, and these gains are not reversed over the following six months, suggesting that it captures a different channel from visual extrapolation of the price path. Beyond price charts, images carry economically relevant information more broadly: \citet{obaid2022} measure investor sentiment from news photos, and \citet{kaczmarek2024} identify firm peers from the images that firms use to depict their operations. Related work interprets economic forecasts from large language models through concept-based representations and feature steering \citep{chen2026brainscan}. We decompose a return-relevant visual representation into sparse features and connect those features to both neighbor selection and portfolio performance through bottom-up discovery, top-down validation, feature steering, and a visual atlas that locates established price-based characteristics among the visual states learned from market data.

The remainder of the paper proceeds as follows. Section~\ref{sec:methodology} describes the sample, the historical-retrieval design, and return-rank alignment. Section~\ref{sec:results} presents the portfolio results, first for the pretrained representation and then for the return-aligned representation, together with model comparisons, ablations, and tradability. Section~\ref{sec:mechanism} examines the economic mechanism through cross-sectional heterogeneity, returns and order flow around portfolio formation, and the salience of retrieved neighbors. Section~\ref{sec:interpretation} introduces a sparse autoencoder and uses it to interpret the return-aligned visual representation, and Section~\ref{sec:conclusion} concludes.

\section{Data and Methodology}\label{sec:methodology}

\subsection{Sample and Price-Chart Construction}\label{sec:sampleconstruction}

We use daily CRSP data for common stocks listed on the NYSE, AMEX, and Nasdaq, excluding ADRs, non-U.S.-incorporated firms, and observations with abnormal trading status. The sample runs from 1993 to 2025 because daily opening, high, and low prices first become available in June 1992. The 1993--2000 period supplies the initial historical library, and 2001--2025 is reserved for out-of-sample evaluation.

For each stock-date observation, we construct a standardized chart from the 20 trading days ending on date $t$. Following the image design of \citet{jiang2023}, the chart contains daily open-high-low-close bars, a 20-day moving average, and trading volume, and prices are adjusted for corporate actions using CRSP returns.\footnote{Because closing prices are constructed from total returns, splits and dividend payments do not appear as artificial price changes in the chart.} Within each image, we normalize the first closing price to one, construct subsequent closing prices recursively from returns, and scale each day's opening, high, and low prices in proportion to that day's closing price. Prices are then scaled within the displayed window, and volume is scaled relative to its within-window maximum. \citet{jiang2023} show that this image scaling, which places every stock's recent price range on a common scale, is a key source of the predictive power of image-based models. A day is drawn in green if its close exceeds its open and in red otherwise; the moving average is blue, and the background is white. The image therefore captures the shape of the recent price path, independent of nominal price and share-volume levels.

Unlike the black-and-white images of \citet{jiang2023}, we render the charts in color. The vision encoders we use are pretrained on natural color images and readily distinguish colors, so color provides an additional visual cue. It separates days that close above their open from days that close below it, and the moving average from the price bars. A black-and-white image conveys these distinctions only through finer details, such as the relative position of the open and close ticks. Figure~\ref{fig:samples} compares our colored chart with the black-and-white design of \citet{jiang2023}. %

\figlink{fig:samples}{Price-Chart Specifications}

The historical library is daily, whereas queries are formed on the last trading day of each week. A daily library gives each weekly query a much richer set of dated historical paths than a weekly library would. We require both the complete price history needed to draw a chart and the complete subsequent return needed for its later use as a neighbor. Table~\ref{tab:summarystats} reports summary statistics for the 4,569,085 out-of-sample stock-week observations. Panel A covers the subsequent five-day return and the retrieval signals, and Panel B covers the stock-level controls used in the cross-sectional tests.

\tablink{tab:summarystats}{Summary Statistics}

\subsection{Historical Visual Retrieval}\label{sec:retrievalmethod}

Our benchmark encoder is DINOv2-S, a self-supervised vision transformer pretrained on a large collection of general, unlabeled images without stock-return labels or a finance-specific objective \citep{oquab2024dinov2}. Let $X_{i,t}$ denote the standardized chart for stock $i$ ending on date $t$. We feed the chart through the frozen encoder and use the final-layer CLS token as a 384-dimensional visual representation,
\begin{equation}
    h_{i,t}=f_{\theta_0}(X_{i,t})\in\mathbb R^{384},
    \qquad
    z_{i,t}=\frac{h_{i,t}}{\lVert h_{i,t}\rVert_2},
    \label{eq:embedding_main}
\end{equation}
where $\theta_0$ denotes the pretrained parameters. The normalized representation $z_{i,t}$ allows us to compare complete price charts in a common visual space.

On each weekly query date, every eligible stock chart is compared with a recursively expanding library of daily historical charts. To prevent look-ahead, a historical observation dated $\tau$ is eligible only after its complete subsequent five-trading-day return has already been realized by the query date $t$. For query $(i,t)$ and eligible historical observation $(j,\tau)$, visual similarity is cosine similarity between their normalized representations,
\begin{equation}
    s\big((i,t),(j,\tau)\big)
    =
    z_{i,t}^{\prime}z_{j,\tau}.
    \label{eq:similarity_main}
\end{equation}
We rank all observations by this similarity and retain the exact top-$K$ historical neighbors,
\begin{equation}
    \mathcal N_K(i,t)
    =
    \operatorname{TopK}_{(j,\tau)}
    s\big((i,t),(j,\tau)\big).
    \label{eq:topk_main}
\end{equation}
The retrieval signal is then the equal-weighted mean of the realized subsequent five-day returns,
\begin{equation}
    \widehat R_{i,t}^{(K)}
    =
    \frac{1}{K}
    \sum_{(j,\tau)\in\mathcal N_K(i,t)}
    R_{j,\tau+1:\tau+5}.
    \label{eq:retrievalsignal_main}
\end{equation}

\subsection{Return-Rank Alignment}\label{sec:rankalignment}

The pretrained representation captures general visual similarity. Two charts can therefore be close because they share visual features that have little to do with what happens next. We fine-tune the encoder before the out-of-sample period so that its geometry better reflects similarity in subsequent returns.

For each training date $t$, we first convert subsequent five-trading-day returns into cross-sectional percentile ranks,
\begin{equation}
    u_{i,t}
    =
    \operatorname{PctRank}_{t}\!\left(R_{i,t+1:t+5}\right)
    \in[0,1].
    \label{eq:returnrank_main}
\end{equation}
Within a training batch, observations with closer return ranks receive greater target affinity. Indexing observations by $a$ and $b$, we define a soft target-neighbor distribution
\begin{equation}
    r_{ab}=-\lvert u_a-u_b\rvert,
    \qquad
    p_{ab}
    =
    \frac{\exp(r_{ab}/\kappa_r)}
    {\sum_{c\neq a}\exp(r_{ac}/\kappa_r)}.
    \label{eq:targetdist_main}
\end{equation}
The model-implied neighbor distribution is constructed from cosine similarities in the current embedding space,
\begin{equation}
    s_{ab}=z_a^{\prime}z_b,
    \qquad
    q_{ab}
    =
    \frac{\exp(s_{ab}/\kappa_z)}
    {\sum_{c\neq a}\exp(s_{ac}/\kappa_z)}.
    \label{eq:modeldist_main}
\end{equation}
Return-rank alignment chooses the encoder parameters to make these two neighborhood distributions similar by minimizing
\begin{equation}
    \mathcal L_{\mathrm{rank}}
    =
    \frac{1}{B}
    \sum_{a=1}^{B}
    D_{\mathrm{KL}}\!\left(p_a\Vert q_a\right).
    \label{eq:rankloss_main}
\end{equation}
The loss therefore reshapes visual similarity so that charts with more similar subsequent-return ranks are pulled toward similar neighborhoods.

We fine-tune DINOv2-S on 1993--1999, use 2000 for validation, and freeze the selected model before the 2001 out-of-sample period begins. Appendix~\ref{app:a_rank_alignment} reports the full training construction and optimization details and shows how the raw embedding geometry changes after alignment.

\section{Historical Visual Retrieval and Stock Returns}\label{sec:results}

In this section, we examine whether historical visual retrieval predicts the cross-section of stock returns. We first ask whether a representation learned without financial supervision already retrieves informative neighbors (Section~\ref{sec:benchmark}), and then adapt the same encoder to return prediction by fine-tuning it so that charts followed by similar returns become close neighbors (FT-DINO). The forecasting rule is identical across the two representations: only the space in which neighbors are selected changes. We evaluate the portfolio performance of the adapted representation (Section~\ref{sec:ftperformance}) and compare FT-DINO with benchmark DINO and a direct image-prediction model (Section~\ref{sec:modelcomparison}). Sections~\ref{sec:ablation}--\ref{sec:alternativehorizons} assess robustness through ablations, tradability and transaction costs, and alternative image windows and return horizons.

\subsection{Historical Visual Retrieval with Pretrained DINO}\label{sec:benchmark}

We begin by examining whether the realized returns of visually similar historical charts predict the current stock's return when the visual representation is learned without financial supervision. If the pretrained representation captures only generic visual similarity unrelated to what follows a chart, the retrieval signal should have no predictive power. If instead general visual features also organize price paths by their subsequent outcomes, stocks with higher retrieval signals should earn higher returns. Following Section~\ref{sec:methodology}, each stock's signal is the average realized subsequent five-day return of its $K$ most similar historical charts (equation~\eqref{eq:retrievalsignal_main}). Each week, we sort stocks into deciles by the signal and form equal-weighted (EW), value-weighted (VW), and capped value-weighted (CVW) high-minus-low (H--L) portfolios, where CVW winsorizes market capitalization at the NYSE 80th percentile \citep{jensen2023}. Table~\ref{tab:benchmark} reports the results for the benchmark DINO signal.

\tablink{tab:benchmark}{DINO Portfolio Performance}

The benchmark signal predicts returns for every value of $K$ and under all three weighting schemes. Under equal weighting, the annualized H--L return rises from 7\% at $K=10$ to 26\% at $K=1{,}000$ and 29\% at $K=3{,}000$, and the Sharpe ratio rises from 1.12 to 2.81 and 2.90. Factor-model alphas are close to raw returns: at $K=1{,}000$, the alpha of 25\% accounts for almost the entire long-short return spread. The retrieved outcomes therefore capture return variation that the standard factors do not price. The gain from broader neighborhoods is consistent with averaging away noise in individual historical outcomes while retaining information common to visually similar cases.

Value-weighted performance is weaker. At $K=1{,}000$, the VW Sharpe ratio is 0.60, about one-fifth of the EW Sharpe ratio, indicating that the benchmark signal is less informative among the largest stocks. This pattern is consistent with price-path information being incorporated more quickly in large, liquid stocks, where arbitrage capital is more abundant \citep{shleifer1997}. Capped value weighting, which limits the influence of the largest firms, raises the Sharpe ratio to 0.87.\footnote{Appendix~\ref{app:b_time} plots performance over time, Appendix~\ref{app:b_library} tests historical-library restrictions, Appendix~\ref{app:b_cases} provides three high-similarity cross-stock pairs, and Appendices~\ref{app:b_weighting}--\ref{app:b_cnn} test neighbor weighting, alternative pretrained encoders, and the comparison with the CNN. Cross-industry retrieval performs about as well as the unrestricted benchmark, every alternative pretrained encoder produces positive performance, and equal weighting performs at least as well as recency weighting. The price configurations that precede similar returns therefore recur across firms and market regimes, and the result does not hinge on one pretrained representation.} Chart similarity measured by a general-purpose image model, pretrained without any financial data, thus carries economically meaningful information about future stock returns.

\subsection{Historical Visual Retrieval with Return-Aligned DINO}\label{sec:ftperformance}

We next examine whether adapting the pretrained representation to the return-prediction task strengthens the signal. Because DINO captures general visual similarity, two charts can be close in the DINO space for reasons unrelated to subsequent returns, which adds noise to the retrieved outcomes. Return-rank alignment (Section~\ref{sec:rankalignment}) addresses this by making charts followed by similar returns closer neighbors, using only pre-2001 data. The forecasting rule is unchanged: FT-DINO forms its signal from the realized returns of its $K$ nearest historical charts exactly as DINO does.

Table~\ref{tab:ftperformance} reports the results. Return-rank alignment sharply increases the predictive content of the retrieved neighborhoods. With only ten neighbors, FT-DINO already delivers an equal-weighted Sharpe ratio of 3.59, more than three times the 1.12 of benchmark DINO. Performance rises with $K$ and stabilizes around $K=1{,}000$. At that neighborhood size, the equal-weighted H--L portfolio earns 85\% per year, with an alpha of 81\% and a Sharpe ratio of 6.37, 2.3 times the benchmark. The CVW portfolio earns 28\% per year with a Sharpe ratio of 2.38, compared with 0.87 for benchmark DINO. Because the forecasting rule is unchanged, the gain suggests that much of the return information in price charts lies in how historical episodes are matched. The smaller CVW spread indicates that, as with the benchmark signal, this information is concentrated among smaller stocks.

\tablink{tab:ftperformance}{FT-DINO Portfolio Performance}

Figure~\ref{fig:ftcum} plots cumulative log returns. The equal-weighted paths rise through most of the out-of-sample period, and broader neighborhoods produce smoother and steeper paths.\footnote{Appendix~\ref{app:c_kweights} extends the neighborhood grid and finds that emphasizing recent neighbors does not improve on equal weighting, and Appendix~\ref{app:c_taskvector} traces the same fine-tuning direction through task-vector steering.}

\figlink{fig:ftcum}{Cumulative Log Returns to FT-DINO Portfolios}

One might worry that alignment merely teaches the encoder to sort stocks by known risk exposures. When we instead fine-tune on FF6-residual returns based on the Fama--French five factors and momentum \citep{fama2015,carhart1997}, Fama--French 12 industry-adjusted returns, or DGTW-adjusted returns \citep{daniel1997}, equal-weighted Sharpe ratios range from 5.82 to 6.55 (Appendix~\ref{app:c_targets}), so the gains do not merely reflect factor, industry, or characteristic exposures.\footnote{Holding the retrieved neighbors fixed and averaging their CAPM or FF6 alphas yields equal-weighted Sharpe ratios of 6.84 and 6.50 (Appendix~\ref{app:c_alpha}).} The signal is also not confined to a particular momentum state: within groups of stocks formed on the signs of their past 250-day and 20-day returns, the FT-DINO 5--1 spread ranges from 43\% to 81\% per year (Appendix~\ref{app:c_momentum}).

\subsection{Model Comparison}\label{sec:modelcomparison}

We next compare historical retrieval with direct image prediction. The CNN follows the 20-day black-and-white design of \citet{jiang2023}, and we train it following their procedure, whereas FT-DINO uses the same 20 days of information but predicts through the outcomes of retrieved historical cases. Table~\ref{tab:modelcomparison} reports the extreme-decile returns and H--L spreads.

\tablink{tab:modelcomparison}{Portfolio Performance: FT-DINO and CNN}

Under equal weighting, FT-DINO produces an annualized spread of 85\% and a Sharpe ratio of 6.37, compared with 61\% and 4.46 for the CNN, a 43\% higher Sharpe ratio. The advantage comes from both legs. FT-DINO's bottom decile earns a lower return and its top decile a higher return than the CNN's. The ranking persists under CVW, where the Sharpe ratios are 2.38 and 1.82. Retrieval therefore identifies both future losers and future winners more sharply than direct prediction. The advantage does not stem from chart color alone: even without color, FT-DINO earns a Sharpe ratio of 5.71 (Section~\ref{sec:ablation}).\footnote{Appendix~\ref{app:b_nonvisual} applies the same retrieval rule to numerical and text representations of the 20-day window. Raw price and volume data yield an equal-weighted Sharpe ratio of 2.47, and a ChronoBERT embedding of the window rendered as text yields 1.42, both below the 2.81 of benchmark DINO and well below the 6.37 of FT-DINO. Numerical similarity performs worse than image-based similarity, consistent with a chart presenting prices, the moving average, and volume jointly on a common time axis.} A linear predictor built on the FT-DINO embedding performs nearly as well as retrieval (Appendix~\ref{app:c_incremental}), so much of the gain over the CNN stems from the return-aligned representation, with the retrieval rule adding incremental information on top of it.\footnote{FT-DINO and the CNN differ in both the encoder, a vision transformer pretrained on natural images versus a CNN trained from scratch, and the forecasting rule. The ridge comparison holds the encoder fixed: after controlling for the FT-DINO ridge portfolio, the CNN, benchmark DINO, and standard factors, FT-DINO retrieval retains an annualized alpha of 21.00\% under EW and 8.64\% under CVW (Appendix Table~\ref{tab:app_ridge}).}

We next examine whether FT-DINO contains information beyond benchmark DINO and the CNN. The signals are far from redundant: averaged over weeks, the cross-sectional correlation of the FT-DINO signal with the CNN signal is about 0.40, and that of the benchmark DINO signal is about 0.16. To test for incremental information formally, we regress each strategy's H--L return on the other image-strategy returns and standard factors. Table~\ref{tab:spanning} reports the results. FT-DINO retains an annualized alpha of 40.95\% ($t$-statistic of 16.53) under EW and 14.66\% ($t$-statistic of 8.50) under CVW, about half of its raw return in each case. By contrast, the CNN's EW alpha falls to 3.7\% ($t$-statistic of 0.74) once FT-DINO and DINO are included, although its CVW alpha of 5.5\% ($t$-statistic of 2.67) remains significant. The relation is largely asymmetric: the retrieval signals absorb most of the CNN's predictive content, whereas the CNN leaves much of FT-DINO's alpha intact.\footnote{Appendix~\ref{app:c_incremental} reaches the same conclusion from Fama--MacBeth regressions and conditional double sorts. Appendix~\ref{app:c_subsample} reports subsample stability and alternative training windows.}

\tablink{tab:spanning}{Factor-Augmented Spanning Regressions}

The two approaches also differ in where they look within the chart. Because the retrieval signal in equation~\eqref{eq:retrievalsignal_main} weights all $K$ neighbors equally, it depends on the query chart only through which historical charts enter the neighborhood, so small changes to individual pixels typically leave the signal unchanged. We therefore measure which pixels matter for the forecast through retrieval strength, the degree to which more similar neighbors receive more weight. Holding the neighborhood $\mathcal N_K(i,t)$ fixed, let neighbor $(j,\tau)$ receive weight proportional to $e^{\lambda s_{j,\tau}}$, where $s_{j,\tau}$ is its similarity to the query in equation~\eqref{eq:similarity_main} and $R_{j,\tau}\equiv R_{j,\tau+1:\tau+5}$ is its subsequent five-day return. Setting $\lambda=0$ recovers the equal-weighted signal, and a first-order expansion around $\lambda=0$ gives
\begin{equation}
    \widehat R_{i,t}^{(K)}(\lambda)
    =
    \widehat R_{i,t}^{(K)}
    +
    \lambda\,g_{i,t}
    +O(\lambda^{2}),
    \qquad
    g_{i,t}
    =
    \frac{1}{K}
    \sum_{(j,\tau)\in\mathcal N_K(i,t)}
    \big(R_{j,\tau}-\widehat R_{i,t}^{(K)}\big)\,s_{j,\tau}.
    \label{eq:forecastsens_main}
\end{equation}
The return-aligned component $g_{i,t}$ rises when the query moves closer to neighbors with above-average subsequent returns and away from those with below-average returns. Forecast sensitivity at a pixel is the magnitude of the gradient of $g_{i,t}$ with respect to that pixel.\footnote{Appendix~\ref{app:c_sensitivity} defines the measure formally and compares it with the sensitivity of the embedding itself and of its similarity to historical charts.}

Figure~\ref{fig:sensitivity} reports the results. The CNN concentrates its sensitivity on the last one or two trading days. Pretrained DINO, by contrast, responds to the entire 20-day window, consistent with the global self-attention of a vision transformer, which relates every region of the image to every other. FT-DINO lies between the two: return alignment shifts its sensitivity toward the most recent days and the region where prices are drawn, while it continues to draw on the full window. FT-DINO also responds strongly to the volume bars across the window, which receive little weight in the CNN. This pattern is consistent with the ablation evidence in Section~\ref{sec:ablation}, where removing volume lowers the FT-DINO Sharpe ratio by 22\%, compared with 14\% for the CNN.\footnote{Appendix~\ref{app:c_temporal} confirms the differences across the chart window with perturbation tests that shuffle the order of days within parts of the window. Appendix~\ref{app:c_transform} reverses the entire path instead: flipping the query chart horizontally or rotating it by 180 degrees eliminates the FT-DINO spread, with equal-weighted Sharpe ratios of $-0.45$ and $-0.07$, whereas benchmark DINO retains about half of its Sharpe ratio.}

\figlink{fig:sensitivity}{Forecast Sensitivity Maps: CNN, DINO, and FT-DINO}

\subsection{Sources of the Retrieval Signal}\label{sec:ablation}

We next examine which parts of the retrieval design and chart specification drive the FT-DINO signal. Figure~\ref{fig:ft_ablation_examples} illustrates the chart-element and color variants, holding the underlying price history fixed.

\figlink{fig:ft_ablation_examples}{Price-Chart Ablations}

Panel A of Table~\ref{tab:ft_ablation} varies the historical library. Restricting retrieval to the previous eight years or to the same industry lowers the Sharpe ratio to 5.77 and 6.06, respectively. Excluding the query stock's own industry leaves it at 6.24, close to the unrestricted 6.37, whereas using only the query stock's own history lowers it to 1.38, a 78\% decline.\footnote{For same-stock retrieval, $K$ is an upper bound. If fewer than $K$ eligible observations are available, we use all of them and average their valid subsequent returns.} Cross-stock histories are the main source of retrieval performance. Part of the decline reflects the much smaller library available from a single stock's past, but excluding only the query stock's industry, which leaves a large library, barely changes performance. The same price configuration appears to carry similar implications for subsequent returns regardless of the firm or industry in which it occurs.

Panels B and C vary the rendered chart. Removing color lowers the Sharpe ratio to 5.71, but reversing the red--green mapping leaves it at 6.29, and randomizing colors still yields 5.92. Removing the moving average has little effect, whereas removing volume lowers the Sharpe ratio to 4.98, a 22\% decline, and removing both yields a similar 4.83.\footnote{Removing volume matters less for direct image prediction. Applied to the CNN with its original weights, without re-training, the same ablation lowers its EW Sharpe ratio from 4.46 to 3.86, a 14\% decline. Under CVW, the FT-DINO Sharpe ratio falls from 2.38 to 1.75, a 26\% decline, whereas the CNN's falls only from 1.82 to 1.75, a 4\% decline.}
The representation relies mainly on price-path geometry, and volume adds to it, consistent with trading volume conveying information about the intensity of the demand behind a price move \citep{lee2000}. Because the encoder is not re-trained on the modified charts, these tests measure only how much the fixed representation relies on each chart element.

\tablink{tab:ft_ablation}{FT-DINO Ablations}

Overall, the signal draws mainly on analogies across stocks and on the geometry of the price path, with volume adding information and color helping as a visual cue, although its conventional red--green mapping plays little role.\footnote{The signal also does not proxy for how much of the image the chart occupies. Within quintiles of the fraction of blank pixels in the chart, the FT-DINO 5--1 spread ranges from 43\% to 86\% per year and is significant in every quintile (Appendix~\ref{app:c_sparsity}).}

\subsection{Tradability and Transaction Costs}\label{sec:tradability}

The large equal-weighted spread raises the question of whether the result is confined to stocks that are difficult to trade. Table~\ref{tab:tradability} applies price and size screens and then deducts transaction costs based on portfolio turnover.

\tablink{tab:tradability}{Tradability and Transaction Costs}

FT-DINO remains economically strong under these restrictions. Among stocks priced at or above \$5, the equal-weighted strategy earns 36\% per year with a Sharpe ratio of 3.94. Excluding the smallest 20\% of stocks yields the same 36\% return and a Sharpe ratio of 3.44. The decline from the full-sample 85\% indicates that low-priced and small stocks contribute a substantial part of the equal-weighted spread, in line with the concentration of predictability among hard-to-arbitrage stocks documented in Section~\ref{sec:econrelevance}. We then charge one-way costs of 10 basis points for stocks above the contemporaneous NYSE 80th market-capitalization percentile and 20 basis points for all other stocks, following the same convention as \citet{jiang2023}. Net of these costs, the weekly strategy earns 55\% per year with a Sharpe ratio of 4.12, retaining about two-thirds of its gross return. The same costs reduce the net returns of benchmark DINO and the CNN to $-6\%$ and 29\%. Return alignment therefore makes the signal considerably more robust to trading costs, although net performance is weaker when larger stocks receive more weight.\footnote{Appendix~\ref{app:c_holding} finds that a four-week holding interval cuts FT-DINO turnover from 600\% to 173\% while preserving positive net performance.}

\subsection{Alternative Image Windows and Return Horizons}\label{sec:alternativehorizons}

The baseline design uses 20 trading days of visual history to predict the subsequent five trading days. We examine whether the result depends on either choice. Figure~\ref{fig:app_windows} shows the 20-, 40-, and 60-day charts, each with the matching moving-average window and otherwise identical conventions.

\figlink{fig:app_windows}{Alternative Price-Chart Windows}

Table~\ref{tab:alternative_horizons} reports the results. For the weekly five-day specification, benchmark DINO Sharpe ratios range from 2.63 to 2.97 across windows, and FT-DINO Sharpe ratios range from 5.86 to 6.37. For the 20-day horizon, queries are formed monthly on the last trading day of each month, and FT-DINO is re-trained on subsequent 20-day return ranks. Predictability remains positive for all three windows, with FT-DINO Sharpe ratios of 2.11 to 2.19 compared with 1.54 to 1.99 for DINO. The lower Sharpe ratios at the monthly horizon are consistent with \citet{jiang2023}, who find that image-based forecasts are most potent in the first week after the chart is observed.\footnote{Appendix~\ref{app:c_horizons} reports the full results across neighborhood sizes.}

\tablink{tab:alternative_horizons}{Alternative Image Windows and Return Horizons}

In summary, historical visual retrieval predicts returns across lookback windows and forecast horizons, with the strongest predictability at short horizons.

\section{Economic Mechanism}\label{sec:mechanism}

Having established that historical visual retrieval predicts returns robustly, we next examine the economic forces behind the FT-DINO spread. We first document where the predictability is strongest (Section~\ref{sec:econrelevance}). We then trace returns and order flow around portfolio formation, which point to temporary price dislocations that partially revert (Section~\ref{sec:dislocation}). Finally, we separate the retrieved neighbors by salience and identify a smaller component whose returns reverse over the following months (Section~\ref{sec:salient}).

\subsection{Limits to Arbitrage}\label{sec:econrelevance}

We begin by examining where FT-DINO predictability is strongest. The cross-sectional pattern helps distinguish between two interpretations of the retrieval spread. If the spread compensates investors for bearing systematic risk, it need not vary with the cost of trading against it. If it instead reflects mispricing, theories of limited arbitrage predict that it should persist mainly where arbitrage is costly and risky \citep{shleifer1997}. Illiquidity and trading costs deter arbitrageurs from exploiting a signal \citep{lesmond2004,sadka2007}, and short-sale constraints combined with investor disagreement allow overpricing to survive \citep{miller1977,diether2002}. \citet{chen2026debt} apply this logic to the debt-equity spread: its return premium is substantially larger among stocks that are less liquid, more costly or risky to short, or subject to greater disagreement, which they interpret as evidence of mispricing sustained by arbitrage frictions. The same test is informative here.

We examine this prediction with sequential double sorts.\footnote{The limits-to-arbitrage score requires securities-lending fees, which are available from 2008 onward.} Each week, stocks are first sorted into quintiles of a limits-to-arbitrage score and then into conditional FT-DINO quintiles within each group. Following \citet{chen2026debt}, the limits-to-arbitrage score averages the weekly percentile ranks of size, Amihud illiquidity, dollar volume, days-to-cover, securities lending fees, and analyst forecast dispersion, to which we add turnover. Each rank is oriented so that higher values indicate greater limits to arbitrage, and averaging across proxies reduces the noise in any single measure.

Table~\ref{tab:econrelevance} reports the results. The FT-DINO spread rises monotonically from 14\% per year in the lowest limits-to-arbitrage quintile to 119\% in the highest, more than an eightfold increase.\footnote{Appendix~\ref{app:d_limits} reports the sorts separately for each proxy. For example, the FT-DINO spread is 127\% in the smallest-stock quintile versus 13\% in the largest, and it rises from 11\% to 135\% across Amihud-illiquidity quintiles. Days-to-cover is the main exception, consistent with the measurement issue emphasized by \citet{asquith2005}.} The strategy is thus most profitable precisely where exploiting it is most costly. This is the pattern that limits-to-arbitrage theories predict for mispricing, and it mirrors the evidence in \citet{chen2026debt}.\footnote{Appendix~\ref{app:d_retail} finds a similar gradient across quintiles of RSVOL, the retail share of trading volume of \citet{laarits2025}, which is concentrated in hard-to-value stocks. Over 2006--2024, the FT-DINO spread rises from 10\% in the lowest RSVOL quintile to 88\% in the highest. An alternative proxy based on Robinhood demand shocks shows no comparable gradient over its short 2018--2020 sample.}

\tablink{tab:econrelevance}{Double Sorts on Limits to Arbitrage}

The predictability of historical visual retrieval is concentrated among stocks that are costly to arbitrage, consistent with a mispricing interpretation rather than compensation for risk. Because short-term reversals are also concentrated in illiquid stocks \citep{avramov2006}, however, this gradient alone does not distinguish retrieval from a reversal sort. We therefore examine the return and order-flow dynamics around portfolio formation directly.

\subsection{Price Dislocation and Correction}\label{sec:dislocation}

We next examine when FT-DINO returns accrue relative to portfolio formation. If the spread reflects temporary, non-fundamental price pressure, high-signal stocks should have been pushed down by selling shortly before formation, and part of the decline should reverse once the pressure subsides \citep{coval2007}. Because such a reversal moves prices back toward fundamentals, it should not be given back at longer horizons. If instead the signal selects stocks whose prices are being pushed away from fundamentals, early gains should reverse over the following months \citep{debondt1985}.

Figure~\ref{fig:returndynamics} reports event-time returns to the equal-weighted FT-DINO H--L portfolio with $K=1{,}000$. Panel (a) shows that the portfolio loses 5.39\% in the formation week: the stocks with the highest retrieval signals are recent losers. The spread then earns 1.64\% over the next five trading days, which recovers 30\% of the formation-week decline, and 2.59\% over the first 20 days. Panel (b) extends the horizon to six months. Returns over days 21 to 60 are small but positive, and those over days 61 to 120 are close to zero, so the cumulative return through day 120 is about 3.1\%. Roughly 58\% of the formation-week decline therefore reverses, and the reversal is not undone. Both legs contribute. High-signal stocks fall 2.73\% in the formation week and rebound 1.18\% over the next five days, whereas low-signal stocks rise 2.67\% and give back 0.46\% (Appendix~\ref{app:d_legs}).

\figlink{fig:returndynamics}{Short- and Long-Run Returns to FT-DINO Portfolios}

We next examine order flow around formation. We obtain daily buyer- and seller-initiated volume from WRDS for 2001--2025, with trades signed by the algorithm of \citet{lee1991}, and measure order imbalance (OIB) for each stock-day as buyer-initiated minus seller-initiated volume divided by their sum. We then compute the difference in average OIB between the top and bottom FT-DINO deciles. Figure~\ref{fig:oib} reports the results. In the formation week, the OIB spread is $-0.99$ percentage points (Panel (a)), and the daily series in Panel (b) shows that net selling builds over the final three days before formation, reaching $-3.48$ percentage points on day 0. Order flow turns to net buying on the first day after formation and remains positive, at 1.29 to 1.50 percentage points per week, through the following four weeks. Selling pressure thus coincides with the price decline, and buying pressure with its partial reversal.\footnote{Because order imbalance co-moves with same-period returns, the formation-week selling partly mirrors the price decline itself. The persistent buying over the following month is less subject to this concern because it occurs after the portfolios are formed.}

\figlink{fig:oib}{Order Imbalance around Portfolio Formation}

The return and order-flow dynamics mirror the evidence in \citet{chan2003}, who finds that returns drift after bad public news but reverse after extreme price moves without news. The roughly 42\% of the decline that does not revert is consistent with the former, so the formation-week decline appears to combine fundamental and non-fundamental components.

One might worry that the pattern restates the short-term reversal effect \citep{jegadeesh1990,lehmann1990}. The evidence suggests otherwise. FT-DINO retains an annualized EW alpha of 40.95\% after controlling for short-term reversal, momentum, the Fama--French five factors, benchmark DINO, and the CNN (Table~\ref{tab:spanning}), and its Fama--MacBeth coefficient remains highly significant with weekly and monthly reversal among the controls (Appendix~\ref{app:c_incremental}).

Taken together, FT-DINO identifies temporary price dislocations driven by trading activity.

\subsection{Salient Retrieval}\label{sec:salient}

We next examine whether the information in the retrieved neighbors depends on how salient they are. Salience theory predicts that payoffs that stand out attract disproportionate weight in decisions \citep{bordalo2012salience}, and stocks whose past returns display salient upsides attract investors, become overvalued, and earn lower subsequent returns \citep{cosemans2021}. Attention-grabbing events similarly attract buying that raises prices temporarily before they reverse \citep{barber2008,da2011}. If some investors form expectations from the most salient historical analogues, the stocks that those analogues point to should experience demand pressure that later reverses.

We measure the salience of each retrieved neighbor with an equally weighted composite of six characteristics: the absolute value of its subsequent five-day return, its abnormal trading volume and abnormal turnover over the same five days, Google search attention to its firm in the preceding calendar month, whether it belongs to the query stock's FF12 industry, and its recency. Each characteristic is percentile-ranked among the query's 1,000 neighbors. We sort the 1,000 neighbors of each query into salience quintiles of 200 neighbors and form a stand-alone retrieval signal from each quintile. The average similarity rank within each quintile is about 500, so we compare the quintiles with the 200 neighbors at similarity ranks 401 to 600, which serve as a similarity-matched benchmark. Figure~\ref{fig:salience} reports the results.

The four least salient quintiles behave like the full-sample signal. In Panel (a), their high-signal stocks decline through the formation week and rebound on the first day afterward, and order flow turns from net selling toward day 0 to persistent net buying over the following three weeks. Their H--L returns remain positive in each of the six months after formation (Panel (b)). The most salient quintile is different. Its high-signal stocks are recent winners: the Q5 H--L return is positive throughout the formation week, and the order-imbalance spread rises into day 0 and then turns to persistent net selling, the mirror image of the other quintiles. The Q5 spread earns about 0.6\% in the first month, well below Q1 to Q4, and then turns negative in each of months 2 through 6. The cumulative loss over months 2 to 6, about 1.8\%, more than offsets the first-month gain.

\figlink{fig:salience}{Returns and Order Imbalance by Neighbor Salience}

On its own, the most salient quintile predicts returns worst. Its EW H--L Sharpe ratio is 1.53, about one-quarter of the 5.77 earned by the similarity-matched benchmark, compared with 3.29 to 5.00 for Q1 to Q4. The sequence of buying, early gains, and later reversal is consistent with demand pressure tied to salient analogues that pushes prices away from fundamentals and subsequently unwinds, in contrast with the main component, whose reversal is not given back. We interpret this evidence as suggestive.\footnote{Two caveats apply. First, the salience composite includes the absolute value of each neighbor's subsequent return, whose signed value the signal averages, so part of the sorting may be mechanical. Second, the Q5 signal selects recent winners, so attention to the stock's own price could generate the same buying and reversal without any role for the historical analogues \citep{barber2008,da2011}. Memory-based models further predict that the most similar, salient, and recent episodes should be the most influential \citep{mullainathan2002,bordalo2020,charles2025}. Appendix~\ref{app:d_recall} finds little support for this structure in the full sample: dropping any band of 200 neighbors by similarity rank barely changes performance, and weighting neighbors by salience, recency, or search attention lowers the Sharpe ratio relative to equal weighting.}

In summary, the retrieval signal captures two forms of price pressure with distinct long-run return patterns. The main component reflects selling pressure that partially reverts, and the reversal is not given back over six months. The smaller salient component reflects buying pressure that reverses over the following months and contributes little to FT-DINO's performance.

\section{Interpreting Historical Visual Retrieval}\label{sec:interpretation}

In this section, we examine what the return-aligned representation captures. Historical retrieval makes each forecast traceable to specific historical charts, but not to the visual features that make those charts comparable.\footnote{Return alignment operates through which historical charts become neighbors: for the same query, historical library, and timing restriction, DINO and FT-DINO retrieve different histories. Figure~\ref{fig:neighbor_compare} in Appendix~\ref{app:c_neighbor} illustrates this for a single query, and Figure~\ref{fig:app_neighbor_composition} documents how return alignment redistributes historical matches across dates and industries over 2001--2025.} The following subsections recover those features. We focus on FT-DINO because it generates our main results.

Section~\ref{sec:saemethod} introduces the sparse autoencoder that decomposes the FT-DINO representation into sparse visual features. The bottom-up analysis in Section~\ref{sec:bottomup} identifies return-related features from real stock charts and evaluates their relevance for retrieval and portfolio performance. The top-down analysis in Section~\ref{sec:topdown} starts from 11 predefined patterns and uses real representative charts to locate them in the learned representation. Section~\ref{sec:visualspace} builds a market-wide visual atlas that organizes historical charts into recurring prototypes and locates literature-based technical patterns within it.%

\subsection{Sparse Autoencoder}\label{sec:saemethod}

FT-DINO represents each chart with a 384-dimensional dense embedding, and its coordinates are not designed to correspond to distinct visual concepts. We therefore use a sparse autoencoder (SAE) to decompose each dense representation into 16,384 sparsely activated visual features that can be examined individually, following recent work that interprets vision-transformer embeddings with sparse representations \citep{olson2025visionsae}. The SAE is trained only on FT-DINO embeddings from 1993--2000 and is fixed before 2001. Appendix~\ref{app:a_sae} reports the training objective and the feature-selection and steering procedures.

\subsection{Bottom-Up Analysis}\label{sec:bottomup}

The bottom-up analysis lets returns nominate the features and then checks whether each nominated feature encodes a recognizable chart pattern. Because each sparse feature activates on only a small fraction of charts, its meaning can be read from the charts that activate it most strongly. The procedure has three steps. First, each year we predict the sign of every stock's subsequent five-day return from its SAE activations, using only earlier data, and rank features by their share of the estimated coefficients. Second, a feature that predicts returns need not correspond to a recognizable pattern, so we inspect, for each of the 100 highest-ranked features, representative real charts from the top 5\% of its positive activations. We retain a feature only when these charts consistently share the same visual characteristic. Third, because several retained features describe variants of the same configuration, we group the 30 surviving features into 11 visual states. The six features in Late Breakdown, for example, all display a flat or weakening upper range that ends in a sharp late-window decline.\footnote{Appendix~\ref{app:e_states} reports the complete classification, and Appendix~\ref{app:e_charts} displays representative charts for every state.}

We next examine whether these visual states matter for portfolio performance. To do so, we hold the FT-DINO signal fixed, remove the stocks that most strongly activate one state at a time, and re-form weekly decile portfolios on the remaining stocks. Table~\ref{tab:saesensitivity} reports the results. Four states produce the largest declines: Isolated Upward Spike, Sustained Uptrend, Late Upside Breakout, and High-Volatility Range. Removing the strongest activators of Isolated Upward Spike lowers the Sharpe ratio from 6.37 to 5.90, a 7\% decline, and the other three states lower it to 6.11, 6.19, and 6.24. These effects are modest. No single visual state accounts for much of FT-DINO's performance, consistent with a signal that draws on many configurations.

The influential states mix abrupt and directional configurations. Isolated Upward Spike, a quiet range interrupted by an extreme upward spike that quickly retraces, resembles the patterns that \citet{jiang2023} isolate when approximating their CNN: future returns tend to be high when a stock closes near the low end of its recent high--low range and after recent increases in volume. Sustained Uptrend and Late Upside Breakout instead capture upward price paths, and High-Volatility Range captures a compressed range with repeated long price bars. By contrast, removing Stable Downtrend, Stepwise Downtrend, Accelerating Downtrend, or Late Breakdown changes the Sharpe ratio by between $-0.03$ and $+0.26$. Removing Accelerating Downtrend raises it to 6.64, suggesting that the signal is less informative for these stocks.

\tablink{tab:saesensitivity}{Portfolio Sensitivity to SAE Visual States}

The exclusion test indicates which states matter for performance, but not whether they matter through the selection of historical neighbors. To examine this channel, we conduct query-only steering on one representative feature from each of the four states: we move only the current query representation away from the feature's SAE decoder direction, hold the historical library and all other settings fixed, and recompute the neighbors and the retrieval signal. Steering thus asks which neighbors the model would retrieve if the query chart displayed less of the feature. Panel A of Table~\ref{tab:steering} reports the results. Sharpe ratios fall as steering becomes stronger, and the declines are large. Steering the High-Volatility Range feature to level $-10$, for example, lowers the Sharpe ratio from 6.37 to 2.32, a 64\% decline. Across the four features, the average Sharpe ratio is 5.61 at level $-3$ and 3.45 at level $-10$. These declines far exceed those from removing the activating stocks, likely because steering alters the neighbors of every query. Since the intervention moves only the query, these results indicate that the selected features shape which historical observations are retrieved and, in turn, the portfolio signal.

\tablink{tab:steering}{Query-Only Steering of SAE Features}

The bottom-up analysis identifies a small set of interpretable visual states: isolated upward spikes, upward price paths, and high-volatility ranges. Removing any single state lowers the FT-DINO Sharpe ratio by at most 7\%, but moving queries away from the corresponding feature directions lowers it by up to 64\%. These features therefore operate through the selection of historical neighbors.

\subsection{Top-Down Analysis}\label{sec:topdown}

The top-down analysis reverses the direction of discovery. Instead of letting returns nominate features, we start from patterns that researchers have already named and look for the features that encode them. We consider 11 patterns. Five follow the standardized construction in \citet{bojstrup2026}: Up Momentum, Down Momentum, Breakout Up, Breakout Down, and Mean Reversion. The other six come from the literature-based characteristics in Appendix Table~\ref{tab:literature_patterns}: discrete winner and loser paths \citep{da2014}, MAX(1) \citep{bali2011}, negative overnight jumps \citep{jiangzhu2017}, and upward and downward paths with positive convexity \citep{gulenwoeppel2026}.

For each pattern, we first generate synthetic reference charts that follow the descriptions and illustrative figures in the original studies. We then retrieve the most similar historical stock charts by FT-DINO cosine similarity, verify by inspection that each displays the target configuration, and retain 50 real charts per pattern. Working with real charts keeps the search within the distribution of charts on which the SAE is trained. We pass the 50 charts through FT-DINO and the SAE and record which sparse features activate consistently across them. A feature that responds consistently across representatives is likely to encode the pattern itself. We then run the check in reverse: for each candidate, we collect the real charts that activate it most strongly and keep the feature whose top charts most consistently reproduce the pattern.

Figure~\ref{fig:topdown_main} illustrates the procedure for Up Momentum. The representative charts in the upper-left panels show a steady advance above a rising moving average. Feature 6946 activates on them, and its highest-activation real-market charts in the upper-right panels display the same configuration. Appendix~\ref{app:f_validation} reports the other ten patterns and the steering-based validation that moves initially inactive queries toward the selected feature direction. Five of the 11 top-down features also pass the bottom-up screen.\footnote{Discrete Winner (feature 16299) is the sole constituent of Sustained Uptrend, Up + Positive Convexity (6232) belongs to Late Upside Breakout, Discrete Loser (9690) to Stable Downtrend, Down Momentum (7123) to Accelerating Downtrend, and Breakout Down (2751) to Late Breakdown.} These results indicate that textbook configurations correspond to distinct, identifiable directions in the return-aligned representation.

\figlink{fig:topdown_main}{Top-Down Feature Validation: Up Momentum}

We next examine whether these directions matter for returns by applying the same query-only steering to the 11 selected features. Panel B of Table~\ref{tab:steering} reports the results, which contrast with those for the bottom-up features in Panel A. Moving the query away from a top-down feature also lowers performance, but by less: averaged across features, the Sharpe ratio is 5.86 at level $-3$ and 4.72 at level $-10$, compared with 5.61 and 3.45 for the bottom-up features. These configurations are thus present in the FT-DINO representation but play a smaller role in its return predictions than the features that returns select.

\subsection{Mapping Literature-Guided Patterns into the Visual Atlas}\label{sec:visualspace}

Beyond the top-down patterns, we use the SAE representation to organize a broader set of technical characteristics from the finance literature. Appendix Table~\ref{tab:literature_patterns} lists 21 characteristics whose constructions in the cited studies imply a visually identifiable price or price-volume configuration. Placing them in the same space as the empirical chart features lets us compare established signals within one representation.

To construct the visual atlas, we encode the 24.2 million out-of-sample 20-day charts from 2001--2025 with FT-DINO and the SAE and cluster the sparse representations into 4,096 prototypes using spherical $k$-means. We project the prototype centers into two dimensions and color each prototype by the average cross-sectional rank of the subsequent five-day return among its observations. Because FT-DINO and the SAE are trained on 1993--2000 data, restricting the atlas to later charts keeps in-sample return outcomes out of the prototype coloring. For each characteristic, we select an illustrative real-market 20-day chart that captures a visual configuration motivated by the cited study and map it to the nearest atlas prototype. Figure~\ref{fig:landmarks_main} displays the resulting landmarks. The atlas connects traditional technical analysis with historical visual retrieval. Any researcher-defined pattern can be located among the visual states learned from market data, and each prototype carries the average subsequent return rank of the charts assigned to it. Established technical patterns can thus be expressed in the coordinates of the learned representation, and learned visual states can be described in the vocabulary of technical analysis.

\figlink{fig:landmarks_main}{Literature-Based Patterns in the Visual Atlas}

\section{Conclusion}\label{sec:conclusion}

This paper develops a historical visual retrieval framework for predicting stock returns from price charts. The framework maps each chart into a learned visual space, retrieves comparable observations from a recursively expanding historical library, and forms a return signal from what followed those observations. Each forecast remains tied to dated historical charts and their realized returns.

We find that general visual similarity contains return information. A representation learned from natural images, without financial supervision, delivers an equal-weighted high-minus-low Sharpe ratio of 2.81. Aligning the representation with subsequent return ranks raises the Sharpe ratio to 6.37, and the aligned signal retains an annualized alpha of about 41\% after controlling for a direct image-prediction CNN, pretrained DINO, and standard factors. The predictability remains economically meaningful under investability screens and transaction costs, particularly for equal-weighted portfolios, extends across chart windows and return horizons, and is concentrated among stocks with greater limits to arbitrage. Around portfolio formation, high-signal stocks are sold down, and about 58\% of the decline reverses without being given back over six months, consistent with temporary price dislocations that partially revert; a smaller component tied to the most salient historical analogues instead reverses over the following months. Cross-stock histories carry most of the signal, and the geometry of the price path is its primary visual input, with volume adding to it.

We also document the visual content behind the return-aligned signal. A sparse autoencoder decomposes the FT-DINO embedding into sparse features, 30 of which form 11 recurring visual states. Portfolio-sensitivity tests and feature steering connect the most important states (isolated upward spikes, upward price paths, and high-volatility ranges) to both neighbor selection and portfolio performance. The top-down analysis locates 11 predefined patterns in the sparse representation, but steering along them affects performance less than steering along the bottom-up features. The visual atlas places 21 characteristics from prior studies within a common map of market states.

Taken together, our evidence suggests that return information in price histories can be organized through the similarity of complete visual states. The representation determines which episodes are comparable, and the outcomes of those episodes determine the forecast. This separation makes the organization of market history an observable component of return prediction. It also offers a natural framework for other financial settings in which current states can be matched to high-dimensional historical experience.

\clearpage
\bibliographystyle{jf}
\bibliography{references}

\input{tables_figures_main}

\clearpage\appendix
\section{Data and Methodological Details}\label{app:a_structure}

\subsection{DINOv2 Pretraining and Retrieval Workflows}\label{app:a_dino_workflows}

DINOv2 is pretrained through self-distillation on unlabeled natural images. A student network and a teacher network receive different augmented views of the same underlying image. The student is updated by gradient descent to match the teacher's representation, while the teacher parameters follow an exponential moving average of the student parameters. The training objective combines agreement in the global image representation with masked-patch prediction, allowing the model to learn transferable visual structure without class labels or financial outcomes. The DINOv2 models used in our analysis are pretrained on LVD-142M, a collection of roughly 142 million general images \citep{oquab2024dinov2}. Figure~\ref{fig:app_dino_pretrain} summarizes the student--teacher design.

\figlink{fig:app_dino_pretrain}{DINOv2 Student--Teacher Pretraining}

For the benchmark retrieval exercise, each standardized stock chart is resized to the DINOv2 input resolution and passed through DINOv2-S. We extract the final-layer CLS token, producing a 384-dimensional representation $h_{i,t}$, and normalize it to unit length as in equation~\eqref{eq:embedding_main}, so that the similarity in equation~\eqref{eq:similarity_main} is the cosine similarity between the query and the historical chart.

The information-timing restriction is imposed at the historical-observation level. Let $R_{j,\tau+1:\tau+5}$ denote the compounded return over the five market trading dates after $\tau$. A historical observation can be searched by a query formed on date $t$ only if that entire return has already been realized. In trading-day notation, the eligible library is
\begin{equation}
    \mathcal L_t
    =
    \left\{(j,\tau):\tau+5\leq t\right\}.
    \label{eq:app_eligible}
\end{equation}
The library begins with daily observations from 1993--2000 and expands recursively throughout 2001--2025. Each weekly query searches all observations in $\mathcal L_t$, ranks them by cosine similarity, and retains the exact top $K$. The signal is the equal-weighted average of the neighbors' realized subsequent returns, as defined in equation~\eqref{eq:retrievalsignal_main}. Panel (a) of Figure~\ref{fig:app_workflows} summarizes the information sequence from chart construction to portfolio formation.

\figlink{fig:app_workflows}{Historical Retrieval and Return-Rank Alignment Workflows}

\subsection{Return-Rank Alignment}\label{app:a_rank_alignment}

Return-rank alignment changes the representation used for neighbor selection while preserving the historical-retrieval rule. For each training date, subsequent five-trading-day returns are converted into within-date percentile ranks (equation~\eqref{eq:returnrank_main}). Within a training batch, the target distribution in equation~\eqref{eq:targetdist_main} assigns greater affinity to observations with closer ranks, and the model-implied distribution in equation~\eqref{eq:modeldist_main} is constructed from cosine similarities between normalized image representations. Self-pairs are excluded from both distributions. The fine-tuning loss in equation~\eqref{eq:rankloss_main} averages, across the $B$ observations in a batch, the Kullback--Leibler divergence from the return-rank target distribution to the model-implied similarity distribution. This soft objective preserves graded information about how close two observations are in relative subsequent performance, which a binary positive--negative classification would discard. The objective is related to continuous-outcome representation learning in \citet{zha2023} and future-aligned cross-sectional asset retrieval in \citet{lee2026retrieval}.

We fine-tune all DINOv2-S parameters on charts from 1993--1999. Each observation retains the return rank computed in the full cross section of its original date, and training batches contain stocks from multiple dates. Optimization uses AdamW with a linear warmup followed by cosine learning-rate decay and one complete pass through the training sample. Hyperparameters and the final checkpoint are selected using the KL loss on the complete 2000 validation sample. The chosen encoder is then frozen before 2001 and used to re-encode both queries and historical-library observations. Panel (b) of Figure~\ref{fig:app_workflows} summarizes this training and deployment sequence.

Figure~\ref{fig:app_raw_space} provides a low-dimensional visualization of the DINO and FT-DINO representation spaces using the same out-of-sample observations and the same projection procedure. The figure visualizes how return-rank alignment reshapes the representation space; the portfolio consequences are evaluated in Appendix~\ref{app:c_structure}.

\figlink{fig:app_raw_space}{Raw Embedding Space: DINO versus FT-DINO}

\subsection{Sparse Autoencoder Training and Interpretation Tests}\label{app:a_sae}

The sparse autoencoder is trained on the raw 384-dimensional FT-DINO representations. We use a JumpReLU SAE with 16,384 latent features \citep{rajamanoharan2024jumprelu}. For an FT-DINO representation $h_{i,t}$, the sparse activation vector is
\begin{equation}
    a_{i,t}
    =
    \operatorname{JumpReLU}_{\kappa}
    \left[W_e(h_{i,t}-b_d)+b_e\right]
    \in\mathbb R^{16{,}384},
    \label{eq:app_sae_encoder}
\end{equation}
and the reconstructed representation is
\begin{equation}
    \widehat h_{i,t}=W_d a_{i,t}+b_d.
    \label{eq:app_sae_decoder}
\end{equation}
The objective combines reconstruction error with an exact sparsity penalty,
\begin{equation}
    \mathcal L_{\mathrm{SAE}}
    =
    \frac{1}{N}\sum_{i,t}
    \left[
       \lVert h_{i,t}-\widehat h_{i,t}\rVert_2^2
       +\lambda_0\lVert a_{i,t}\rVert_0
    \right].
    \label{eq:app_sae_loss}
\end{equation}
The SAE is trained on FT-DINO embeddings from 1993 through 2000 and is fixed before the 2001 out-of-sample period. Each column of $W_d$ is the decoder direction associated with one sparse feature. We interpret the sparse features ex post using the stock charts that activate them.

We identify return-related sparse features using a rolling procedure that avoids using future outcomes to select features for an earlier test year, following the same procedure as in \citet{chen2026brainscan}. On each weekly query date, define $y_{i,t}=\mathbf 1\{R_{i,t+1:t+5}>0\}$. For a test year $T$, the training sample ends in $T-3$, years $T-2$ and $T-1$ select the $L_2$ penalty, and the selected specification is re-estimated through $T-1$. To make estimation tractable, we project standardized SAE activations onto 500 truncated-SVD components. We then back-project the fitted coefficients to the original SAE coordinates for feature-level attribution. We estimate an $L_2$-penalized logistic regression in that component space and project the estimated coefficients back to the original SAE features. Feature $f$'s importance in test year $T$ is its share of the absolute back-projected coefficients,
\begin{equation}
    \omega_{f,T}
    =
    \frac{|\widehat\beta_{f,T}|}
    {\sum_g|\widehat\beta_{g,T}|}.
    \label{eq:app_feature_importance}
\end{equation}
We aggregate these year-specific importance measures across test years ex post and inspect the 100 highest-ranked features.

Visual consistency is evaluated using representative charts from the top 5\% of positive activations, where a feature's visual characteristics are most consistent. A feature is retained when these charts share the same visual characteristic. This screen retains 30 features. GPT-5.6 is used to generate a concise description of each retained feature's representative charts and to group visually similar features into 11 recurring visual states. Appendix~\ref{app:e_structure} reports the full state map and representative real-market charts.

We use two diagnostics to assess their relevance to portfolio performance and historical retrieval. The portfolio-sensitivity test holds the original FT-DINO retrieval signal fixed. For each stock $i$ in week $t$, we first percentile-rank each positive SAE feature activation within the weekly cross section. For visual state $C$, cluster activation is defined as the maximum activation rank among its constituent features,
\begin{equation}
    S_{i,C,t}
    =
    \max_{f\in C} q_{i,f,t},
    \label{eq:app_cluster_activation}
\end{equation}
where $q_{i,f,t}$ is the cross-sectional percentile rank of feature $f$'s positive activation. We remove the stocks with the highest $S_{i,C,t}$, subject to a cap of 10\% of the weekly cross section. If fewer than 10\% of stocks activate the state, all active stocks are removed. Thus,
\begin{equation}
    M_{C,t}
    =
    \min\left\{\lfloor0.1N_t\rfloor,N^{\mathrm{act}}_{C,t}\right\},
    \label{eq:app_sae_removal}
\end{equation}
where $N^{\mathrm{act}}_{C,t}$ is the number of stocks with positive activation in at least one constituent feature of state $C$. 
We then re-form equal-weighted decile portfolios using the unchanged FT-DINO retrieval signal.
 
The second diagnostic perturbs the representation used for neighbor selection. If $d_f$ is the decoder direction of feature $f$, steering level $\ell$ replaces the raw query representation with
\begin{equation}
    h_{i,t}^{(f,\ell)}=h_{i,t}+\ell d_f,
    \label{eq:app_query_steering}
\end{equation}
while leaving all historical-library representations unchanged. We re-normalize the steered query and repeat exact cosine-similarity retrieval under the original eligibility rule. Because only the query moves, differences in retrieved neighbors can be attributed directly to movement along the selected representation direction.

The top-down analysis uses the same SAE but reverses the direction of discovery. We begin with 11 patterns. For each pattern, we write code, with the assistance of GPT-5.6, that generates synthetic reference charts following the descriptions and illustrative figures in the corresponding studies. We then encode the synthetic references with FT-DINO, retrieve the historical stock charts with the highest embedding cosine similarity from the full historical chart library, manually verify that each displays the target configuration, and retain 50 real charts per pattern.
We encode these charts with FT-DINO and pass them through the SAE. A sparse feature is retained as a candidate if it activates consistently across the 50 representative charts.

We then rank real stock charts by candidate activation and select the feature whose highest-activation examples most consistently reproduce the intended pattern. Appendix~\ref{app:f_structure} reports the selected real-image validations.

\section{Historical Visual Retrieval with Pretrained DINO}\label{app:b_structure}

\subsection{Portfolio Performance over Time}\label{app:b_time}

Figure~\ref{fig:dino_cum} plots cumulative log high-minus-low returns for benchmark DINO across neighborhood sizes and portfolio weighting schemes. Equal-weighted cumulative performance generally becomes smoother as $K$ increases, consistent with the improvement in Sharpe ratios reported in the main table. The VW paths are more volatile and weaker, while capped value weighting reduces the influence of the largest firms.

\figlink{fig:dino_cum}{Cumulative Log Returns to Benchmark DINO Portfolios}

\subsection{Historical-Library Restrictions}\label{app:b_library}

Table~\ref{tab:app_dino_universe} varies the set of historical observations eligible for retrieval while leaving the representation and portfolio construction unchanged. At $K=1{,}000$, limiting the library to the preceding eight years lowers the EW Sharpe ratio from 2.81 to 2.48. Restricting retrieval to the same Fama--French 12 industry yields a Sharpe ratio of 2.36, whereas excluding the query stock's industry leaves performance almost unchanged at 2.80. By contrast, same-stock retrieval produces a negative spread. Cross-stock and cross-industry historical observations account for most of the benchmark retrieval performance.

\tablink{tab:app_dino_universe}{Historical-Library Restrictions}

\subsection{High-Similarity Cross-Stock Cases}\label{app:b_cases}

The historical-library restrictions show that benchmark DINO draws most of its predictive content from cross-stock histories. We next provide observation-level illustrations of the kinds of cross-stock analogies that the representation retrieves. We identify cross-company chart pairs with very high cosine similarity and then examine corporate events near the beginning of the corresponding chart windows. Figure~\ref{fig:app_high_similarity_cases} presents three representative pairs.

The first pair links Allison Transmission in 2022 with D.R. Horton in 2004. Both periods begin with strong earnings and favorable forward-looking signals, after which both charts display closely similar upward price-adjustment paths within the window. The second links The Hanover Insurance Group in 2016 with FedEx in 2007, when external operating headwinds compressed profitability; both charts then display closely similar downward price-adjustment paths within the window. The third links Ford in 2017 with Home Depot in 2010, when weak headline sales coincided with stronger underlying operating performance; both charts then display closely similar upward price-adjustment paths within the window. These cases illustrate how visual retrieval can connect episodes from different firms and periods that share similar price adjustments and economically comparable circumstances.

\figlink{fig:app_high_similarity_cases}{Illustrative High-Similarity Cross-Stock Cases}

\subsection{Neighbor Weighting}\label{app:b_weighting}

The baseline signal assigns equal weight to the realized returns of the retrieved neighbors. Table~\ref{tab:app_dino_recency} instead applies exponentially declining weights based on each neighbor's age. Short half-lives materially weaken the signal: with a one-year half-life, the EW Sharpe ratio is 1.16, compared with 2.81 under equal weighting. Performance approaches the equal-weight benchmark as the half-life lengthens, showing that older retrieved observations continue to contribute useful information.

\tablink{tab:app_dino_recency}{Neighbor Weighting: Equal versus Recency-Weighted}

\subsection{Alternative Pretrained Encoders}\label{app:b_encoders}

We next apply the same historical-retrieval procedure to three alternative pretrained encoders: the larger DINOv2-L model, a masked autoencoder (MAE) \citep{he2022mae}, and CLIP \citep{radford2021clip}. Table~\ref{tab:encoderspecs} summarizes their specifications. The encoders differ in pretraining objective, training data, model size, and embedding dimension, but none of them is pretrained with stock-return labels or a finance-specific objective.

\tablink{tab:encoderspecs}{Pretrained Vision Encoders}

Table~\ref{tab:encodercompare} applies the same charts, cosine-similarity search, expanding historical library, and neighbor-return averaging to each encoder. At $K=1{,}000$, the EW Sharpe ratios are 2.81 for DINOv2-S, 3.17 for DINOv2-L, 3.26 for MAE, and 2.91 for CLIP. Historical retrieval therefore does not hinge on a particular pretraining objective or embedding dimension.

\tablink{tab:encodercompare}{Portfolio Performance across Pretrained Encoders}

\subsection{Numerical and Text Representations}\label{app:b_nonvisual}

We next examine whether the chart image matters or whether the same 20-day history performs as well in numerical or text form. We construct two non-visual representations of each window and apply the same retrieval rule to each. The first, Raw OHLCV, stacks the daily open, high, low, and close prices, the 20-day moving average, and volume into a 120-dimensional vector, with each variable standardized by its 1993--2000 mean and standard deviation. The second writes the chart-normalized values used to render the images (prices scaled by the high--low range of the window and volume by its within-window maximum) as structured text, one line per day, and embeds the text with ChronoBERT \citep{he2025chrono}, a language model trained only on historically available text. Its December 31, 2000 checkpoint, which we hold fixed throughout the out-of-sample period, cannot draw on information from after 2000. We use the mean of its last-layer token representations, a 768-dimensional vector. Each representation is normalized to unit length, and retrieval proceeds exactly as for the image encoders, with the same eligibility rule and $K=1{,}000$.

Table~\ref{tab:app_nonvisual} reports the results with $K=1{,}000$. Raw OHLCV earns an equal-weighted Sharpe ratio of 2.47 and ChronoBERT 1.42, both below the 2.81 of benchmark DINO and far below the 6.37 of FT-DINO, which is 2.6 times that of Raw OHLCV. The ranking is the same under capped value weighting, where the Sharpe ratios are 0.70 and 0.23, compared with 0.87 and 2.38.

Numerical and text representations thus perform worse than image-based retrieval, even before return alignment. One explanation is that a chart presents several sequences jointly. Prices, the moving average, and volume share a common time axis, so their alignment within and across days is visible at once, and a vision encoder can learn how the series move together. A stacked numerical vector instead treats each variable on each day as a separate coordinate, and cosine similarity compares two windows coordinate by coordinate, without regard to how the coordinates relate to one another. A text rendering similarly presents the window as a sequence of numbers for a language model to parse. The image format adds to retrieval performance, and aligning the visual representation with subsequent returns adds substantially more.

\tablink{tab:app_nonvisual}{Numerical, Text, and Image Representations}

\subsection{Comparison with CNN}\label{app:b_cnn}

Before introducing return alignment, Table~\ref{tab:spanningnoft} asks whether benchmark DINO contains information beyond the CNN. In EW portfolios, DINO retains an annualized alpha of 11.04\% with a $t$-statistic of 3.47 after controlling for the CNN and standard factors. The CNN also retains substantial incremental information relative to DINO. The pretrained retrieval signal and the direct CNN signal are related but not redundant. The main-text spanning tests then show how this relation changes once return-aligned retrieval is introduced.

\tablink{tab:spanningnoft}{DINO--CNN Factor-Augmented Spanning Regressions}

\section{Additional Results for Return-Aligned Retrieval}\label{app:c_structure}

\subsection{Historical-Neighbor Composition}\label{app:c_neighbor}

Figure~\ref{fig:neighbor_compare} illustrates how return alignment changes which historical charts are retrieved. For a fixed query chart, it displays the three nearest eligible historical charts under benchmark DINO and FT-DINO, together with randomly selected eligible charts for reference. The query, the historical library, and the timing restriction are identical, yet the two models retrieve different histories. Return alignment thus operates directly through which prior cases become neighbors. Economically, it shifts the notion of comparability from episodes that look alike to episodes in a space where proximity reflects similar subsequent relative performance.

\figlink{fig:neighbor_compare}{Retrieved Historical Neighbors: DINO versus FT-DINO}

Figure~\ref{fig:app_neighbor_composition} extends this comparison to the full 2001--2025 period and summarizes the composition of the top-1,000 historical neighborhoods. Panel (a) reports the temporal distribution of retrieved neighbors. Panel (b) reports Fama--French 12 industry preferences after adjusting for each industry's contemporaneous supply in the eligible library. The difference panels are FT-DINO minus DINO. These full-sample diagnostics show systematic changes in neighbor selection across dates and industries.

\figlink{fig:app_neighbor_composition}{Historical-Neighbor Composition for Benchmark DINO and FT-DINO}

\subsection{Neighborhood Size and Neighbor Weighting}\label{app:c_kweights}

Figure~\ref{fig:app_sharpe_k} compares DINO and FT-DINO Sharpe ratios over a broad grid of neighborhood sizes. Return alignment improves performance throughout the displayed range under both EW and CVW. For FT-DINO, the gains from increasing $K$ are largest at relatively small neighborhoods and flatten once the neighborhood contains roughly 1,000 observations.

\figlink{fig:app_sharpe_k}{Sharpe Ratios across Neighborhood Sizes: DINO versus FT-DINO}

Table~\ref{tab:app_ft_recency} repeats the neighbor-weighting test after return alignment. A one-year half-life lowers the EW Sharpe ratio from 6.37 to 4.61; longer half-lives gradually recover the equal-weight performance. As in the benchmark representation, aggressive recency weighting does not improve the signal. The gains from return alignment persist even when older historical episodes continue to receive substantial weight.

\tablink{tab:app_ft_recency}{FT-DINO Neighbor Weighting: Equal versus Recency-Weighted}

\subsection{Cross-Sectional Predictability and Incremental Information}\label{app:c_incremental}

Table~\ref{app:fmb} reports weekly Fama--MacBeth regressions of subsequent five-day stock returns on the image signals and standard characteristics. All image signals and controls are ranked within the weekly cross section before estimation. FT-DINO has a coefficient of 0.0111 with a $t$-statistic of 22.73 when entered with the controls. When DINO and CNN are included jointly, the FT-DINO coefficient remains 0.0092 with a $t$-statistic of 22.52. DINO and CNN also retain positive coefficients in the joint specification, but the DINO coefficient falls from 0.0032 in Model 1 to 0.0010 once FT-DINO is added in Model 5, a 69\% decline. The stock-level tests therefore support the portfolio evidence that return alignment adds information beyond the benchmark representation and the CNN.

\tablink{app:fmb}{Fama--MacBeth Regressions}

Table~\ref{tab:app_cnn_double_sorts} provides a nonparametric comparison with the CNN. Stocks are first sorted into CNN quintiles and then into conditional DINO or FT-DINO quintiles. The DINO 5--1 spread is positive in every CNN quintile, ranging from 9\% to 17\% per year. The FT-DINO spread is about three to four times as large, ranging from 32\% to 63\%, and is strongly significant in every CNN quintile. The retrieval signals therefore retain cross-sectional variation within portfolios formed on the direct CNN prediction.

\tablink{tab:app_cnn_double_sorts}{DINO--CNN and FT-DINO--CNN Double Sorts}

A nearest-neighbor forecast is nonlinear because it conditions on local observations in the embedding space. Table~\ref{tab:app_ridge} asks whether the same representations contain similar information when used in a simple global linear predictor. Panel A compares historical retrieval with ridge-regression predictions from the corresponding DINO and FT-DINO embeddings. Ridge signals are predictive for both representations, showing that return-relevant information is present in the embeddings themselves. For FT-DINO, the ridge portfolio performs nearly as well as retrieval, with an EW Sharpe ratio of 5.99 versus 6.37, and slightly better under CVW, at 2.43 versus 2.38, so much of FT-DINO's predictive content resides in the representation itself. Panels B and C then include the ridge portfolios in spanning regressions. In Panel B, benchmark DINO retrieval adds little once FT-DINO retrieval is included: its alpha turns negative, at $-9.68\%$ in EW and $-3.40\%$ in CVW. FT-DINO retrieval retains a positive annualized alpha of 21.00\% in EW and 8.64\% in CVW even after controlling for FT-DINO ridge, CNN, benchmark DINO retrieval, and standard factors. The return-aligned representation and the historical-retrieval rule therefore each contribute information to the final strategy.

\tablink{tab:app_ridge}{Historical Retrieval versus Ridge Prediction}

\subsection{Input Sensitivity Measures}\label{app:c_sensitivity}

The forecast-sensitivity maps in Figure~\ref{fig:sensitivity} are one of three pixel-level measures that differ in which part of the representation they track. Let $x$ denote a query chart, $z(x)\in\mathbb R^{384}$ its normalized CLS embedding, and $x_p$ the three color channels of pixel $p$. All three measures build on the embedding Jacobian $J_p(x)=\partial z(x)/\partial x_p\in\mathbb R^{384\times 3}$. Embedding sensitivity aggregates changes in all embedding directions, $S_{\mathrm{Emb}}(p)=\lVert J_p(x)\rVert_F$. Retrieval, however, depends on the embedding only through its similarity to historical charts, $s_j(x)=z(x)^{\prime}z_j$, whose pixel gradient is $J_p(x)^{\prime}z_j$. Matching sensitivity averages the squared similarity gradients over a set $\mathcal B$ of 64 embeddings drawn at random from the eligible historical library,
\begin{equation}
    S_{\mathrm{Match}}(p)
    =
    \frac{1}{|\mathcal B|}
    \sum_{j\in\mathcal B}
    \big\lVert J_p(x)^{\prime}z_j\big\rVert_2^{2}.
    \label{eq:app_match_sens}
\end{equation}
Similarity alone does not distinguish historical charts followed by high returns from those followed by low returns. Forecast sensitivity weights the similarity gradients by the demeaned subsequent returns of the top-$K$ neighbors, which yields the pixel gradient of the return-aligned component $g$ in equation~\eqref{eq:forecastsens_main},
\begin{equation}
    S_{\mathrm{Forecast}}(p)
    =
    \Big\lVert
    \frac{1}{K}
    \sum_{j\in\mathcal N_K}
    \big(R_j-\bar R\big)\,J_p(x)^{\prime}z_j
    \Big\rVert_1,
    \qquad K=1{,}000,
    \label{eq:app_forecast_sens}
\end{equation}
where $\bar R$ is the equal-weighted retrieval signal.

Figure~\ref{fig:app_sensitivity_measures} reports the three measures for benchmark DINO and FT-DINO. For FT-DINO, the three maps are closely aligned. Each concentrates on the most recent days of the price region and on the volume bars, so return alignment makes the pixels that move the representation the same pixels that move similarity and the forecast. Benchmark DINO behaves differently. Its embedding and forecast maps spread across the full 20-day window, whereas matching sensitivity concentrates on the most recent day. The pixels that most change DINO's similarity to a typical historical chart are thus not the ones that drive its return forecast, which draws on the full sequence of the window. This pattern is consistent with the perturbation evidence in Appendix~\ref{app:c_temporal}, where reordering the earlier part of the window lowers the DINO Sharpe ratio by 30\%.

\figlink{fig:app_sensitivity_measures}{Three Measures of Input Sensitivity: DINO and FT-DINO}
\subsection{Temporal Structure Perturbations}\label{app:c_temporal}

Figure~\ref{fig:sensitivity} suggests that the CNN, DINO, and FT-DINO draw on different parts of the chart window. Table~\ref{tab:app_temporal} tests this directly. We divide the 20-day window into four consecutive five-day blocks, B1 through B4, where B4 contains the most recent days. For each block, we randomly permute the daily observations within the block and redraw the query chart, which removes the ordering of information inside that block while preserving its content. The shuffle moves each daily candlestick together with its volume, so the shuffled chart need not preserve the cumulative price change within the block. In a separate test, we permute the order of the first three blocks while keeping B4 fixed. The perturbations apply only to the query charts: no model is re-trained, and the historical retrieval library is unchanged.

Shuffling within B4 lowers the CNN's EW Sharpe ratio from 4.46 to 2.32, a 48\% decline, and FT-DINO's from 6.37 to 4.37, a 31\% decline, but lowers benchmark DINO's by only 6\%. Reordering the first three blocks produces the opposite pattern. It lowers the DINO Sharpe ratio from 2.81 to 1.96, a 30\% decline, whereas the CNN and FT-DINO decline by only 2\% and 6\%. Shuffling within any of the earlier blocks changes performance little for all three models, and the CVW results in Panel B are similar. The CNN therefore relies on the most recent days, benchmark DINO on the sequence of the full window, and FT-DINO on both, with greater weight on recent days.

\tablink{tab:app_temporal}{Temporal Structure Perturbations}

\subsection{Transformed Queries}\label{app:c_transform}

The temporal perturbations reorder days within parts of the window. A stronger placebo reverses the entire price path. We flip each query chart horizontally, which reverses the order of time, and rotate it by 180 degrees, which reverses both time and the direction of price changes. The historical library and the retrieval rule are unchanged. Figure~\ref{fig:app_transform} illustrates the two transformations and the nearest neighbor that each model retrieves for them. If a representation encodes the direction and sequence of the price path, a transformed query should no longer resemble the original and should retrieve neighbors with different subsequent returns.

Panel A of Table~\ref{tab:app_transform} measures how far each transformation moves the query in the representation space. For a randomly selected query chart, we compute the cosine similarity between the transformed copy and the original and record the rank that this similarity would attain among the original query's similarities to the historical library. Under DINO, the flipped and rotated copies would rank 524,306th and 1,198,188th. Under FT-DINO, the flipped copy would rank 586,020th, and the rotated copy 16,901,476th, 14 times as far as under DINO. Return alignment thus makes the representation especially sensitive to the direction of price changes.

Panel B reports portfolio performance. For benchmark DINO, flipping or rotating the query lowers the equal-weighted Sharpe ratio from 2.81 to 1.53 and 1.38, about half of the original. For FT-DINO, both transformations eliminate the spread: the equal-weighted Sharpe ratios are $-0.45$ and $-0.07$, compared with 6.37, and the capped value-weighted Sharpe ratios are 0.04. The FT-DINO signal therefore depends on the direction and time order of the price path, whereas part of the DINO signal comes from visual features that survive these transformations.

\figlink{fig:app_transform}{Transformed Queries and Retrieved Neighbors}

\tablink{tab:app_transform}{Transformed Queries}

\subsection{Subsample Stability and Alternative Training Windows}\label{app:c_subsample}

Table~\ref{app:subsample} divides the 2001--2025 out-of-sample period into three nonoverlapping intervals. FT-DINO remains strong in each period. The EW Sharpe ratios are 8.49 in 2001--2009, 7.91 in 2010--2017, and 4.41 in 2018--2025. Predictability weakens in the most recent period, a 48\% decline in the Sharpe ratio relative to 2001--2009, but remains economically large.

\tablink{app:subsample}{Subsample Portfolio Performance}

The original FT-DINO model is frozen before 2001. We next re-estimate FT-DINO using a later training window. Table~\ref{tab:app_oos_periods} trains FT-DINO-NEW on 1993--2007 and validates on 2008--2009, so the common out-of-sample evaluation begins in 2010. Over 2010--2025, FT-DINO-NEW has an EW return of 73\% and a Sharpe ratio of 6.79, compared with 62\% and 5.51 for the original FT-DINO over the same period. The later training window also performs strongly in both 2010--2017 and 2018--2025.

\tablink{tab:app_oos_periods}{Alternative Training Windows and Out-of-Sample Performance}

\subsection{Alternative Fine-Tuning Targets}\label{app:c_targets}

The baseline alignment target is the cross-sectional rank of the raw subsequent return. Table~\ref{tab:app_adjusted_ft} replaces that target with three measures designed to remove different common components: FF6-residual returns based on the Fama--French five factors and momentum \citep{fama2015,carhart1997}, Fama--French 12 industry-adjusted returns, and DGTW-adjusted returns \citep{daniel1997}. All models otherwise use the same encoder, historical-retrieval procedure, and portfolio construction. At $K=1{,}000$, the EW Sharpe ratios are 5.82, 6.55, and 6.27 for the three adjusted-return targets, compared with 6.37 for the raw-return target. Return alignment therefore remains effective when the supervision focuses more narrowly on firm-specific relative performance.

\tablink{tab:app_adjusted_ft}{Alternative Fine-Tuning Targets}

\subsection{Factor-Adjusted Neighbor Returns}\label{app:c_alpha}

The alternative fine-tuning targets change what the encoder learns. A complementary test holds the retrieved neighbors fixed and changes the outcome that the signal averages. For the $K=1{,}000$ neighbors of each query, we replace each neighbor's raw subsequent five-day return with its CAPM alpha or its FF6 alpha relative to the Fama--French five factors and momentum \citep{fama2015,carhart1997}. Factor loadings are estimated by rolling 36-month OLS regressions, updated monthly. If the signal reflects the neighbors' exposures to priced factors, removing these components should weaken it.

Table~\ref{tab:app_alpha} reports the results. Factor adjustment leaves the signal intact and slightly strengthens it. For FT-DINO, the equal-weighted Sharpe ratio is 6.84 with CAPM alphas and 6.50 with FF6 alphas, compared with 6.37 with raw returns, and the capped value-weighted Sharpe ratios are 2.55 and 2.47, compared with 2.38. For benchmark DINO, the equal-weighted Sharpe ratio rises from 2.81 to 3.31 and 3.34, an increase of about 18\%. The return information in the retrieved neighbors therefore lies mainly in the parts of their outcomes that standard factors do not explain.

\tablink{tab:app_alpha}{Factor-Adjusted Neighbor Returns}

\subsection{Time-Series Momentum States}\label{app:c_momentum}

Because the retrieval signal is built from the recent price path, a natural concern is that it merely identifies stocks in a particular momentum state. Table~\ref{tab:app_momentum} examines this possibility. Each week, we group stocks by the sign of their own past returns, in the spirit of time-series momentum \citep{moskowitz2012}, and then sort them into FT-DINO quintiles within each group. The first two columns split stocks by the sign of the past 250-day return. The remaining four columns interact it with the sign of the past 20-day return, which yields the Bull $(+,+)$, Correction $(+,-)$, Bear $(-,-)$, and Rebound $(-,+)$ states of \citet{goulding2023}.

The FT-DINO spread is large and significant in every state. The 5--1 spread is 43\% per year among stocks with positive long-term momentum and 74\% among those with negative long-term momentum, and it ranges from 43\% in the Bull state to 81\% in the Bear state, with $t$-statistics above 15 throughout. The spread is about 1.7 to 1.9 times as large among long-term losers as among long-term winners, consistent with the concentration of predictability among hard-to-arbitrage stocks in Section~\ref{sec:econrelevance}. The signal is also not confined to recent losers: the spread is 43\% in the Bull state and 72\% in the Rebound state, in which the past 20-day return is positive. Retrieval therefore captures information beyond the direction of a stock's own past returns.

\tablink{tab:app_momentum}{FT-DINO and Time-Series Momentum States}

\subsection{Retrieval-Library and Chart Ablations}\label{app:c_ablation}

Table~\ref{tab:app_ft_full_ablation} reports the complete $K$ grid underlying the main-text ablation table. Panel A changes the historical retrieval library, Panel B changes color information, and Panel C changes chart elements. The cross-industry and same-industry specifications remain strong across broad neighborhoods, while same-stock retrieval is much weaker. Removing the conventional red-green color mapping reduces performance less than removing volume, and removing the moving average has little effect. These patterns persist across the reported neighborhood sizes.

\tablink{tab:app_ft_full_ablation}{FT-DINO Ablations: Retrieval Library and Chart Design}

\subsection{Chart Sparsity}\label{app:c_sparsity}

Charts differ in how much of the image the price and volume marks occupy. Because each chart is scaled to its own high--low range, a path dominated by a few large moves leaves much of the image blank, whereas a path with steady, comparable daily ranges fills more of it. A concern is that retrieval merely sorts stocks by this image density. Table~\ref{tab:app_sparsity} examines this possibility. Each week, stocks are first sorted into quintiles by the fraction of white pixels in their original $64\times60$ chart images and then, within each quintile, into conditional quintiles of the retrieval signal.

The FT-DINO 5--1 spread is positive and highly significant in every white-pixel quintile, ranging from 43\% per year in the densest charts to 86\% in the sparsest, with $t$-statistics between 16.78 and 20.28. The benchmark DINO spread is also positive in every quintile but rises more steeply, from 5\% to 41\%. Both signals are strongest among the sparsest charts, consistent with the importance of abrupt, localized price and volume events in the sparse-feature analysis (Section~\ref{sec:bottomup}). Retrieval performance is therefore not an artifact of image density.

\tablink{tab:app_sparsity}{Double Sorts on Chart Sparsity}

\subsection{Alternative Image Windows and Return Horizons}\label{app:c_horizons}

Table~\ref{tab:app_window_compare} expands the weekly five-day-horizon comparison across the full $K$ grid for D20, D40, and D60 charts. At large neighborhoods, all three image windows produce similar benchmark DINO performance, while FT-DINO more than doubles the Sharpe ratio for every window. Historical visual retrieval therefore does not hinge on a particular lookback window.

\tablink{tab:app_window_compare}{Alternative Image Windows: DINO versus FT-DINO}

Table~\ref{tab:app_r20} changes the outcome horizon to the subsequent 20 trading days. Queries are formed monthly on the last market trading day of each month, and FT-DINO is retrained using cross-sectional ranks of the subsequent 20-day return. Performance remains positive over the full neighborhood grid. At $K=1{,}000$, FT-DINO Sharpe ratios are 2.13, 2.11, and 2.19 for D20, D40, and D60, respectively. The corresponding DINO Sharpe ratios are 1.54, 1.99, and 1.86.

\tablink{tab:app_r20}{20-Day Return Horizon: DINO versus FT-DINO}

\subsection{Holding Horizons and Transaction Costs}\label{app:c_holding}

We hold the D20R5 signal fixed and evaluate rebalancing intervals from one to four weeks. Table~\ref{tab:app_holding_costs} reports the resulting turnover and portfolio performance. Under EW, FT-DINO turnover falls from 600\% at H5 to 329\%, 228\%, and 173\% at H10, H15, and H20. Gross performance declines with the slower rebalancing schedule, but net Sharpe ratios remain 1.55, 2.06, and 1.78 at the three longer intervals. The result illustrates the trade-off between exploiting the short-horizon signal more frequently and reducing turnover.

\tablink{tab:app_holding_costs}{Holding Horizons and Transaction Costs}

\subsection{Task-Vector Steering}\label{app:c_taskvector}

Return alignment can also be viewed as a direction in model-parameter space. Let $\theta_{\mathrm{DINO}}$ and $\theta_{\mathrm{FT\text{-}DINO}}$ denote the pretrained and fine-tuned parameter vectors and define the task vector
\begin{equation}
    \Delta\theta
    =
    \theta_{\mathrm{FT\text{-}DINO}}
    -
    \theta_{\mathrm{DINO}}.
    \label{eq:app_task_vector}
\end{equation}
We construct steered encoders
\begin{equation}
    \theta(\lambda)
    =
    \theta_{\mathrm{DINO}}+\lambda\Delta\theta.
    \label{eq:app_task_steering}
\end{equation}
Thus, $\lambda=0$ reproduces DINO, $\lambda=1$ reproduces FT-DINO, values between zero and one interpolate between them, and values above one extrapolate in the same fine-tuning direction. Negative values move the model in the direction opposite to fine-tuning.

Table~\ref{tab:app_task_vector} shows a smooth improvement as the model moves from DINO toward the return-aligned parameters. At $K=1{,}000$, the EW Sharpe ratio rises from 2.81 at $\lambda=0$ to 5.15 at $\lambda=0.5$ and 6.37 at $\lambda=1$. Extrapolating to $\lambda=1.5$ leaves performance high at 6.15 but does not improve on the fine-tuned model. Moving in the opposite direction to $\lambda=-0.5$ lowers the Sharpe ratio to 2.30. The monotonic improvement from $\lambda=0$ to $\lambda=1$ links the performance gain to the return-aligned fine-tuning direction.

\tablink{tab:app_task_vector}{Task-Vector Steering}

\section{Economic Mechanism}\label{app:d_structure}

\subsection{Limits to Arbitrage}\label{app:d_limits}

The main text combines several arbitrage-friction proxies into a single oriented score following \citet{chen2026debt}. Tables~\ref{tab:app_dino_limits} and \ref{tab:app_ft_limits} report the underlying sequential double sorts separately for size, Amihud illiquidity, dollar volume, securities lending fees, analyst forecast dispersion, turnover, and days-to-cover. Proxy quintiles retain their natural ordering; greater limits to arbitrage correspond to smaller size, lower dollar volume and turnover, and higher illiquidity, lending fees, forecast dispersion, and days-to-cover.

The benchmark DINO signal is generally stronger in the more constrained parts of the market. For example, the annualized DINO 5--1 spread is 42\% among the smallest stocks versus 5\% among the largest, and it rises from 4\% in the most liquid Amihud quintile to 46\% in the least liquid quintile.

\tablink{tab:app_dino_limits}{DINO and Limits to Arbitrage}

The pattern becomes more pronounced after return alignment. The FT-DINO spread is 127\% in the smallest-stock quintile versus 13\% in the largest-stock quintile, and it rises from 11\% to 135\% across Amihud-illiquidity quintiles. The dollar-volume, lending-fee, forecast-dispersion, and turnover sorts similarly place the largest spreads on the side associated with tighter arbitrage constraints. Days-to-Cover is the main exception: the FT-DINO spread declines from 104\% in the lowest-DTC quintile to 32\% in the highest-DTC quintile. This pattern is consistent with the measurement issue emphasized by \citet{asquith2005}, who characterize short-sale constraints as jointly reflecting shorting demand and the supply of lendable shares; observed short interest, which enters the numerator of DTC, therefore need not vary monotonically with the underlying constraint. Overall, the component-level evidence is broadly consistent with the composite pattern reported in the main text.

\tablink{tab:app_ft_limits}{FT-DINO and Limits to Arbitrage}

\subsection{Retail Share of Trading Volume}\label{app:d_retail}

Tables~\ref{tab:app_dino_retail} and \ref{tab:app_ft_retail} examine two proxies for retail participation. Panel A uses RSVOL, the retail share of trading volume of \citet{laarits2025}, with retail-initiated trades identified by the algorithm of \citet{boehmer2021}, and is available from 2006 through 2024. \citet{laarits2025} show that high-RSVOL stocks are hard to value: they have more intangible capital, longer-duration cash flows, and higher mispricing scores, and their prices are less sensitive to earnings news. Because valuation uncertainty makes trading against a dislocation risky \citep{shleifer1997}, a mispricing interpretation predicts a stronger spread among these stocks. For DINO, the annualized conditional 5--1 spread rises from 1\% in the lowest RSVOL quintile to 34\% in the highest. For FT-DINO, the corresponding spread rises from 10\% to 88\%. The concentration of predictability in high-RSVOL stocks is present in the pretrained representation and becomes stronger after return alignment, with the high-RSVOL spread rising from 34\% to 88\%.

\tablink{tab:app_dino_retail}{DINO and Retail Trading}

Panel B uses the Robinhood demand shock over the July 2018--September 2020 period. FT-DINO produces positive conditional spreads in all five shock quintiles, but unlike RSVOL, the spreads show no clear gradient across shock quintiles, at 41\% in the lowest and 29\% in the highest. Because this proxy covers only about two years, we interpret the retail evidence as suggestive.

\tablink{tab:app_ft_retail}{FT-DINO and Retail Trading}

\subsection{Long and Short Legs}\label{app:d_legs}

Figure~\ref{fig:app_legs} decomposes the event-time H--L returns in Figure~\ref{fig:returndynamics} into the top and bottom FT-DINO deciles. Both legs contribute to the formation-week decline and to the subsequent correction. High-signal stocks fall 2.73\% in the formation week and then earn 1.18\% over days 1 to 5 and between 0.37\% and 0.51\% in each of the next three five-day blocks. Low-signal stocks rise 2.67\% in the formation week and give back 0.46\% over days 1 to 5, after which their returns are close to zero. The correction is therefore larger and more persistent on the long side, whereas the short side mainly contributes the first-week reversal.

\figlink{fig:app_legs}{Event-Time Returns to the Top and Bottom FT-DINO Deciles}

\subsection{Neighbor Similarity and Weighting}\label{app:d_recall}

Memory-based models of belief formation imply that the most similar, salient, and recent past episodes should carry the most weight in expectations \citep{mullainathan2002,bordalo2020}, and \citet{charles2025} finds that recalled episodes move prices. We examine whether the retrieval signal displays this structure. Table~\ref{tab:app_similarity} forms stand-alone signals from each band of 200 neighbors by similarity rank. Performance is nearly flat across bands: the EW Sharpe ratio is 6.00 for the 200 most similar neighbors and 5.63 for the 200 least similar, a decline of only 6\%, and every band falls short of the 6.37 obtained from all 1,000 neighbors. Breadth thus matters more than closeness.

\tablink{tab:app_similarity}{FT-DINO Performance by Neighbor Similarity Rank}

Table~\ref{tab:app_memory_weights} instead reweights the 1,000 neighbors by characteristics that make past episodes memorable. Every such scheme lowers performance relative to equal weighting. Weighting by the composite salience measure lowers the EW Sharpe ratio from 6.37 to 5.50, a 14\% decline, time-decay weighting lowers it to 5.98, and weighting by historical or current Google search attention lowers it to 2.68 and 3.72. Weighting by abnormal volume or turnover lowers it to about 1.3. The retrieved neighbors that are most informative about subsequent returns are therefore not those that a memory-based investor would weight most heavily.

\tablink{tab:app_memory_weights}{FT-DINO Performance under Alternative Neighbor Weights}

\section{Sparse-Feature Interpretation}\label{app:e_structure}

\subsection{Visual-State Classification}\label{app:e_states}

Table~\ref{app:saestates} reports the complete mapping from the 30 visually consistent SAE features to the 11 visual states. The states range from directional configurations such as Sustained Uptrend and Stable Downtrend to more localized patterns such as Peak Pullback, Late Upside Breakout, Late Breakdown, and Isolated Upward Spike. The labels summarize recurring visual configurations in the representative high-activation charts.

\tablink{app:saestates}{SAE Visual States and Constituent Features}

\subsection{Representative Charts for Visual States}\label{app:e_charts}

Figures~\ref{fig:sae_states_c01_c03}--\ref{fig:sae_states_c09_c11} display representative high-activation real-market charts for all 11 states. The figures are organized by state. Within a state, the constituent features can emphasize different variants of the same broad visual configuration. We therefore interpret each state from recurring patterns across multiple high-activation observations.

\figrangelink{fig:sae_states_c01_c03}{fig:sae_states_c09_c11}

\subsection{Literature-Based Visual Characteristics}\label{app:e_literature}

Table~\ref{tab:literature_patterns} lists 21 visual characteristics motivated by the cited finance literature. For each characteristic, we select an illustrative real-market 20-day chart that captures a visual configuration motivated by the cited study and map it to the nearest atlas prototype. Figure~\ref{fig:landmarks_main} displays the resulting locations.

\tablink{tab:literature_patterns}{Literature-Based Visual Characteristics}

\section{Top-Down Analysis}\label{app:f_structure}

\subsection{Top-Down Feature Validation}\label{app:f_validation}

The top-down procedure starts from 11 patterns and searches for sparse features that activate on real representative charts of each pattern. Up Momentum is reported in the main text. The remaining ten validations are shown here: Down Momentum (feature 7123), Breakout Up (13482), Breakout Down (2751), Mean Reversion (8565), Discrete Winner (16299), Discrete Loser (9690), MAX(1) (10461), Negative Overnight Jump (4232), Up + Positive Convexity (6232), and Down + Positive Convexity (2940). For each selected feature, the displayed high-activation real-market charts check whether the feature captures the intended configuration beyond the representative charts used to select it.

\figrangelink{fig:app_down_momentum}{fig:app_down_pos_convexity}

Across the 11 patterns, validation combines three steps: activation on real representative charts identifies candidate sparse features, high-activation real charts evaluate whether the same visual configuration appears more broadly in market data, and query steering examines how movement along the selected feature direction affects the nearest real-market neighbors. Together, these tests provide evidence that the target patterns are represented by sparse directions with consistent behavior in the learned visual space.

\input{tables_figures_appendix}

\end{document}

%% file: tables_figures_main.tex
\clearpage

\begin{figure}[p]
\centering
\caption{Price-Chart Specifications}
\label{fig:samples}
\exhibitnote{This figure compares the price-chart designs for the same 20-trading-day window. Panel (a) shows the black-and-white price image of \citet{jiang2023}, and Panel (b) shows the colored price image used in this paper. Both designs display daily open-high-low-close bars, a 20-day moving average, and trading volume. Within each image, the first closing price is normalized to one, subsequent closing prices are constructed from CRSP returns, daily open, high, and low prices are scaled in proportion to the closing price, and volume is scaled relative to its within-window maximum. In Panel (b), days that close above their open are green, other days are red, and the moving average is blue. Both designs therefore abstract from nominal price and share-volume levels.}
\begin{subfigure}[t]{0.42\textwidth}
  \centering
  \includegraphics[width=\linewidth]{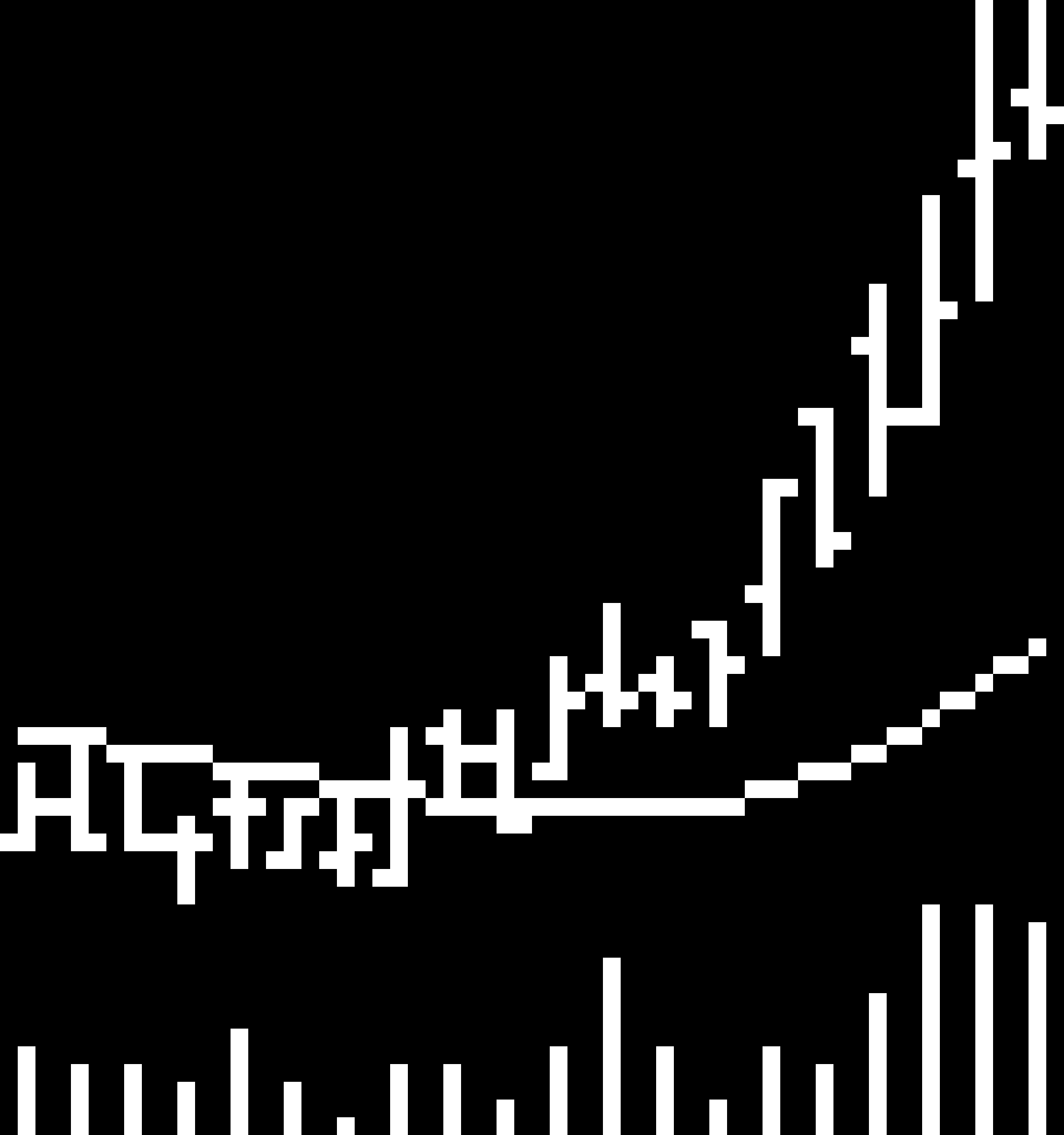}
  \caption{\citet{jiang2023} price-image}
\end{subfigure}\hfill
\begin{subfigure}[t]{0.42\textwidth}
  \centering
  \includegraphics[width=\linewidth]{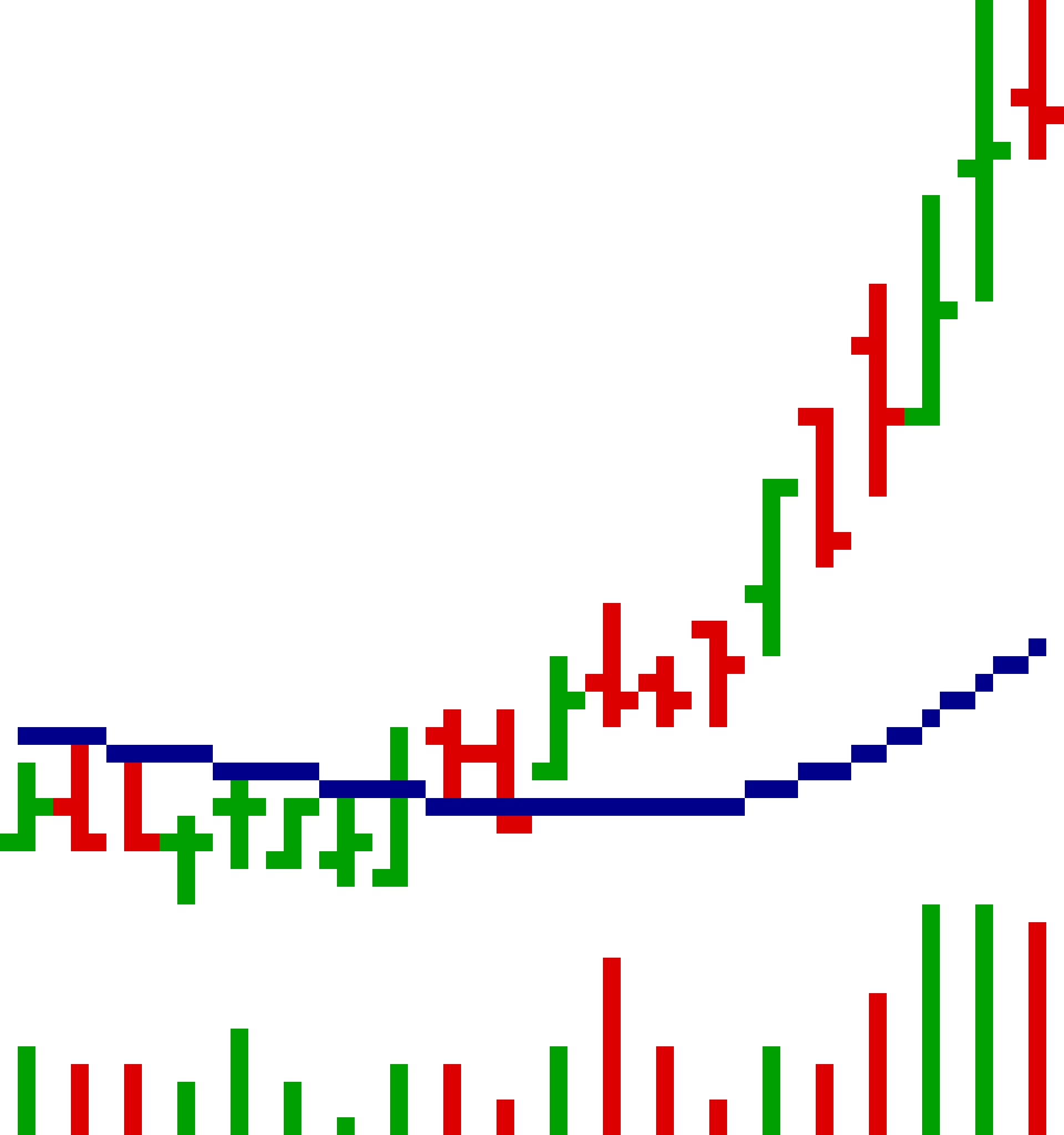}
  \caption{Our colored price-image}
\end{subfigure}
\end{figure}

\clearpage
\begin{figure}[p]
\centering
\caption{Cumulative Log Returns to FT-DINO Portfolios}
\label{fig:ftcum}
\exhibitnote{This figure plots cumulative log returns to FT-DINO high-minus-low portfolios from 2001 to 2025. FT-DINO denotes DINOv2-S after return-rank alignment on 1993--1999 charts, with 2000 used for validation and the encoder frozen before 2001. Each week, stocks are sorted into deciles by the signal and held over the subsequent five trading days; H--L is the return on decile 10 minus decile 1. Panels (a), (b), and (c) report equal-weighted, value-weighted, and capped value-weighted (CVW) portfolios, where capped value weighting winsorizes market capitalization at the contemporaneous NYSE 80th percentile following \citet{jensen2023}. Each line corresponds to one neighborhood size, $K\in\{5,10,30,100,500,1{,}000,3{,}000\}$. The dashed horizontal line marks zero.}
\begin{subfigure}{0.67\textwidth}
  \centering
  \includegraphics[width=\linewidth]{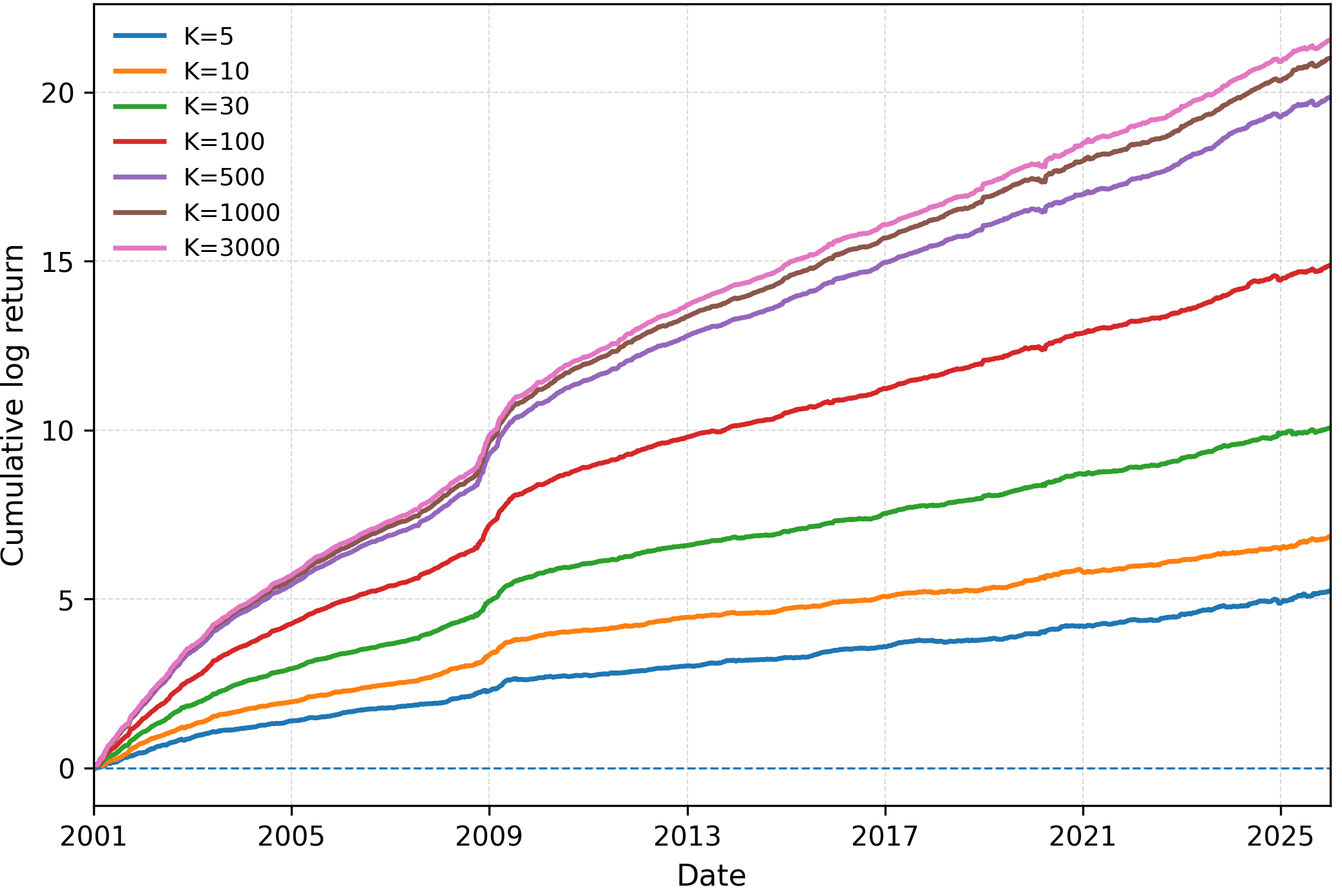}
  \caption{EW H--L}
\end{subfigure}

\vspace{0.2cm}

\begin{subfigure}{0.67\textwidth}
  \centering
  \includegraphics[width=\linewidth]{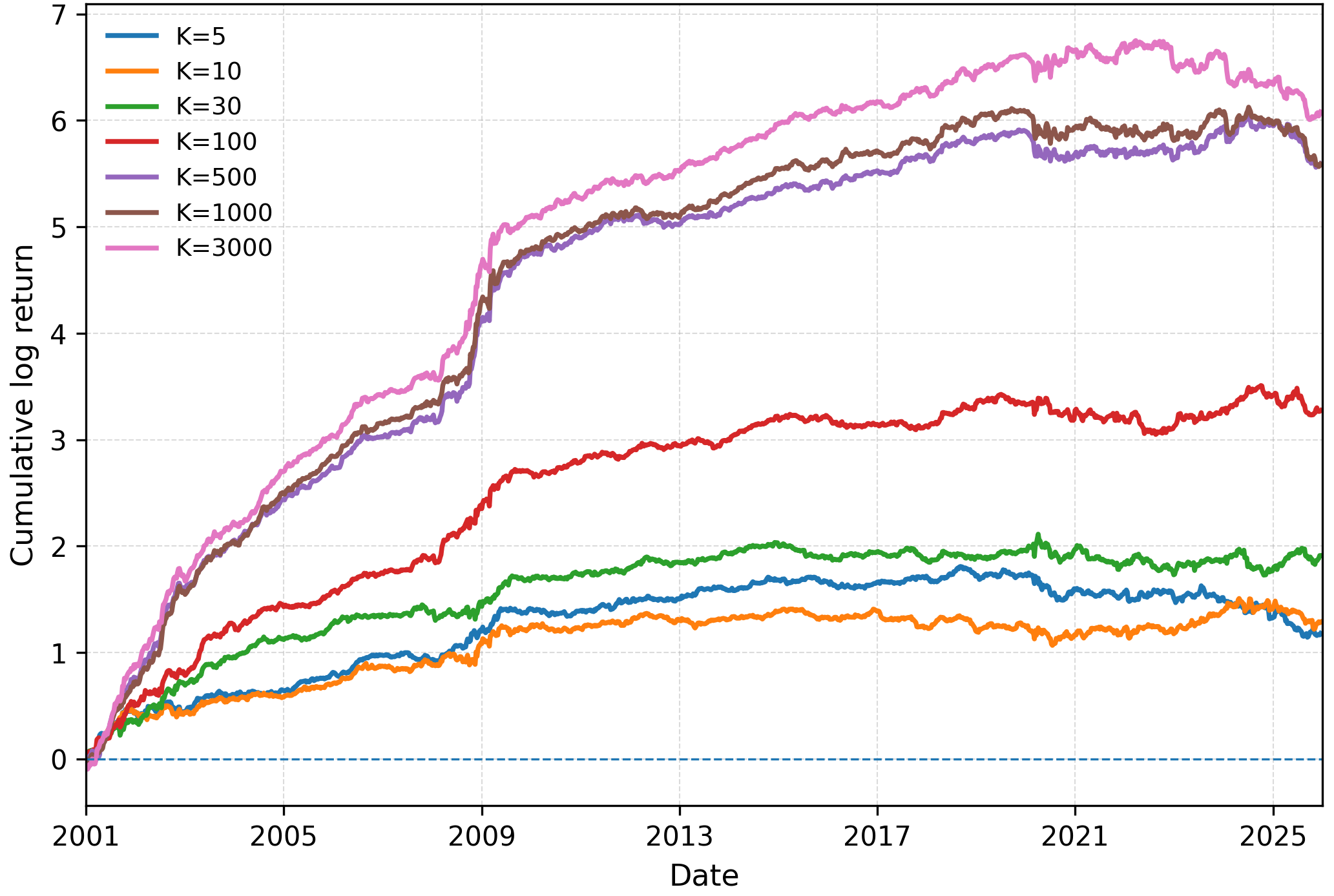}
  \caption{VW H--L}
\end{subfigure}

\end{figure}

\clearpage
\begin{figure}[p]
\centering
\ContinuedFloat
\caption[]{Cumulative Log Returns to FT-DINO Portfolios (continued)}
\exhibitnote{This figure continues Figure~\ref{fig:ftcum}. Panel (c) reports CVW portfolios.}
\begin{subfigure}{0.67\textwidth}
  \centering
  \includegraphics[width=\linewidth]{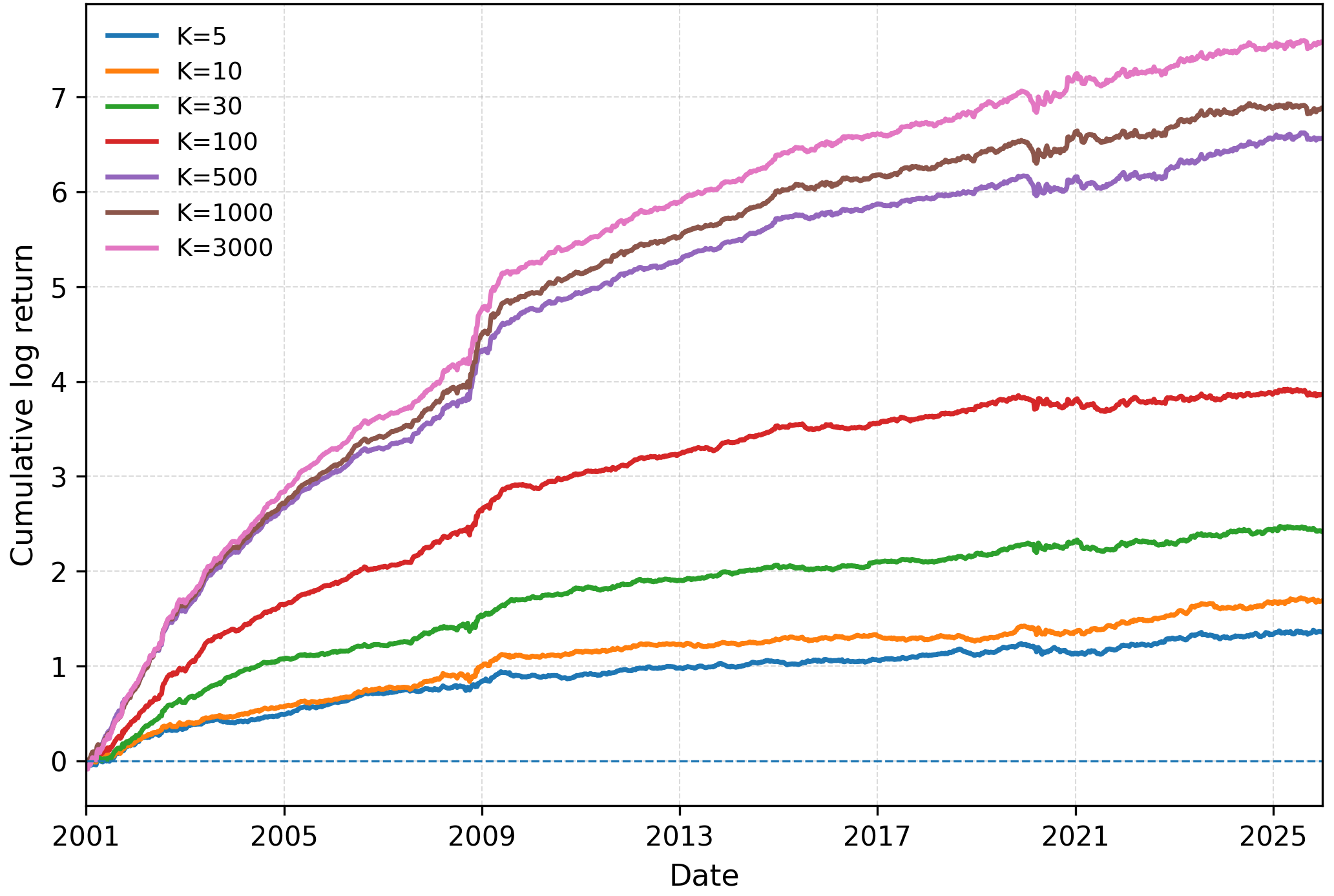}
  \caption{CVW H--L}
\end{subfigure}
\end{figure}

\clearpage
\begin{figure}[p]
\centering
\caption{Forecast Sensitivity Maps: CNN, DINO, and FT-DINO}
\label{fig:sensitivity}
\exhibitnote{This figure plots the average forecast sensitivity of each model to each pixel of the 20-day price chart, averaged across randomly sampled query charts. For DINO and FT-DINO, forecast sensitivity is the $L_1$ norm of the pixel gradient of the return-aligned component $g$ of the similarity-weighted retrieval forecast in equation~\eqref{eq:forecastsens_main}, holding the top-$K$ neighborhood fixed with $K=1{,}000$. For the CNN, it is the absolute pixel gradient of the five-seed ensemble's predicted probability of a positive return. For all three models, each query's sensitivity map is normalized to sum to one before averaging across the same 100 randomly sampled charts. The horizontal axis is the trading day within the chart window, and the bottom of each panel corresponds to the volume bars. Each panel has its own color scale, and brighter colors indicate greater sensitivity. Appendix Figure~\ref{fig:app_sensitivity_measures} compares this measure with embedding and matching sensitivity for DINO and FT-DINO.}
\includegraphics[width=\linewidth]{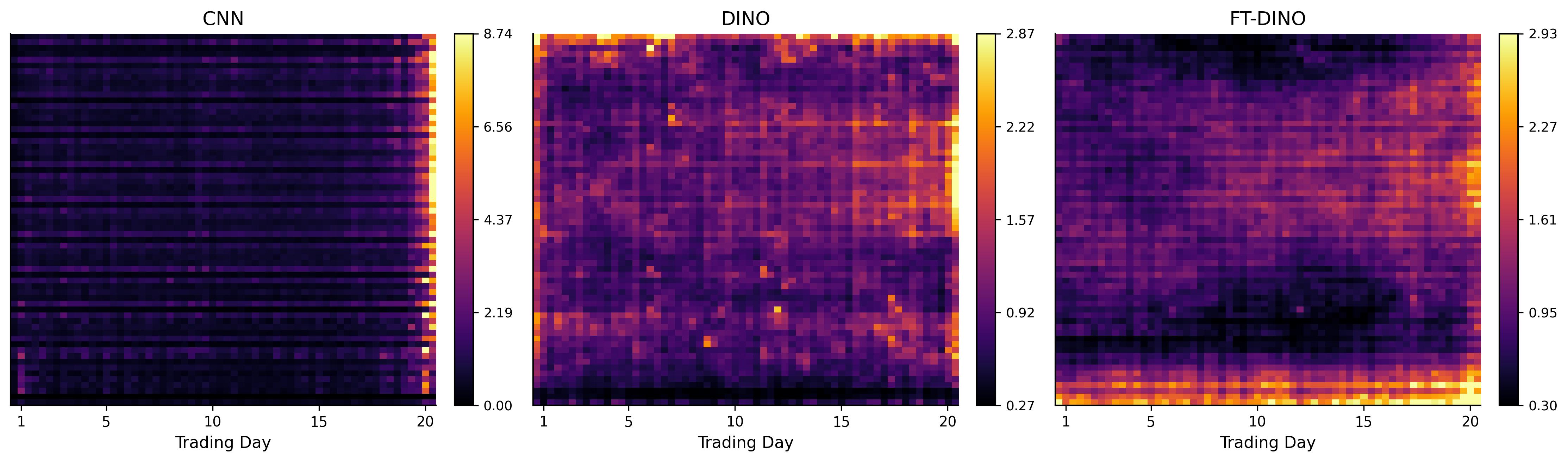}
\end{figure}

\clearpage
\begin{figure}[p]
\centering
\caption{Price-Chart Ablations}
\label{fig:ft_ablation_examples}
\exhibitnote{This figure illustrates the chart variants used in the FT-DINO ablation tests in Table~\ref{tab:ft_ablation}. Panel (a) shows the chart-element ablations, which remove the moving average, remove the volume bars, remove both, or plot the moving average on the price scale. Panel (b) shows the color ablations, which remove color, reverse the conventional red--green mapping of down and up days, or assign colors at random. Each rendering uses the same underlying 20-day price history and changes only the indicated chart element or color specification.}
\begin{subfigure}[t]{0.96\textwidth}
  \centering
  \includegraphics[width=\linewidth,height=0.30\textheight,keepaspectratio]{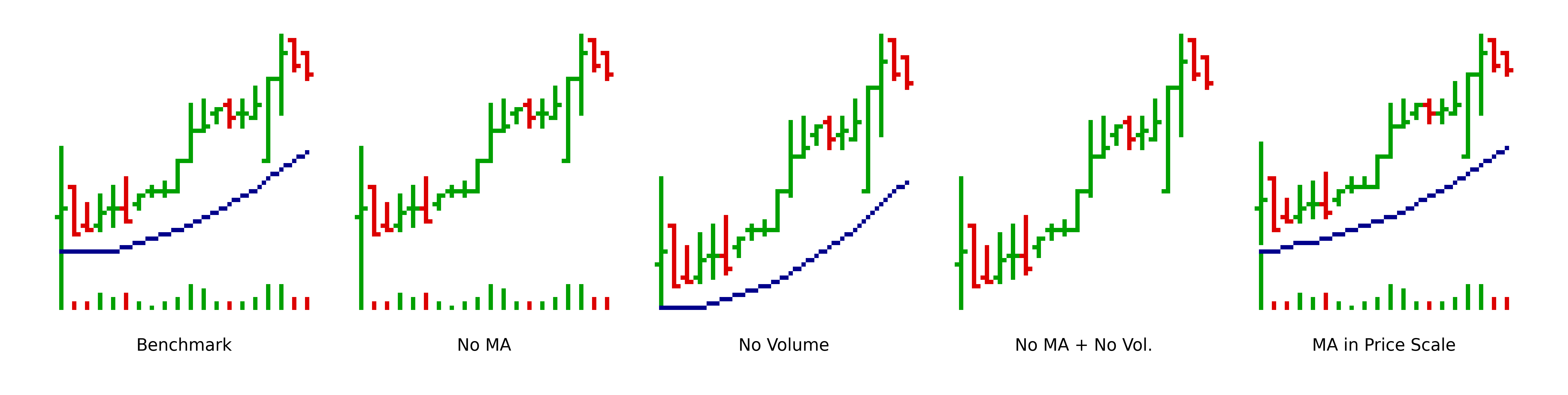}
  \caption{Chart-element ablations}
\end{subfigure}

\vspace{0.45cm}

\begin{subfigure}[t]{0.96\textwidth}
  \centering
  \includegraphics[width=\linewidth,height=0.30\textheight,keepaspectratio]{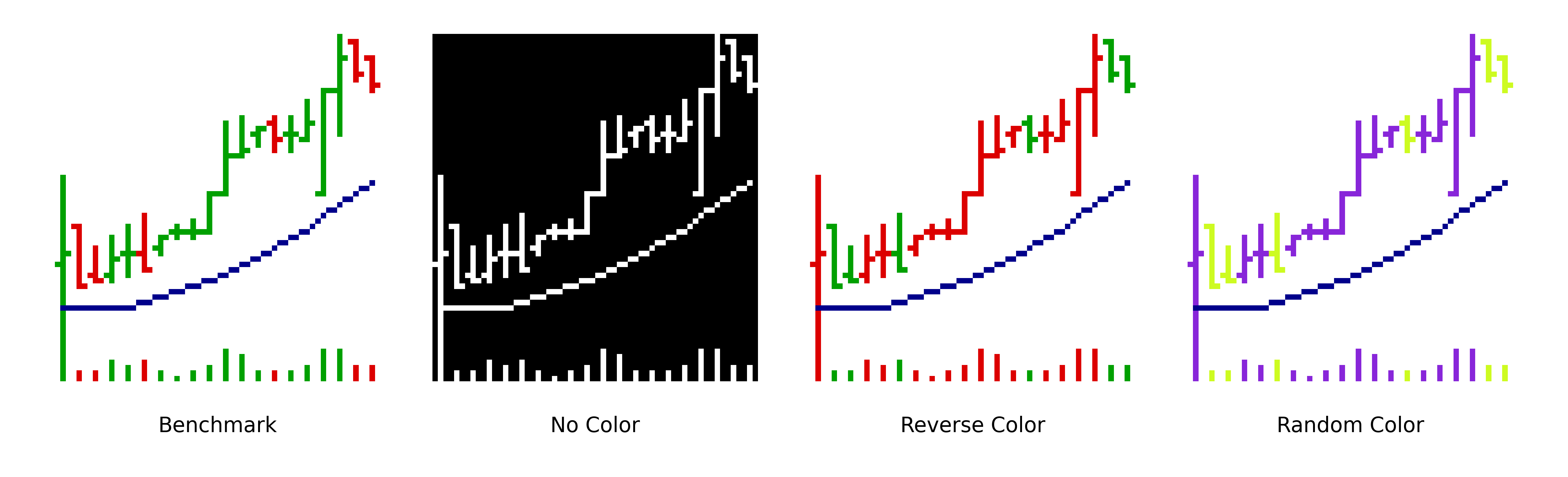}
  \caption{Color ablations}
\end{subfigure}
\end{figure}

\clearpage
\begin{figure}[p]
\centering
\caption{Alternative Price-Chart Windows}
\label{fig:app_windows}
\exhibitnote{This figure shows representative D20, D40, and D60 price images, in Panels (a), (b), and (c), used in Table~\ref{tab:alternative_horizons}. The three designs display the prior 20, 40, and 60 trading days, respectively, with open-high-low-close bars, trading volume, and a moving average whose window matches the image length (MA20, MA40, and MA60). Normalization, color conventions, and the remaining plotting rules are the same as in the baseline D20 design. The panels use a common height and a fixed horizontal width per trading day.}
\begin{subfigure}{\textwidth}
  \centering
  \includegraphics[height=0.22\textheight]{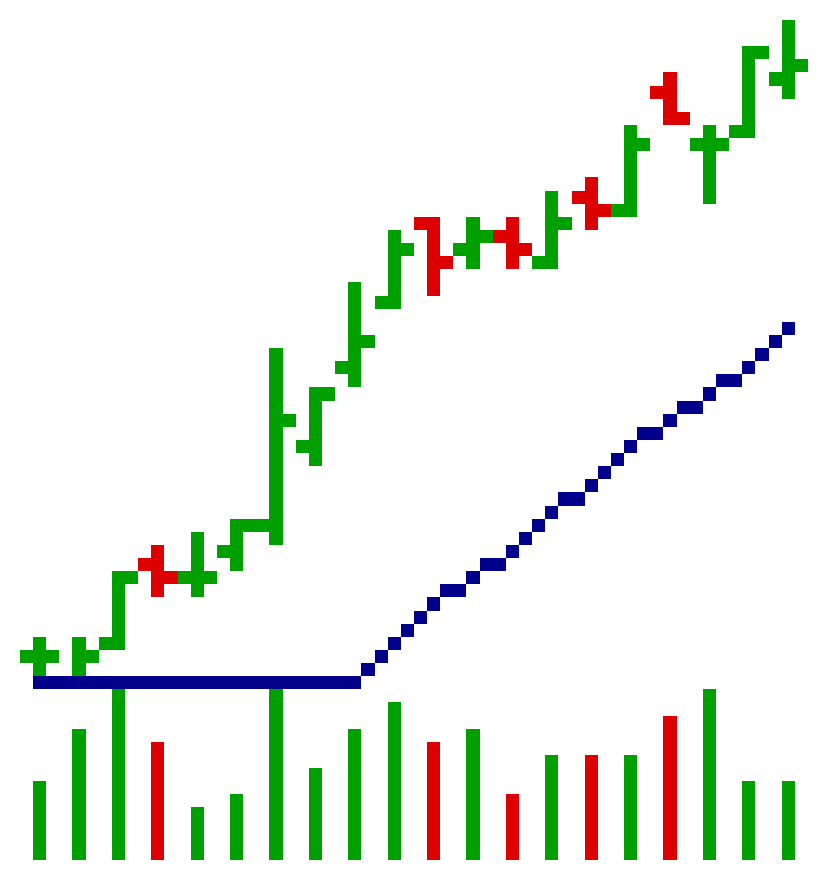}
  \caption{D20}
\end{subfigure}

\vspace{0.3cm}

\begin{subfigure}{\textwidth}
  \centering
  \includegraphics[height=0.22\textheight]{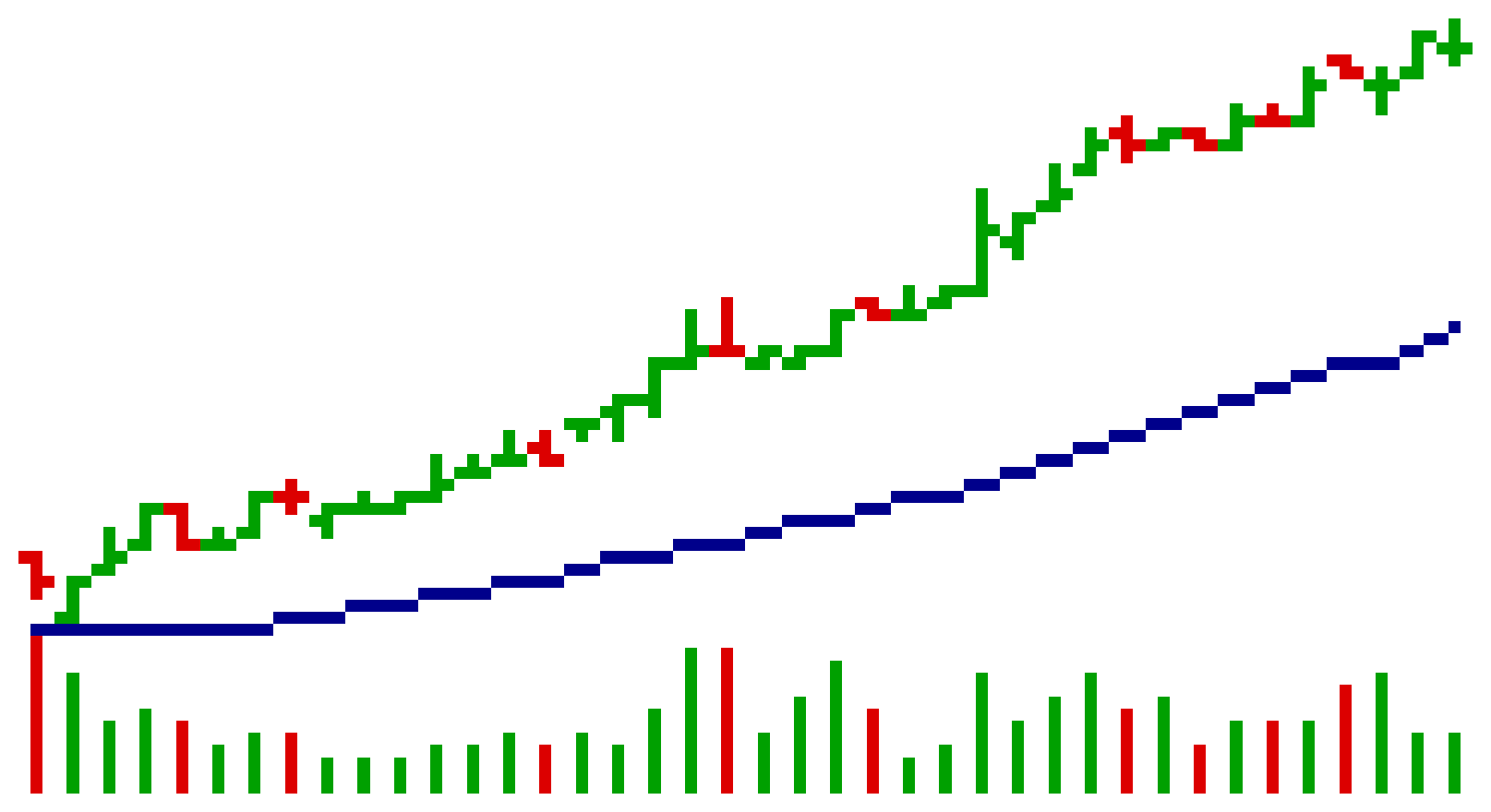}
  \caption{D40}
\end{subfigure}

\vspace{0.3cm}

\begin{subfigure}{\textwidth}
  \centering
  \includegraphics[height=0.22\textheight]{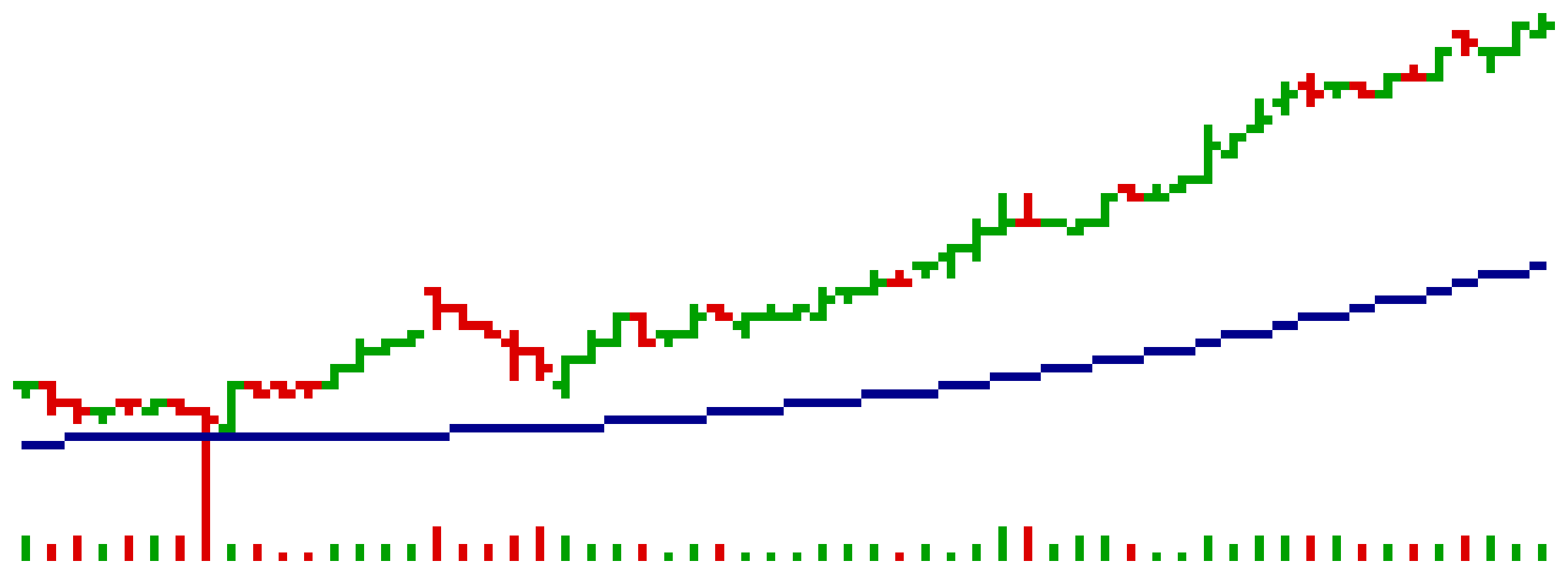}
  \caption{D60}
\end{subfigure}
\end{figure}

\clearpage
\begin{figure}[p]
\centering
\caption{Short- and Long-Run Returns to FT-DINO Portfolios}
\label{fig:returndynamics}
\exhibitnote{This figure plots event-time returns to the equal-weighted FT-DINO high-minus-low portfolio with $K=1{,}000$ over 2001--2025. Each week, stocks are sorted into deciles by the FT-DINO retrieval signal formed at the end of the formation week (day 0), and H--L is the return on decile 10 minus decile 1. Panel (a) reports H--L returns over five-trading-day blocks from the formation week (days $-4$ to 0) through day 20. Panel (b) reports H--L returns over 20-trading-day blocks from day 1 through day 120. Returns are in percent and are not annualized. Appendix Figure~\ref{fig:app_legs} reports the two legs separately.}
\begin{subfigure}[t]{0.49\textwidth}
  \centering
  \adjustbox{trim={0pt 0pt {.5\width} 0pt},clip,width=\linewidth}{\includegraphics{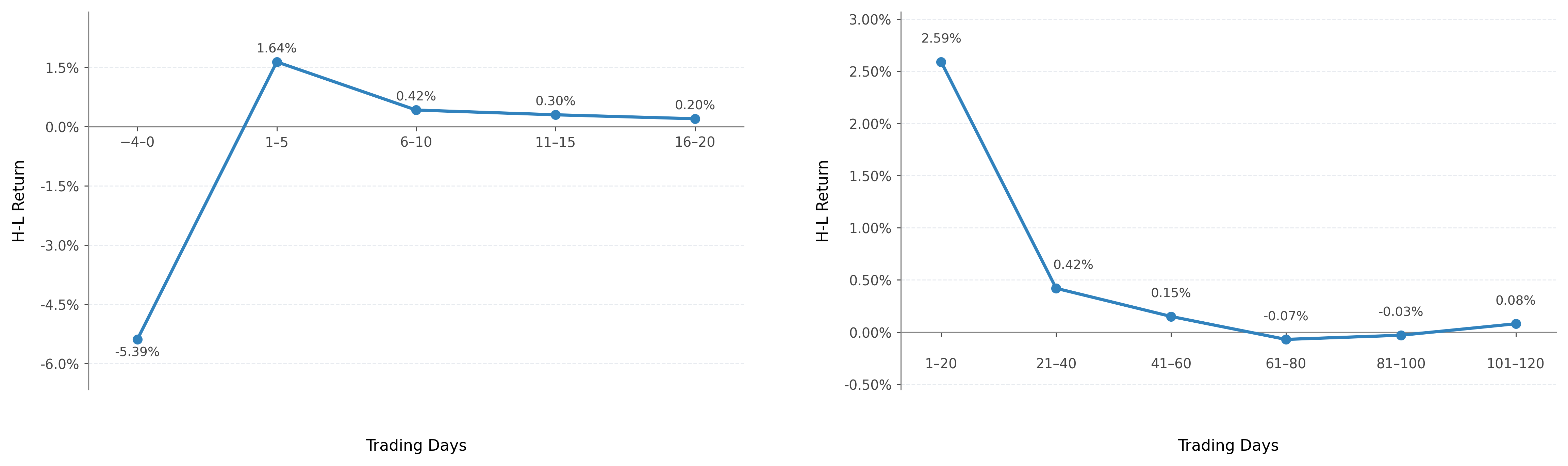}}
  \caption{Days $-4$ to 20}
\end{subfigure}\hfill
\begin{subfigure}[t]{0.49\textwidth}
  \centering
  \adjustbox{trim={{.5\width} 0pt 0pt 0pt},clip,width=\linewidth}{\includegraphics{usfigs/ftdino_full1000_5weeks_6months.png}}
  \caption{Days 1 to 120}
\end{subfigure}
\end{figure}

\clearpage
\begin{figure}[p]
\centering
\caption{Order Imbalance around Portfolio Formation}
\label{fig:oib}
\exhibitnote{This figure plots the order-imbalance (OIB) spread between the top and bottom FT-DINO deciles ($K=1{,}000$) around portfolio formation. For each stock and day, OIB is buyer-initiated minus seller-initiated volume divided by their sum, with trades signed by the algorithm of \citet{lee1991}. Buyer- and seller-initiated volume are obtained from WRDS for 2001--2025. The spread is the average OIB of decile 10 minus that of decile 1, in percentage points. Panel (a) reports averages over five-trading-day blocks from the formation week (days $-4$ to 0) through day 20. Panel (b) reports daily values from day $-4$ through day 20, where day 0 ($t$) is the formation date.}
\begin{subfigure}[t]{0.49\textwidth}
  \centering
  \adjustbox{trim={{.5\width} 0pt 0pt 0pt},clip,width=\linewidth}{\includegraphics{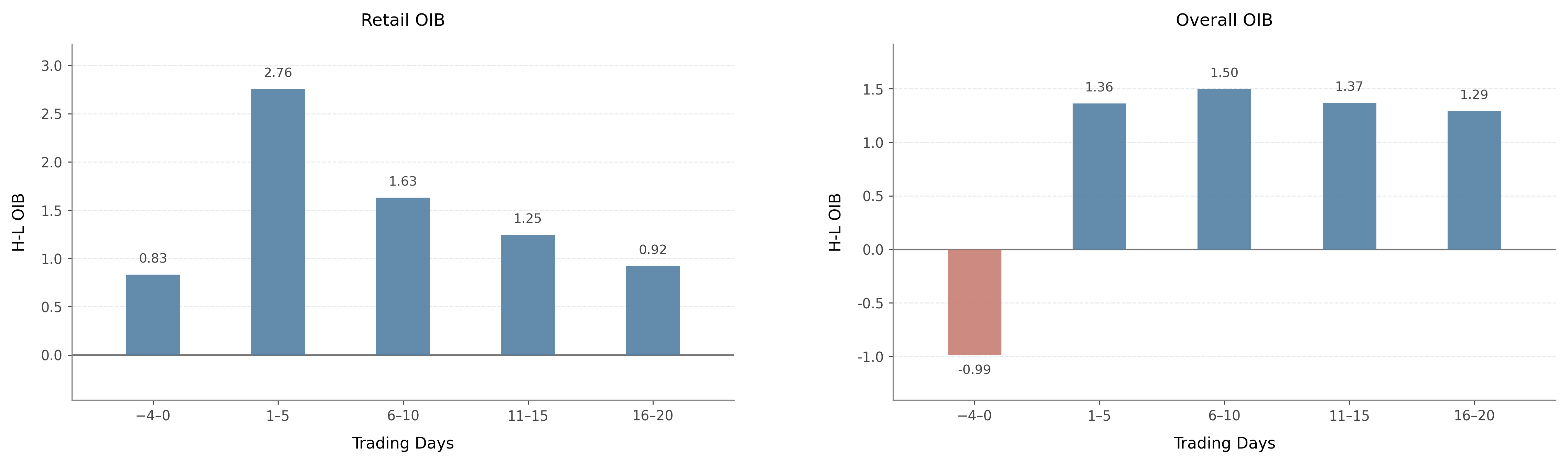}}
  \caption{Five-day blocks}
\end{subfigure}\hfill
\begin{subfigure}[t]{0.49\textwidth}
  \centering
  \adjustbox{trim={{.5\width} 0pt 0pt 0pt},clip,width=\linewidth}{\includegraphics{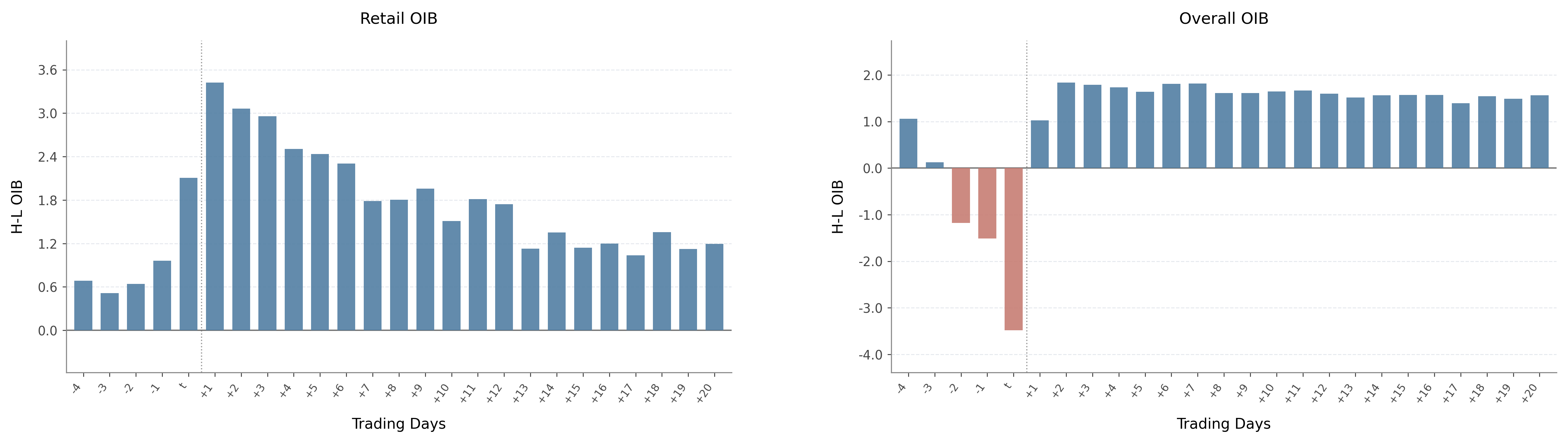}}
  \caption{Daily}
\end{subfigure}
\end{figure}

\clearpage
\begin{figure}[p]
\centering
\caption{Returns and Order Imbalance by Neighbor Salience}
\label{fig:salience}
\exhibitnote{This figure reports returns and order imbalance for stand-alone FT-DINO signals formed from neighbors sorted by salience. For each query, the 1,000 retrieved neighbors are sorted into salience quintiles of 200 neighbors based on an equally weighted composite of six characteristics, each percentile-ranked among the query's 1,000 neighbors: the absolute subsequent five-day return, abnormal trading volume, abnormal turnover, Google search attention in the calendar month preceding the neighbor's formation date, same FF12 industry, and recency; Q5 is the most salient. Each quintile forms a separate retrieval signal. The dashed line in Panel (a) is a similarity-matched benchmark formed from the neighbors at similarity ranks 401--600, chosen because the average similarity rank within each quintile is about 500. Panel (a) plots the daily H--L return (upper plot, in percent) and the daily OIB spread (lower plot, in percentage points) from day $-4$ through day 15, where day 0 is the formation date. Panel (b) plots H--L returns over 20-trading-day blocks from day 1 through day 120.}
\begin{subfigure}{\textwidth}
  \centering
  \adjustbox{trim={0pt 0pt 0pt {.065\height}},clip,height=0.40\textheight}{\includegraphics{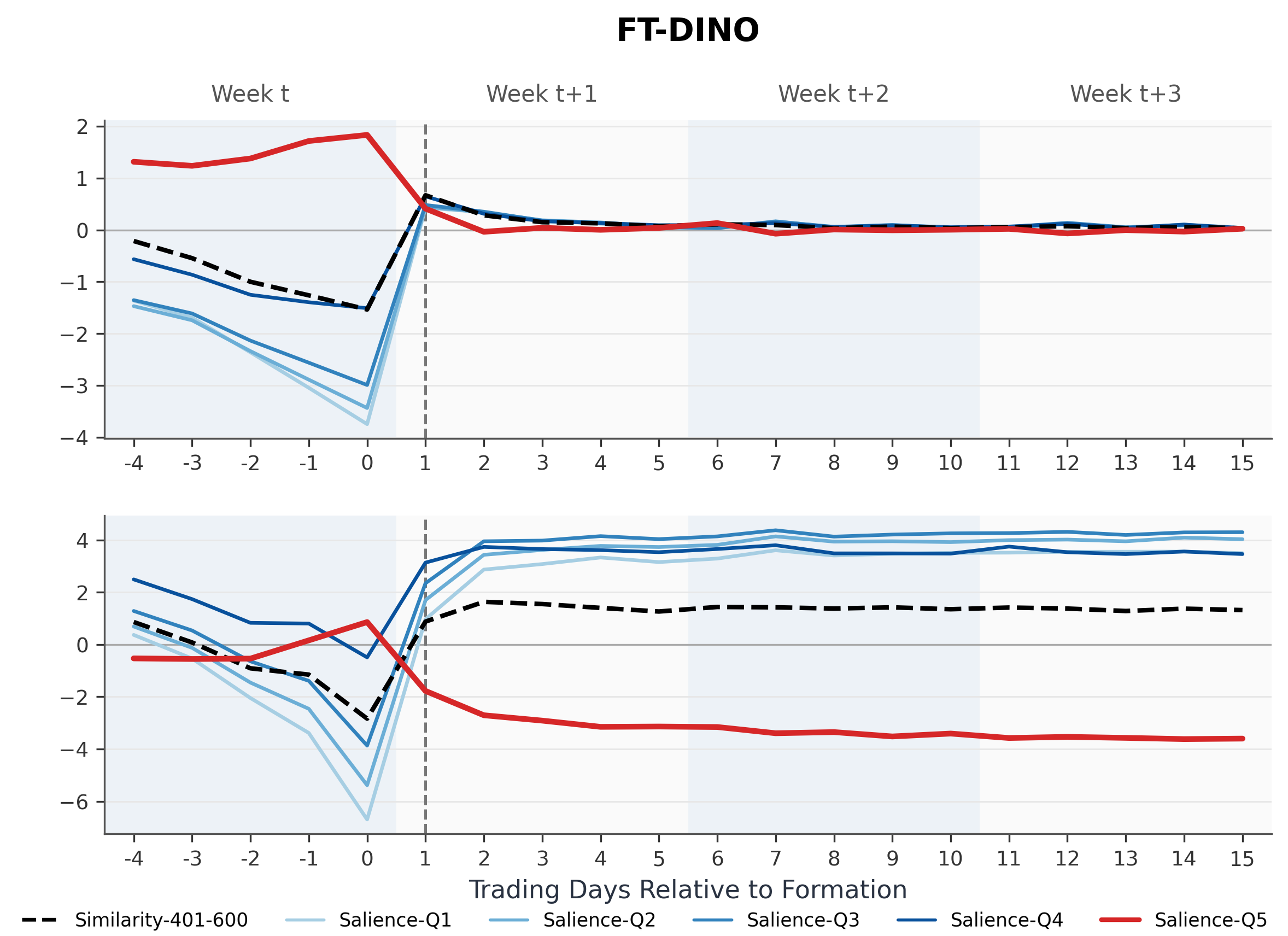}}
  \caption{Daily returns and order imbalance}
\end{subfigure}

\vspace{0.2cm}

\begin{subfigure}{\textwidth}
  \centering
  \includegraphics[height=0.28\textheight]{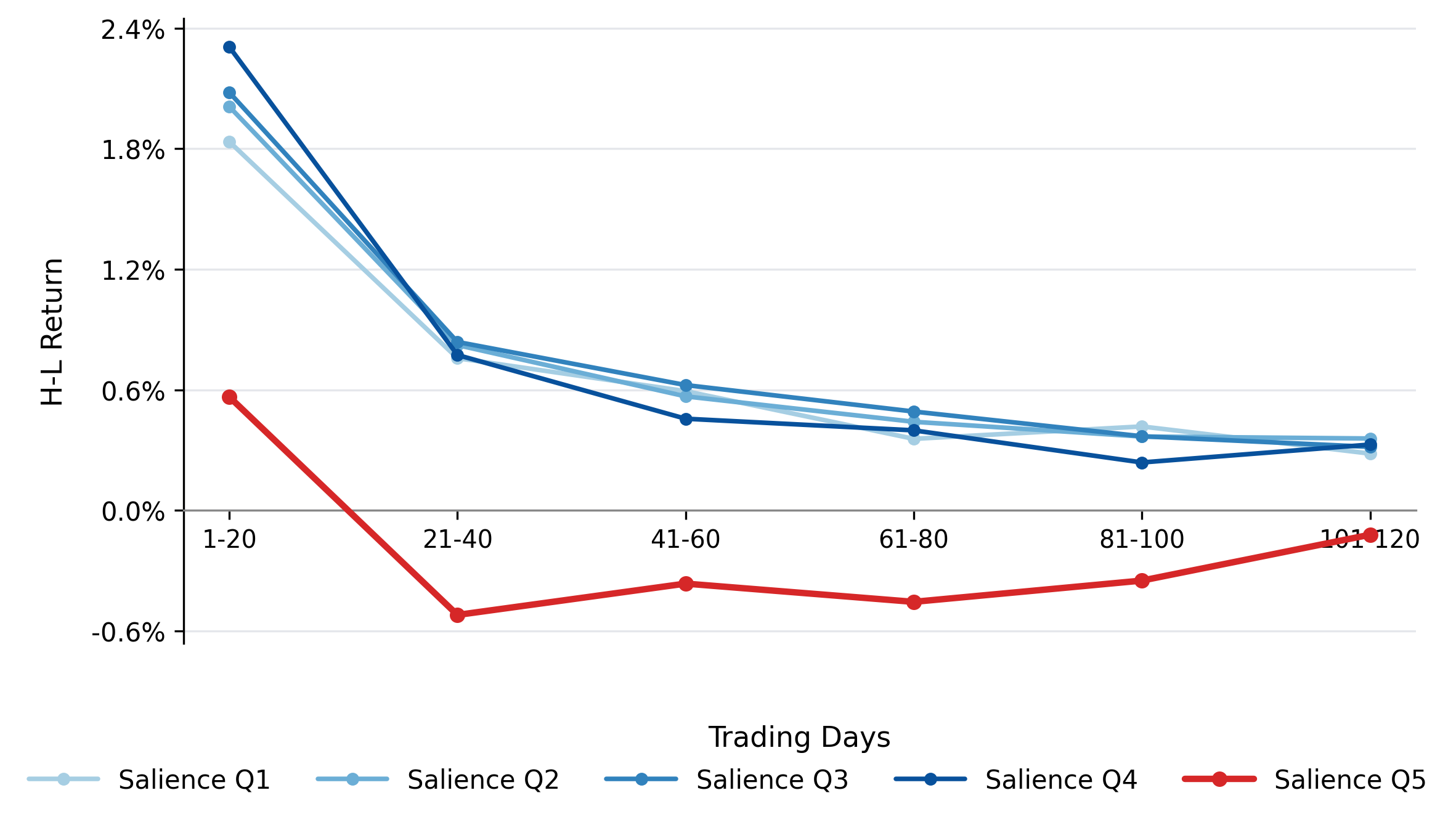}
  \caption{Returns over six months}
\end{subfigure}
\end{figure}

\clearpage
\begin{figure}[p]
\centering
\caption{Top-Down Feature Validation: Up Momentum}
\label{fig:topdown_main}
\exhibitnote{This figure reports the top-down validation of FT-DINO SAE feature 6946 for Up Momentum. The upper-left panels (Real Representatives) show two of the 50 real-market 20-day charts retained as representative examples of Up Momentum, following \citet{bojstrup2026}; feature 6946 activates on the representative charts. The upper-right panels (Highest Activations) show the two real-market charts with the highest activation of feature 6946. The bottom row reports query steering: starting from a real-market query on which the feature is initially inactive (Original), the query representation is moved along the feature's SAE decoder direction by two, three, and five times the feature's maximum activation, and each panel shows the resulting nearest real-market chart. The bar beneath each steering panel indicates the feature's activation relative to its maximum. All charts follow the colored 20-day design in Figure~\ref{fig:samples}.}
\includegraphics[width=\textwidth,height=0.73\textheight,keepaspectratio]{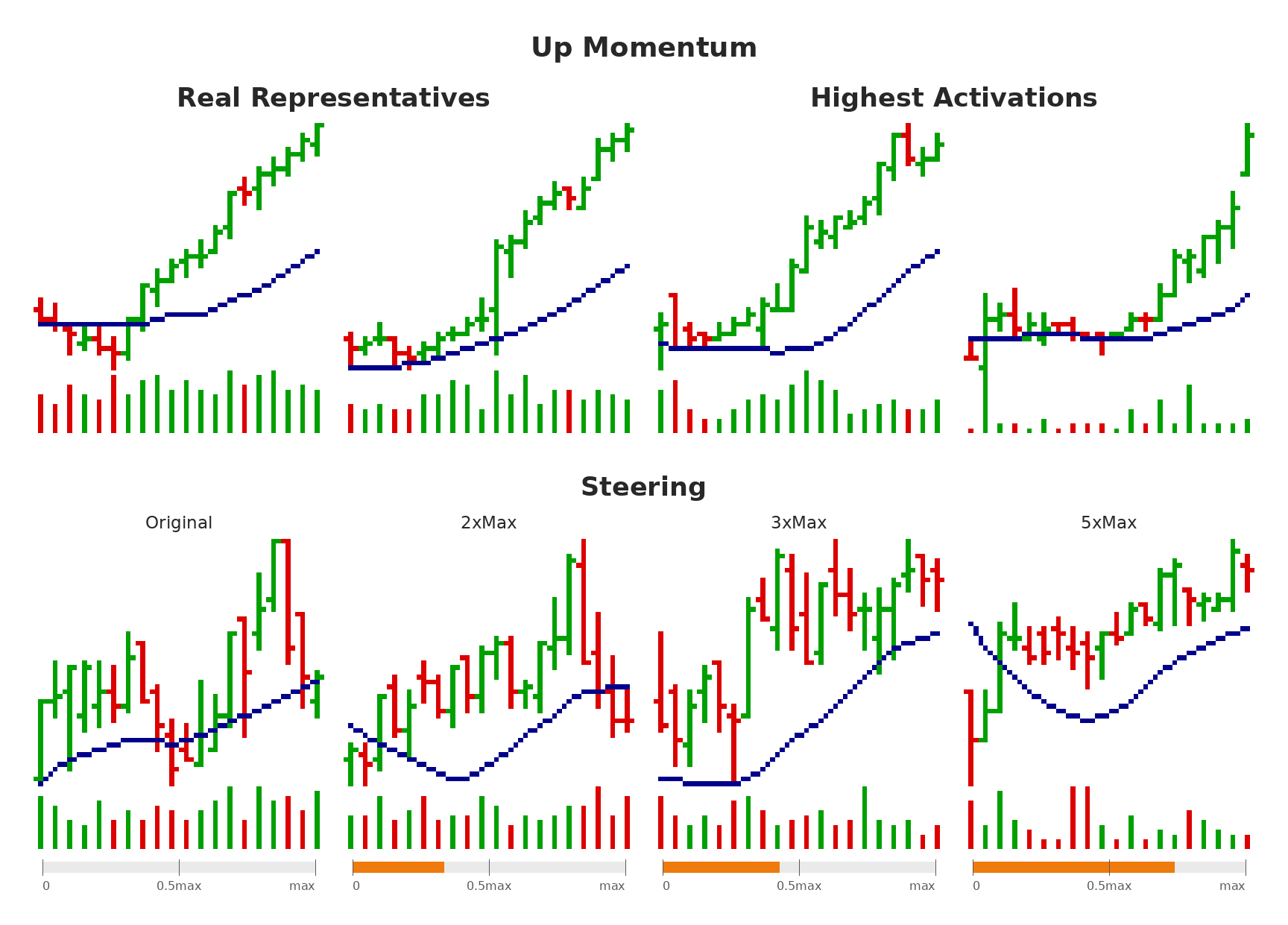}
\end{figure}

\clearpage
\begin{figure}[p]
\centering
\caption{Literature-Guided Patterns in the Visual Atlas}
\label{fig:landmarks_main}
\exhibitnote{This figure locates the literature-based visual characteristics in Table~\ref{tab:literature_patterns} within the visual atlas. To construct the atlas, the 24.2 million out-of-sample 20-day charts from 2001--2025 are encoded with FT-DINO and the SAE, and the sparse representations are clustered into 4,096 prototypes using spherical $k$-means. Each dot is a prototype center projected into two dimensions and colored by the average cross-sectional rank of the subsequent five-day return among the observations assigned to it, with green indicating higher and red indicating lower average ranks. For each characteristic, we select an illustrative real-market 20-day chart that captures a visual configuration motivated by the cited study and map it to the nearest atlas prototype. The enlarged marker identifies that prototype, and the inset displays the corresponding chart. Labels identify the characteristics listed in Table~\ref{tab:literature_patterns}.}
\includegraphics[width=\textwidth,keepaspectratio]{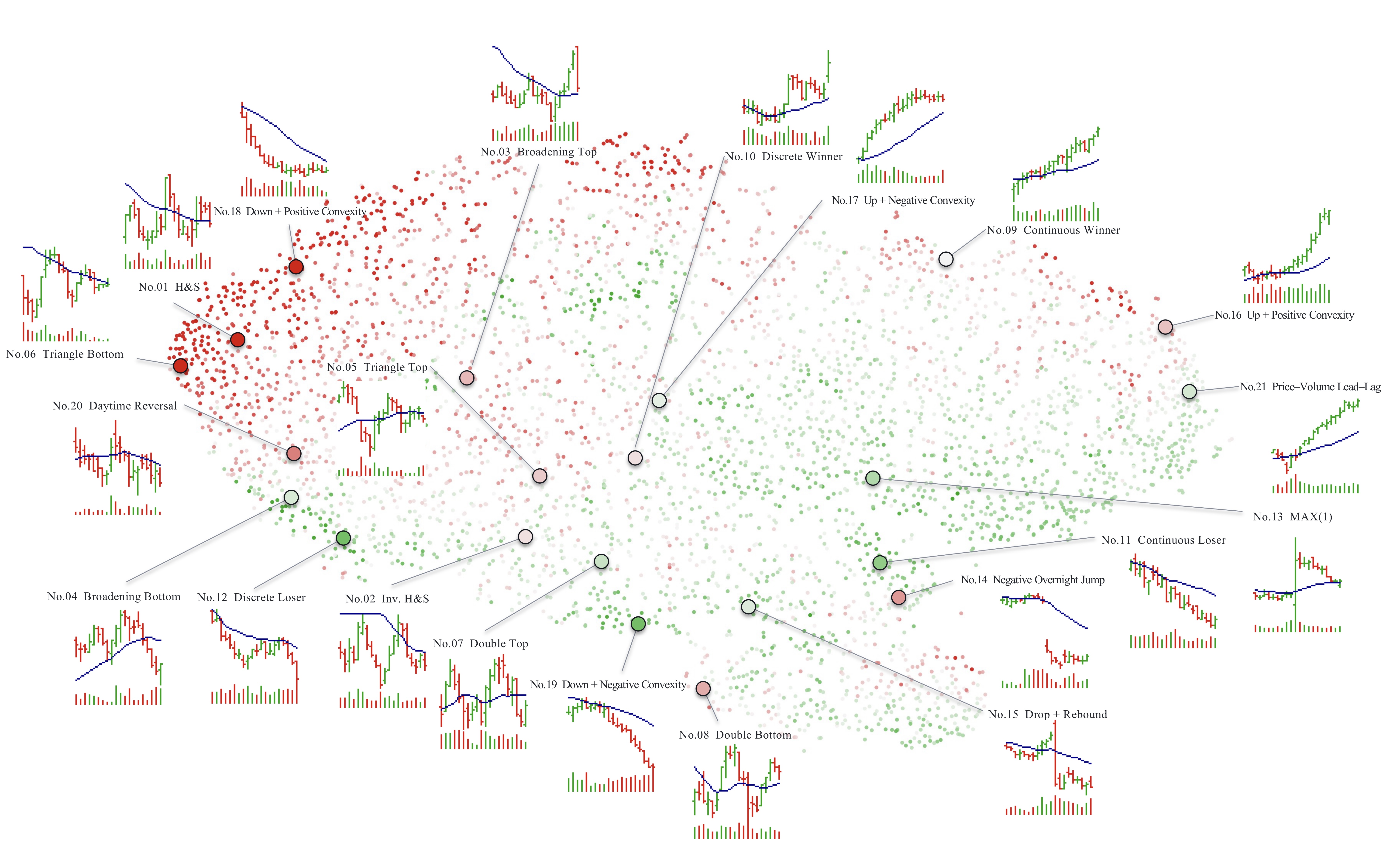}
\end{figure}

\clearpage

\begin{table}[H]
\centering
\caption{Summary Statistics}
\label{tab:summarystats}
\tablenote{This table reports summary statistics for the stock-week sample. The sample consists of CRSP common stocks listed on the NYSE, AMEX, and Nasdaq, and the out-of-sample period is 2001 to 2025. Observations are formed on the last trading day of each week. Panel A reports the subsequent five-trading-day raw return (SubseqRet) and the retrieval signals: Sig20, Sig40, and Sig60 are benchmark DINO signals based on 20-, 40-, and 60-day charts, respectively, and FTSig20 is the FT-DINO signal based on 20-day charts. Each signal is the equal-weighted average of the realized subsequent five-day returns of the $K=1{,}000$ retrieved historical neighbors. Panel B reports the stock-level control variables following \citet{jiang2023}: momentum (MOM), short-term reversal (STR), weekly short-term reversal (WSTR), market beta (Beta), return volatility (RetVol), the ratio of price to its 52-week high (High52), the bid-ask spread (BASpread), log dollar volume (DolVol), zero-trading days (ZeroTrade), price delay (PriceDelay), log market capitalization (Size), and Amihud illiquidity (Illiq). All variables are reported before any cross-sectional ranking. P10--P90 denote percentiles.}
\begin{minipage}{\textwidth}
\centering
\renewcommand{\arraystretch}{0.88}
\panelheading{Panel A. Subsequent Returns and Retrieval Signals}
\begin{tabular*}{\linewidth}{@{\extracolsep{\fill}}lrrrrrrrr@{}}
\toprule
Variable & N & Mean & Std. Dev. & P10 & P25 & P50 & P75 & P90 \\
\midrule
SubseqRet & 4,569,085 & 0.002 & 0.097 & -0.076 & -0.032 & 0.000 & 0.032 & 0.076 \\
Sig20 & 4,569,085 & 0.002 & 0.005 & -0.003 & -0.000 & 0.002 & 0.005 & 0.008 \\
Sig40 & 4,569,085 & 0.002 & 0.006 & -0.004 & -0.001 & 0.002 & 0.005 & 0.008 \\
Sig60 & 4,569,085 & 0.002 & 0.005 & -0.004 & -0.001 & 0.002 & 0.005 & 0.008 \\
FTSig20 & 4,569,085 & 0.002 & 0.007 & -0.005 & -0.001 & 0.002 & 0.005 & 0.009 \\
\bottomrule
\end{tabular*}

\vspace{8pt}
\panelheading{Panel B. Control Variables}
\begin{tabular*}{\linewidth}{@{\extracolsep{\fill}}lrrrrrrrr@{}}
\toprule
Variable & N & Mean & Std. Dev. & P10 & P25 & P50 & P75 & P90 \\
\midrule
MOM & 4,569,085 & 0.117 & 0.823 & -0.531 & -0.233 & 0.041 & 0.305 & 0.681 \\
STR & 4,569,085 & 0.010 & 0.190 & -0.156 & -0.065 & 0.004 & 0.071 & 0.162 \\
WSTR & 4,569,085 & 0.003 & 0.096 & -0.076 & -0.031 & 0.000 & 0.032 & 0.076 \\
Beta & 4,569,085 & 1.110 & 0.573 & 0.445 & 0.737 & 1.065 & 1.427 & 1.821 \\
RetVol & 4,569,085 & 0.031 & 0.032 & 0.011 & 0.016 & 0.024 & 0.038 & 0.058 \\
High52 & 4,569,085 & 0.739 & 0.241 & 0.358 & 0.601 & 0.815 & 0.934 & 0.984 \\
BASpread & 4,569,085 & 0.006 & 0.011 & 0.000 & 0.001 & 0.002 & 0.006 & 0.015 \\
DolVol & 4,569,085 & 18.368 & 2.402 & 15.101 & 16.601 & 18.471 & 20.129 & 21.473 \\
ZeroTrade & 4,569,085 & 0.108 & 1.359 & 0.000 & 0.000 & 0.000 & 0.000 & 0.000 \\
PriceDelay & 4,569,085 & 0.338 & 0.275 & 0.055 & 0.116 & 0.251 & 0.502 & 0.797 \\
Size & 4,569,085 & 20.433 & 2.053 & 17.835 & 19.058 & 20.380 & 21.776 & 23.103 \\
Illiq & 4,569,085 & 0.184 & 2.319 & 0.000 & 0.001 & 0.004 & 0.035 & 0.212 \\
\bottomrule
\end{tabular*}
\end{minipage}

\end{table}
\clearpage

\begin{table}[H]
\centering
\caption{DINO Portfolio Performance}
\label{tab:benchmark}
\tablenote{This table reports the performance of weekly decile portfolios sorted on the benchmark DINO retrieval signal. The sample consists of CRSP common stocks listed on the NYSE, AMEX, and Nasdaq, and the out-of-sample period is 2001 to 2025. DINO denotes the pretrained DINOv2-S encoder \citep{oquab2024dinov2}, used without any finance-specific training. For each stock, on the last trading day of each week, the retrieval signal is the equal-weighted average of the realized subsequent five-trading-day returns of its $K$ most similar historical charts, where similarity is the cosine similarity between normalized chart representations and the recursively expanding daily library contains only observations whose subsequent five-day returns are fully realized by the query date. $K$ is the number of retrieved neighbors. Each week, stocks are sorted into deciles by the signal and held over the subsequent five trading days; H--L is the return on decile 10 minus decile 1. EW, VW, and CVW denote equal-weighted, value-weighted, and capped value-weighted portfolios, respectively; capped value weighting winsorizes market capitalization at the contemporaneous NYSE 80th percentile following \citet{jensen2023}. Ret is 52 times the mean weekly H--L return, Alpha is 52 times the intercept from a regression of weekly H--L returns on the Fama--French five factors, momentum, and short-term reversal, and SR is the mean weekly H--L return divided by its standard deviation and multiplied by $\sqrt{52}$.}
\begin{threeparttable}
\begin{tabular*}{\textwidth}{@{\extracolsep{\fill}}c ccc ccc ccc@{}}
\toprule
& \multicolumn{3}{c}{EW} & \multicolumn{3}{c}{VW} & \multicolumn{3}{c}{CVW} \\
\cmidrule(lr){2-4}\cmidrule(lr){5-7}\cmidrule(lr){8-10}
$K$ & Ret & Alpha & SR & Ret & Alpha & SR & Ret & Alpha & SR \\
\midrule
10   & 0.07 & 0.07 & 1.12 & 0.02 & 0.02 & 0.20 & 0.03 & 0.03 & 0.54 \\
30   & 0.11 & 0.11 & 1.83 & 0.05 & 0.05 & 0.51 & 0.03 & 0.03 & 0.63 \\
100  & 0.16 & 0.16 & 2.15 & 0.04 & 0.03 & 0.36 & 0.04 & 0.04 & 0.70 \\
500  & 0.23 & 0.22 & 2.74 & 0.06 & 0.04 & 0.57 & 0.06 & 0.04 & 0.79 \\
1{,}000 & 0.26 & 0.25 & 2.81 & 0.07 & 0.05 & 0.60 & 0.07 & 0.06 & 0.87 \\
3{,}000 & 0.29 & 0.27 & 2.90 & 0.08 & 0.06 & 0.64 & 0.08 & 0.06 & 0.83 \\
\bottomrule
\end{tabular*}
\end{threeparttable}
\end{table}
\clearpage

\begin{table}[H]\caption{FT-DINO Portfolio Performance}
\label{tab:ftperformance}
\tablenote{This table reports the performance of weekly decile portfolios sorted on the FT-DINO retrieval signal. The sample consists of CRSP common stocks listed on the NYSE, AMEX, and Nasdaq, and the out-of-sample period is 2001 to 2025. FT-DINO denotes DINOv2-S after return-rank alignment on 1993--1999 charts, with 2000 used for validation and the encoder frozen before 2001. For each stock, on the last trading day of each week, the retrieval signal is the equal-weighted average of the realized subsequent five-trading-day returns of its $K$ most similar historical charts, where similarity is the cosine similarity between normalized chart representations and the recursively expanding daily library contains only observations whose subsequent five-day returns are fully realized by the query date. $K$ is the number of retrieved neighbors, and the DINO row reports benchmark DINO at $K=1{,}000$ for comparison. Each week, stocks are sorted into deciles by the signal and held over the subsequent five trading days; H--L is the return on decile 10 minus decile 1. EW, VW, and CVW denote equal-weighted, value-weighted, and capped value-weighted portfolios, respectively; capped value weighting winsorizes market capitalization at the contemporaneous NYSE 80th percentile following \citet{jensen2023}. Ret is 52 times the mean weekly H--L return, Alpha is 52 times the intercept from a regression of weekly H--L returns on the Fama--French five factors, momentum, and short-term reversal, and SR is the mean weekly H--L return divided by its standard deviation and multiplied by $\sqrt{52}$.}

\centering
\begin{threeparttable}
\begin{tabular*}{\textwidth}{@{\extracolsep{\fill}}c ccc ccc ccc@{}}
\toprule
& \multicolumn{3}{c}{EW} & \multicolumn{3}{c}{VW} & \multicolumn{3}{c}{CVW} \\
\cmidrule(lr){2-4}\cmidrule(lr){5-7}\cmidrule(lr){8-10}
$K$ & Ret & Alpha & SR & Ret & Alpha & SR & Ret & Alpha & SR \\
\midrule
10   & 0.28 & 0.26 & 3.59 & 0.06 & 0.05 & 0.51 & 0.07 & 0.06 & 1.14 \\
30   & 0.41 & 0.39 & 4.59 & 0.08 & 0.07 & 0.72 & 0.10 & 0.08 & 1.46 \\
100  & 0.60 & 0.57 & 5.47 & 0.14 & 0.11 & 1.06 & 0.16 & 0.13 & 1.83 \\
500  & 0.81 & 0.77 & 6.27 & 0.24 & 0.19 & 1.50 & 0.27 & 0.23 & 2.42 \\
1{,}000 & 0.85 & 0.81 & 6.37 & 0.24 & 0.19 & 1.45 & 0.28 & 0.24 & 2.38 \\
3{,}000 & 0.88 & 0.83 & 6.31 & 0.26 & 0.21 & 1.48 & 0.31 & 0.27 & 2.45 \\
\midrule
DINO & 0.26 & 0.25 & 2.81 & 0.07 & 0.05 & 0.60 & 0.07 & 0.06 & 0.87 \\
\bottomrule
\end{tabular*}
\end{threeparttable}
\end{table}
\clearpage

\begin{table}[H]
\centering
\caption{Portfolio Performance: FT-DINO and CNN}
\label{tab:modelcomparison}
\tablenote{This table compares decile portfolios sorted on FT-DINO and CNN signals. The sample consists of CRSP common stocks listed on the NYSE, AMEX, and Nasdaq, and the out-of-sample period is 2001 to 2025. FT-DINO denotes DINOv2-S after return-rank alignment on 1993--1999 charts, with 2000 used for validation and the encoder frozen before 2001; the FT-DINO signal uses $K=1{,}000$ neighbors. CNN denotes the 20-day black-and-white price-image CNN of \citet{jiang2023}, trained following their procedure, which predicts returns directly from the image using the same 20 trading days of price and volume information. Each week, stocks are sorted into deciles by each signal and held over the subsequent five trading days. Group 1 (Low) and Group 10 (High) are the lowest- and highest-signal deciles, and 10--1 Spread is Group 10 minus Group 1. EW denotes equal weighting, and CVW denotes capped value weighting, which winsorizes market capitalization at the contemporaneous NYSE 80th percentile following \citet{jensen2023}. Ret is the annualized mean return, and SR is the annualized Sharpe ratio.}
\begin{threeparttable}
\setlength{\tabcolsep}{3.2pt}
\begin{tabular*}{\textwidth}{@{\extracolsep{\fill}}l cc cc cc cc@{}}
\toprule
& \multicolumn{4}{c}{EW} & \multicolumn{4}{c}{CVW} \\
\cmidrule(lr){2-5}\cmidrule(lr){6-9}
& \multicolumn{2}{c}{FT-DINO} & \multicolumn{2}{c}{CNN}
& \multicolumn{2}{c}{FT-DINO} & \multicolumn{2}{c}{CNN} \\
\cmidrule(lr){2-3}\cmidrule(lr){4-5}
\cmidrule(lr){6-7}\cmidrule(lr){8-9}
Group & Ret & SR & Ret & SR & Ret & SR & Ret & SR \\
\midrule
1 (Low)      & -0.28 & -1.15 & -0.23 & -1.01 & -0.03 & -0.22 & -0.03 & -0.21 \\
10 (High)    &  0.57 &  1.87 &  0.39 &  1.53 &  0.25 &  0.89 &  0.19 &  0.80 \\
\midrule
10--1 Spread &  0.85 &  6.37 &  0.61 &  4.46 &  0.28 &  2.38 &  0.22 &  1.82 \\
\bottomrule
\end{tabular*}
\end{threeparttable}
\end{table}
\clearpage

\begin{table}[H]
\centering
\caption{Factor-Augmented Spanning Regressions}
\label{tab:spanning}
\tablenote{This table reports weekly factor-augmented spanning regressions of image-based high-minus-low portfolio returns over 2001--2025. In each column, the dependent variable is the weekly H--L return of the strategy named in the column header, formed under the indicated weighting scheme (EW or CVW). The regressors are the H--L returns of the other image strategies (FT-DINO, benchmark DINO, and CNN) under the same weighting scheme, together with the Fama--French five factors, momentum, and short-term reversal (Factor Controls). FT-DINO and DINO use $K=1{,}000$, and CNN uses its out-of-sample prediction following \citet{jiang2023}. Alpha is 52 times the weekly intercept. $t$-statistics based on Newey--West standard errors \citep{newey1987} are reported in parentheses.}
\begin{threeparttable}
\setlength{\tabcolsep}{4.0pt}
\begin{tabular*}{\textwidth}{@{\extracolsep{\fill}}lcccc@{}}
\toprule
& \multicolumn{2}{c}{EW} & \multicolumn{2}{c}{CVW} \\
\cmidrule(lr){2-3}\cmidrule(lr){4-5}
Variable & FT-DINO & CNN & FT-DINO & CNN \\
\midrule
Alpha & 0.410 & 0.037 & 0.147 & 0.055 \\
      & (16.53) & (0.74) & (8.50) & (2.67) \\[2pt]
FT-DINO & & 0.659 & & 0.483 \\
        & & (12.23) & & (10.21) \\[2pt]
DINO & 0.563 & $-$0.050 & 0.580 & 0.036 \\
     & (6.91) & ($-$0.65) & (9.84) & (0.34) \\[2pt]
CNN & 0.467 & & 0.365 & \\
    & (10.95) & & (8.59) & \\[2pt]
Factor Controls & Yes & Yes & Yes & Yes \\
\midrule
Adj. $R^2$ & 0.627 & 0.502 & 0.523 & 0.367 \\
\bottomrule
\end{tabular*}
\end{threeparttable}
\end{table}
\clearpage

\begin{table}[H]
\centering
\caption{FT-DINO Ablations}
\label{tab:ft_ablation}
\tablenote{This table reports FT-DINO equal-weighted high-minus-low portfolio performance over 2001--2025 with $K=1{,}000$ under alternative retrieval libraries and chart designs. Each week, stocks are sorted into deciles by the signal and held over the subsequent five trading days; H--L is the return on decile 10 minus decile 1. Panel A varies the historical retrieval library: Rolling 8Y retains only observations from the preceding eight years, Same-Ind. restricts retrieval to the query stock's Fama--French 12 industry, Cross-Ind. excludes that industry, and Same-Stock retains only the query stock's own history. Panel B varies color: No Color renders the chart in black and white, Reverse Color swaps the red--green mapping of down and up days, and Random Color assigns colors at random. Panel C varies chart elements: No MA removes the moving average, No Volume removes the volume bars, No MA + No Vol. removes both, and MA in Price Scale plots the moving average on the price scale. Baseline is the unrestricted FT-DINO specification with the original colored chart. Each alternative changes only the indicated component. Figure~\ref{fig:ft_ablation_examples} illustrates the chart variants. Ret is the annualized mean H--L return, and SR is the annualized Sharpe ratio.}
\begin{threeparttable}
\begin{tabular*}{\textwidth}{@{\extracolsep{\fill}}lcc@{}}
\toprule
Specification & Ret & SR \\
\midrule
\multicolumn{3}{@{}l}{\textit{Panel A. Historical Retrieval Library}} \\
Baseline       & 0.85 & 6.37 \\
Rolling 8Y     & 0.78 & 5.77 \\
Same-Ind.      & 0.79 & 6.06 \\
Cross-Ind.     & 0.85 & 6.24 \\
Same-Stock     & 0.17 & 1.38 \\
\addlinespace[6pt]
\multicolumn{3}{@{}l}{\textit{Panel B. Color Information}} \\
Baseline       & 0.85 & 6.37 \\
No Color       & 0.64 & 5.71 \\
Reverse Color  & 0.85 & 6.29 \\
Random Color   & 0.78 & 5.92 \\
\addlinespace[6pt]
\multicolumn{3}{@{}l}{\textit{Panel C. Chart Elements}} \\
Baseline           & 0.85 & 6.37 \\
No MA              & 0.83 & 6.36 \\
No Volume          & 0.62 & 4.98 \\
No MA + No Vol.    & 0.61 & 4.83 \\
MA in Price Scale  & 0.85 & 6.43 \\
\bottomrule
\end{tabular*}
\end{threeparttable}
\end{table}
\clearpage

\begin{table}[H]
\centering
\caption{Tradability and Transaction Costs}
\label{tab:tradability}
\tablenote{This table reports high-minus-low portfolio performance for FT-DINO, benchmark DINO, and CNN under tradability screens and transaction costs over 2001--2025. FT-DINO and DINO use $K=1{,}000$. Full sample, gross reports unrestricted gross performance. Close $\geq \$5$ restricts the sample to stocks with a CRSP closing price of at least \$5 on the portfolio-formation date. Drop bottom size 20\% removes the bottom 20 percent of stocks by formation-date market capitalization. Net cost, R5 deducts transaction costs from the weekly strategy. Prior target weights are first drifted by realized returns and then traded to the current target weights. Transaction costs are charged on the full absolute weight changes, covering both purchases and sales, at 10 basis points for stocks above the contemporaneous NYSE 80th market-capitalization percentile and 20 basis points otherwise. Costs are deducted on both the long and short legs, and the initial establishment of the portfolio is also charged. EW denotes equal weighting, and CVW denotes capped value weighting, which winsorizes market capitalization at the contemporaneous NYSE 80th percentile following \citet{jensen2023}. Ret is the annualized mean H--L return, and SR is the annualized Sharpe ratio.}
\begin{threeparttable}
\setlength{\tabcolsep}{2.3pt}
\begin{tabular*}{\textwidth}{@{\extracolsep{\fill}}l cc cc cc cc cc cc@{}}
\toprule
& \multicolumn{6}{c}{EW} & \multicolumn{6}{c}{CVW} \\
\cmidrule(lr){2-7}\cmidrule(lr){8-13}
& \multicolumn{2}{c}{FT-DINO} & \multicolumn{2}{c}{DINO} & \multicolumn{2}{c}{CNN}
& \multicolumn{2}{c}{FT-DINO} & \multicolumn{2}{c}{DINO} & \multicolumn{2}{c}{CNN} \\
\cmidrule(lr){2-3}\cmidrule(lr){4-5}\cmidrule(lr){6-7}
\cmidrule(lr){8-9}\cmidrule(lr){10-11}\cmidrule(lr){12-13}
Test & Ret & SR & Ret & SR & Ret & SR & Ret & SR & Ret & SR & Ret & SR \\
\midrule
Full sample, gross      & 0.85 & 6.37 & 0.26 & 2.81 & 0.61 & 4.46 & 0.28 & 2.38 & 0.07 & 0.87 & 0.22 & 1.82 \\
Close $\geq \$5$        & 0.36 & 3.94 & 0.11 & 1.56 & 0.30 & 3.00 & 0.20 & 1.89 & 0.06 & 0.71 & 0.17 & 1.53 \\
Drop bottom size 20\%   & 0.36 & 3.44 & 0.11 & 1.46 & 0.30 & 2.69 & 0.21 & 1.97 & 0.06 & 0.76 & 0.18 & 1.61 \\
Net cost, R5            & 0.55 & 4.12 & -0.06 & -0.63 & 0.29 & 2.11 & 0.00 & 0.01 & -0.20 & -2.39 & -0.07 & -0.62 \\
\bottomrule
\end{tabular*}
\end{threeparttable}
\end{table}
\clearpage

\begin{table}[H]
\centering
\caption{Alternative Image Windows and Return Horizons}
\label{tab:alternative_horizons}
\tablenote{This table reports equal-weighted high-minus-low portfolio performance with $K=1{,}000$ across price-image windows and return horizons for benchmark DINO and FT-DINO. D20, D40, and D60 use the prior 20, 40, and 60 trading days, respectively, with the corresponding moving-average window (Figure~\ref{fig:app_windows}). For R5, signals are formed weekly from the realized subsequent five-trading-day returns of retrieved neighbors, and portfolios are held for five trading days over 2001--2025. For R20, signals are formed monthly on the last trading day of each month from subsequent 20-trading-day returns, and portfolios are held for 20 trading days; FT-DINO is re-trained for this specification using cross-sectional ranks of subsequent 20-trading-day returns as the fine-tuning target. Appendix Tables~\ref{tab:app_window_compare} and~\ref{tab:app_r20} report the full $K$ grid. Ret is the annualized mean H--L return, and SR is the annualized Sharpe ratio.}
\begin{threeparttable}
\begin{tabular*}{\textwidth}{@{\extracolsep{\fill}}ll cc cc cc@{}}
\toprule
& & \multicolumn{2}{c}{D20} & \multicolumn{2}{c}{D40} & \multicolumn{2}{c}{D60} \\
\cmidrule(lr){3-4}\cmidrule(lr){5-6}\cmidrule(lr){7-8}
Return horizon & Model & Ret & SR & Ret & SR & Ret & SR \\
\midrule
R5  & DINO    & 0.26 & 2.81 & 0.31 & 2.97 & 0.26 & 2.63 \\
    & FT-DINO & 0.85 & 6.37 & 0.93 & 6.25 & 0.89 & 5.86 \\
\addlinespace[4pt]
R20 & DINO    & 0.13 & 1.54 & 0.20 & 1.99 & 0.19 & 1.86 \\
    & FT-DINO & 0.30 & 2.13 & 0.36 & 2.11 & 0.38 & 2.19 \\
\bottomrule
\end{tabular*}
\end{threeparttable}
\end{table}
\clearpage

\begin{table}[H]
\centering
\caption{Double Sorts on Limits to Arbitrage}
\label{tab:econrelevance}
\tablenote{This table reports equal-weighted sequential $5\times5$ double sorts on a composite limits-to-arbitrage score and the FT-DINO retrieval signal ($K=1{,}000$) from 2008 onward, when securities-lending fees become available. Each week, stocks are first sorted into quintiles (Q1--Q5, columns) of the score and then, within each quintile, into conditional quintiles (rows) of the FT-DINO signal. Following \citet{chen2026debt}, the score is the average of the available weekly percentile ranks of size, Amihud illiquidity, dollar volume, days-to-cover, securities lending fees, analyst forecast dispersion, and turnover. Each component rank is oriented so that a higher score indicates greater limits to arbitrage. FT-DINO 5--1 is the return on the top minus the bottom conditional FT-DINO quintile within each score quintile. Returns are annualized, and $t$-statistics are reported in parentheses. Appendix Table~\ref{tab:app_ft_limits} reports the sorts for each component, and Appendix Table~\ref{tab:app_ft_retail} reports the sorts on retail-trading proxies.}
\begin{threeparttable}
\setlength{\tabcolsep}{4.0pt}
\renewcommand{\arraystretch}{0.96}
\begin{tabular*}{\textwidth}{@{\extracolsep{\fill}}l rrrrr@{}}
\toprule
& Q1 & Q2 & Q3 & Q4 & Q5 \\
\midrule
FT-DINO Q1 & 0.04 & 0.01 & -0.10 & -0.19 & -0.40 \\
            & (0.83) & (0.09) & (-1.56) & (-2.71) & (-6.18) \\
FT-DINO Q2 & 0.08 & 0.09 & 0.07 & 0.05 & -0.00 \\
            & (1.80) & (1.91) & (1.36) & (0.83) & (-0.03) \\
FT-DINO Q3 & 0.09 & 0.11 & 0.10 & 0.14 & 0.15 \\
            & (2.26) & (2.41) & (2.00) & (2.53) & (2.77) \\
FT-DINO Q4 & 0.12 & 0.13 & 0.15 & 0.20 & 0.30 \\
            & (2.66) & (2.70) & (2.87) & (3.60) & (5.49) \\
FT-DINO Q5 & 0.18 & 0.21 & 0.29 & 0.47 & 0.80 \\
            & (3.60) & (3.84) & (4.57) & (6.73) & (11.60) \\
\midrule
FT-DINO 5--1 & 0.14 & 0.21 & 0.38 & 0.65 & 1.19 \\
              & (7.55) & (9.83) & (13.37) & (18.11) & (24.48) \\
\bottomrule
\end{tabular*}
\end{threeparttable}
\end{table}
\clearpage

\begin{table}[H]
\centering
\caption{Portfolio Sensitivity to SAE Visual States}
\label{tab:saesensitivity}
\tablenote{This table reports how FT-DINO equal-weighted high-minus-low performance changes when the stocks that most strongly activate each SAE visual state are removed, over 2001--2025 with $K=1{,}000$. The SAE is trained on FT-DINO representations. For each stock and week, each constituent feature's positive activation is percentile-ranked within the weekly cross section, and cluster activation is the maximum rank among the state's constituent features. Each week, we remove the stocks with the highest cluster activations, up to 10 percent of the cross section; if fewer than 10 percent of stocks activate the state, all active stocks are removed. The remaining stocks are re-sorted into deciles using the unchanged FT-DINO signal. SR$_f$ is the post-removal annualized Sharpe ratio, and $\Delta\mathrm{SR}=\mathrm{SR}_f-6.37$, where 6.37 is the unrestricted FT-DINO Sharpe ratio. States are ordered from the largest decline to the largest increase.}
\begin{minipage}{\textwidth}
\centering
\begin{tabular*}{\linewidth}{@{\extracolsep{\fill}}c l cc c l cc@{}}
\toprule
Cluster & Visual State & SR$_f$ & $\Delta$SR & Cluster & Visual State & SR$_f$ & $\Delta$SR \\
\midrule
C10 & Isolated Upward Spike & 5.90 & -0.47 & C03 & Downtrend Recovery & 6.30 & -0.07 \\
C01 & Sustained Uptrend & 6.11 & -0.27 & C06 & Stepwise Downtrend & 6.35 & -0.03 \\
C04 & Late Upside Breakout & 6.19 & -0.18 & C08 & Late Breakdown & 6.39 & 0.02 \\
C09 & High-Volatility Range & 6.24 & -0.13 & C11 & High-Price Plateau & 6.43 & 0.06 \\
C02 & Peak Pullback & 6.29 & -0.09 & C05 & Stable Downtrend & 6.49 & 0.11 \\
& & & & C07 & Accelerating Downtrend & 6.64 & 0.26 \\
\bottomrule
\end{tabular*}
\end{minipage}
\end{table}
\clearpage

\begin{table}[H]
\centering
\caption{Query-Only Steering of SAE Features}
\label{tab:steering}
\tablenote{This table reports how query-only steering along selected SAE feature directions changes FT-DINO equal-weighted high-minus-low performance over 2001--2025 with $K=1{,}000$. The SAE is trained on FT-DINO representations. Panel A reports bottom-up features: for each of the four visual states with the largest portfolio-exclusion effects in Table~\ref{tab:saesensitivity} (C01, C04, C09, and C10), steering uses the highest-ranked return-related feature within that state. Panel B reports top-down features: for each of the 11 predefined patterns, steering uses the feature selected in the top-down validation (Section~\ref{sec:topdown}). At steering level $\ell$, the FT-DINO query representation $h_{i,t}$ is replaced with $h_{i,t}+\ell d_f$, where $d_f$ is the SAE decoder direction of feature $f$; the steered query is re-normalized and exact cosine-similarity retrieval is repeated under the original eligibility rule. Historical-library representations are unchanged, so differences arise only through the query. Negative levels move the query away from the feature direction, and level zero reproduces the FT-DINO signal. Entries are annualized Sharpe ratios.}
\begin{threeparttable}
\panelheading{Panel A. Bottom-Up Features}
\begin{tabular*}{\textwidth}{@{\extracolsep{\fill}}p{7.6cm} ccccc@{}}
\toprule
& \multicolumn{5}{c}{Steering level} \\
\cmidrule(lr){2-6}
Feature & $-10$ & $-5$ & $-3$ & $-1$ & $0$ \\
\midrule
C01 Sustained Uptrend: Feature 16299 & 4.97 & 6.10 & 6.38 & 6.39 & 6.37 \\
C04 Late Upside Breakout: Feature 10529 & 3.30 & 5.08 & 5.59 & 6.21 & 6.37 \\
C09 High-Volatility Range: Feature 4500 & 2.32 & 3.36 & 4.63 & 6.11 & 6.37 \\
C10 Isolated Upward Spike: Feature 12695 & 3.20 & 5.22 & 5.85 & 6.18 & 6.37 \\
\bottomrule
\end{tabular*}

\vspace{8pt}
\panelheading{Panel B. Top-Down Features}
\begin{tabular*}{\textwidth}{@{\extracolsep{\fill}}p{7.6cm} ccccc@{}}
\toprule
& \multicolumn{5}{c}{Steering level} \\
\cmidrule(lr){2-6}
Feature & $-10$ & $-5$ & $-3$ & $-1$ & $0$ \\
\midrule
Up Momentum: Feature 6946 & 4.17 & 5.69 & 6.08 & 6.23 & 6.37 \\
Down Momentum: Feature 7123 & 4.06 & 4.67 & 6.02 & 6.41 & 6.37 \\
Breakout Up: Feature 13482 & 5.27 & 5.95 & 6.43 & 6.36 & 6.37 \\
Breakout Down: Feature 2751 & 4.58 & 5.11 & 5.54 & 6.21 & 6.37 \\
Mean Reversion: Feature 8565 & 4.61 & 5.49 & 5.66 & 6.13 & 6.37 \\
Discrete Winner: Feature 16299 & 4.97 & 6.10 & 6.38 & 6.39 & 6.37 \\
Discrete Loser: Feature 9690 & 5.27 & 5.93 & 6.35 & 6.38 & 6.37 \\
MAX(1): Feature 10461 & 4.59 & 5.13 & 5.32 & 6.02 & 6.37 \\
Negative Overnight Jump: Feature 4232 & 5.04 & 5.35 & 5.62 & 6.35 & 6.37 \\
Up + Positive Convexity: Feature 6232 & 4.15 & 4.70 & 5.35 & 6.08 & 6.37 \\
Down + Positive Convexity: Feature 2940 & 5.16 & 5.32 & 5.69 & 6.22 & 6.37 \\
\bottomrule
\end{tabular*}
\end{threeparttable}
\end{table}
\clearpage

%% file: tables_figures_appendix.tex
\clearpage
\section*{Appendix Figures}\addcontentsline{toc}{section}{Appendix Figures}

\setappfig{A}
\subsection*{Appendix A: Data and Methodological Details}
\begin{figure}[H]
\centering
\caption{DINOv2 Student--Teacher Pretraining}
\label{fig:app_dino_pretrain}
\exhibitnote{This figure provides a schematic of the student--teacher self-distillation framework used to pretrain DINOv2 ViT-g \citep{oquab2024dinov2}. A student network and a teacher network receive augmented views of the same unlabeled image. The student is updated by gradient descent to match self-distillation targets produced by the teacher, while the teacher parameters follow an exponential moving average of the student parameters. The main training objectives combine global self-distillation with masked-patch prediction. The ViT-g model is pretrained on LVD-142M, a collection of roughly 142 million general images; no stock charts or stock-return labels enter this stage.}
\includegraphics[width=0.85\textwidth]{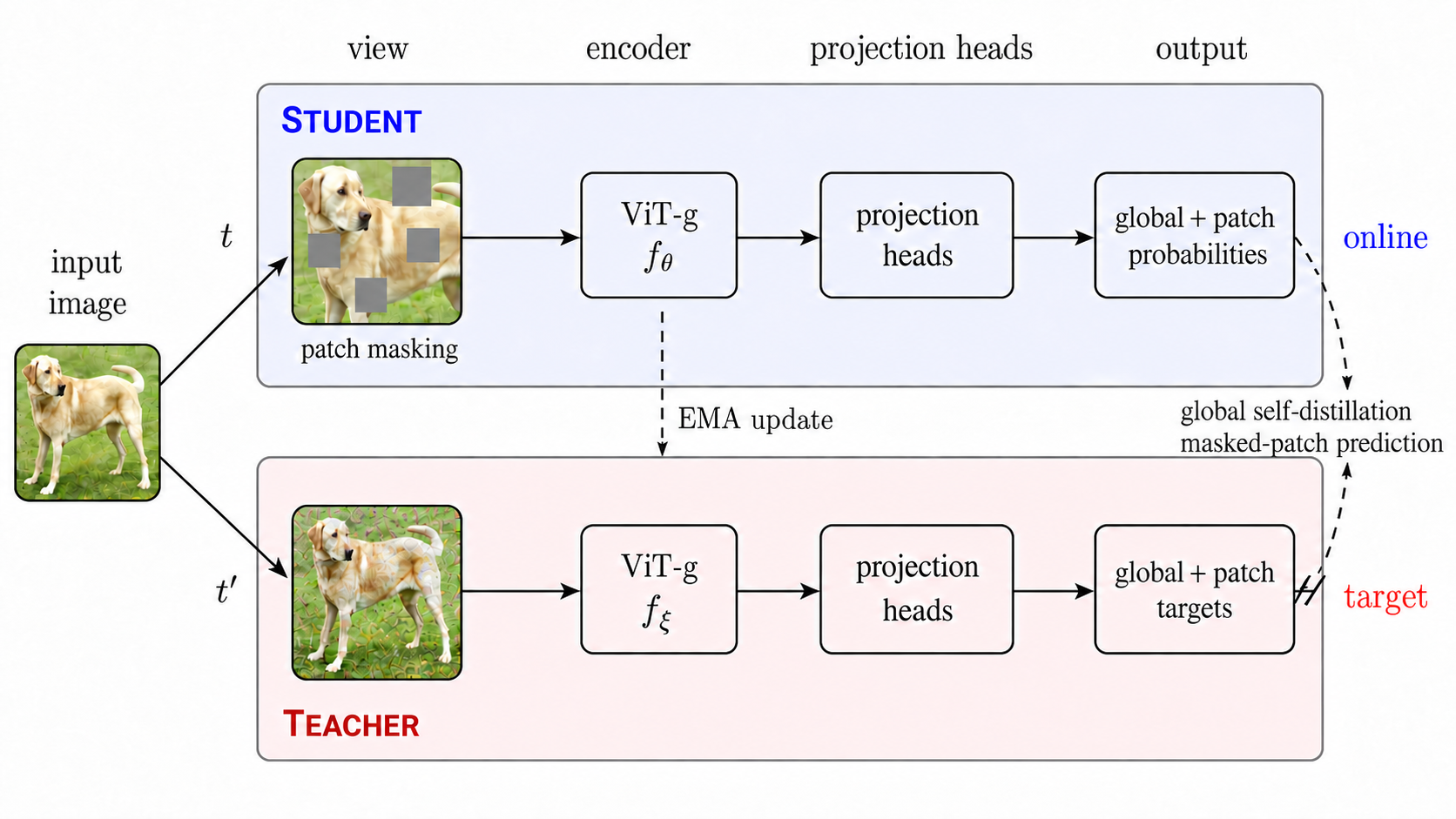}
\end{figure}

\begin{figure}[H]
\centering
\caption{Historical Retrieval and Return-Rank Alignment Workflows}
\label{fig:app_workflows}
\exhibitnote{This figure summarizes the two procedures used in the paper. Panel (a) shows historical visual retrieval: each stock's 20-day chart is encoded into a normalized representation, compared by cosine similarity with a recursively expanding daily library that contains only observations whose subsequent five-day returns are fully realized by the query date, and the equal-weighted average realized return of the top-$K$ neighbors forms the signal used for weekly decile portfolios. Panel (b) shows return-rank alignment: subsequent five-day returns on 1993--1999 training dates are converted into within-date percentile ranks, DINOv2-S is fine-tuned to minimize the Kullback--Leibler divergence between the return-rank target neighbor distribution and the model-implied similarity distribution, the checkpoint is selected on the 2000 validation sample, and the frozen encoder re-encodes both queries and the historical library from 2001 onward.}
\begin{subfigure}[t]{0.70\textwidth}
\centering
\includegraphics[width=\linewidth]{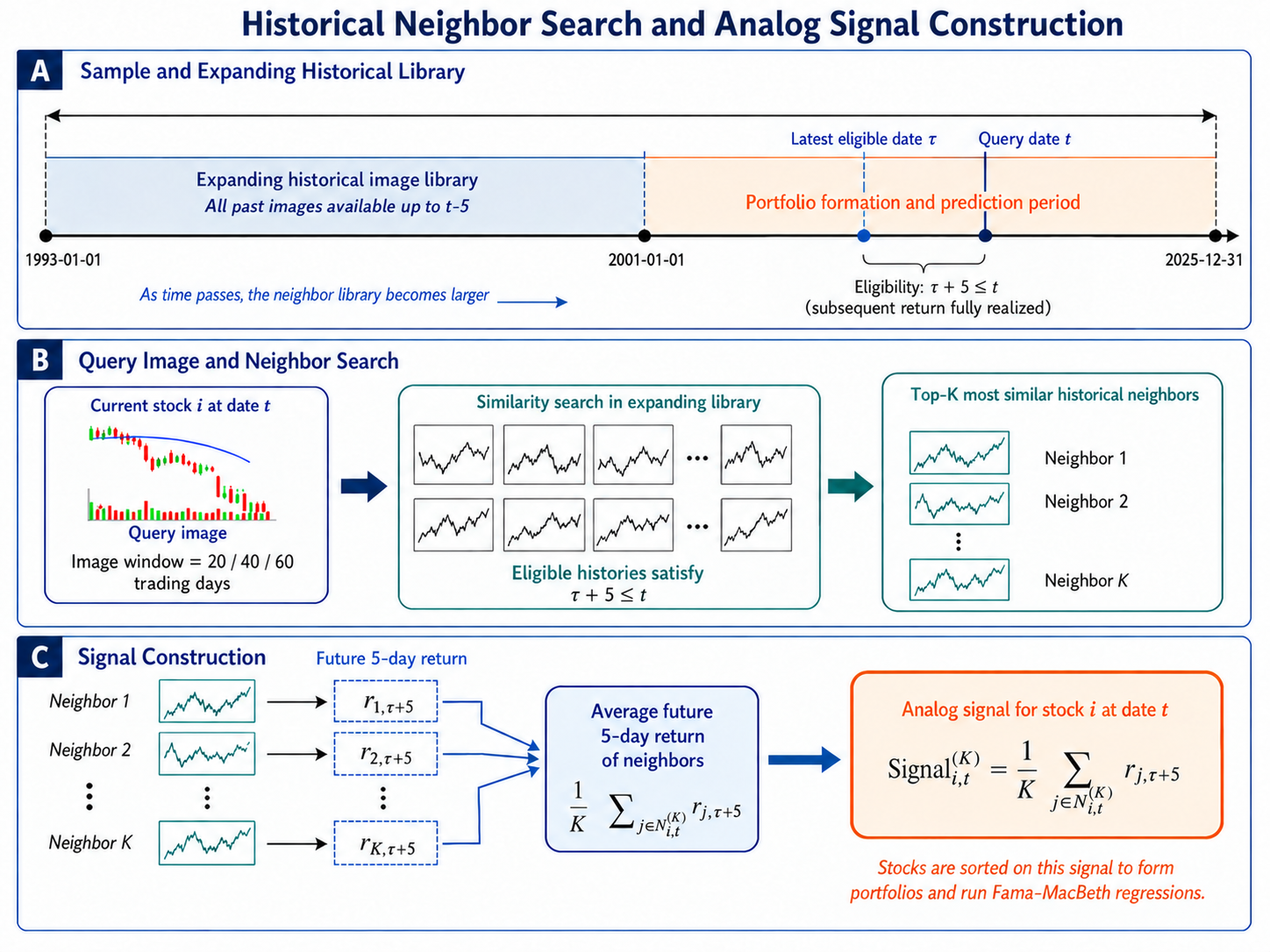}
\caption{Historical visual retrieval procedure}
\end{subfigure}

\vspace{0.28cm}

\begin{subfigure}[t]{0.70\textwidth}
\centering
\includegraphics[width=\linewidth]{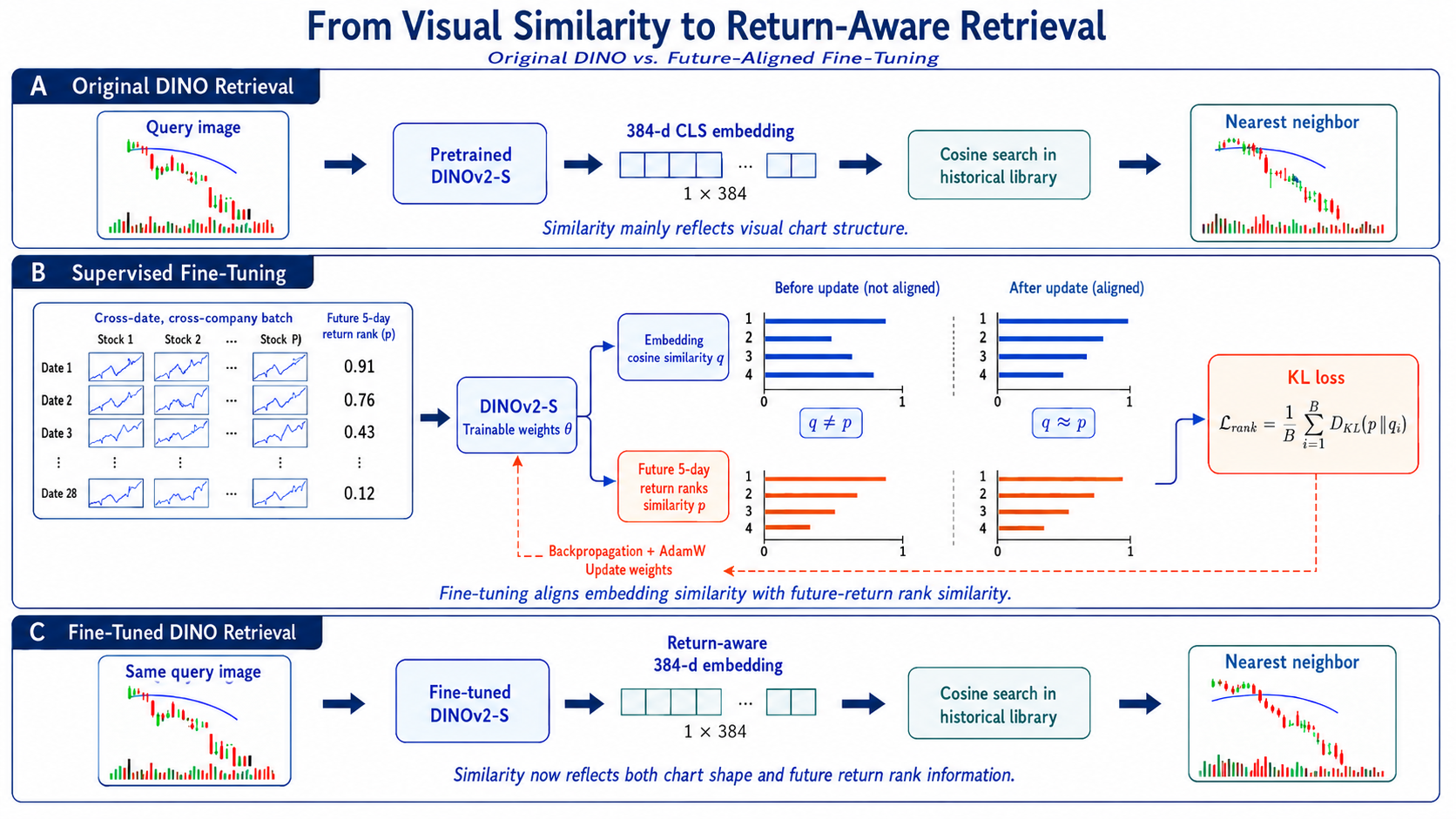}
\caption{Return-rank alignment procedure}
\end{subfigure}
\label{fig:pipeline}
\label{fig:finetune}

\end{figure}
\clearpage
\begin{figure}[H]
\centering
\caption{Raw Embedding Space: DINO versus FT-DINO}
\label{fig:app_raw_space}
\exhibitnote{This figure visualizes the benchmark DINO representation space in Panel (a) and the FT-DINO representation space in Panel (b). Both panels use the same out-of-sample observations and the same two-dimensional projection and plotting procedure, so differences between the panels reflect only the change in representation induced by return-rank alignment. The portfolio consequences of this change are evaluated in Appendix~\ref{app:c_structure}.}
\begin{subfigure}{0.62\textwidth}
  \centering
  \includegraphics[width=\linewidth]{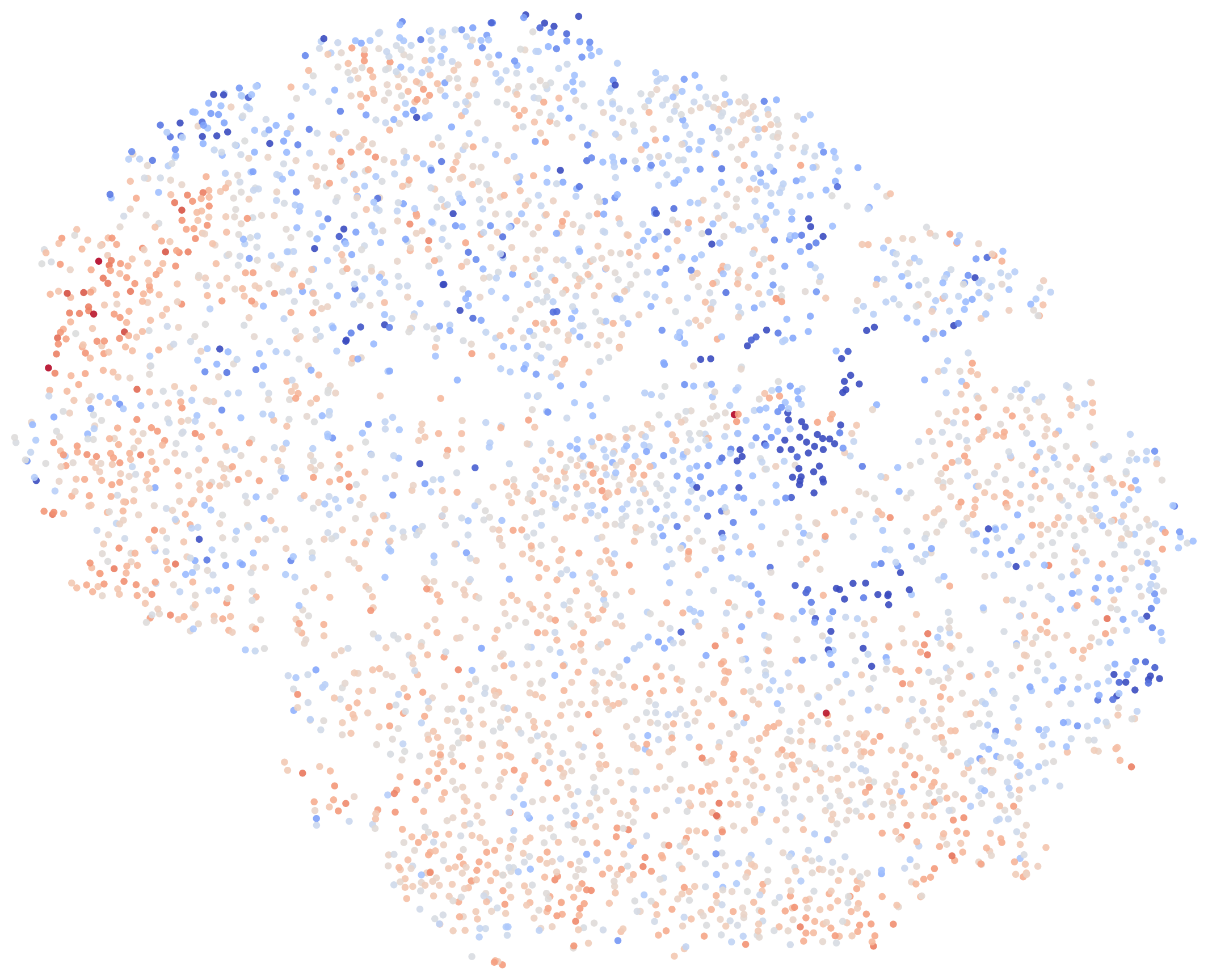}
  \caption{DINO}
\end{subfigure}

\vspace{0.2cm}

\begin{subfigure}{0.62\textwidth}
  \centering
  \includegraphics[width=\linewidth]{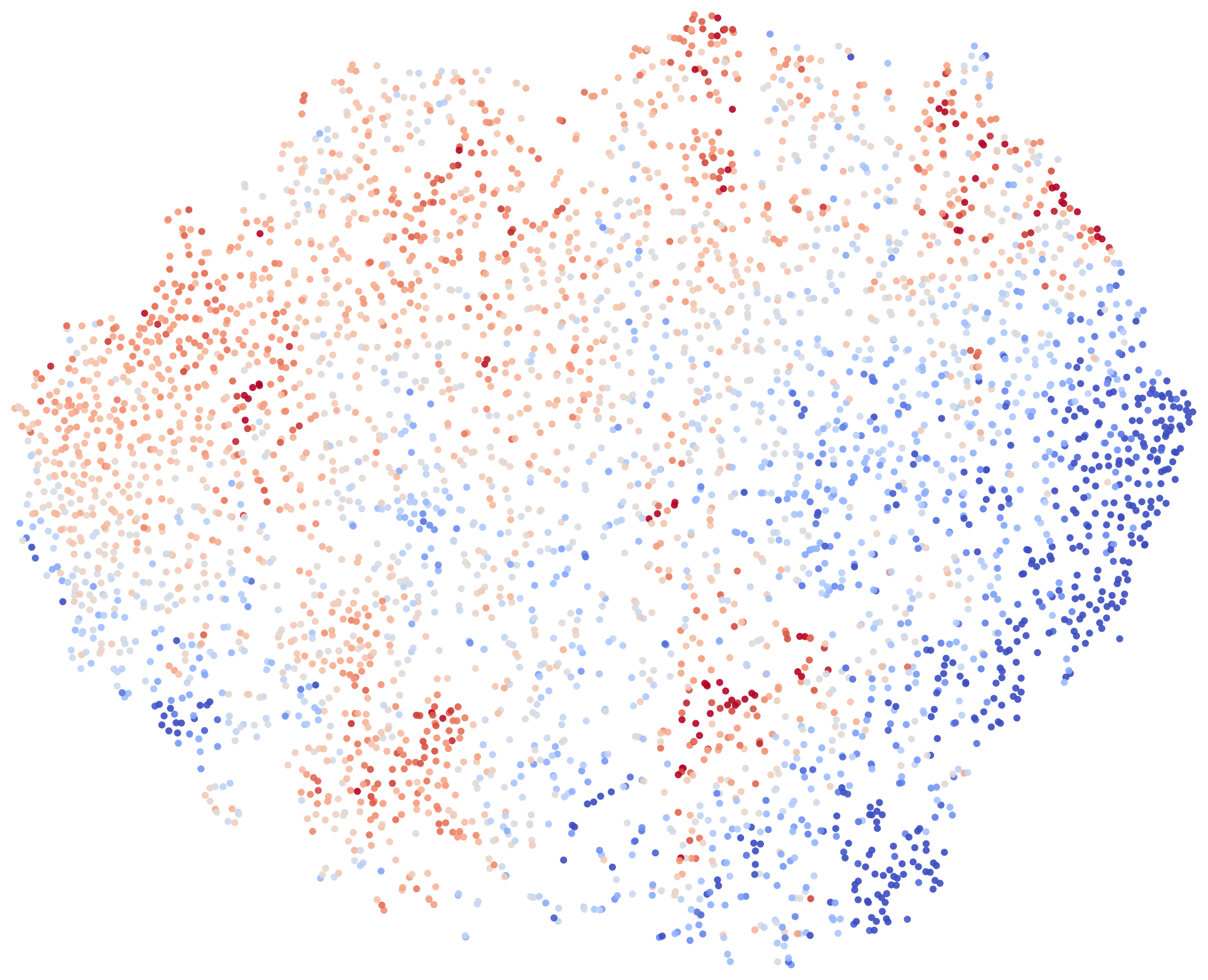}
  \caption{FT-DINO}
\end{subfigure}
\end{figure}

\clearpage
\setappfig{B}
\subsection*{Appendix B: Historical Visual Retrieval with Pretrained DINO}
\begin{figure}[H]
\centering
\caption{Cumulative Log Returns to Benchmark DINO Portfolios}
\label{fig:dino_cum}
\exhibitnote{This figure plots cumulative log returns to benchmark DINO high-minus-low portfolios from 2001 to 2025. DINO denotes the pretrained DINOv2-S encoder \citep{oquab2024dinov2}. Each week, stocks are sorted into deciles by the signal and held over the subsequent five trading days; H--L is the return on decile 10 minus decile 1. Panels (a), (b), and (c) report equal-weighted, value-weighted, and capped value-weighted (CVW) portfolios, where capped value weighting winsorizes market capitalization at the contemporaneous NYSE 80th percentile following \citet{jensen2023}. Each line corresponds to one neighborhood size, $K\in\{5,10,30,100,500,1{,}000,3{,}000\}$. The dashed horizontal line marks zero.}
\begin{subfigure}{0.67\textwidth}
  \centering
  \includegraphics[width=\linewidth]{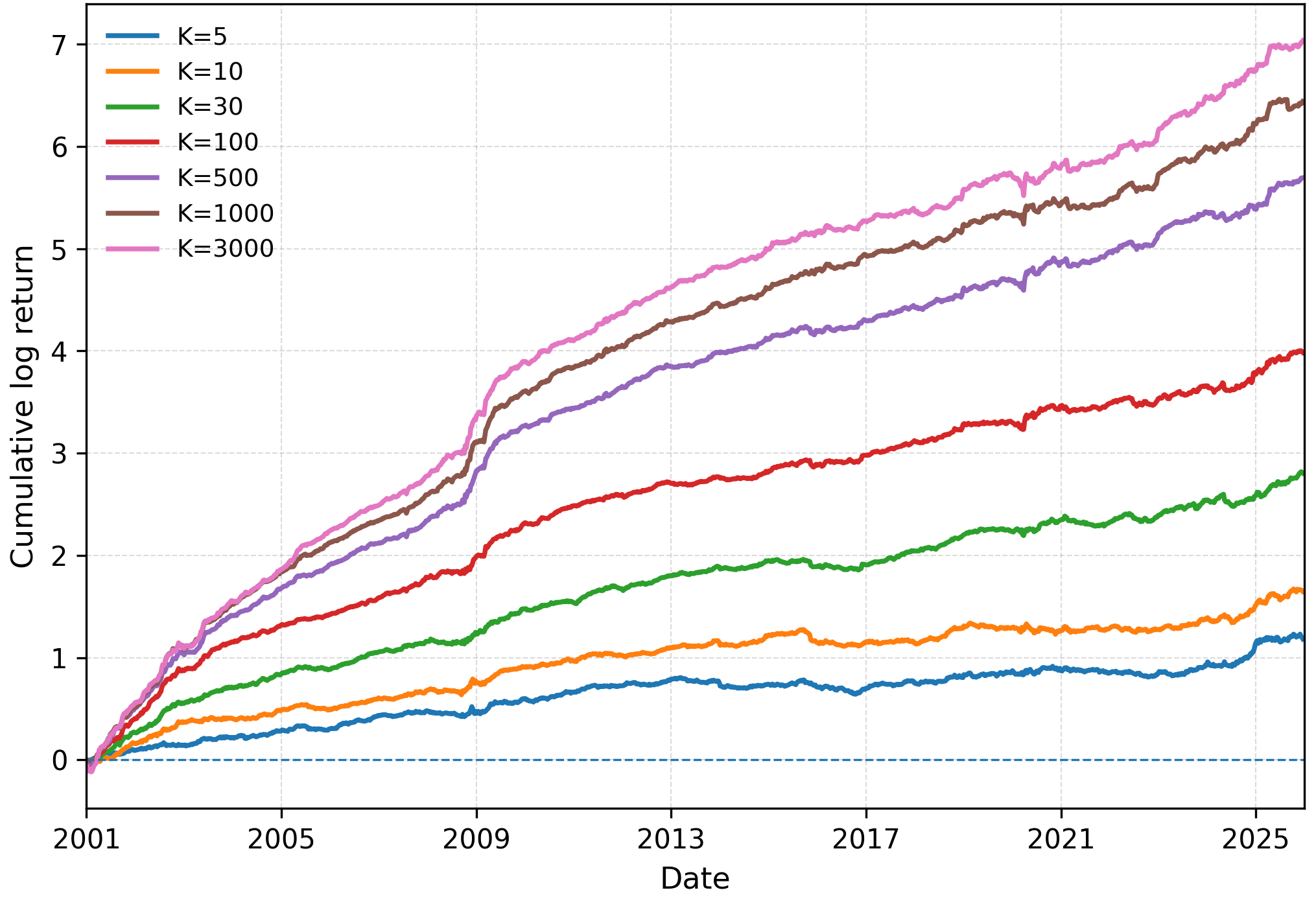}
  \caption{EW H--L}
\end{subfigure}

\vspace{0.2cm}

\begin{subfigure}{0.67\textwidth}
  \centering
  \includegraphics[width=\linewidth]{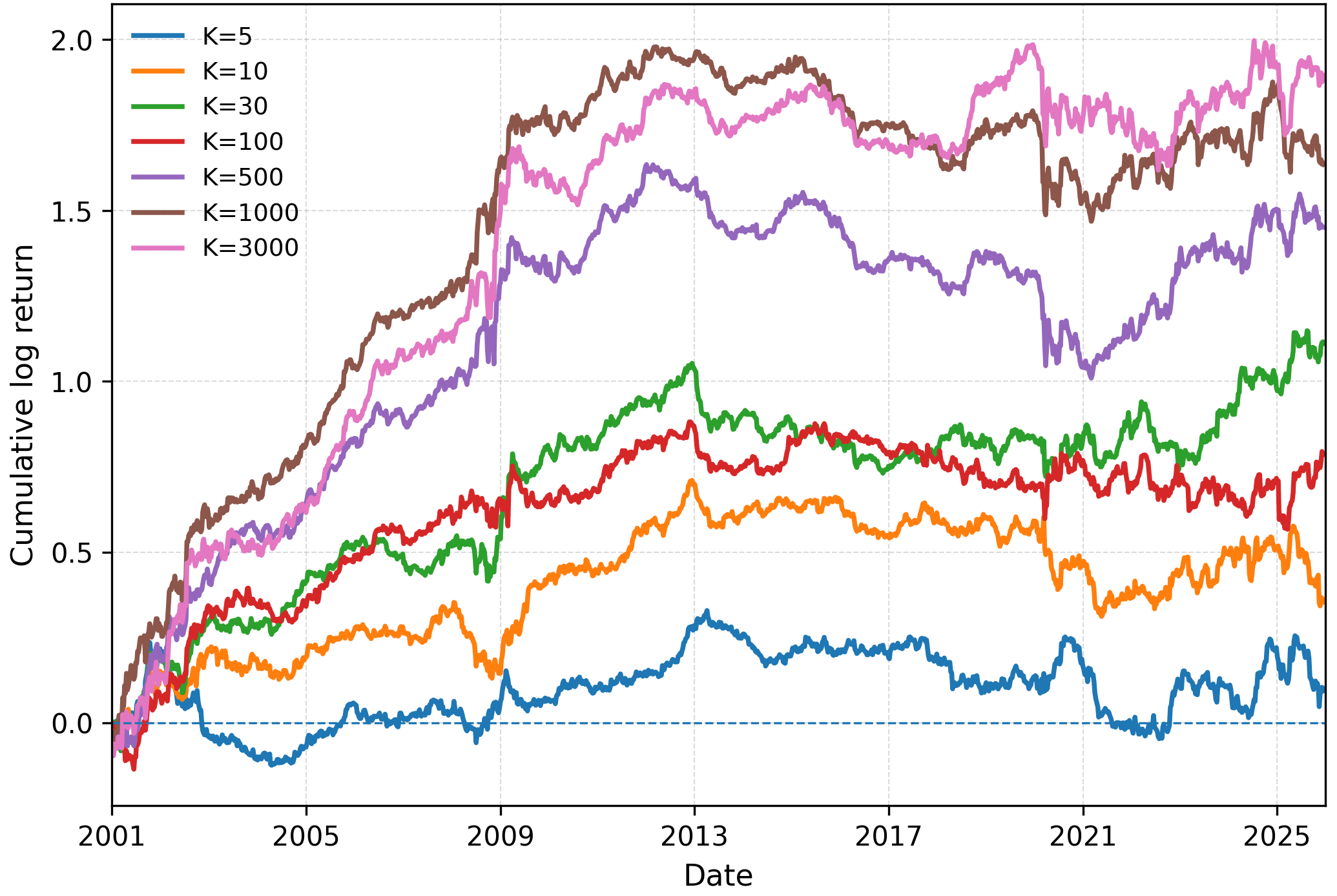}
  \caption{VW H--L}
\end{subfigure}
\end{figure}

\clearpage
\begin{figure}[H]
\centering
\ContinuedFloat
\caption[]{Cumulative Log Returns to Benchmark DINO Portfolios (continued)}
\exhibitnote{This figure continues Figure~\ref{fig:dino_cum}. Panel (c) reports CVW portfolios.}
\begin{subfigure}{0.67\textwidth}
  \centering
  \includegraphics[width=\linewidth]{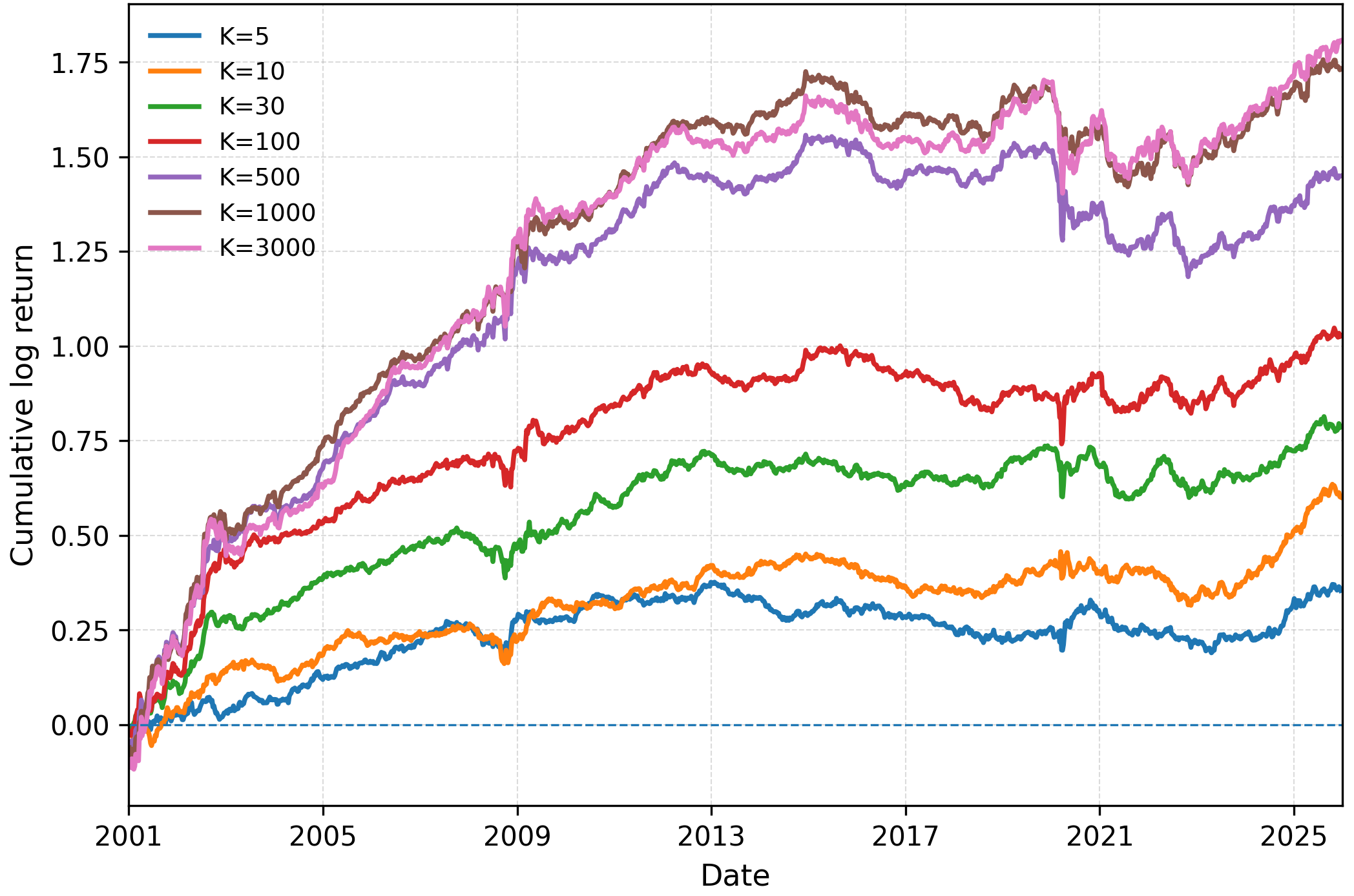}
  \caption{CVW H--L}
\end{subfigure}
\end{figure}

\clearpage
\begin{figure}[H]
\centering
\caption{High-Similarity Cross-Stock Cases}
\label{fig:app_high_similarity_cases}
\exhibitnote{This figure presents three cross-stock pairs with very high cosine similarity under benchmark DINO. Each panel pairs charts from different firms and periods and reports the cosine similarity between their normalized DINO representations. We first identify high-similarity cross-company pairs and then examine corporate events near the beginning of each displayed chart window. Panel (a) pairs Allison Transmission (ALSN) with D.R. Horton (DHI). On October 26, 2022, Allison reported third-quarter 2022 results and raised its full-year 2022 guidance midpoints. On November 10, 2004, D.R. Horton reported fourth-quarter results, a record \$4.6 billion year-end backlog, and fiscal-2005 guidance. Following these announcements, both charts display similar upward price-adjustment paths over the displayed windows. Panel (b) pairs The Hanover Insurance Group (THG) with FedEx (FDX). On July 28, 2016, Hanover reported second-quarter results, including a 97.3\% combined ratio with 4.5 points of catastrophe losses. On December 20, 2007, FedEx reported second-quarter results in which operating income declined 7\% year over year, while management cited high fuel prices and weak U.S. economic growth as pressures on profitability. Following these announcements, both charts display similar downward price-adjustment paths over the displayed windows.}

\begin{subfigure}[t]{0.96\textwidth}
\centering

\begin{minipage}[c]{0.32\linewidth}
    \centering
    \includegraphics[
        width=0.70\linewidth,
        height=0.14\textheight,
        keepaspectratio
    ]{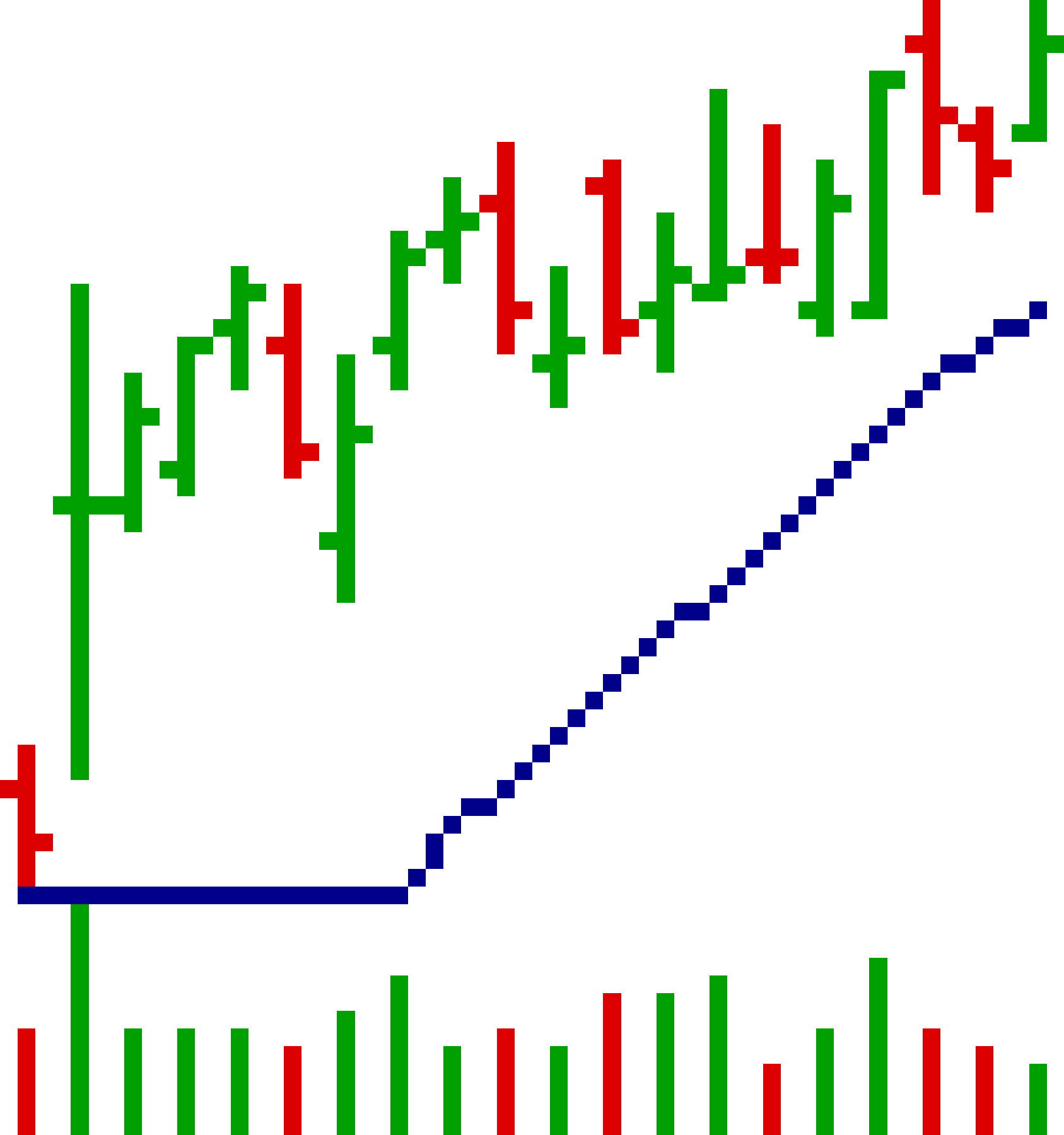}

    \vspace{2pt}
    {\small ALSN}\\
    {\small 2022-10-26 to 2022-11-22}
\end{minipage}
\hfill
\begin{minipage}[c]{0.22\linewidth}
    \centering
    {\small Cosine similarity}\\[2pt]
    \(\longleftrightarrow\)\\[2pt]
    {\small 0.99632609}
\end{minipage}
\hfill
\begin{minipage}[c]{0.32\linewidth}
    \centering
    \includegraphics[
        width=0.70\linewidth,
        height=0.14\textheight,
        keepaspectratio
    ]{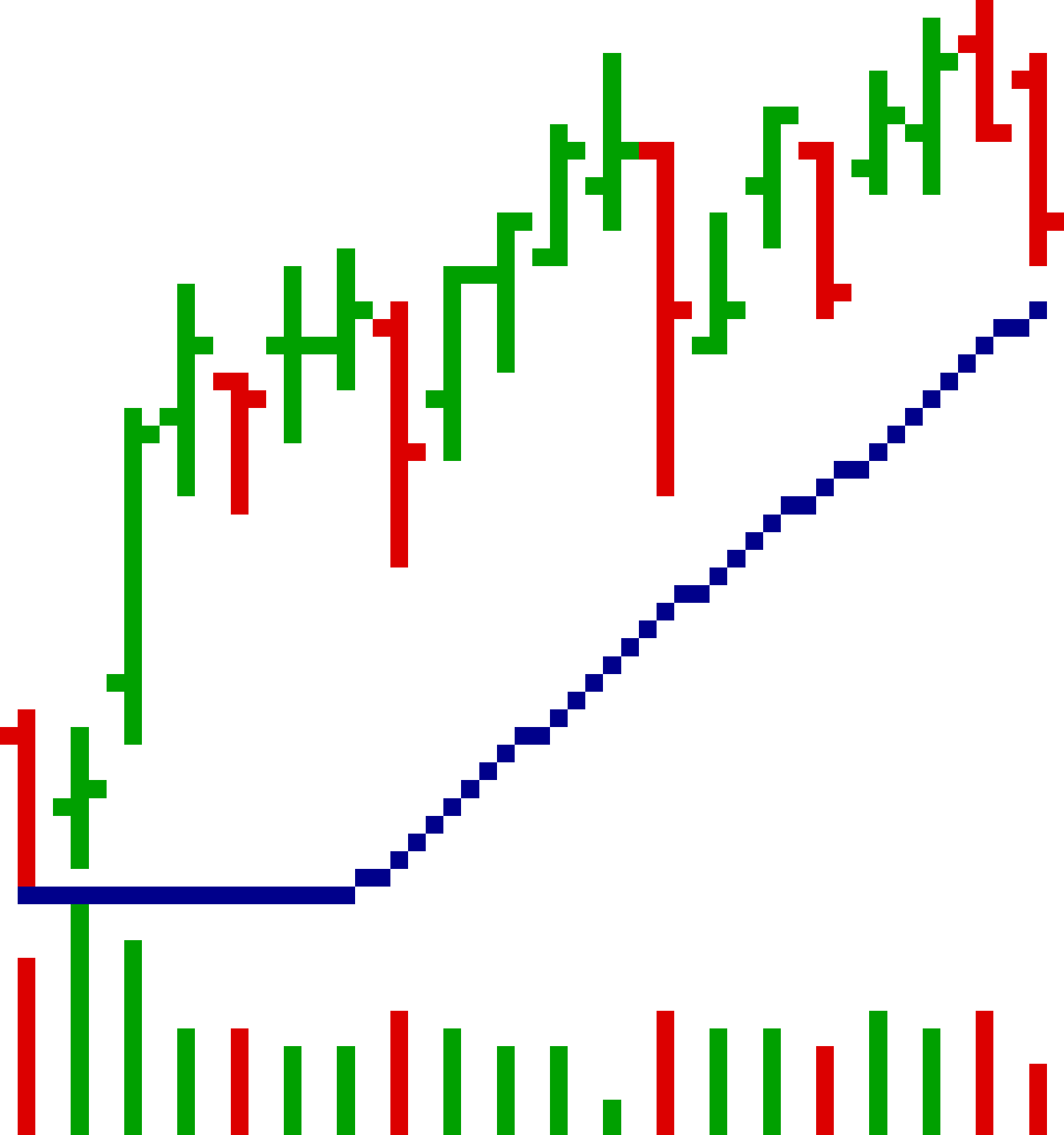}

    \vspace{2pt}
    {\small DHI}\\
    {\small 2004-11-10 to 2004-12-08}
\end{minipage}

\caption{Positive Outlook}
\end{subfigure}

\vspace{0.22cm}

\begin{subfigure}[t]{0.96\textwidth}
\centering

\begin{minipage}[c]{0.32\linewidth}
    \centering
    \includegraphics[
        width=0.70\linewidth,
        height=0.14\textheight,
        keepaspectratio
    ]{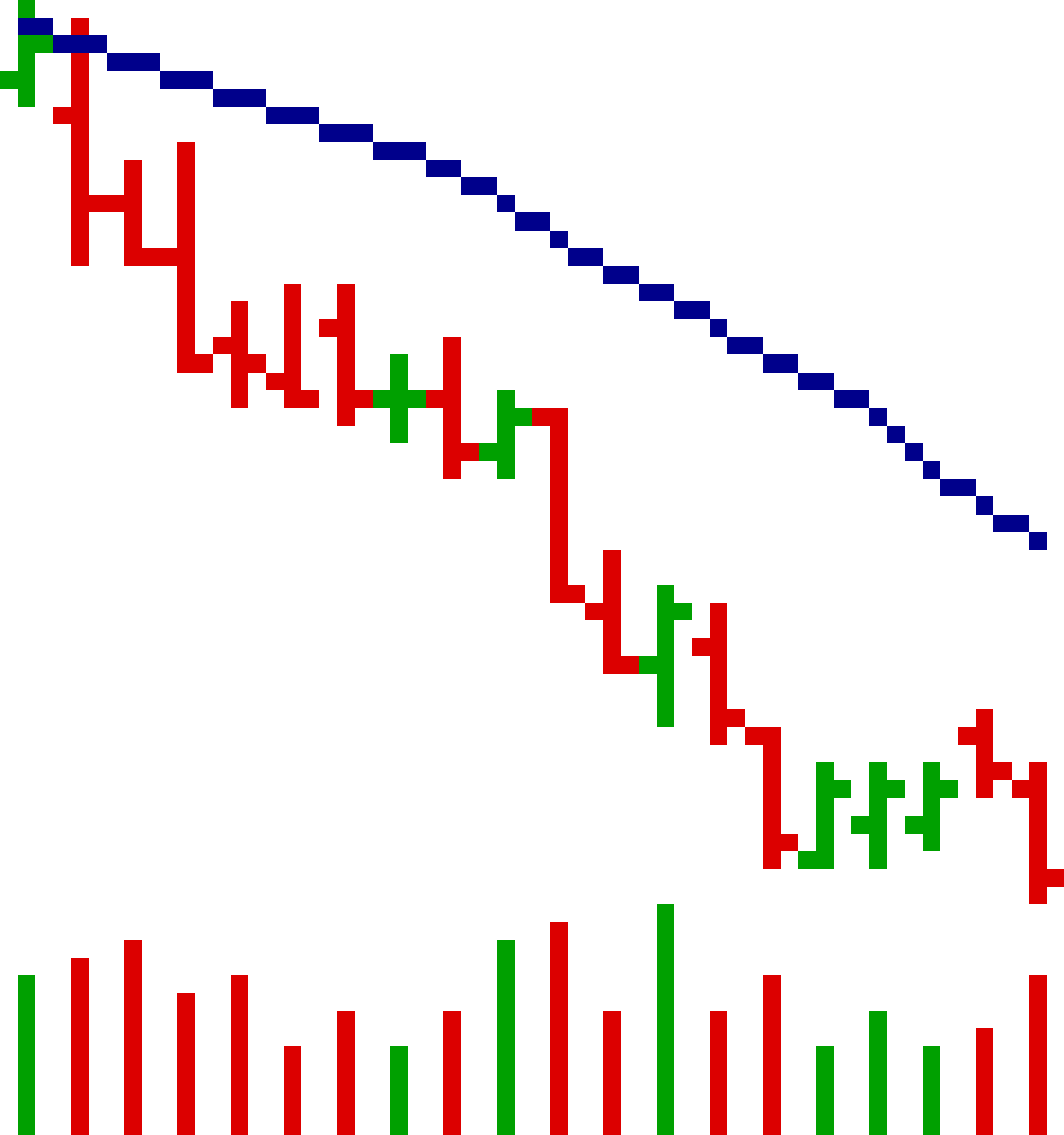}

    \vspace{2pt}
    {\small THG}\\
    {\small 2016-07-28 to 2016-08-24}
\end{minipage}
\hfill
\begin{minipage}[c]{0.22\linewidth}
    \centering
    {\small Cosine similarity}\\[2pt]
    \(\longleftrightarrow\)\\[2pt]
    {\small 0.98418027}
\end{minipage}
\hfill
\begin{minipage}[c]{0.32\linewidth}
    \centering
    \includegraphics[
        width=0.70\linewidth,
        height=0.14\textheight,
        keepaspectratio
    ]{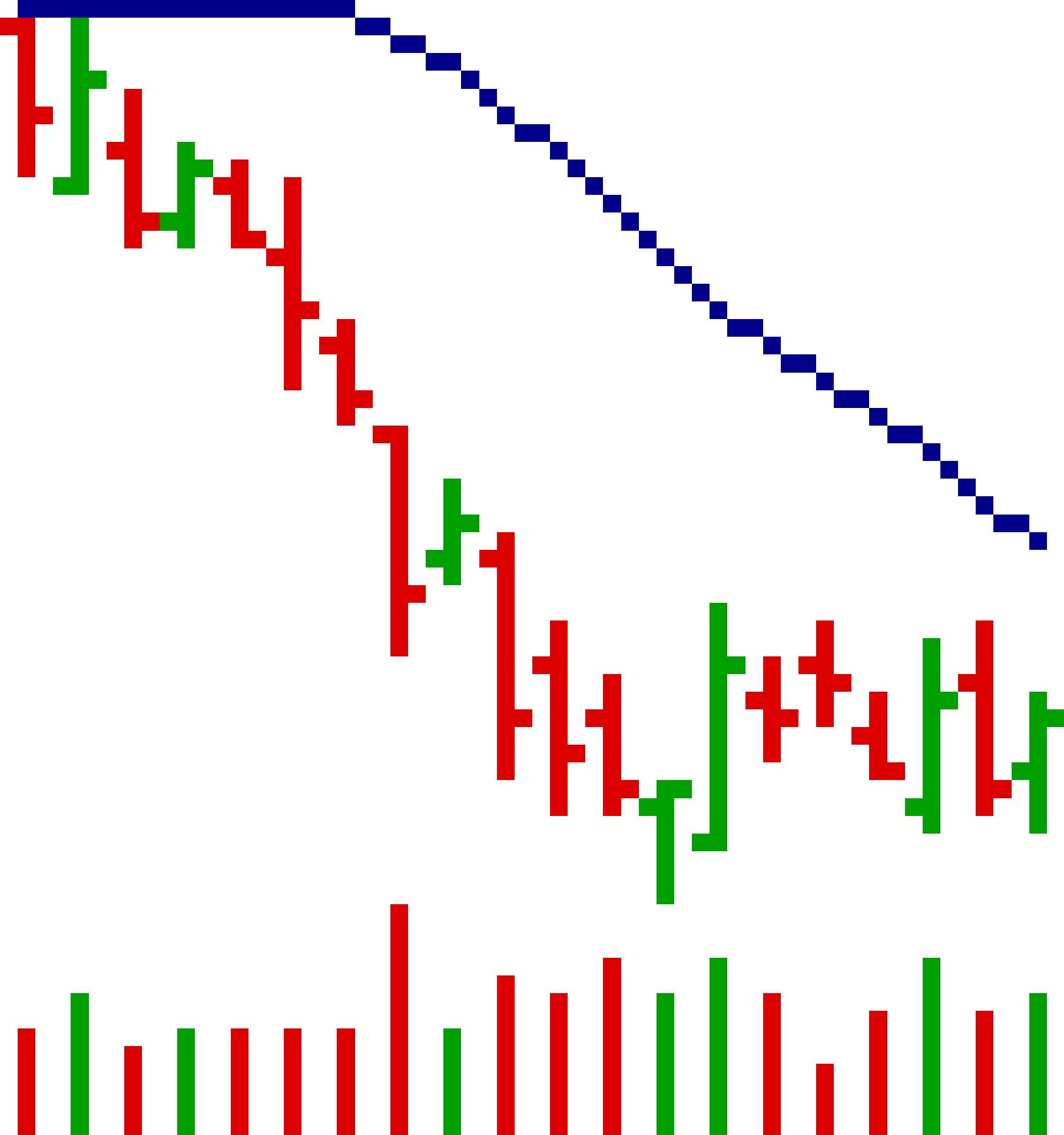}

    \vspace{2pt}
    {\small FDX}\\
    {\small 2007-12-20 to 2008-01-18}
\end{minipage}

\caption{Profitability Pressure}
\end{subfigure}

\end{figure}

\clearpage
\begin{figure}[H]
\centering
\ContinuedFloat
\caption[]{High-Similarity Cross-Stock Cases (continued)}
\exhibitnote{This figure continues Figure~\ref{fig:app_high_similarity_cases}. Panel (c) pairs Ford Motor Company (F) with Home Depot (HD). On September 1, 2017, Ford reported that August U.S. vehicle sales declined 2.1\% year over year, while F-Series sales increased 15.0\%. On February 23, 2010, Home Depot reported fourth-quarter fiscal-2009 results: sales declined 0.3\% and U.S. comparable-store sales declined 1.1\%, while adjusted earnings from continuing operations increased 22.9\%. In both cases, weak headline sales coincided with stronger performance in economically important underlying measures, and both charts display similar upward price-adjustment paths over the displayed windows.}

\begin{subfigure}[t]{0.96\textwidth}
\centering

\begin{minipage}[c]{0.32\linewidth}
    \centering
    \includegraphics[
        width=0.70\linewidth,
        height=0.14\textheight,
        keepaspectratio
    ]{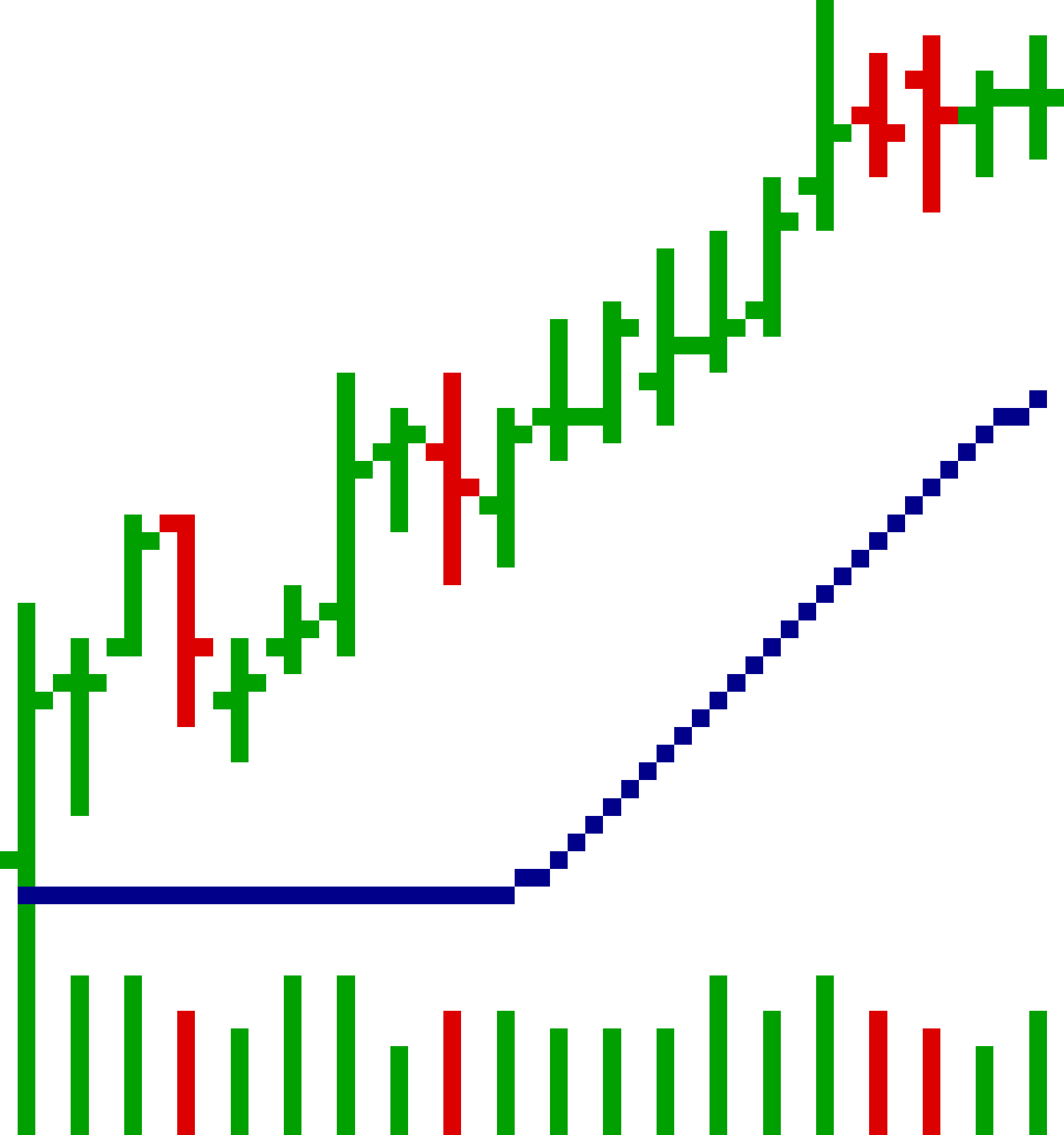}

    \vspace{2pt}
    {\small Ford (F)}\\
    {\small 2017-09-01 to 2017-09-29}
\end{minipage}
\hfill
\begin{minipage}[c]{0.22\linewidth}
    \centering
    {\small Cosine similarity}\\[2pt]
    \(\longleftrightarrow\)\\[2pt]
    {\small 0.99668866}
\end{minipage}
\hfill
\begin{minipage}[c]{0.32\linewidth}
    \centering
    \includegraphics[
        width=0.70\linewidth,
        height=0.14\textheight,
        keepaspectratio
    ]{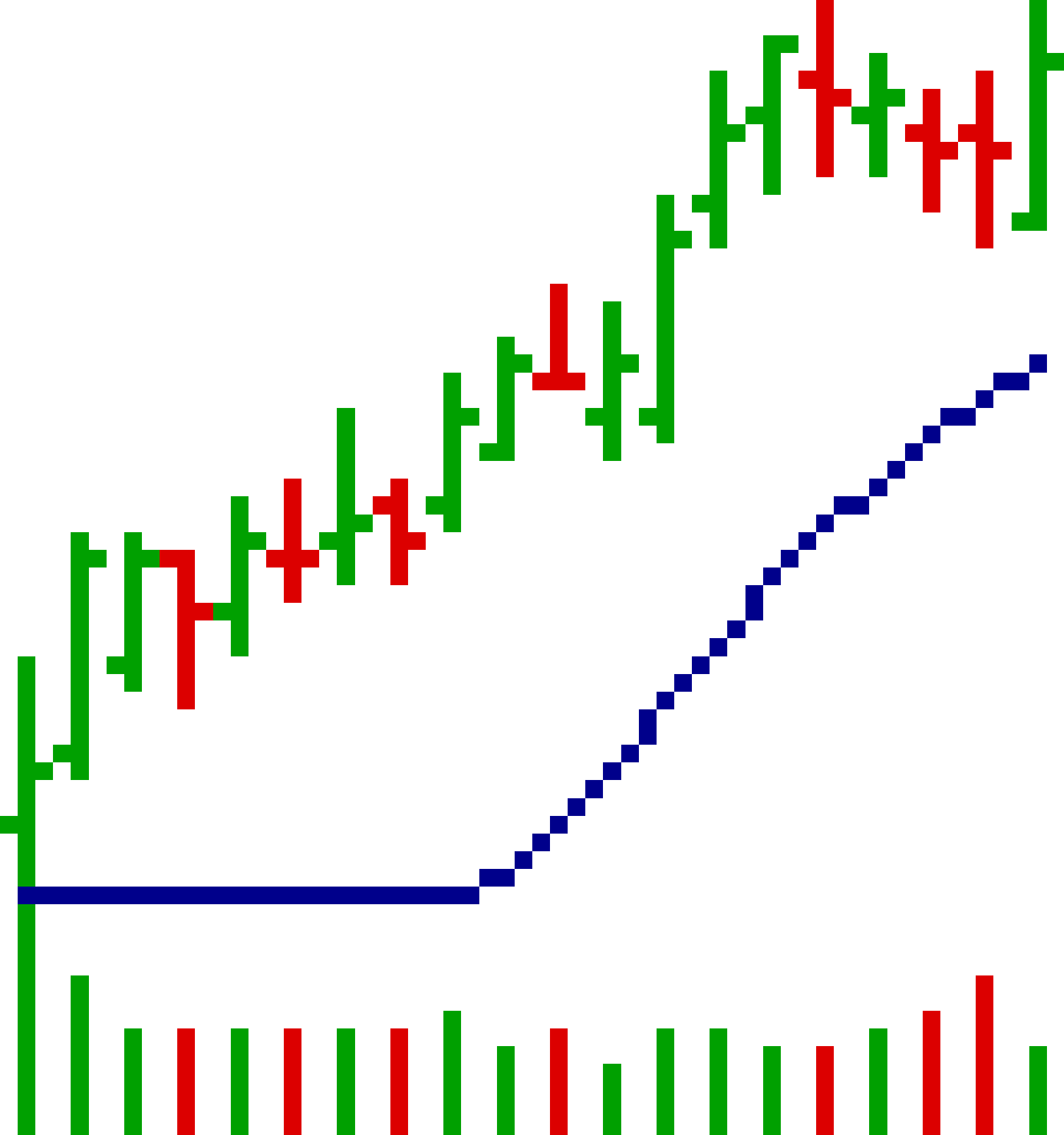}

    \vspace{2pt}
    {\small Home Depot (HD)}\\
    {\small 2010-02-23 to 2010-03-22}
\end{minipage}

\caption{Hidden Strength}
\end{subfigure}

\end{figure}

\clearpage
\setappfig{C}
\subsection*{Appendix C: Additional Results for Return-Aligned Retrieval}
\begin{figure}[H]
\centering
\caption{Retrieved Historical Neighbors: DINO versus FT-DINO}
\label{fig:neighbor_compare}
\exhibitnote{This figure illustrates historical retrieval for a single query chart (IRDM, January 6, 2023), shown in the first column of each row. The first row reports the three nearest eligible historical charts under benchmark DINO, the second row reports the three nearest eligible historical charts under FT-DINO, and the third row reports three randomly selected eligible historical charts for reference. The title of each chart reports the ticker, the chart end date, and the realized subsequent five-trading-day return. The query, the historical library, and the information-timing restriction are identical across rows, so differences between the first two rows reflect only the change in representation. Figure~\ref{fig:app_neighbor_composition} summarizes the temporal and industry composition of neighbors over the full 2001--2025 out-of-sample period.}
\includegraphics[width=\textwidth,height=0.72\textheight,keepaspectratio]{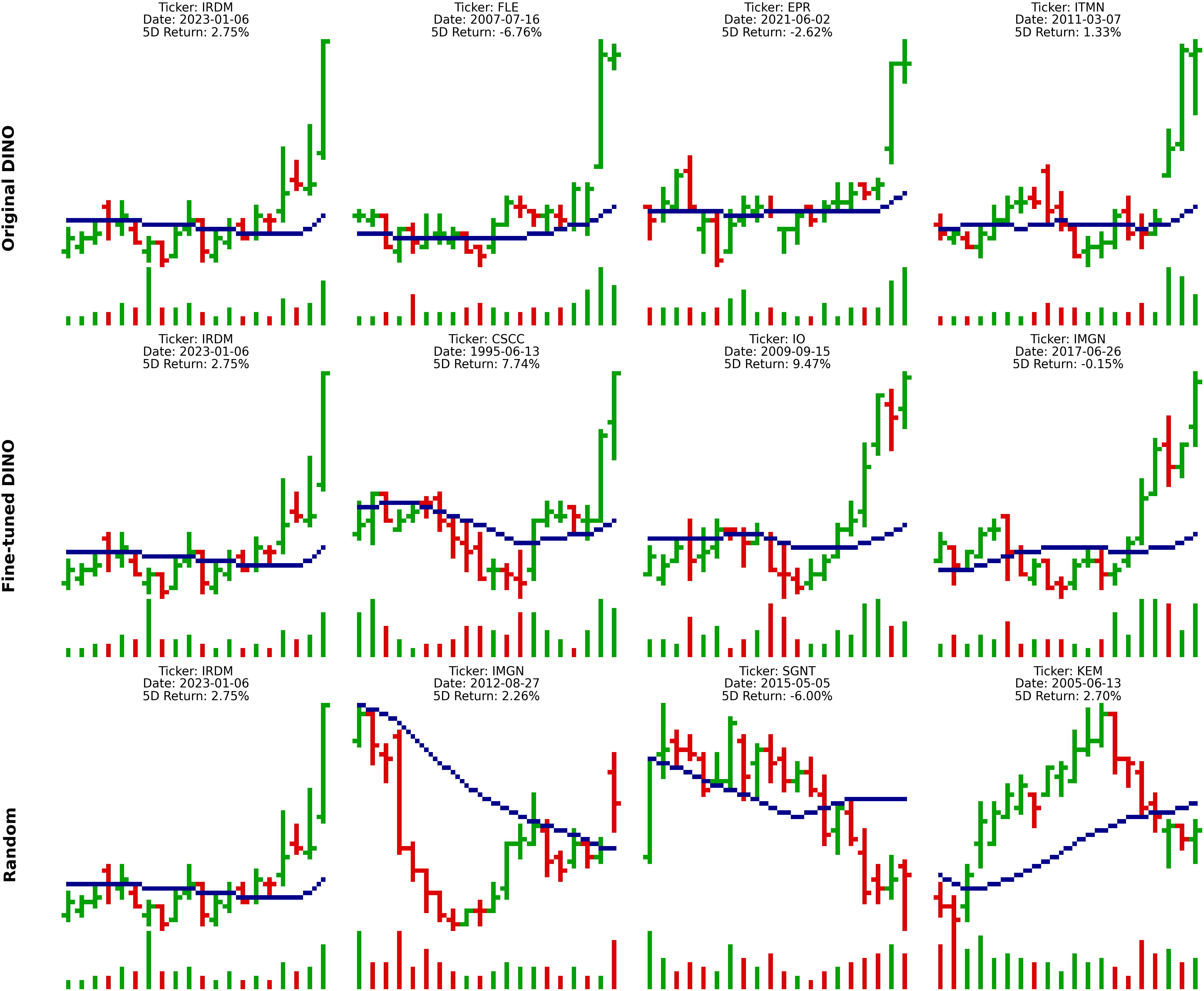}
\end{figure}

\clearpage
\begin{figure}[H]
\centering
\caption{Historical-Neighbor Composition for Benchmark DINO and FT-DINO}
\label{fig:app_neighbor_composition}
\exhibitnote{This figure summarizes the composition of the top-1,000 historical neighborhoods for benchmark DINO and FT-DINO over the 2001--2025 out-of-sample period. Panel (a) reports the temporal distribution: each heatmap cell shows the share of a query week's top-1,000 neighbors whose chart end date falls in the indicated neighbor week, with query dates on the horizontal axis. Panel (b) reports supply-adjusted Fama–French 12 neighbor-industry preferences. For each query week and query industry, observed neighbor-industry shares are divided by the corresponding shares in the eligible historical library and renormalized across neighbor industries. The resulting probabilities are averaged equally across query weeks. The DINO and FT-DINO heatmaps display deviations from the equal-industry benchmark of 1/12. In each panel, the left and middle heatmaps report DINO and FT-DINO, and the right heatmap reports FT-DINO minus DINO.}
\begin{subfigure}[t]{0.96\textwidth}
\centering
\includegraphics[width=\linewidth,height=0.34\textheight,keepaspectratio]{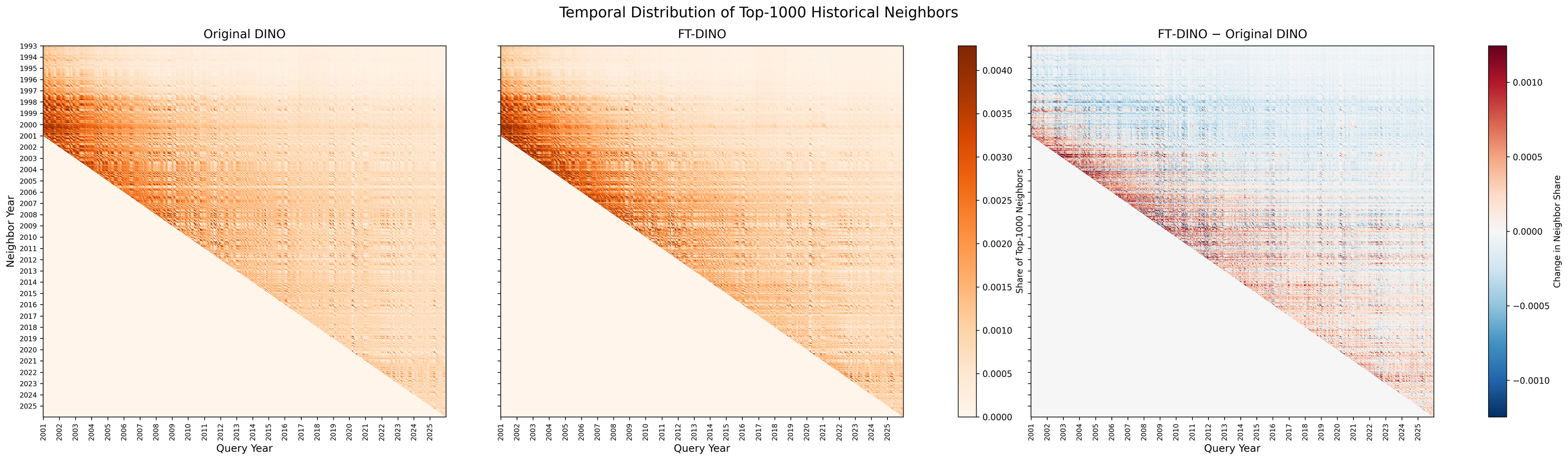}
\caption{Temporal distribution of retrieved historical neighbors}
\label{fig:app_neighbor_time}
\end{subfigure}

\vspace{0.35cm}

\begin{subfigure}[t]{0.96\textwidth}
\centering
\includegraphics[width=\linewidth,height=0.34\textheight,keepaspectratio]{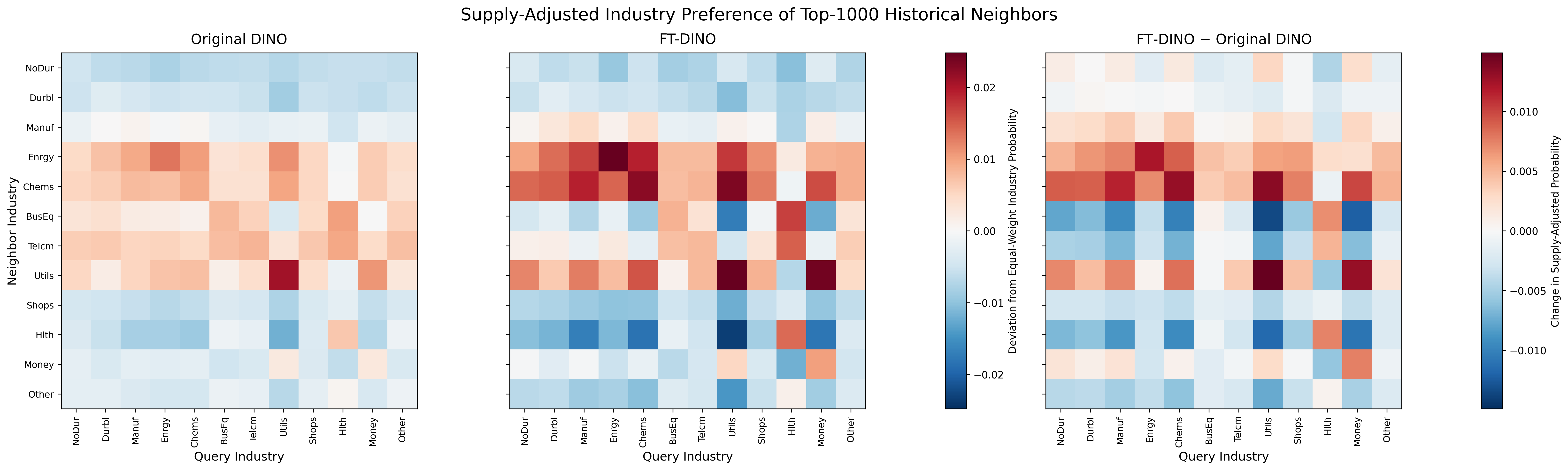}
\caption{Supply-adjusted industry distribution of retrieved historical neighbors}
\label{fig:app_neighbor_industry}
\end{subfigure}
\end{figure}

\clearpage
\begin{figure}[H]
\centering
\caption{Sharpe Ratios across Neighborhood Sizes: DINO versus FT-DINO}
\label{fig:app_sharpe_k}
\exhibitnote{This figure reports annualized Sharpe ratios of high-minus-low portfolios sorted on benchmark DINO and FT-DINO retrieval signals over 2001--2025 for neighborhood sizes from $K=5$ to $K=20{,}000$. Each week, stocks are sorted into deciles by the signal and held over the subsequent five trading days; H--L is the return on decile 10 minus decile 1. Panel (a) uses equal-weighted portfolios, and Panel (b) uses capped value-weighted (CVW) portfolios, which winsorize market capitalization at the contemporaneous NYSE 80th percentile following \citet{jensen2023}. The horizontal axis is not to scale.}
\begin{subfigure}{0.75\textwidth}
  \centering
  \includegraphics[width=\linewidth]{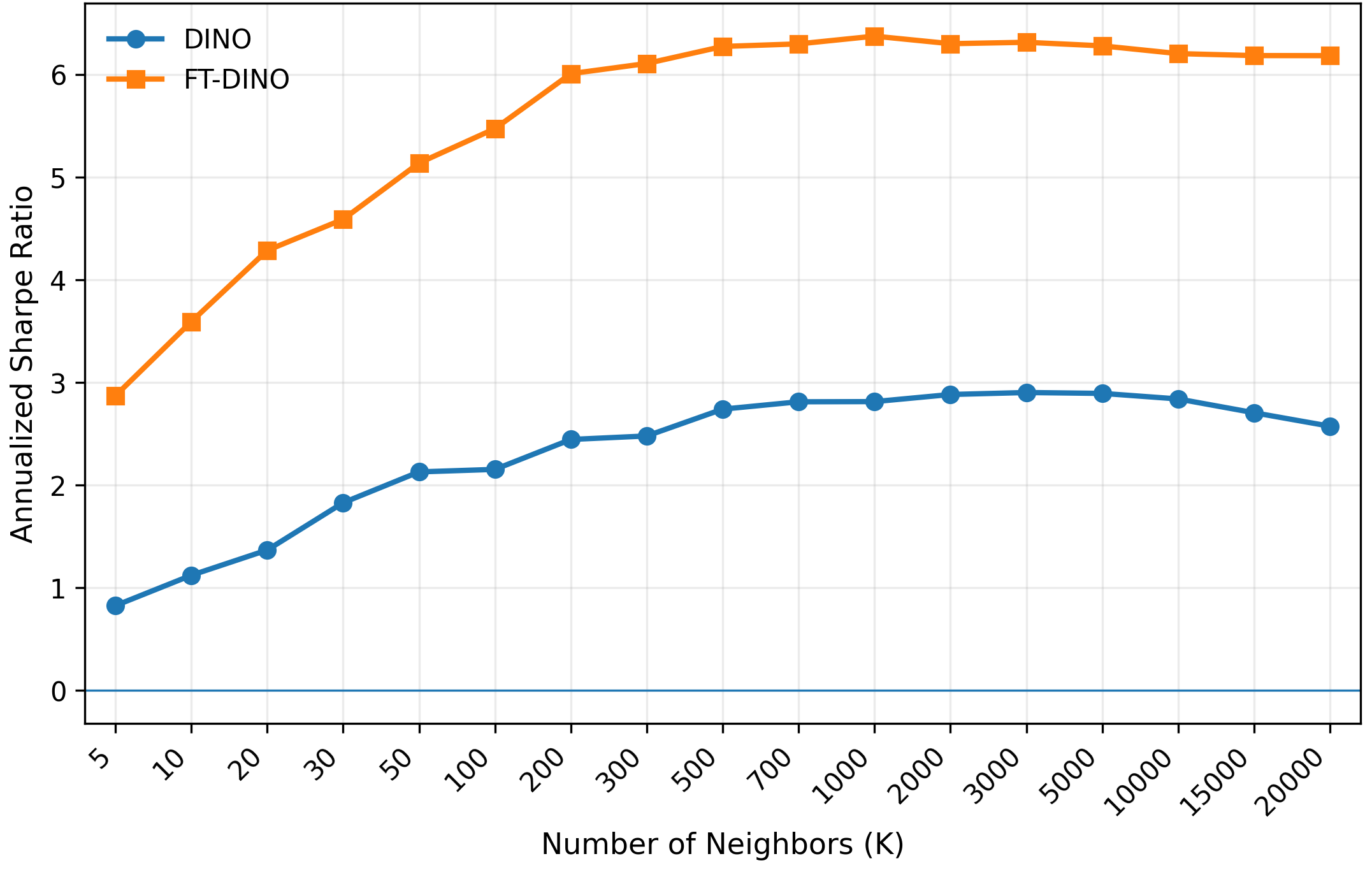}
  \caption{EW H--L}
\end{subfigure}

\vspace{0.1cm}

\begin{subfigure}{0.75\textwidth}
  \centering
  \includegraphics[width=\linewidth]{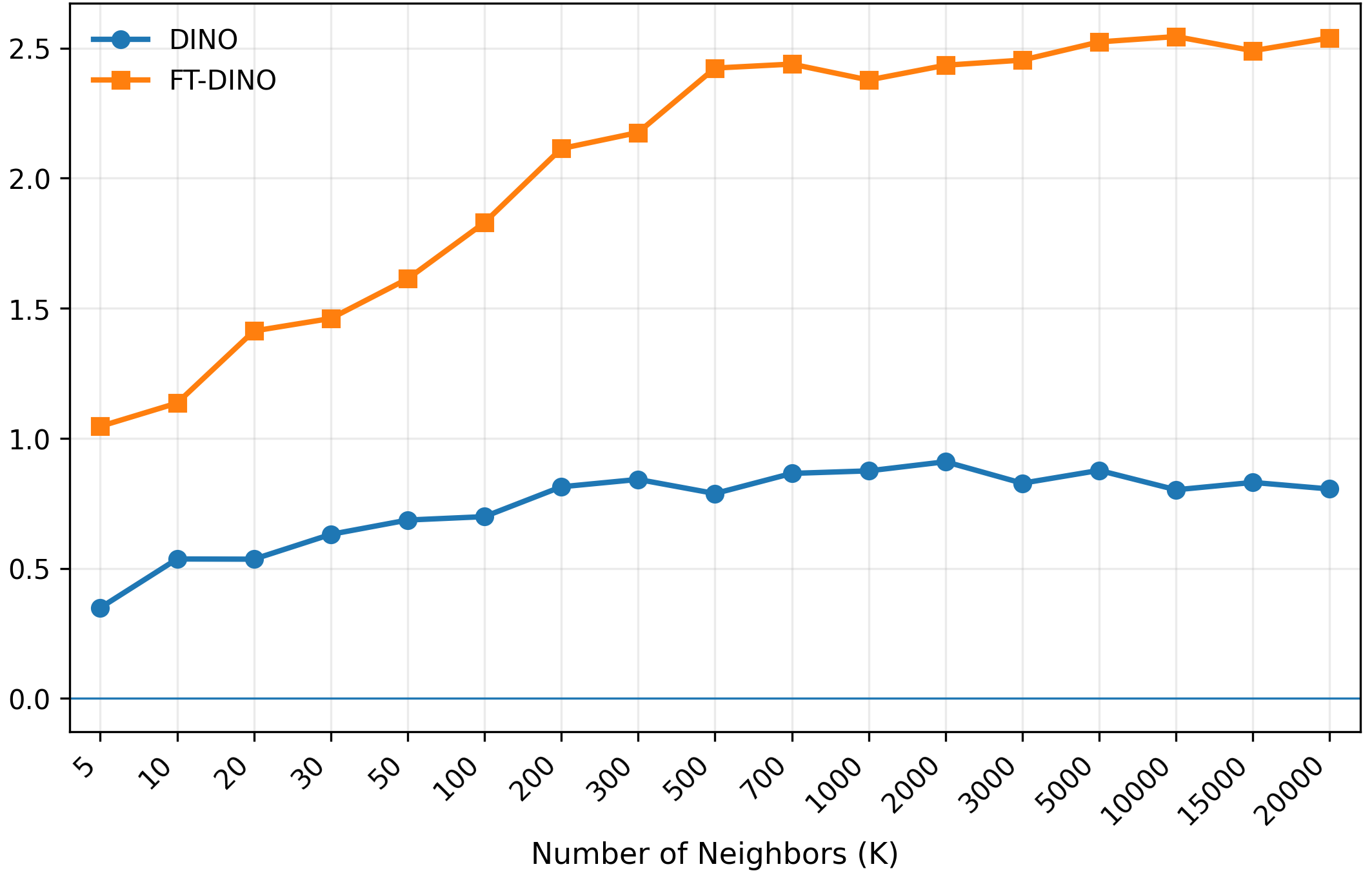}
  \caption{CVW H--L}
\end{subfigure}
\end{figure}

\clearpage
\begin{figure}[H]
\centering
\caption{Three Measures of Input Sensitivity: DINO and FT-DINO}
\label{fig:app_sensitivity_measures}
\exhibitnote{This figure plots three pixel-level sensitivity measures for benchmark DINO (Panel (a)) and FT-DINO (Panel (b)), averaged across the same 100 randomly sampled query charts as in Figure~\ref{fig:sensitivity}. Embedding sensitivity is the Frobenius norm of the Jacobian of the normalized CLS embedding with respect to each pixel. Matching sensitivity averages the squared pixel gradients of cosine similarity to 64 embeddings drawn at random from the eligible historical library, as in equation~\eqref{eq:app_match_sens}. Forecast sensitivity is the $L_1$ norm of the pixel gradient of the return-aligned component of the similarity-weighted retrieval forecast over the fixed top-$K$ neighborhood with $K=1{,}000$, as in equation~\eqref{eq:app_forecast_sens}. For each measure, each query's sensitivity map is normalized to sum to one before averaging across queries. The horizontal axis is the trading day within the chart window, and the bottom of each panel corresponds to the volume bars. Each panel has its own color scale, and brighter colors indicate greater sensitivity.}
\begin{subfigure}{\textwidth}
  \centering
  \includegraphics[width=\linewidth]{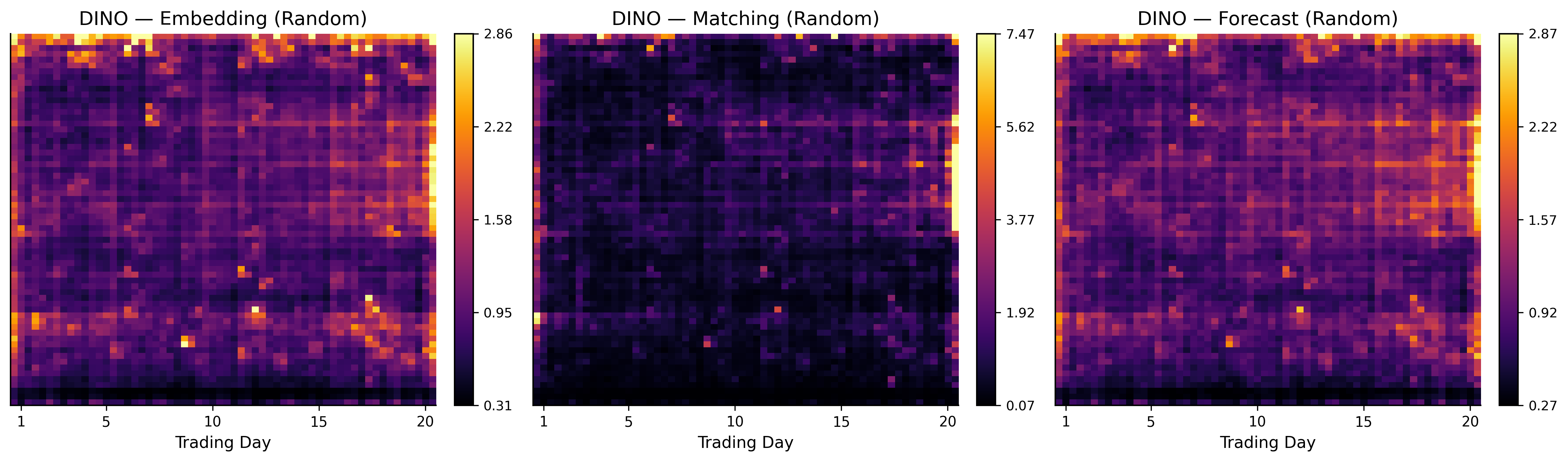}
  \caption{DINO}
\end{subfigure}

\vspace{0.3cm}

\begin{subfigure}{\textwidth}
  \centering
  \includegraphics[width=\linewidth]{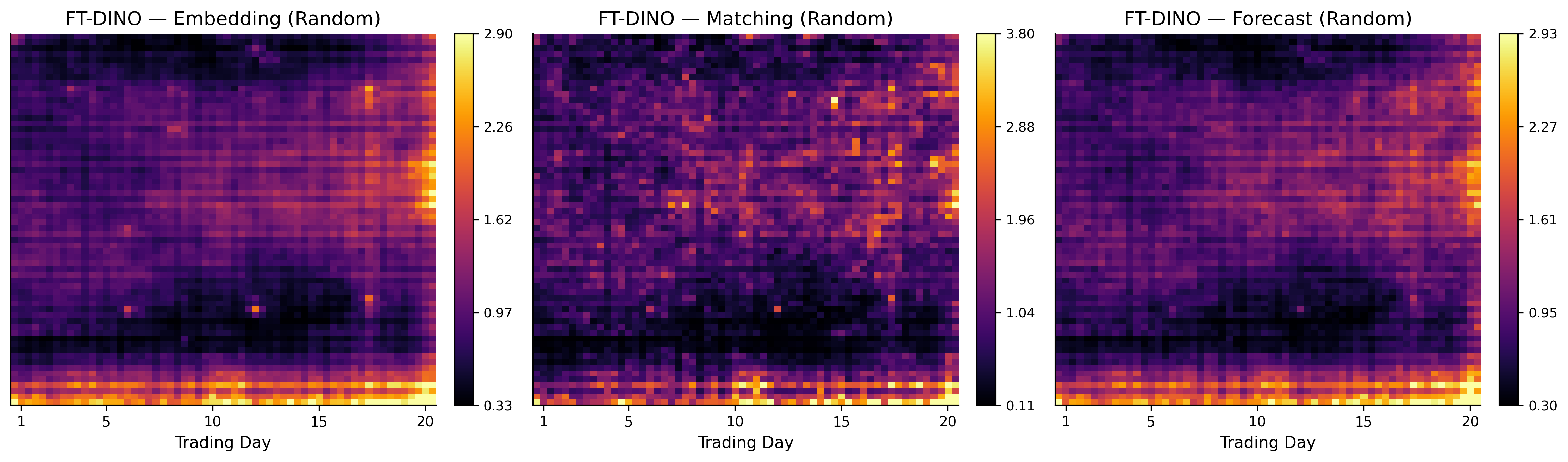}
  \caption{FT-DINO}
\end{subfigure}
\end{figure}

\clearpage
\begin{figure}[H]
\centering
\caption{Transformed Queries and Retrieved Neighbors}
\label{fig:app_transform}
\exhibitnote{This figure illustrates the query transformations in Appendix~\ref{app:c_transform} for one query chart. The first row shows the original query, its horizontal flip, which reverses the order of time, and its 180-degree rotation, which reverses both time and the direction of price changes. The second and third rows show the nearest eligible historical chart retrieved for each query under benchmark DINO and FT-DINO. The historical library and the retrieval rule are unchanged; only the query is transformed.}
\includegraphics[width=\textwidth,height=0.6\textheight,keepaspectratio]{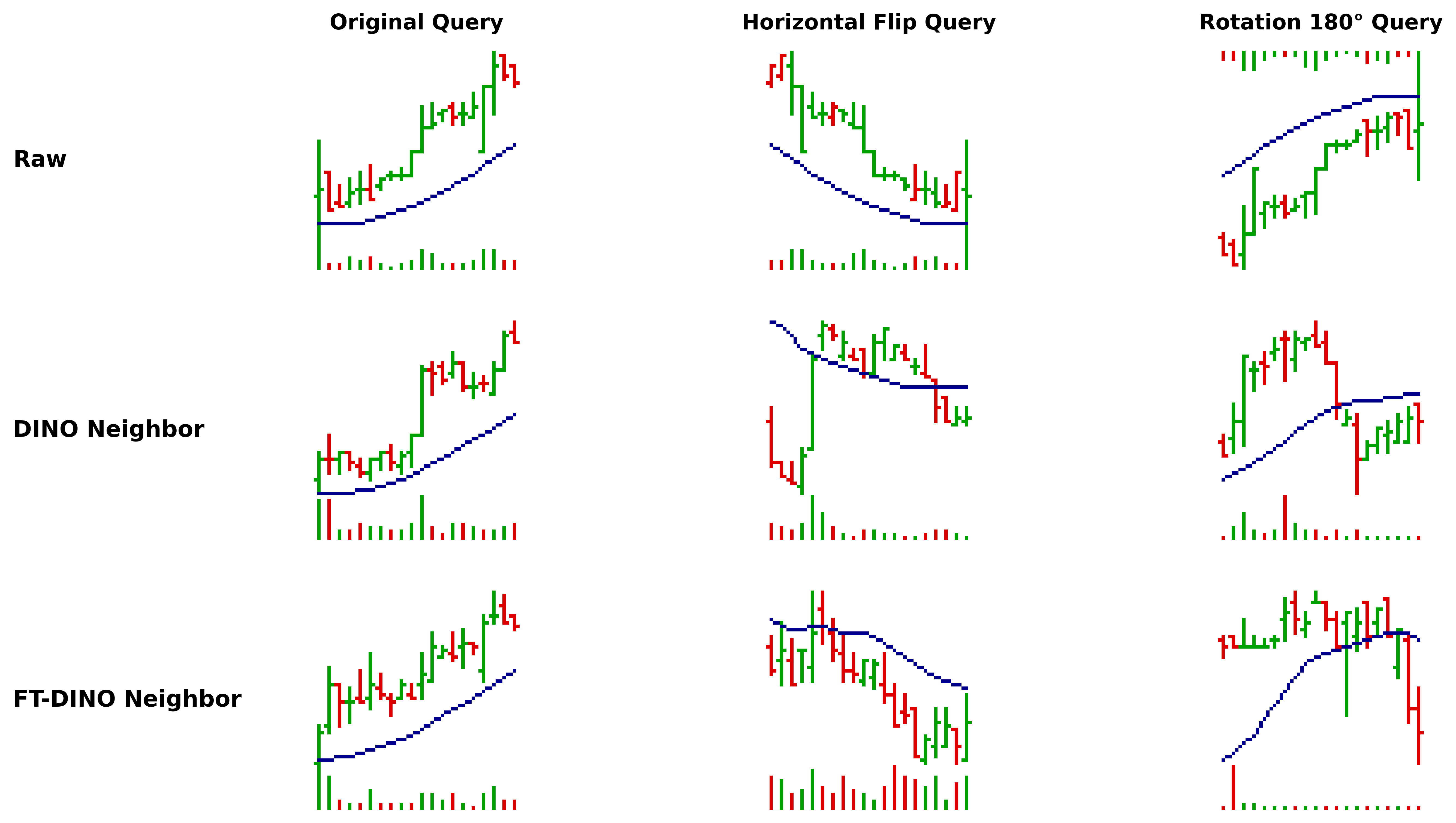}
\end{figure}

\clearpage
\setappfig{D}
\subsection*{Appendix D: Economic Mechanism}
\begin{figure}[H]
\centering
\caption{Event-Time Returns to the Top and Bottom FT-DINO Deciles}
\label{fig:app_legs}
\exhibitnote{This figure decomposes the event-time H--L returns in Figure~\ref{fig:returndynamics} into the two legs. Panel (a) reports the equal-weighted return of the bottom FT-DINO decile ($K=1{,}000$), and Panel (b) reports that of the top decile, over five-trading-day blocks from the formation week (days $-4$ to 0) through day 20. Returns are in percent and are not annualized.}
\begin{subfigure}[t]{0.49\textwidth}
  \centering
  \includegraphics[width=\linewidth]{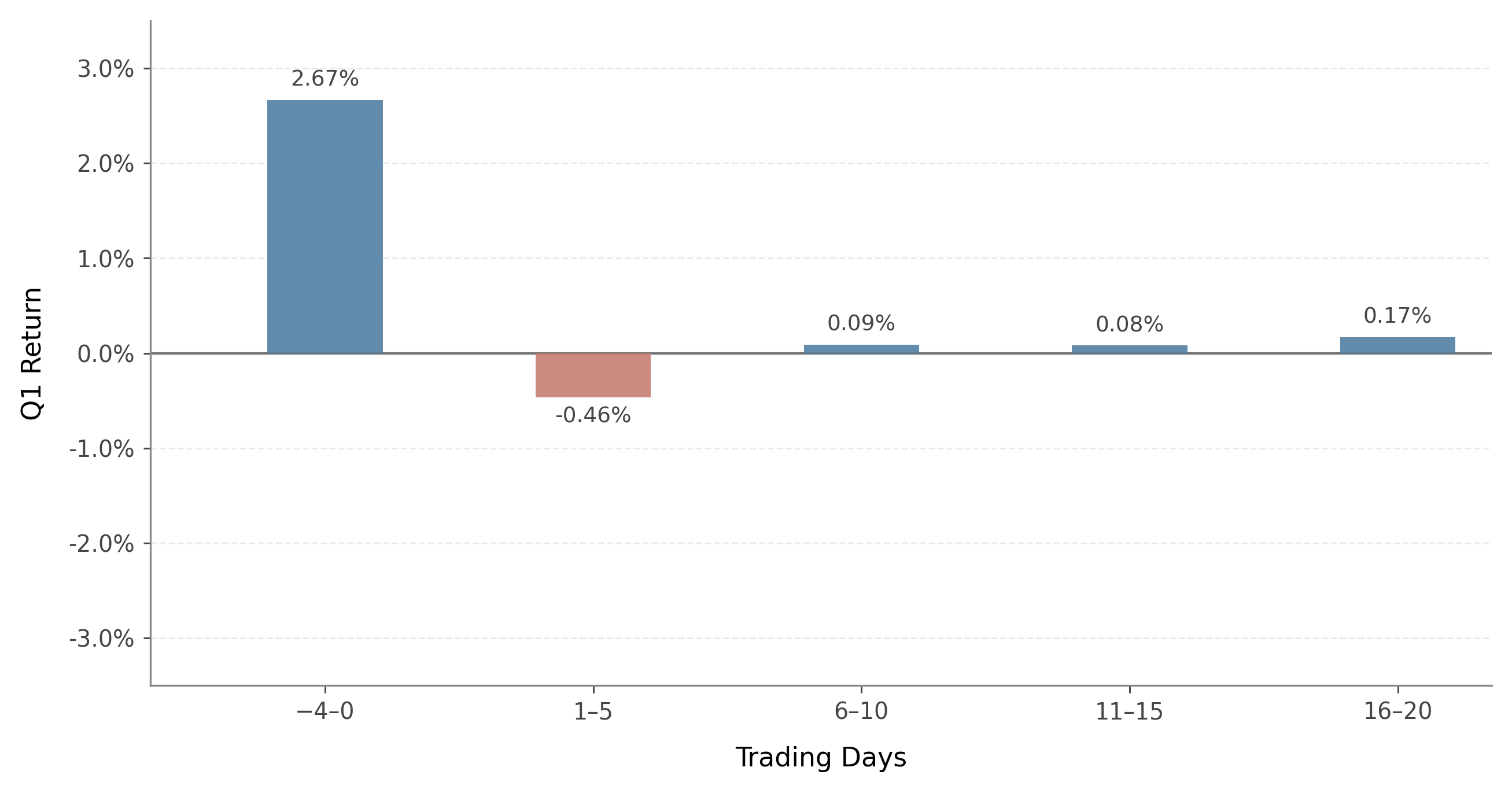}
  \caption{Bottom decile}
\end{subfigure}\hfill
\begin{subfigure}[t]{0.49\textwidth}
  \centering
  \includegraphics[width=\linewidth]{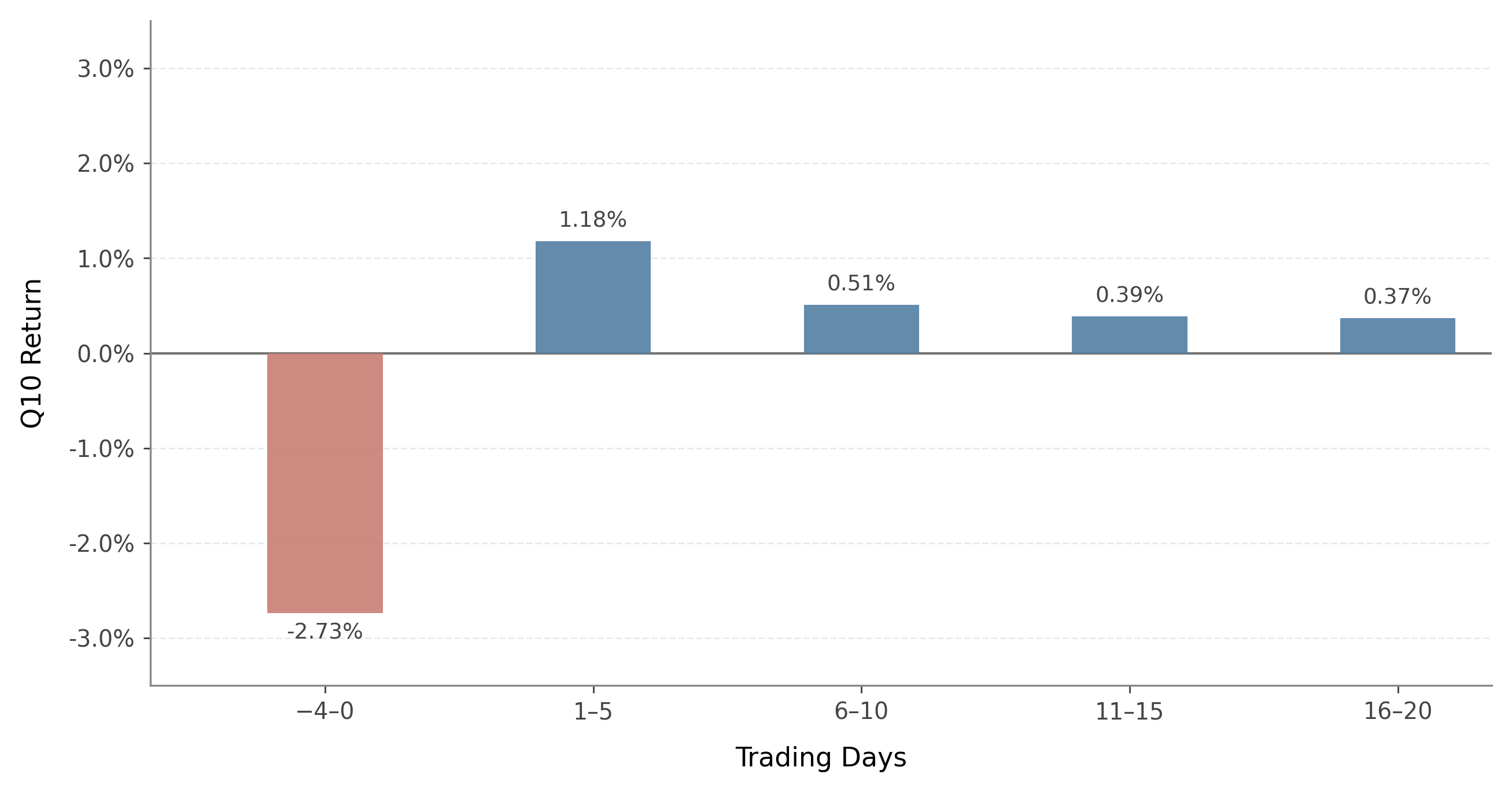}
  \caption{Top decile}
\end{subfigure}
\end{figure}

\clearpage
\setappfig{E}
\subsection*{Appendix E: Sparse-Feature Interpretation}

\begin{figure}[H]
\centering
\caption{Representative Charts for SAE Visual States C01--C03}
\label{fig:sae_states_c01_c03}
\exhibitnote{This figure displays representative real-market charts for the SAE features assigned to visual states C01 (Sustained Uptrend), C02 (Peak Pullback), and C03 (Downtrend Recovery). Each card corresponds to one constituent feature and reports its identifier, a short description of the visual configuration shared by its high-activation charts, and a representative 20-day chart with strong activation. The SAE is a JumpReLU sparse autoencoder with 16,384 latent features trained on FT-DINO representations from 1993--2000. The 30 visually consistent features are drawn from the 100 features with the highest return-related importance and are grouped into 11 visual states, as described in Table~\ref{app:saestates}.}
\begin{subfigure}[t]{0.96\textwidth}
\centering
\includegraphics[width=\linewidth,height=0.22\textheight,keepaspectratio]{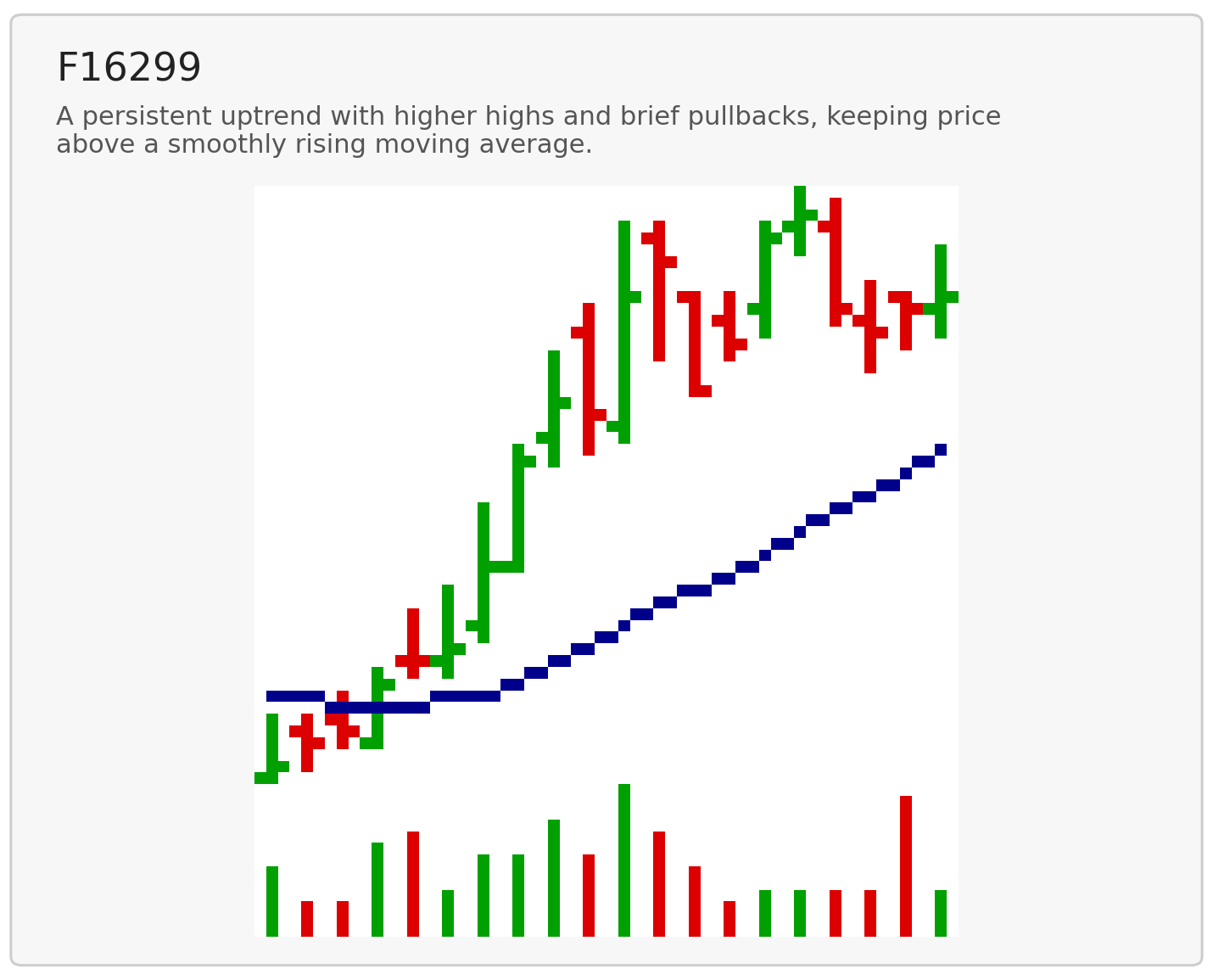}
\caption{C01: Sustained Uptrend.}
\end{subfigure}

\vspace{0.22cm}
\begin{subfigure}[t]{0.96\textwidth}
\centering
\includegraphics[width=\linewidth,height=0.22\textheight,keepaspectratio]{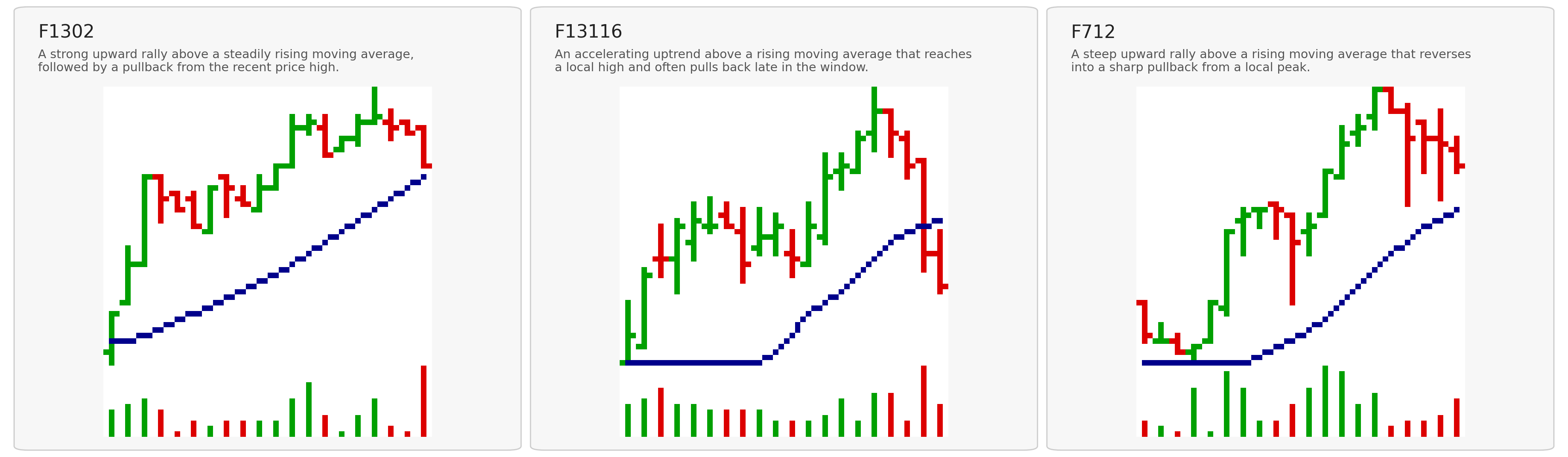}
\caption{C02: Peak Pullback.}
\end{subfigure}

\vspace{0.22cm}
\begin{subfigure}[t]{0.96\textwidth}
\centering
\includegraphics[width=\linewidth,height=0.22\textheight,keepaspectratio]{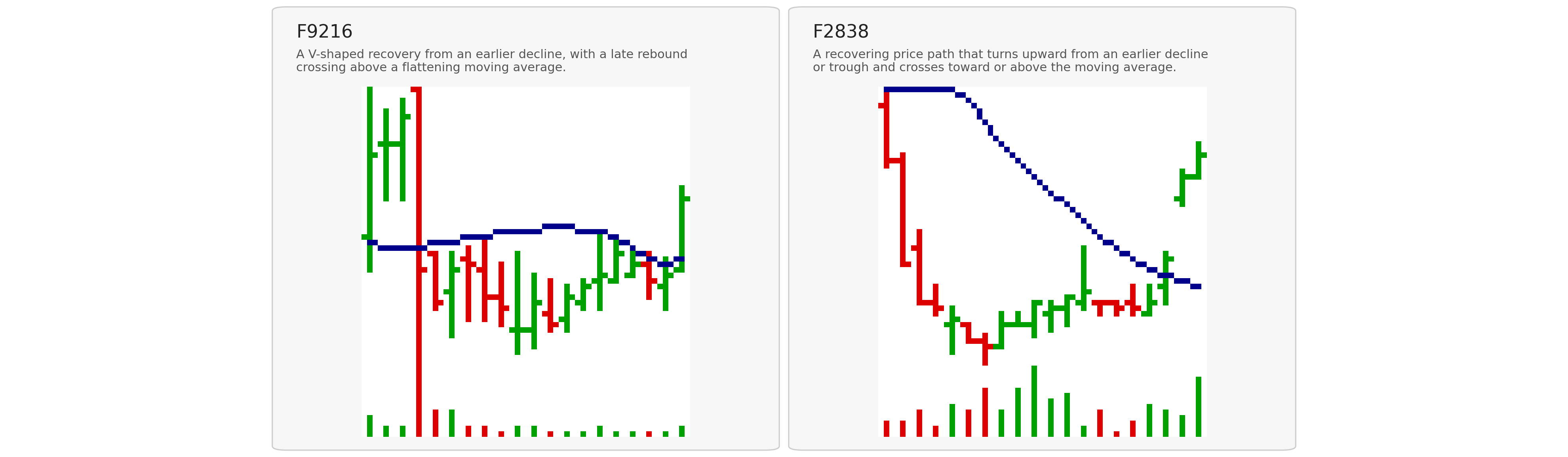}
\caption{C03: Downtrend Recovery.}
\end{subfigure}
\end{figure}

\clearpage
\begin{figure}[H]
\centering
\caption{Representative Charts for SAE Visual States C04--C06}
\label{fig:sae_states_c04_c06}
\exhibitnote{This figure displays representative real-market charts for the SAE features assigned to visual states C04 (Late Upside Breakout), C05 (Stable Downtrend), and C06 (Stepwise Downtrend). Each card corresponds to one constituent feature and reports its identifier, a short description of the visual configuration shared by its high-activation charts, and a representative 20-day chart with strong activation. The SAE is a JumpReLU sparse autoencoder with 16,384 latent features trained on FT-DINO representations from 1993--2000. The 30 visually consistent features are drawn from the 100 features with the highest return-related importance and are grouped into 11 visual states, as described in Table~\ref{app:saestates}.}
\begin{subfigure}[t]{0.96\textwidth}
\centering
\includegraphics[width=\linewidth,height=0.22\textheight,keepaspectratio]{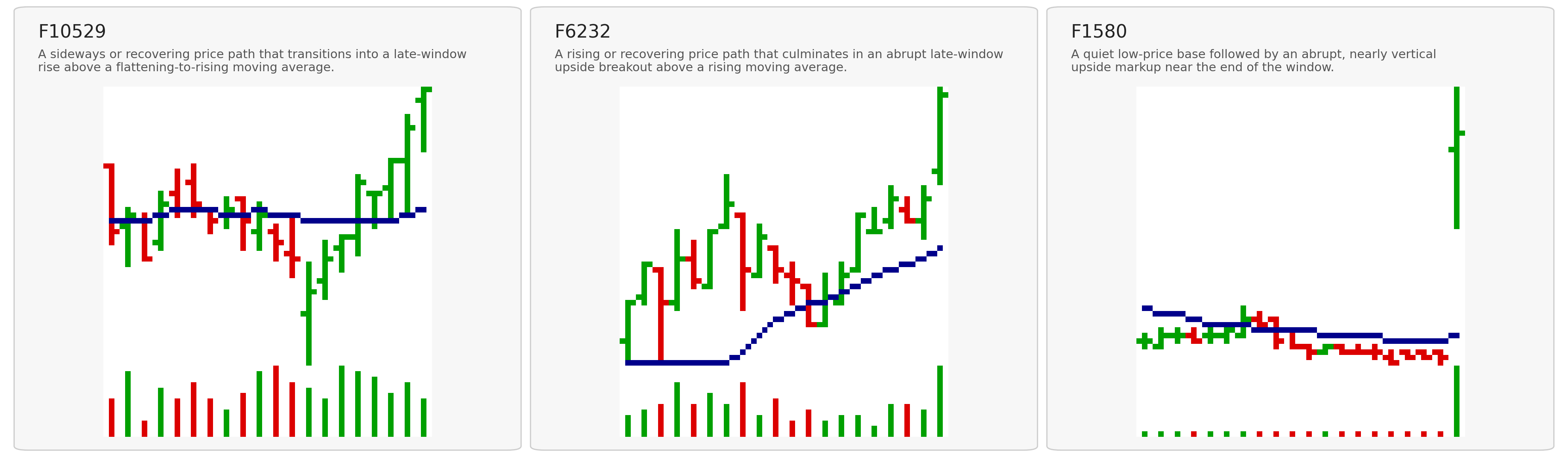}
\caption{C04: Late Upside Breakout.}
\end{subfigure}

\vspace{0.22cm}
\begin{subfigure}[t]{0.96\textwidth}
\centering
\includegraphics[width=\linewidth,height=0.22\textheight,keepaspectratio]{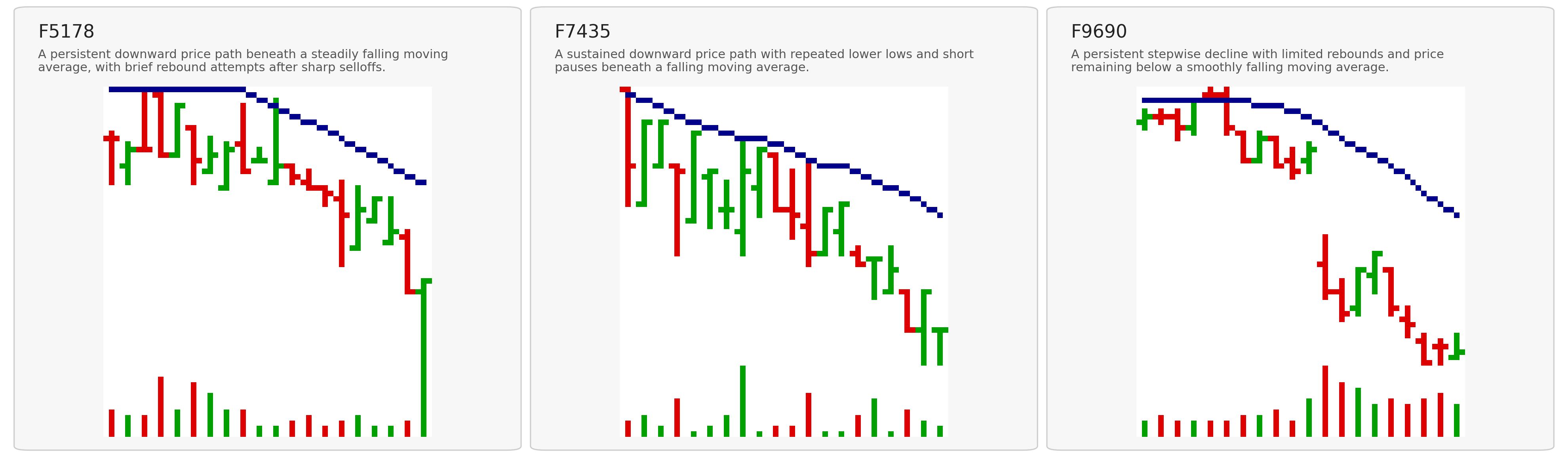}
\caption{C05: Stable Downtrend.}
\end{subfigure}

\vspace{0.22cm}
\begin{subfigure}[t]{0.96\textwidth}
\centering
\includegraphics[width=\linewidth,height=0.22\textheight,keepaspectratio]{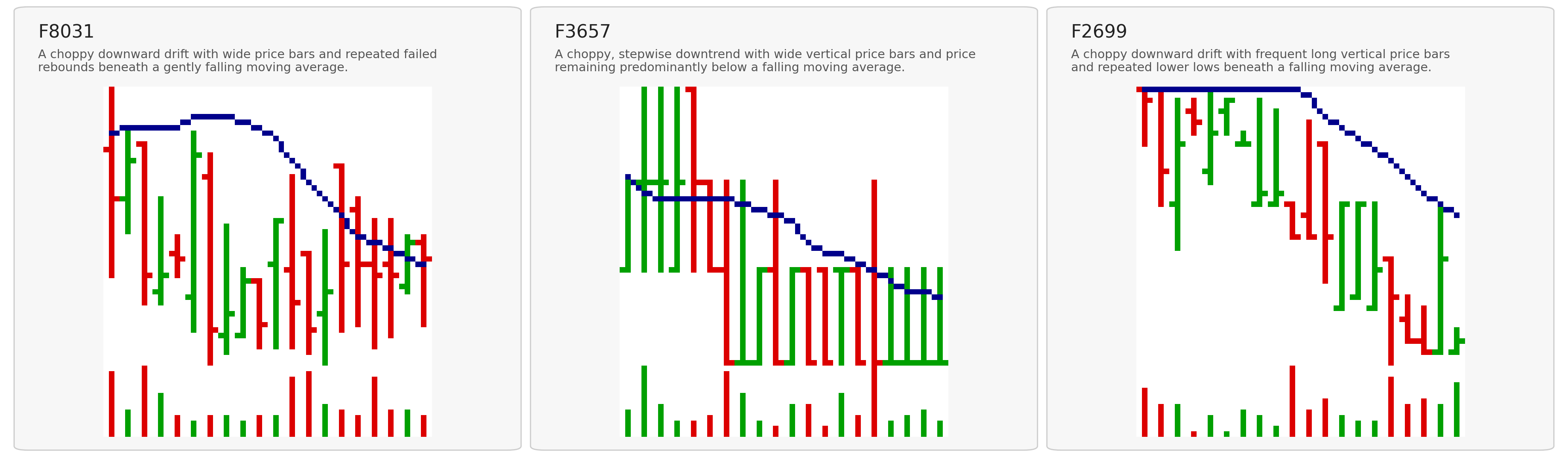}
\caption{C06: Stepwise Downtrend.}
\end{subfigure}
\end{figure}

\clearpage
\begin{figure}[H]
\centering
\caption{Representative Charts for SAE Visual States C07--C08}
\label{fig:sae_states_c07_c08}
\exhibitnote{This figure displays representative real-market charts for the SAE features assigned to visual states C07 (Accelerating Downtrend) and C08 (Late Breakdown, displayed in two parts). Each card corresponds to one constituent feature and reports its identifier, a short description of the visual configuration shared by its high-activation charts, and a representative 20-day chart with strong activation. The SAE is a JumpReLU sparse autoencoder with 16,384 latent features trained on FT-DINO representations from 1993--2000. The 30 visually consistent features are drawn from the 100 features with the highest return-related importance and are grouped into 11 visual states, as described in Table~\ref{app:saestates}.}
\begin{subfigure}[t]{0.96\textwidth}
\centering
\includegraphics[width=\linewidth,height=0.22\textheight,keepaspectratio]{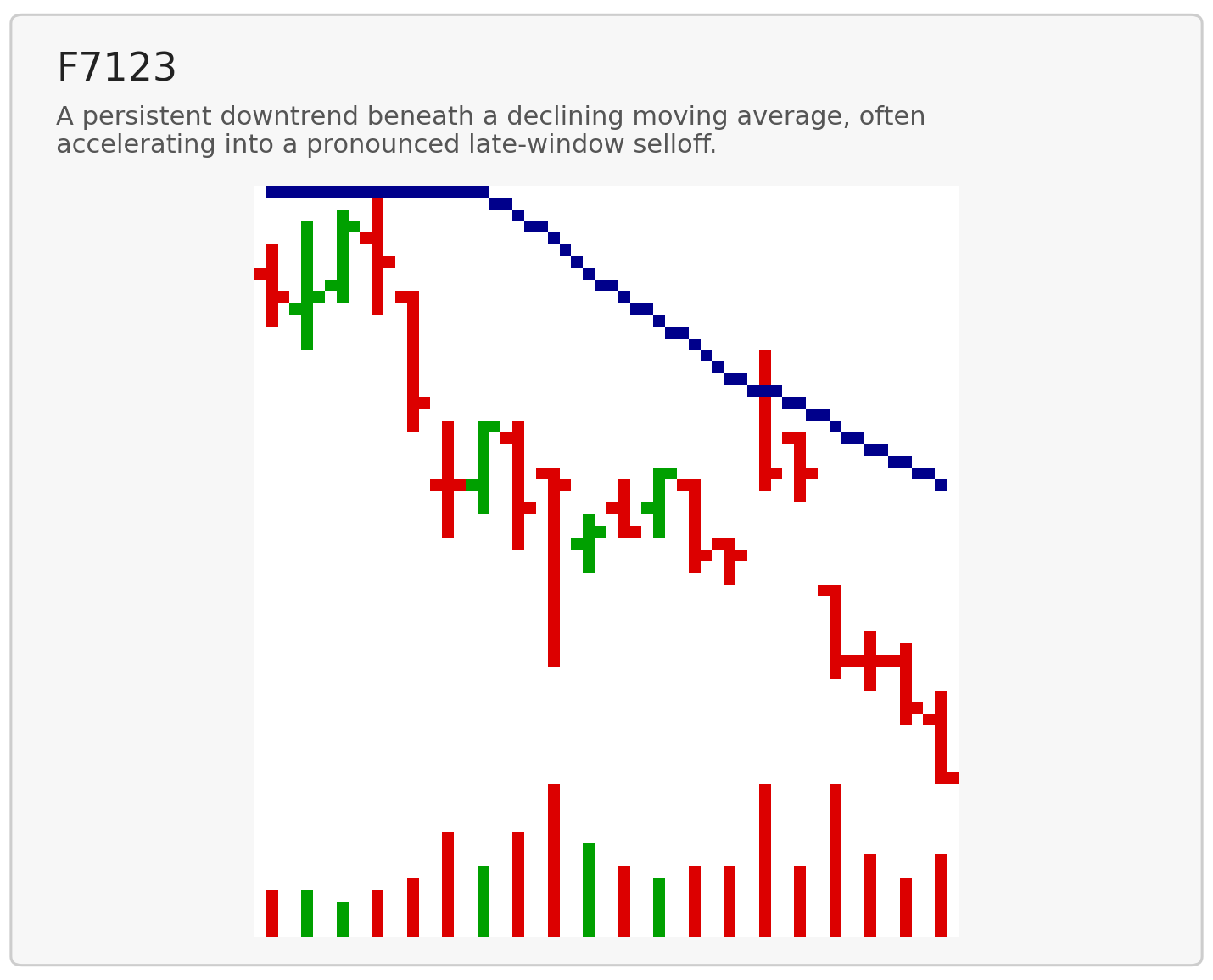}
\caption{C07: Accelerating Downtrend.}
\end{subfigure}

\vspace{0.22cm}
\begin{subfigure}[t]{0.96\textwidth}
\centering
\includegraphics[width=\linewidth,height=0.22\textheight,keepaspectratio]{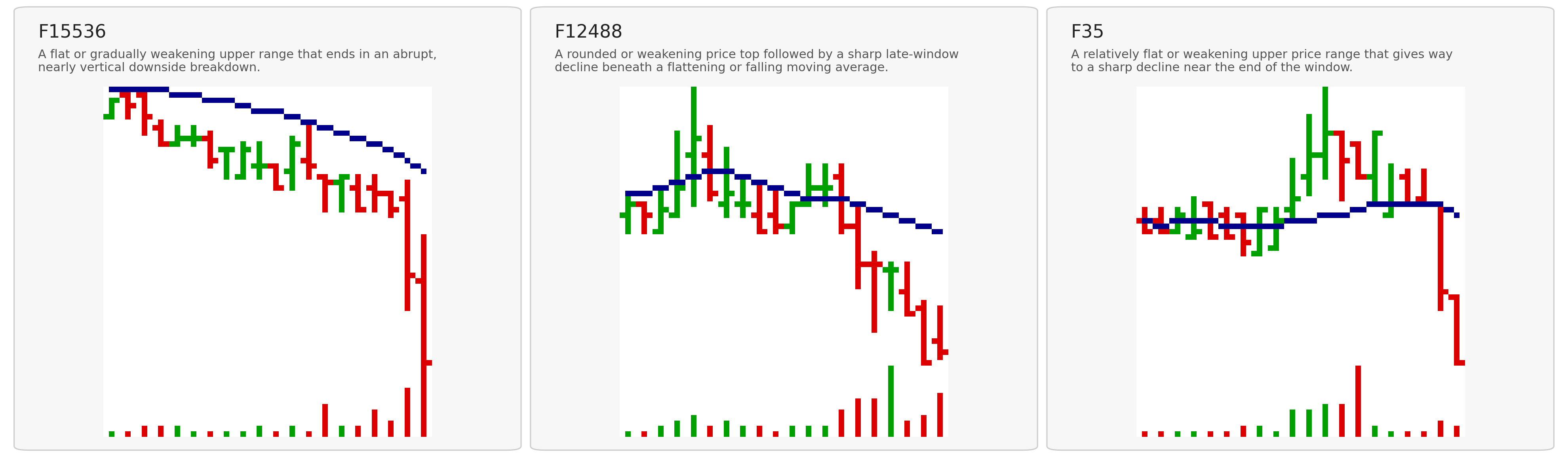}
\caption{C08: Late Breakdown, Part 1.}
\end{subfigure}

\vspace{0.22cm}
\begin{subfigure}[t]{0.96\textwidth}
\centering
\includegraphics[width=\linewidth,height=0.22\textheight,keepaspectratio]{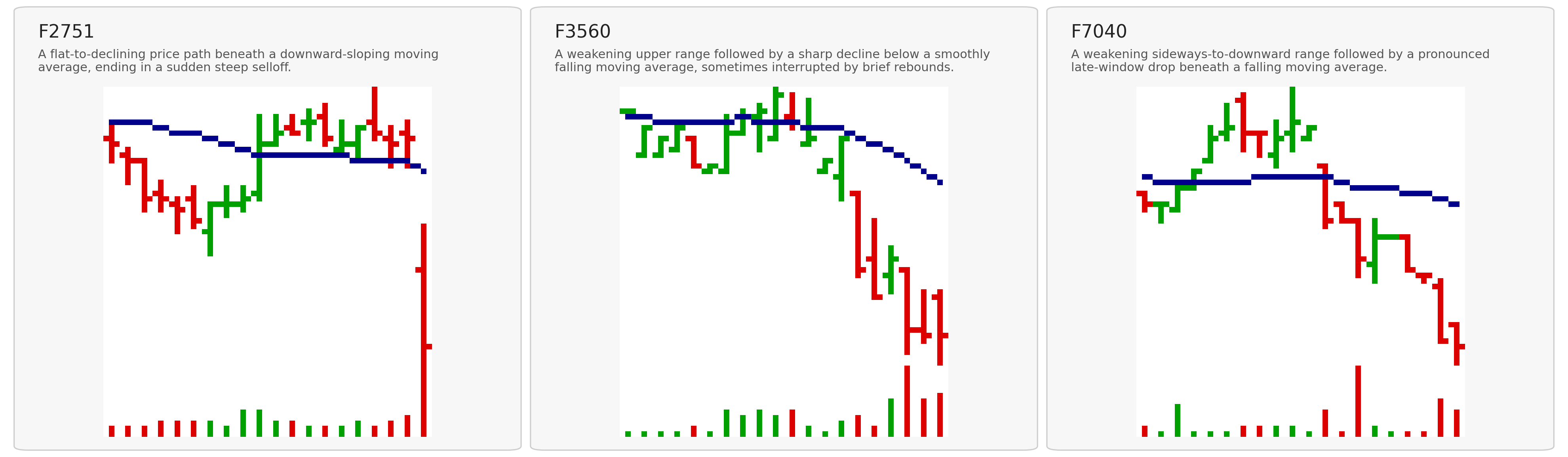}
\caption{C08: Late Breakdown, Part 2.}
\end{subfigure}
\end{figure}

\clearpage
\begin{figure}[H]
\centering
\caption{Representative Charts for SAE Visual States C09--C11}
\label{fig:sae_states_c09_c11}
\exhibitnote{This figure displays representative real-market charts for the SAE features assigned to visual states C09 (High-Volatility Range), C10 (Isolated Upward Spike), and C11 (High-Price Plateau). Each card corresponds to one constituent feature and reports its identifier, a short description of the visual configuration shared by its high-activation charts, and a representative 20-day chart with strong activation. The SAE is a JumpReLU sparse autoencoder with 16,384 latent features trained on FT-DINO representations from 1993--2000. The 30 visually consistent features are drawn from the 100 features with the highest return-related importance and are grouped into 11 visual states, as described in Table~\ref{app:saestates}.}
\begin{subfigure}[t]{0.96\textwidth}
\centering
\includegraphics[width=\linewidth,height=0.22\textheight,keepaspectratio]{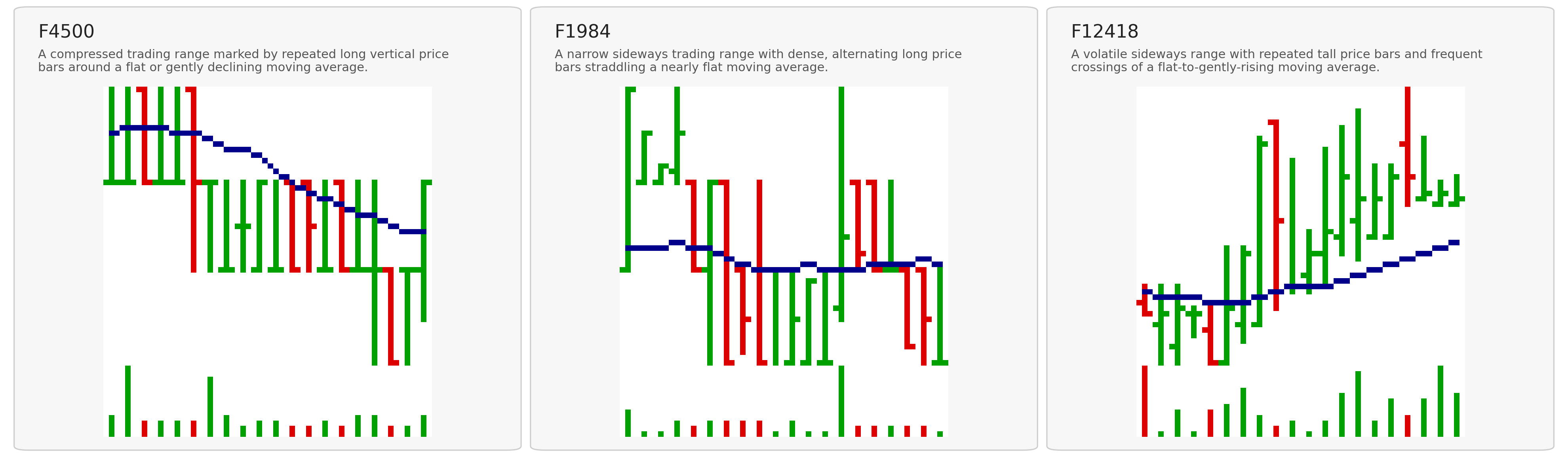}
\caption{C09: High-Volatility Range.}
\end{subfigure}

\vspace{0.22cm}
\begin{subfigure}[t]{0.96\textwidth}
\centering
\includegraphics[width=\linewidth,height=0.22\textheight,keepaspectratio]{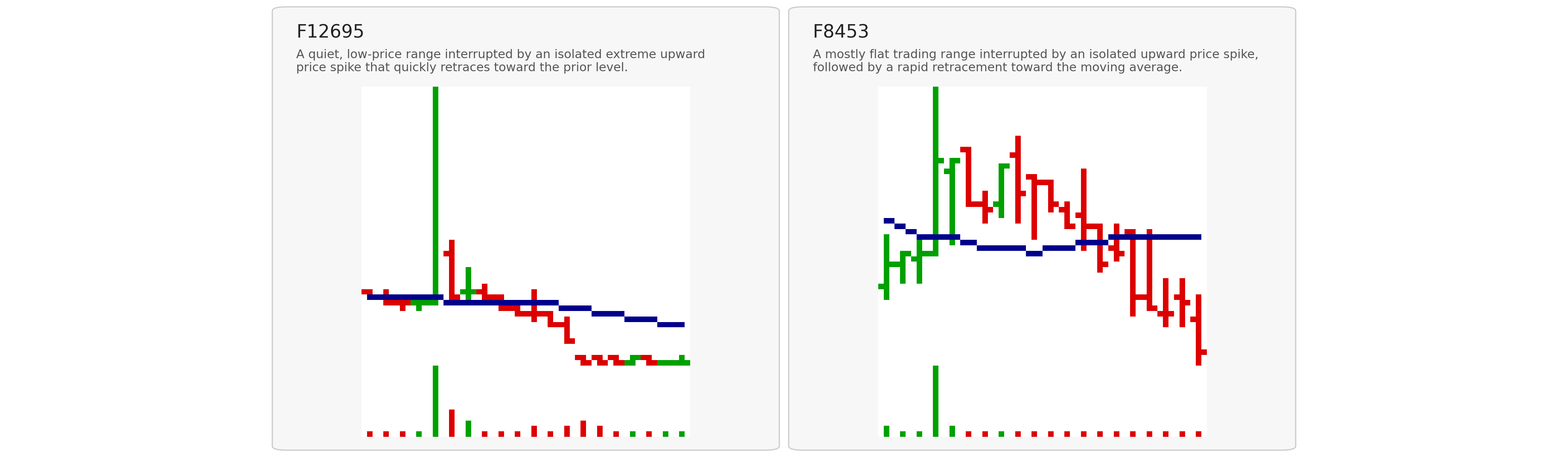}
\caption{C10: Isolated Upward Spike.}
\end{subfigure}

\vspace{0.22cm}
\begin{subfigure}[t]{0.96\textwidth}
\centering
\includegraphics[width=\linewidth,height=0.22\textheight,keepaspectratio]{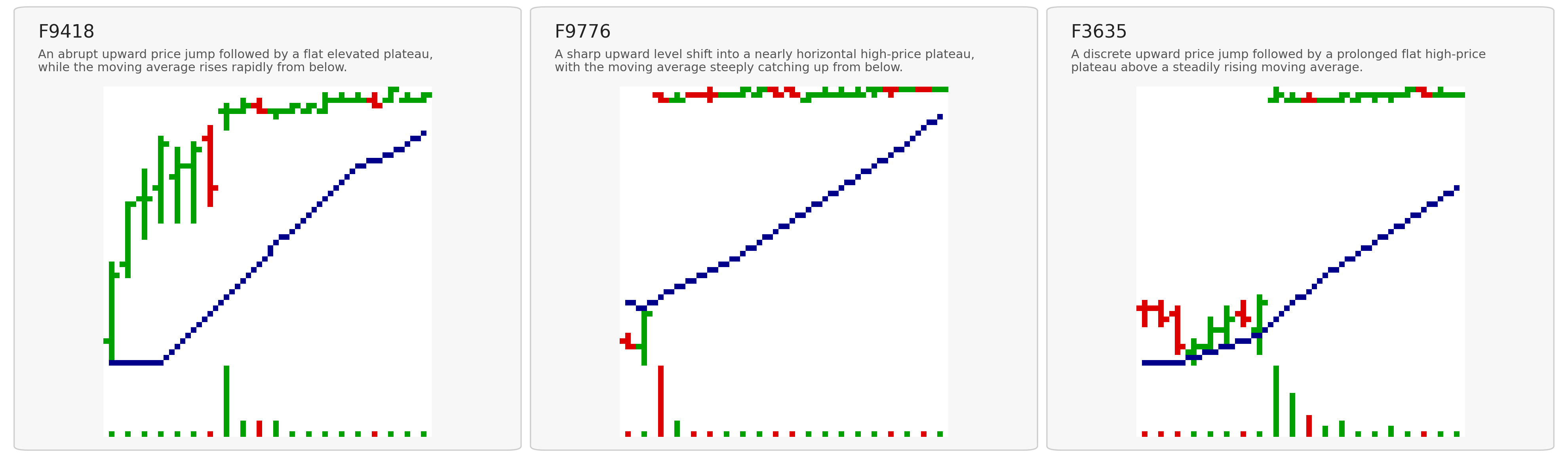}
\caption{C11: High-Price Plateau.}
\end{subfigure}
\end{figure}

\clearpage
\setappfig{F}
\subsection*{Appendix F: Top-Down Analysis}
\begin{figure}[H]
\centering
\caption{Down Momentum}
\label{fig:app_down_momentum}
\exhibitnote{This figure reports the top-down validation of FT-DINO SAE feature 7123 for Down Momentum. The upper-left panels (Real Representatives) show two of the 50 real-market 20-day charts retained as representative examples of Down Momentum, following \citet{bojstrup2026}; feature 7123 activates on the representative charts. The upper-right panels (Highest Activations) show the two real-market charts with the highest activation of feature 7123. The bottom row reports query steering: starting from a real-market query on which the feature is initially inactive (Original), the query representation is moved along the feature's SAE decoder direction by two, three, and five times the feature's maximum activation, and each panel shows the resulting nearest real-market chart. The bar beneath each steering panel indicates the feature's activation relative to its maximum. All charts follow the colored 20-day design in Figure~\ref{fig:samples}.}
\includegraphics[width=0.72\textwidth]{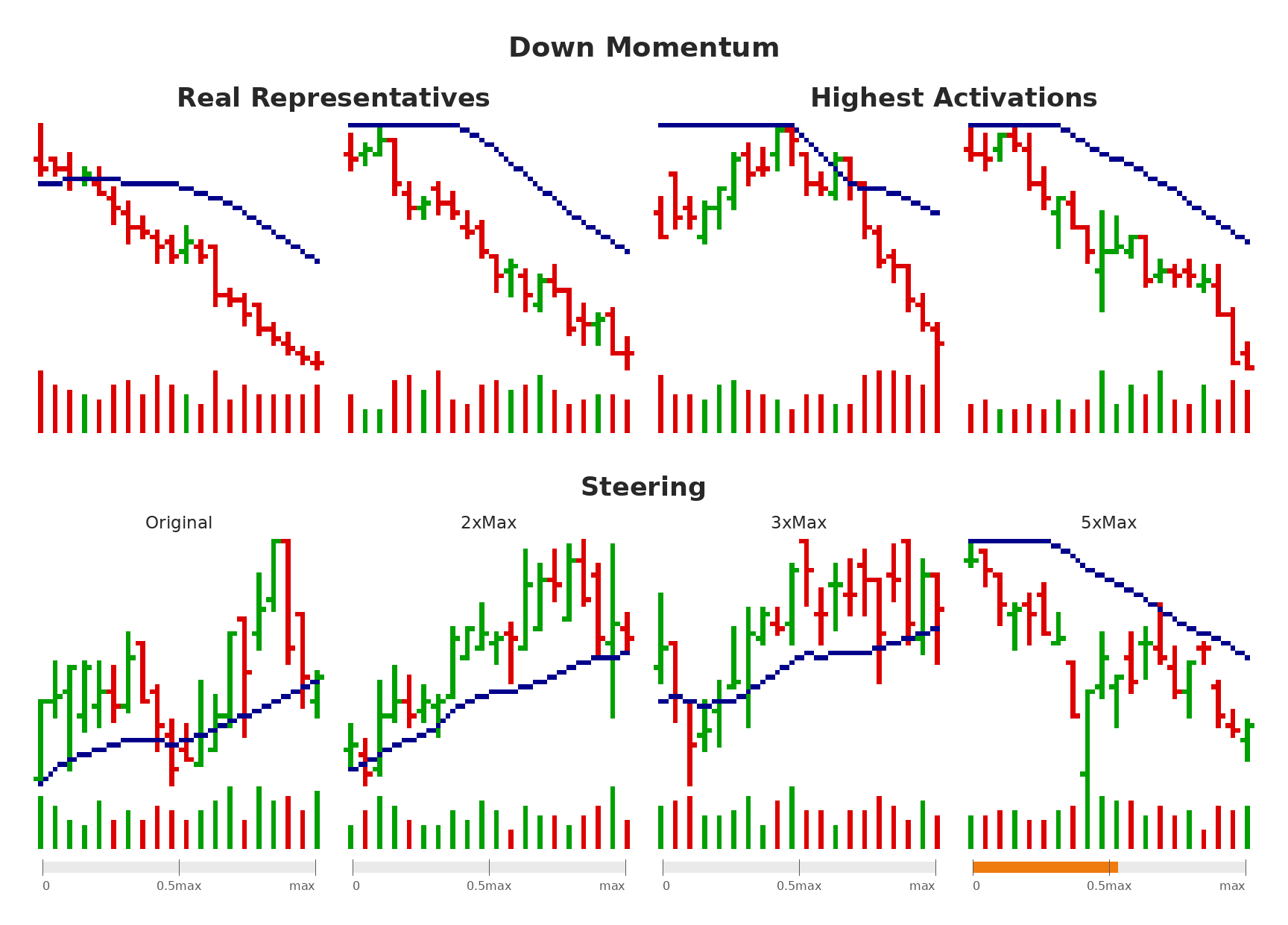}
\end{figure}

\vspace{0.50cm}

\begin{figure}[H]
\centering
\caption{Breakout Up}
\label{fig:app_break_up}
\exhibitnote{This figure reports the top-down validation of FT-DINO SAE feature 13482 for Breakout Up. The upper-left panels (Real Representatives) show two of the 50 real-market 20-day charts retained as representative examples of Breakout Up, following \citet{bojstrup2026}; feature 13482 activates on the representative charts. The upper-right panels (Highest Activations) show the two real-market charts with the highest activation of feature 13482. The bottom row reports query steering: starting from a real-market query on which the feature is initially inactive (Original), the query representation is moved along the feature's SAE decoder direction by two, three, and five times the feature's maximum activation, and each panel shows the resulting nearest real-market chart. The bar beneath each steering panel indicates the feature's activation relative to its maximum. All charts follow the colored 20-day design in Figure~\ref{fig:samples}.}
\includegraphics[width=0.72\textwidth]{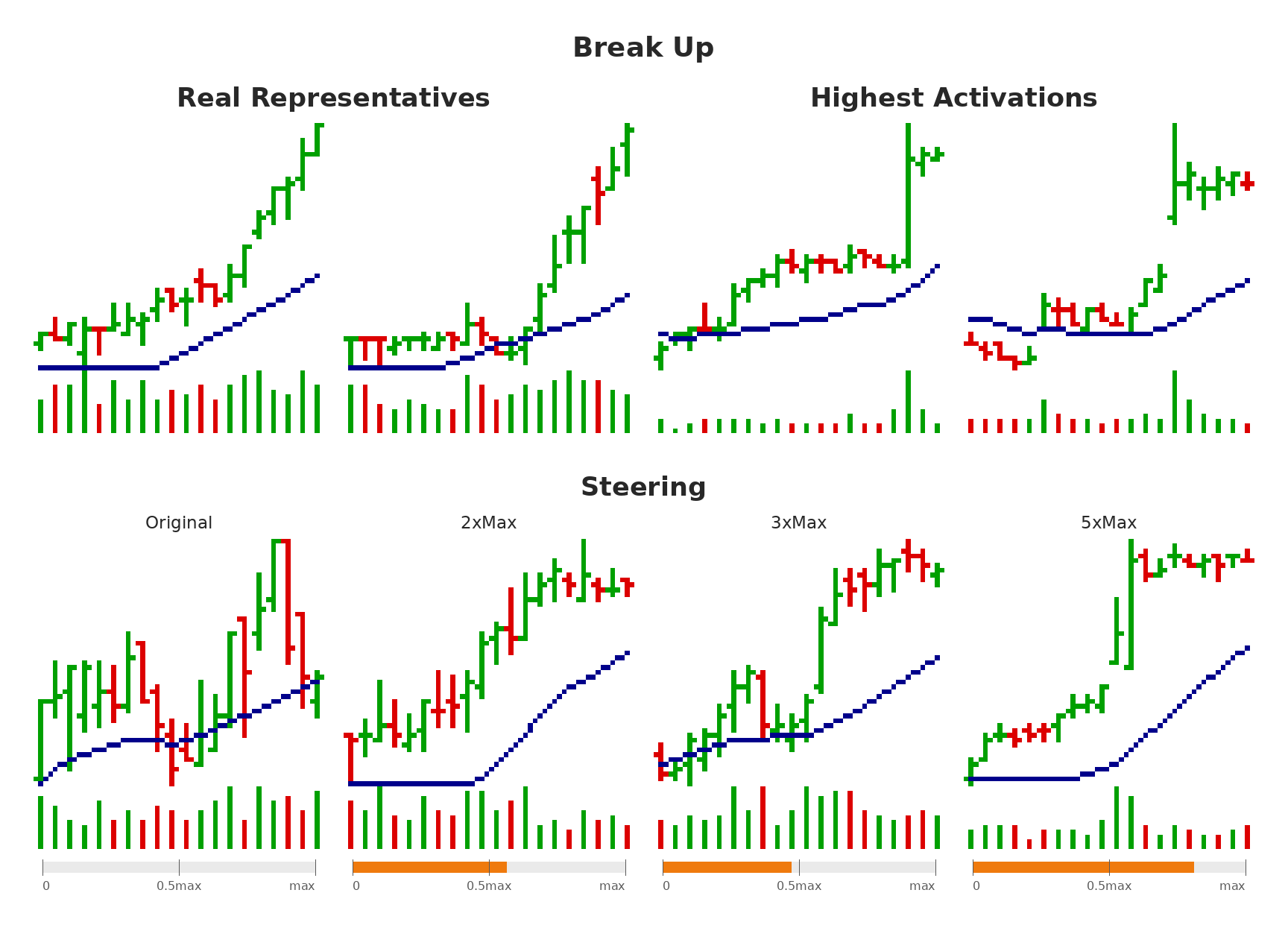}
\end{figure}

\clearpage
\begin{figure}[H]
\centering
\caption{Breakout Down}
\label{fig:app_break_down}
\exhibitnote{This figure reports the top-down validation of FT-DINO SAE feature 2751 for Breakout Down. The upper-left panels (Real Representatives) show two of the 50 real-market 20-day charts retained as representative examples of Breakout Down, following \citet{bojstrup2026}; feature 2751 activates on the representative charts. The upper-right panels (Highest Activations) show the two real-market charts with the highest activation of feature 2751. The bottom row reports query steering: starting from a real-market query on which the feature is initially inactive (Original), the query representation is moved along the feature's SAE decoder direction by two, three, and five times the feature's maximum activation, and each panel shows the resulting nearest real-market chart. The bar beneath each steering panel indicates the feature's activation relative to its maximum. All charts follow the colored 20-day design in Figure~\ref{fig:samples}.}
\includegraphics[width=0.72\textwidth]{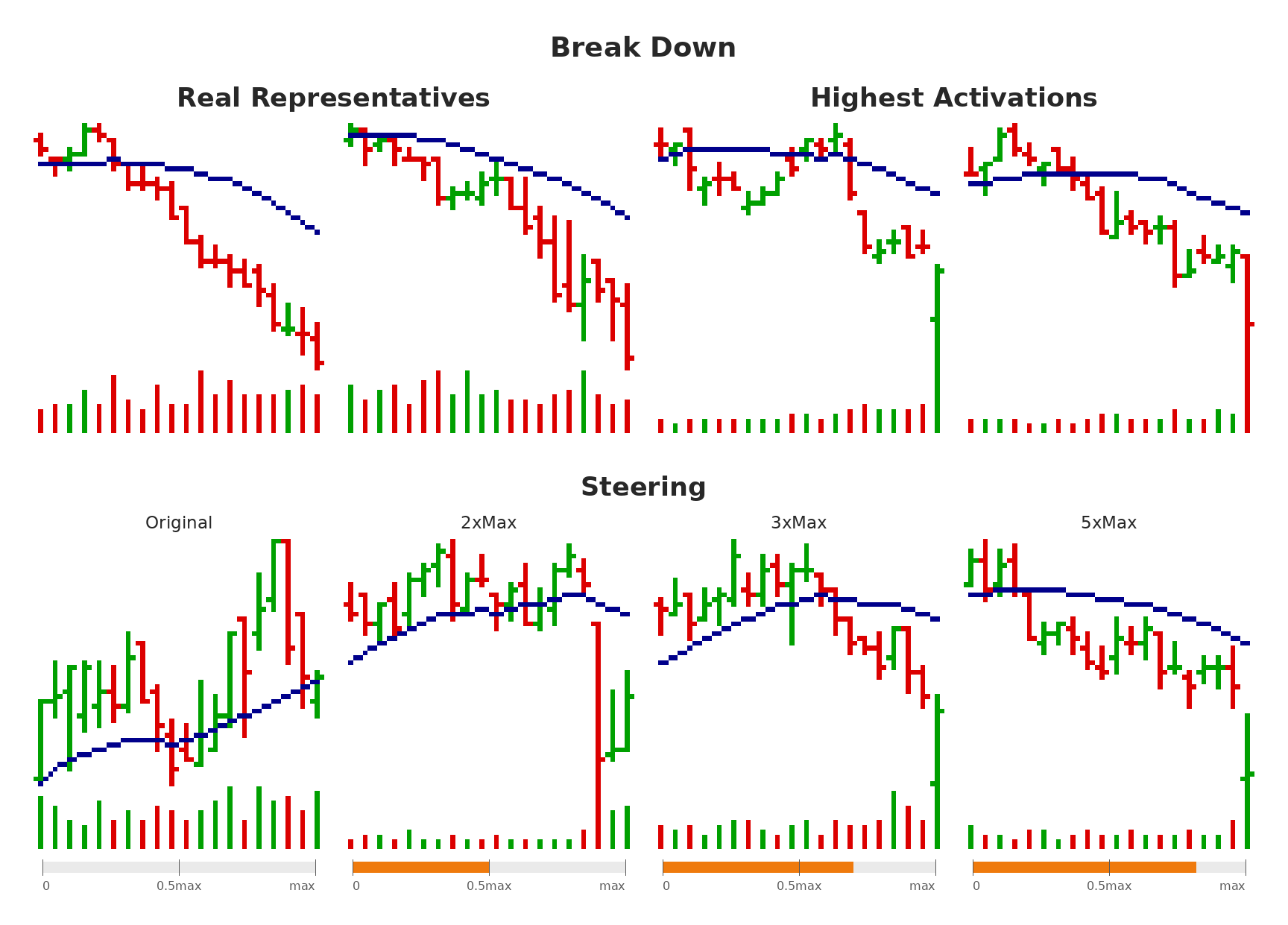}
\end{figure}

\vspace{0.50cm}

\begin{figure}[H]
\centering
\caption{Mean Reversion}
\label{fig:app_mean_reversion}
\exhibitnote{This figure reports the top-down validation of FT-DINO SAE feature 8565 for Mean Reversion. The upper-left panels (Real Representatives) show two of the 50 real-market 20-day charts retained as representative examples of Mean Reversion, following \citet{bojstrup2026}; feature 8565 activates on the representative charts. The upper-right panels (Highest Activations) show the two real-market charts with the highest activation of feature 8565. The bottom row reports query steering: starting from a real-market query on which the feature is initially inactive (Original), the query representation is moved along the feature's SAE decoder direction by two, three, and five times the feature's maximum activation, and each panel shows the resulting nearest real-market chart. The bar beneath each steering panel indicates the feature's activation relative to its maximum. All charts follow the colored 20-day design in Figure~\ref{fig:samples}.}
\includegraphics[width=0.72\textwidth]{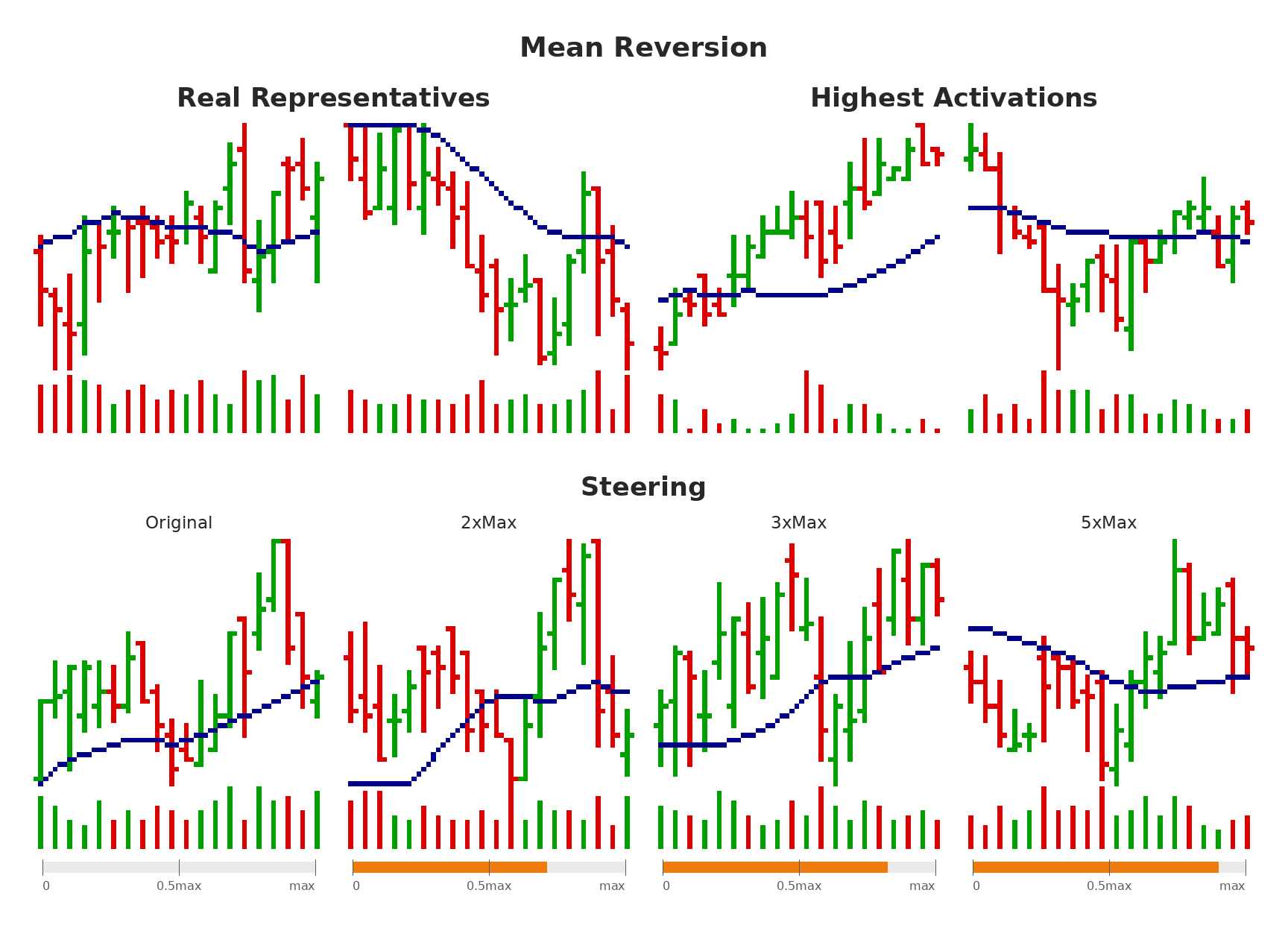}
\end{figure}

\clearpage
\begin{figure}[H]
\centering
\caption{Discrete Winner}
\label{fig:app_discrete_winner}
\exhibitnote{This figure reports the top-down validation of FT-DINO SAE feature 16299 for Discrete Winner. The upper-left panels (Real Representatives) show two of the 50 real-market 20-day charts retained as representative examples of Discrete Winner, following \citet{da2014}; feature 16299 activates on the representative charts. The upper-right panels (Highest Activations) show the two real-market charts with the highest activation of feature 16299. The bottom row reports query steering: starting from a real-market query on which the feature is initially inactive (Original), the query representation is moved along the feature's SAE decoder direction by two, three, and five times the feature's maximum activation, and each panel shows the resulting nearest real-market chart. The bar beneath each steering panel indicates the feature's activation relative to its maximum. All charts follow the colored 20-day design in Figure~\ref{fig:samples}.}
\includegraphics[width=0.72\textwidth]{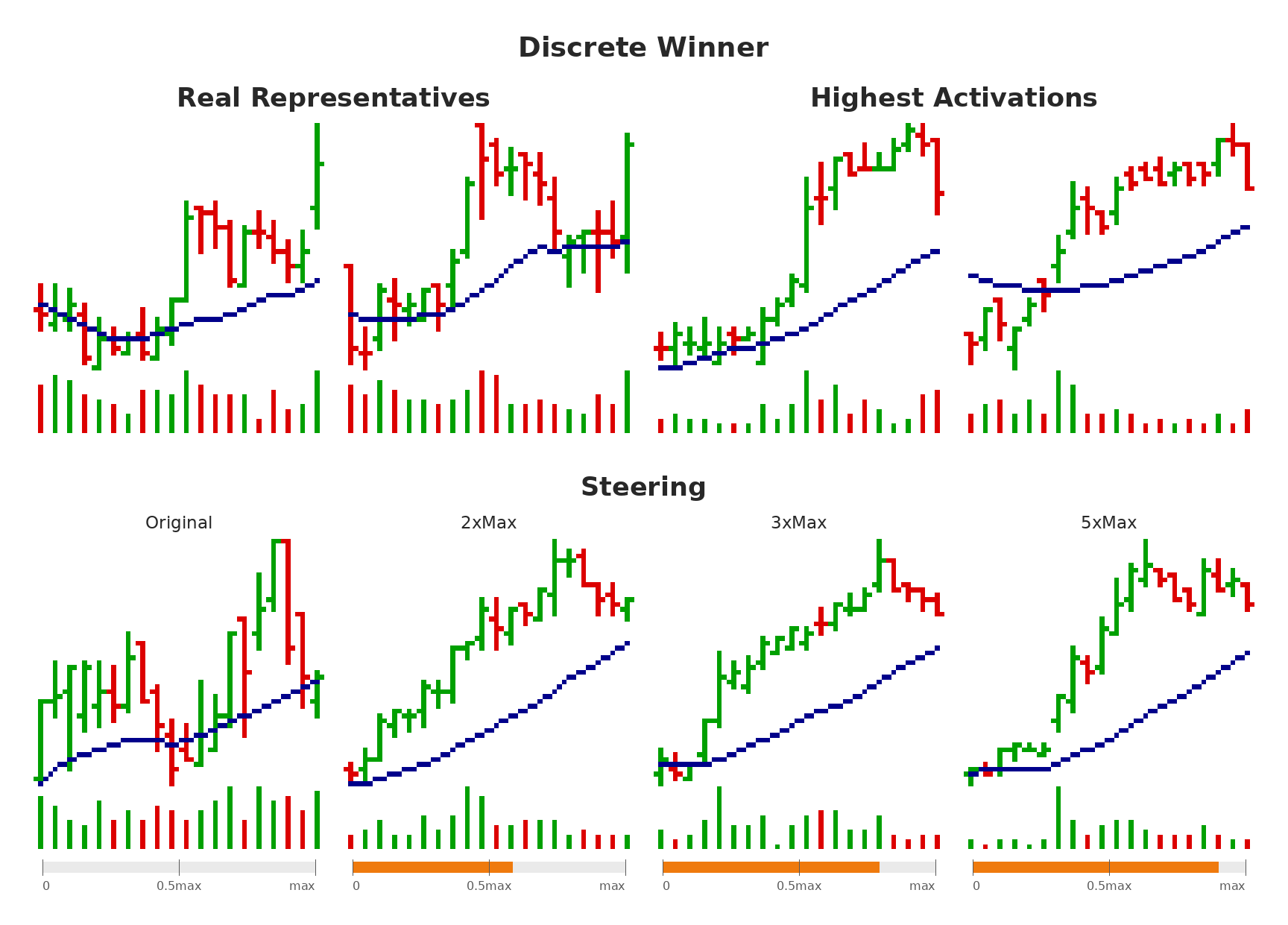}
\end{figure}

\vspace{0.50cm}

\begin{figure}[H]
\centering
\caption{Discrete Loser}
\label{fig:app_discrete_loser}
\exhibitnote{This figure reports the top-down validation of FT-DINO SAE feature 9690 for Discrete Loser. The upper-left panels (Real Representatives) show two of the 50 real-market 20-day charts retained as representative examples of Discrete Loser, following \citet{da2014}; feature 9690 activates on the representative charts. The upper-right panels (Highest Activations) show the two real-market charts with the highest activation of feature 9690. The bottom row reports query steering: starting from a real-market query on which the feature is initially inactive (Original), the query representation is moved along the feature's SAE decoder direction by two, three, and five times the feature's maximum activation, and each panel shows the resulting nearest real-market chart. The bar beneath each steering panel indicates the feature's activation relative to its maximum. All charts follow the colored 20-day design in Figure~\ref{fig:samples}.}
\includegraphics[width=0.72\textwidth]{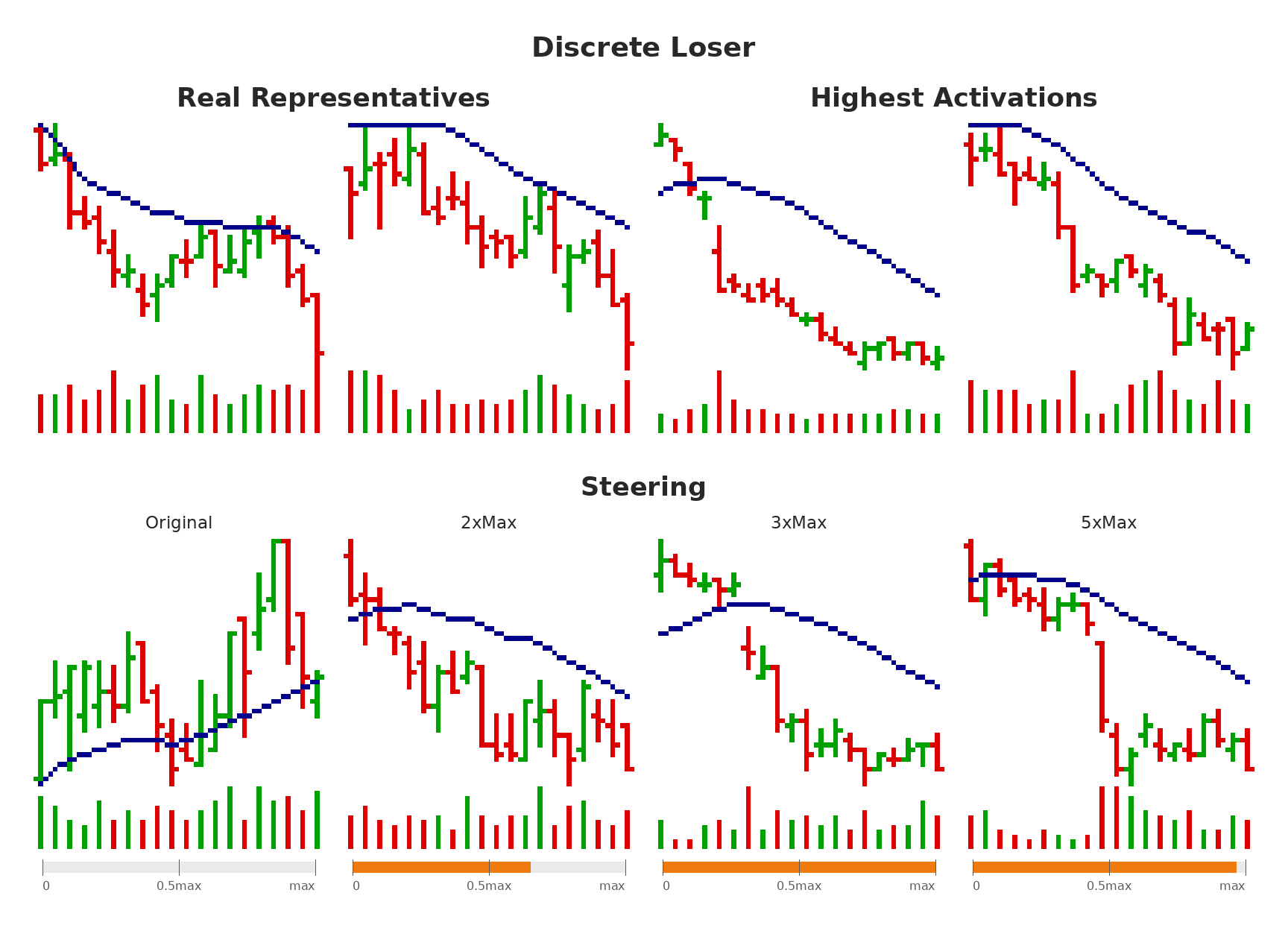}
\end{figure}

\clearpage
\begin{figure}[H]
\centering
\caption{MAX(1)}
\label{fig:app_max1}
\exhibitnote{This figure reports the top-down validation of FT-DINO SAE feature 10461 for MAX(1). The upper-left panels (Real Representatives) show two of the 50 real-market 20-day charts retained as representative examples of MAX(1), following \citet{bali2011}; feature 10461 activates on the representative charts. The upper-right panels (Highest Activations) show the two real-market charts with the highest activation of feature 10461. The bottom row reports query steering: starting from a real-market query on which the feature is initially inactive (Original), the query representation is moved along the feature's SAE decoder direction by two, three, and five times the feature's maximum activation, and each panel shows the resulting nearest real-market chart. The bar beneath each steering panel indicates the feature's activation relative to its maximum. All charts follow the colored 20-day design in Figure~\ref{fig:samples}.}
\includegraphics[width=0.72\textwidth]{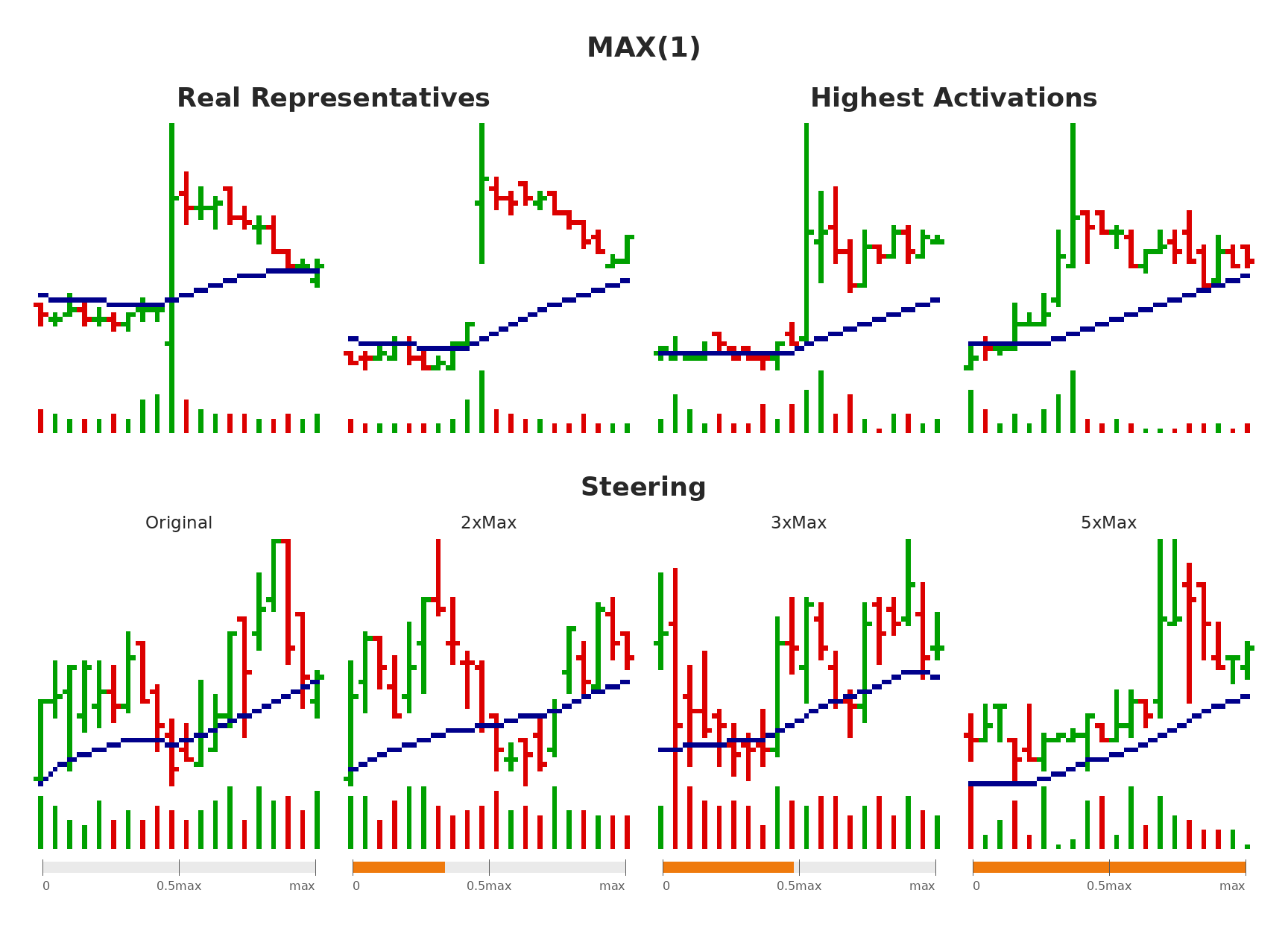}
\end{figure}

\vspace{0.50cm}

\begin{figure}[H]
\centering
\caption{Negative Overnight Jump}
\label{fig:app_neg_overnight}
\exhibitnote{This figure reports the top-down validation of FT-DINO SAE feature 4232 for Negative Overnight Jump. The upper-left panels (Real Representatives) show two of the 50 real-market 20-day charts retained as representative examples of Negative Overnight Jump, following \citet{jiangzhu2017}; feature 4232 activates on the representative charts. The upper-right panels (Highest Activations) show the two real-market charts with the highest activation of feature 4232. The bottom row reports query steering: starting from a real-market query on which the feature is initially inactive (Original), the query representation is moved along the feature's SAE decoder direction by two, three, and five times the feature's maximum activation, and each panel shows the resulting nearest real-market chart. The bar beneath each steering panel indicates the feature's activation relative to its maximum. All charts follow the colored 20-day design in Figure~\ref{fig:samples}.}
\includegraphics[width=0.72\textwidth]{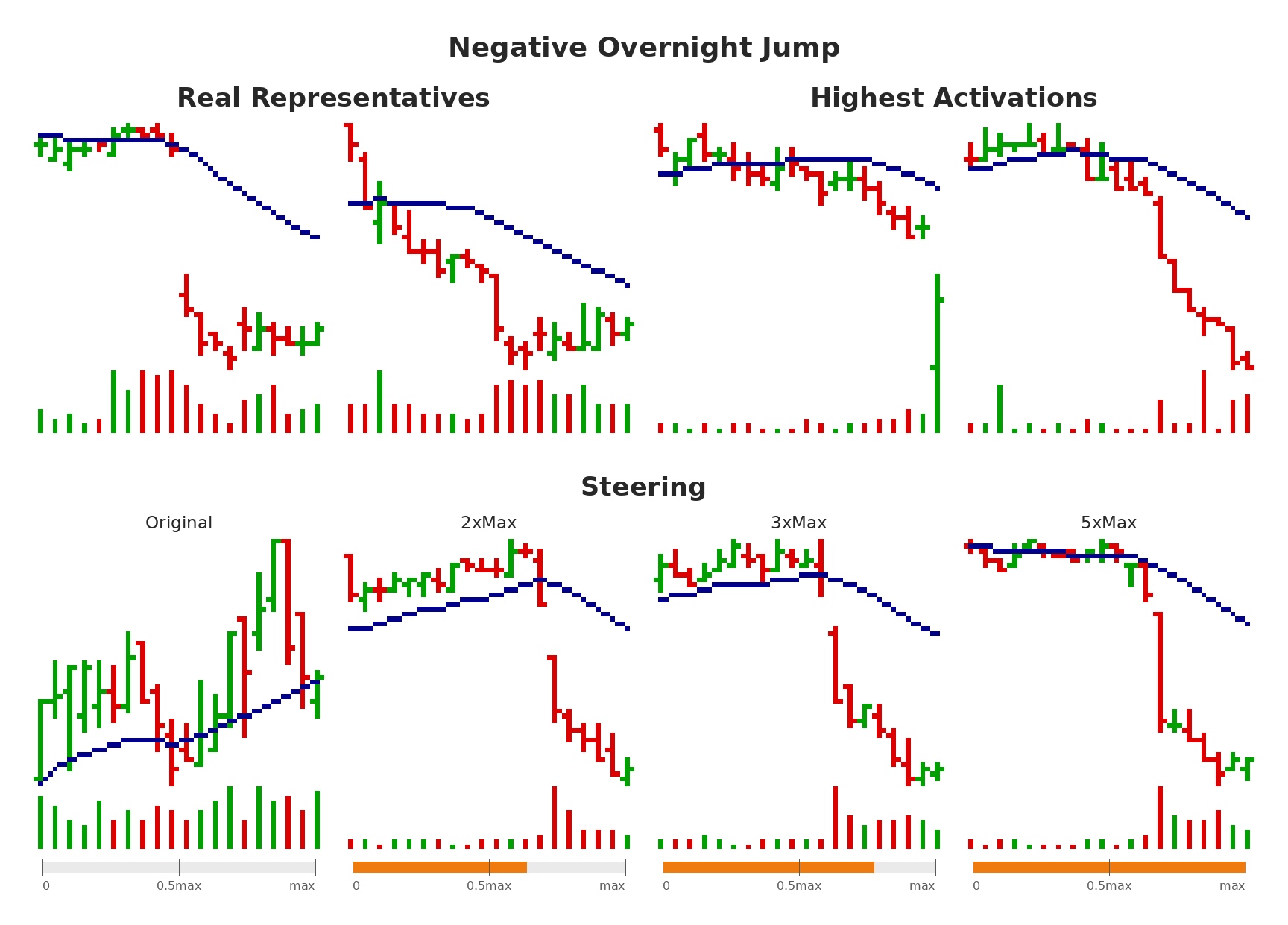}
\end{figure}

\clearpage
\begin{figure}[H]
\centering
\caption{Up + Positive Convexity}
\label{fig:app_up_pos_convexity}
\exhibitnote{This figure reports the top-down validation of FT-DINO SAE feature 6232 for Up + Positive Convexity. The upper-left panels (Real Representatives) show two of the 50 real-market 20-day charts retained as representative examples of Up + Positive Convexity, following \citet{gulenwoeppel2026}; feature 6232 activates on the representative charts. The upper-right panels (Highest Activations) show the two real-market charts with the highest activation of feature 6232. The bottom row reports query steering: starting from a real-market query on which the feature is initially inactive (Original), the query representation is moved along the feature's SAE decoder direction by two, three, and five times the feature's maximum activation, and each panel shows the resulting nearest real-market chart. The bar beneath each steering panel indicates the feature's activation relative to its maximum. All charts follow the colored 20-day design in Figure~\ref{fig:samples}.}
\includegraphics[width=0.72\textwidth]{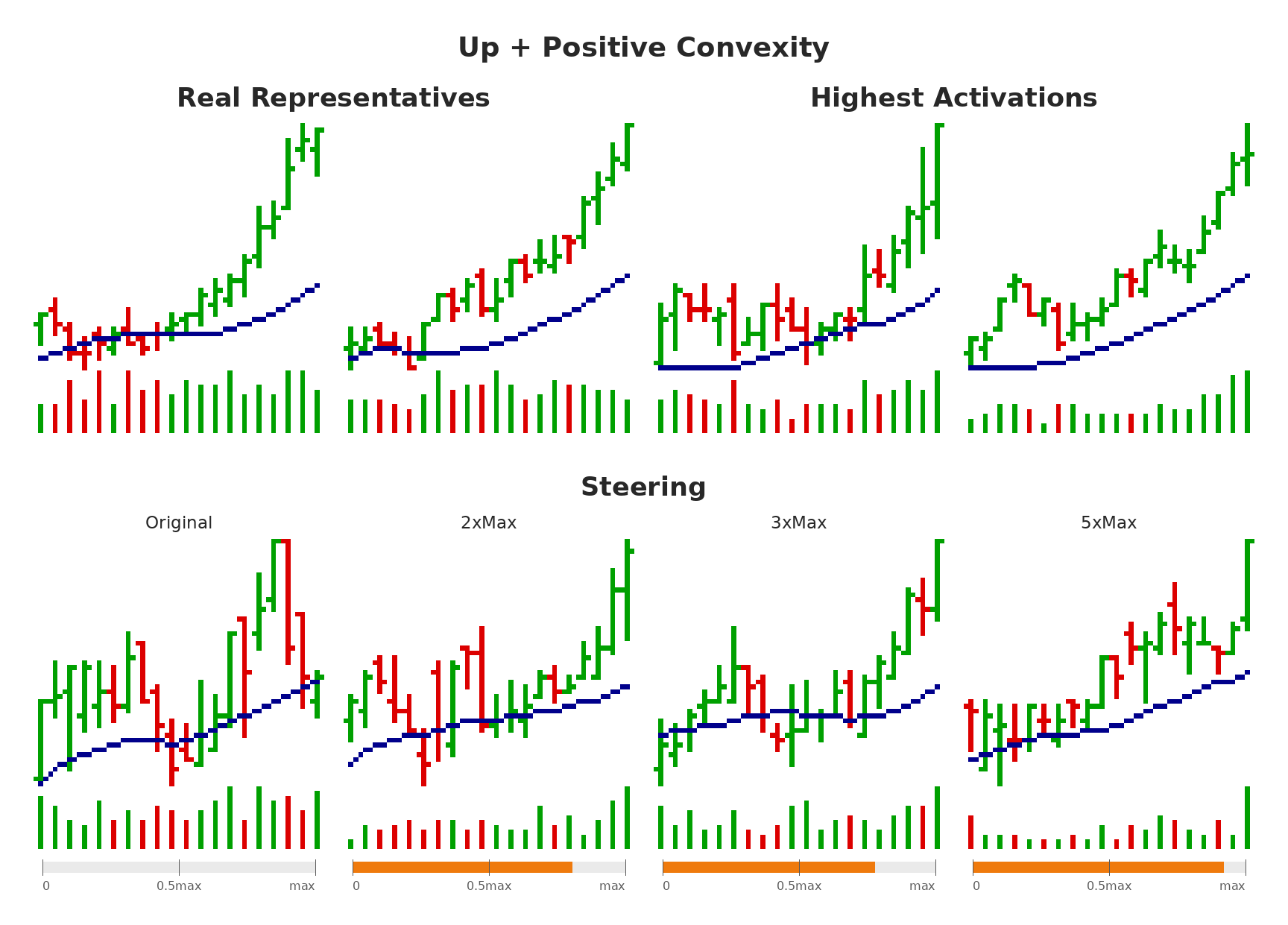}
\end{figure}

\vspace{0.50cm}

\begin{figure}[H]
\centering
\caption{Down + Positive Convexity}
\label{fig:app_down_pos_convexity}
\exhibitnote{This figure reports the top-down validation of FT-DINO SAE feature 2940 for Down + Positive Convexity. The upper-left panels (Real Representatives) show two of the 50 real-market 20-day charts retained as representative examples of Down + Positive Convexity, following \citet{gulenwoeppel2026}; feature 2940 activates on the representative charts. The upper-right panels (Highest Activations) show the two real-market charts with the highest activation of feature 2940. The bottom row reports query steering: starting from a real-market query on which the feature is initially inactive (Original), the query representation is moved along the feature's SAE decoder direction by two, three, and five times the feature's maximum activation, and each panel shows the resulting nearest real-market chart. The bar beneath each steering panel indicates the feature's activation relative to its maximum. All charts follow the colored 20-day design in Figure~\ref{fig:samples}.}
\includegraphics[width=0.72\textwidth]{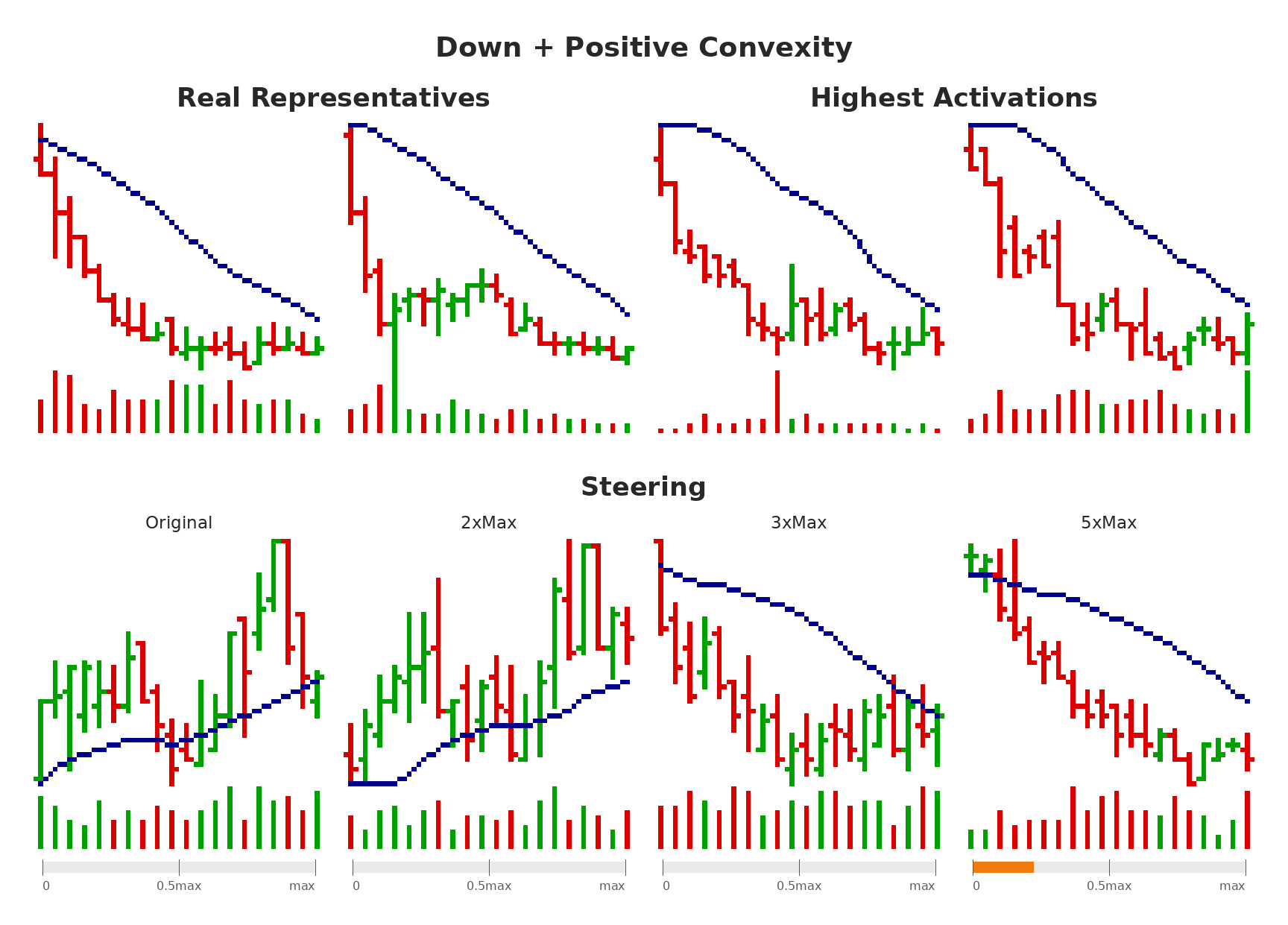}
\end{figure}

\clearpage\section*{Appendix Tables}\addcontentsline{toc}{section}{Appendix Tables}
\setapptab{B}
\subsection*{Appendix B: Historical Visual Retrieval with Pretrained DINO}

\begin{table}[H]\caption{Historical-Library Restrictions}
\label{tab:app_dino_universe}
\tablenote{This table reports benchmark DINO equal-weighted high-minus-low portfolio performance over 2001--2025 under alternative historical retrieval libraries. For each stock, on the last trading day of each week, the retrieval signal is the equal-weighted average of the realized subsequent five-trading-day returns of its $K$ most similar historical charts, where similarity is the cosine similarity between normalized chart representations and the recursively expanding daily library contains only observations whose subsequent five-day returns are fully realized by the query date. Each week, stocks are sorted into deciles by the signal and held over the subsequent five trading days; H--L is the return on decile 10 minus decile 1. Benchmark uses all eligible historical observations; Rolling 8Y retains only observations from the preceding eight years; Same-Ind. restricts retrieval to the query stock's Fama--French 12 industry; Cross-Ind. excludes that industry; and Same-Stock retains only the query stock's own eligible history. The representation, retrieval rule, and portfolio construction are otherwise unchanged. Ret is the annualized mean H--L return, and SR is the annualized Sharpe ratio.}

\apptablesize
\centering
\begin{tabular*}{\textwidth}{@{\extracolsep{\fill}}c cc cc cc cc cc@{}}
            \toprule
            & \multicolumn{2}{c}{Benchmark}
            & \multicolumn{2}{c}{Rolling 8Y}
            & \multicolumn{2}{c}{Same-Ind.}
            & \multicolumn{2}{c}{Cross-Ind.}
            & \multicolumn{2}{c}{Same-Stock} \\
            \cmidrule(lr){2-3}
            \cmidrule(lr){4-5}
            \cmidrule(lr){6-7}
            \cmidrule(lr){8-9}
            \cmidrule(lr){10-11}

            $K$
            & Ret & SR
            & Ret & SR
            & Ret & SR
            & Ret & SR
            & Ret & SR \\
            \midrule
            5
            & 0.05 & 0.83
            & 0.05 & 0.76
            & 0.07 & 1.09
            & 0.04 & 0.74
            & -0.03 & -0.38 \\

            10
            & 0.07 & 1.12
            & 0.06 & 0.96
            & 0.07 & 1.14
            & 0.05 & 0.84
            & -0.03 & -0.36 \\

            30
            & 0.11 & 1.83
            & 0.08 & 1.24
            & 0.09 & 1.50
            & 0.11 & 1.74
            & -0.05 & -0.49 \\

            100
            & 0.16 & 2.15
            & 0.15 & 2.03
            & 0.14 & 1.95
            & 0.15 & 2.12
            & -0.06 & -0.55 \\

            500
            & 0.23 & 2.74
            & 0.20 & 2.30
            & 0.19 & 2.27
            & 0.23 & 2.76
            & -0.09 & -0.63 \\

            1{,}000
            & 0.26 & 2.81
            & 0.22 & 2.48
            & 0.21 & 2.36
            & 0.25 & 2.80
            & -0.10 & -0.64 \\

            3{,}000
            & 0.29 & 2.90
            & 0.23 & 2.28
            & 0.23 & 2.41
            & 0.27 & 2.76
            & -0.10 & -0.64 \\
            \bottomrule
\end{tabular*}
\end{table}
\clearpage

\begin{table}[H]\caption{Neighbor Weighting: Equal versus Recency-Weighted}
\label{tab:app_dino_recency}
\tablenote{This table compares equal and recency-weighted aggregation of benchmark DINO neighbors with $K=1{,}000$ over 2001--2025. Equal assigns the same weight to each retrieved neighbor's realized subsequent five-day return. H1Y--H25Y apply exponentially declining weights according to the age of the historical observation, with half-lives of 1, 2, 5, 10, 15, 20, and 25 years. The retrieved neighbor set and portfolio construction are unchanged. Each week, stocks are sorted into deciles by the signal and held over the subsequent five trading days; H--L is the return on decile 10 minus decile 1. EW, VW, and CVW denote equal-weighted, value-weighted, and capped value-weighted portfolios, respectively; capped value weighting winsorizes market capitalization at the contemporaneous NYSE 80th percentile following \citet{jensen2023}. Ret is the annualized mean H--L return, and SR is the annualized Sharpe ratio.}

\apptablesize
\centering
\begin{tabular*}{\textwidth}{@{\extracolsep{\fill}}c cc cc cc@{}}
            \toprule
            & \multicolumn{2}{c}{EW}
            & \multicolumn{2}{c}{VW}
            & \multicolumn{2}{c}{CVW} \\
            \cmidrule(lr){2-3}
            \cmidrule(lr){4-5}
            \cmidrule(lr){6-7}

            Neighbor weighting
            & Ret & SR
            & Ret & SR
            & Ret & SR \\
            \midrule

            Equal
            & 0.26 & 2.81
            & 0.07 & 0.60
            & 0.07 & 0.87 \\

            H1Y
            & 0.10 & 1.16
            & 0.00 & 0.03
            & 0.01 & 0.13 \\

            H2Y
            & 0.14 & 1.69
            & 0.03 & 0.24
            & 0.02 & 0.32 \\

            H5Y
            & 0.22 & 2.60
            & 0.06 & 0.53
            & 0.06 & 0.75 \\

            H10Y
            & 0.24 & 2.77
            & 0.07 & 0.58
            & 0.07 & 0.84 \\

            H15Y
            & 0.25 & 2.72
            & 0.07 & 0.58
            & 0.07 & 0.83 \\

            H20Y
            & 0.25 & 2.73
            & 0.06 & 0.52
            & 0.07 & 0.81 \\

            H25Y
            & 0.25 & 2.75
            & 0.06 & 0.53
            & 0.07 & 0.82 \\

            \bottomrule
\end{tabular*}
\end{table}
\clearpage

\begin{table}[H]\caption{Pretrained Vision Encoders}
\label{tab:encoderspecs}
\tablenote{This table summarizes the pretrained vision encoders used in the robustness tests. For each model, it reports the pretraining objective, how the image embedding is extracted from a price chart, and the embedding dimension and training data. DINOv2-S and DINOv2-L are self-supervised on LVD-142M \citep{oquab2024dinov2}, MAE is a masked autoencoder self-supervised on ImageNet-1K \citep{he2022mae}, and CLIP is pretrained with image--text contrastive learning on approximately 400 million internet image--text pairs \citep{radford2021clip}. None of the encoders is trained with stock-return labels or a finance-specific objective.}

\apptablesize
\centering
\begin{threeparttable}

\setlength{\tabcolsep}{3pt}
\renewcommand{\arraystretch}{1.08}
\begin{tabularx}{\textwidth}{@{}>{\raggedright\arraybackslash}p{2.0cm}>{\raggedright\arraybackslash}p{3.15cm}YY@{}}
\toprule
Model & Pretraining Objective & Embedding Extraction & Output and Features \\
\midrule
\mbox{DINOv2-S} & Self-supervised visual representation learning & Resize price images to $518\times518$ and directly extract the final-layer CLS token. & 384 dimensions; emphasizes global image structure and cross-image similarity. \\
\midrule
\mbox{DINOv2-L} & Self-supervised visual representation learning & Use the same preprocessing as DINOv2-S and directly extract the final-layer CLS token. & 1,024 dimensions; same pretraining objective with a larger model. \\
\midrule
MAE & Masked image reconstruction & Use a ViT-Base encoder and take its output CLS token as the image embedding. & 768 dimensions; learns visual structure by reconstructing masked patches. \\
\midrule
CLIP & Image--text contrastive learning & Use a ViT-B/32 image encoder and extract the global image embedding after visual projection. & 512 dimensions; images lie in a shared semantic space aligned with text. \\
\bottomrule
\end{tabularx}
\end{threeparttable}

\end{table}
\clearpage

\begin{table}[H]\caption{Portfolio Performance across Pretrained Encoders}
\label{tab:encodercompare}
\tablenote{This table reports equal-weighted high-minus-low portfolio performance over 2001--2025 when the retrieval signal is computed from four pretrained vision encoders described in Table~\ref{tab:encoderspecs}. All encoders process the same price charts, and cosine similarity, the recursively expanding historical library, the eligibility rule, and the equal-weighted averaging of neighbors' realized subsequent five-day returns are unchanged. Each week, stocks are sorted into deciles by the signal and held over the subsequent five trading days; H--L is the return on decile 10 minus decile 1. $K$ is the number of retrieved neighbors. Ret is the annualized mean H--L return, and SR is the annualized Sharpe ratio.}

\apptablesize
\centering
\begin{threeparttable}
\begin{tabular*}{\textwidth}{@{\extracolsep{\fill}}c cc cc cc cc@{}}
\toprule
& \multicolumn{2}{c}{DINOv2-S} & \multicolumn{2}{c}{DINOv2-L} & \multicolumn{2}{c}{MAE} & \multicolumn{2}{c}{CLIP} \\
\cmidrule(lr){2-3}\cmidrule(lr){4-5}\cmidrule(lr){6-7}\cmidrule(lr){8-9}
$K$ & Ret & SR & Ret & SR & Ret & SR & Ret & SR \\
\midrule
5    & 0.05 & 0.83 & 0.07 & 1.09 & 0.06 & 0.95 & 0.04 & 0.63 \\
10   & 0.07 & 1.12 & 0.08 & 1.17 & 0.07 & 1.02 & 0.06 & 1.00 \\
30   & 0.11 & 1.83 & 0.12 & 1.77 & 0.12 & 1.85 & 0.09 & 1.32 \\
100  & 0.16 & 2.15 & 0.18 & 2.42 & 0.18 & 2.51 & 0.13 & 1.79 \\
500  & 0.23 & 2.74 & 0.25 & 3.04 & 0.28 & 3.09 & 0.24 & 2.72 \\
1{,}000 & 0.26 & 2.81 & 0.28 & 3.17 & 0.32 & 3.26 & 0.28 & 2.91 \\
3{,}000 & 0.29 & 2.90 & 0.31 & 3.01 & 0.34 & 3.16 & 0.32 & 2.79 \\
\bottomrule
\end{tabular*}
\end{threeparttable}

\end{table}
\clearpage
\begin{table}[H]\caption{Numerical, Text, and Image Representations}
\label{tab:app_nonvisual}
\tablenote{This table reports high-minus-low portfolio performance over 2001--2025 when the retrieval signal is computed from different representations of the same 20-day price and volume history, with $K=1{,}000$. Raw OHLCV stacks the daily open, high, low, and close prices, the 20-day moving average, and volume into a 120-dimensional vector, with each variable standardized by its 1993--2000 mean and standard deviation. ChronoBERT embeds the same variables, written as structured text with one line per day, with prices normalized by the high--low range of the window and volume scaled by its within-window maximum, using the model of \citet{he2025chrono} pretrained on text available before 2001; the representation is the mean of the last-layer token representations. DINO and FT-DINO use the price charts. Every representation is normalized to unit length, and the eligibility rule, cosine-similarity search, and equal-weighted averaging of neighbors' realized subsequent five-day returns are identical. Ret is the annualized mean H--L return, and SR is the annualized Sharpe ratio, under equal weighting (EW) and capped value weighting (CVW).}
\apptablesize
\centering
\begin{tabular*}{\textwidth}{@{\extracolsep{\fill}}l cc cc@{}}
\toprule
& \multicolumn{2}{c}{EW} & \multicolumn{2}{c}{CVW} \\
\cmidrule(lr){2-3}\cmidrule(lr){4-5}
Representation & Ret & SR & Ret & SR \\
\midrule
Raw OHLCV & 0.19 & 2.47 & 0.05 & 0.70 \\
ChronoBERT & 0.10 & 1.42 & 0.01 & 0.23 \\
\addlinespace[2pt]
DINO & 0.26 & 2.81 & 0.07 & 0.87 \\
FT-DINO & 0.85 & 6.37 & 0.28 & 2.38 \\
\bottomrule
\end{tabular*}
\end{table}
\clearpage

\begin{table}[H]\caption{DINO--CNN Factor-Augmented Spanning Regressions}
\label{tab:spanningnoft}
\tablenote{This table reports weekly factor-augmented spanning regressions of benchmark DINO and CNN high-minus-low portfolio returns over 2001--2025, before introducing return alignment. In each column, the dependent variable is the weekly H--L return of the strategy named in the column header, formed under the indicated weighting scheme (EW or CVW). The regressors are the other image strategy's H--L return under the same weighting scheme, together with the Fama--French five factors, momentum, and short-term reversal (Factor Controls). Benchmark DINO uses $K=1{,}000$, and CNN denotes the 20-day black-and-white price-image CNN of \citet{jiang2023}. Alpha is 52 times the weekly intercept, and $t$-statistics are reported in parentheses.}

\apptablesize
\centering
\begin{threeparttable}
\begin{tabular*}{\textwidth}{@{\extracolsep{\fill}}lcccc@{}}
\toprule
& \multicolumn{2}{c}{EW} & \multicolumn{2}{c}{CVW} \\
\cmidrule(lr){2-3}\cmidrule(lr){4-5}
Variable & DINO & CNN & DINO & CNN \\
\midrule
Alpha & 0.1104 & 0.4436 & 0.0202 & 0.1522 \\
      & (3.47) & (14.26) & (1.17) & (6.91) \\[2pt]
DINO & & 0.4633 & & 0.3835 \\
     & & (5.49) & & (3.52) \\[2pt]
CNN & 0.2463 & & 0.2130 & \\
    & (6.07) & & (4.79) & \\[2pt]
Factor Controls & Yes & Yes & Yes & Yes \\
\midrule
Adj. $R^2$ & 0.1635 & 0.2815 & 0.1286 & 0.2327 \\
\bottomrule
\end{tabular*}
\end{threeparttable}
\end{table}

\clearpage
\setapptab{C}
\subsection*{Appendix C: Additional Return-Aligned Retrieval Results}

\begin{table}[H]\caption{FT-DINO Neighbor Weighting: Equal versus Recency-Weighted}
\label{tab:app_ft_recency}
\tablenote{This table compares equal and recency-weighted aggregation of FT-DINO neighbors with $K=1{,}000$ over 2001--2025. Equal assigns the same weight to each retrieved neighbor's realized subsequent five-day return. H1Y--H25Y apply exponentially declining weights according to the age of the historical observation, with half-lives of 1, 2, 5, 10, 15, 20, and 25 years. The retrieved neighbor set and portfolio construction are unchanged. Each week, stocks are sorted into deciles by the signal and held over the subsequent five trading days; H--L is the return on decile 10 minus decile 1. EW, VW, and CVW denote equal-weighted, value-weighted, and capped value-weighted portfolios, respectively; capped value weighting winsorizes market capitalization at the contemporaneous NYSE 80th percentile following \citet{jensen2023}. Ret is the annualized mean H--L return, and SR is the annualized Sharpe ratio.}

\apptablesize
\centering
\begin{tabular*}{\textwidth}{@{\extracolsep{\fill}}c cc cc cc@{}}
            \toprule
            & \multicolumn{2}{c}{EW}
            & \multicolumn{2}{c}{VW}
            & \multicolumn{2}{c}{CVW} \\
            \cmidrule(lr){2-3}
            \cmidrule(lr){4-5}
            \cmidrule(lr){6-7}

            Neighbor weighting
            & Ret & SR
            & Ret & SR
            & Ret & SR \\
            \midrule

            Equal
            & 0.85 & 6.37
            & 0.24 & 1.45
            & 0.28 & 2.38 \\

            H1Y
            & 0.55 & 4.61
            & 0.17 & 1.10
            & 0.18 & 1.65 \\

            H2Y
            & 0.67 & 5.36
            & 0.18 & 1.18
            & 0.22 & 1.99 \\

            H5Y
            & 0.80 & 5.98
            & 0.23 & 1.43
            & 0.26 & 2.29 \\

            H10Y
            & 0.83 & 6.21
            & 0.23 & 1.41
            & 0.27 & 2.31 \\

            H15Y
            & 0.85 & 6.26
            & 0.23 & 1.43
            & 0.27 & 2.32 \\

            H20Y
            & 0.85 & 6.29
            & 0.24 & 1.47
            & 0.28 & 2.37 \\

            H25Y
            & 0.85 & 6.33
            & 0.23 & 1.45
            & 0.28 & 2.38 \\

            \bottomrule
\end{tabular*}
\end{table}

\clearpage
\begin{table}[H]\caption{Fama--MacBeth Regressions}
\label{app:fmb}
\tablenote{This table reports weekly Fama--MacBeth regressions over 2001--2025. The dependent variable is the stock's subsequent five-trading-day return. Models 1--6 include the image signals shown in each column: DINO and FT-DINO are retrieval signals with $K=1{,}000$, and CNN is the out-of-sample prediction of the 20-day price-image CNN of \citet{jiang2023}. All models include the control variables following \citet{jiang2023}: MOM, STR, WSTR, Beta, RetVol, High52, BASpread, DolVol, ZeroTrade, PriceDelay, Size, and Illiq (defined in Table~\ref{tab:summarystats}). The image signals and controls are ranked within each weekly cross section before estimation. Coefficients are time-series averages of the weekly cross-sectional estimates, Adj. $R^2$ is the average adjusted $R^2$, and $t$-statistics are reported in parentheses. Summary statistics in Table~\ref{tab:summarystats} are reported before these cross-sectional transformations.}

\apptablesize
\centering
\begin{threeparttable}
\begin{tabular*}{\textwidth}{@{\extracolsep{\fill}}lcccccc@{}}
\toprule
Variable & Model 1 & Model 2 & Model 3 & Model 4 & Model 5 & Model 6 \\
\midrule
FT-DINO & & 0.0111 & & 0.0093 & 0.0109 & 0.0092 \\
        & & (22.73) & & (22.41) & (22.90) & (22.52) \\[2pt]
DINO & 0.0032 & & 0.0024 & & 0.0010 & 0.0008 \\
     & (13.18) & & (11.14) & & (5.15) & (4.52) \\[2pt]
CNN & & & 0.0086 & 0.0052 & & 0.0052 \\
    & & & (20.03) & (16.39) & & (16.44) \\[2pt]
Other Controls & Yes & Yes & Yes & Yes & Yes & Yes \\
\midrule
Adj. $R^2$ & 0.0506 & 0.0522 & 0.0518 & 0.0528 & 0.0523 & 0.0529 \\
\bottomrule
\end{tabular*}
\end{threeparttable}
\end{table}
\clearpage
\begin{table}[H]\caption{DINO--CNN and FT-DINO--CNN Double Sorts}
\label{tab:app_cnn_double_sorts}
\tablenote{This table reports equal-weighted sequential $5\times5$ double sorts over 2001--2025. Each week, stocks are first sorted into quintiles (columns) of the CNN prediction of \citet{jiang2023} and then, within each CNN quintile, into conditional quintiles (rows) of the benchmark DINO retrieval signal in Panel A or the FT-DINO retrieval signal in Panel B, both with $K=1{,}000$. Portfolios are held over the subsequent five trading days. Cells report annualized returns. The DINO 5--1 and FT-DINO 5--1 rows report the return on the top minus the bottom conditional retrieval quintile within each CNN quintile, and the CNN 5--1 column reports the difference between CNN quintiles 5 and 1. $t$-statistics are reported in parentheses.}

\apptablesize
\centering
\label{tab:app_dino_cnn}
\label{tab:app_ft_cnn}
\begin{minipage}{\textwidth}
\centering

\setlength{\tabcolsep}{1.8pt}
\renewcommand{\arraystretch}{0.78}

\panelheading{Panel A. DINO--CNN Double Sort}
\begin{tabular*}{\linewidth}{@{\extracolsep{\fill}}lrrrrrr@{}}
\toprule
& CNN Q1 & CNN Q2 & CNN Q3 & CNN Q4 & CNN Q5 & CNN 5--1 \\
\midrule
\multirow[c]{2}{*}{DINO Q1} & -0.21 & 0.05 & 0.11 & 0.15 & 0.28 & 0.49 \\
& (-3.32) & (0.84) & (2.03) & (3.01) & (5.51) & (14.49) \\[0.15em]
\multirow[c]{2}{*}{DINO Q2} & -0.15 & 0.07 & 0.11 & 0.18 & 0.29 & 0.43 \\
& (-2.59) & (1.31) & (2.18) & (3.68) & (5.96) & (14.31) \\[0.15em]
\multirow[c]{2}{*}{DINO Q3} & -0.12 & 0.06 & 0.13 & 0.17 & 0.27 & 0.39 \\
& (-2.09) & (1.14) & (2.55) & (3.59) & (5.64) & (13.25) \\[0.15em]
\multirow[c]{2}{*}{DINO Q4} & -0.08 & 0.10 & 0.15 & 0.21 & 0.31 & 0.39 \\
& (-1.40) & (1.83) & (3.10) & (4.34) & (6.22) & (12.88) \\[0.15em]
\multirow[c]{2}{*}{DINO Q5} & -0.04 & 0.14 & 0.20 & 0.27 & 0.45 & 0.49 \\
& (-0.66) & (2.54) & (3.61) & (4.88) & (7.71) & (12.54) \\
\midrule
\multirow[c]{2}{*}{DINO 5--1} & 0.17 & 0.10 & 0.09 & 0.12 & 0.17 & \\
& (7.17) & (4.66) & (5.02) & (6.22) & (8.21) & \\
\bottomrule
\end{tabular*}

\panelheading{Panel B. FT-DINO--CNN Double Sort}
\begin{tabular*}{\linewidth}{@{\extracolsep{\fill}}lrrrrrr@{}}
\toprule
& CNN Q1 & CNN Q2 & CNN Q3 & CNN Q4 & CNN Q5 & CNN 5--1 \\
\midrule
\multirow[c]{2}{*}{FT-DINO Q1} & -0.50 & -0.08 & 0.01 & 0.08 & 0.13 & 0.63 \\
& (-6.97) & (-1.37) & (0.25) & (1.49) & (2.71) & (14.01) \\[0.15em]
\multirow[c]{2}{*}{FT-DINO Q2} & -0.17 & 0.04 & 0.09 & 0.13 & 0.19 & 0.36 \\
& (-2.83) & (0.74) & (1.86) & (2.87) & (4.32) & (11.00) \\[0.15em]
\multirow[c]{2}{*}{FT-DINO Q3} & -0.07 & 0.07 & 0.11 & 0.16 & 0.25 & 0.32 \\
& (-1.33) & (1.35) & (2.38) & (3.43) & (5.28) & (12.40) \\[0.15em]
\multirow[c]{2}{*}{FT-DINO Q4} & 0.00 & 0.13 & 0.14 & 0.20 & 0.31 & 0.31 \\
& (0.04) & (2.51) & (2.94) & (3.96) & (6.14) & (12.07) \\[0.15em]
\multirow[c]{2}{*}{FT-DINO Q5} & 0.13 & 0.26 & 0.33 & 0.42 & 0.71 & 0.58 \\
& (2.11) & (4.39) & (5.60) & (6.95) & (10.25) & (13.10) \\
\midrule
\multirow[c]{2}{*}{FT-DINO 5--1} & 0.63 & 0.34 & 0.32 & 0.35 & 0.58 & \\
& (19.72) & (15.31) & (14.66) & (13.03) & (14.81) & \\
\bottomrule
\end{tabular*}

\end{minipage}
\end{table}

\clearpage

\begin{table}[H]\caption{Historical Retrieval versus Ridge Prediction}
\label{tab:app_ridge}
\tablenote{This table compares historical retrieval with ridge prediction over 2001--2025. Retrieval signals average the realized subsequent five-day returns of the $K=1{,}000$ nearest historical charts, whereas ridge signals are ridge-regression predictions of subsequent five-day returns estimated directly on the same embeddings. Panel A reports H--L decile portfolio performance under EW, VW, and CVW weighting. Panels B and C report weekly spanning regressions in which the dependent variable is the H--L return of DINO retrieval and FT-DINO retrieval, respectively. Column (1) regresses it on the ridge portfolio built from the same representation, column (2) adds the Fama--French five factors, momentum, and short-term reversal (Factor Controls), column (3) adds the CNN portfolio, and column (4) adds the other retrieval portfolio. Returns, alphas, and Sharpe ratios are annualized, Observations is the number of weeks, and $t$-statistics are reported in parentheses.}

\singlespacing
\apptablesize
\centering
\begin{minipage}{\textwidth}
\centering

\setlength{\tabcolsep}{3pt}
\renewcommand{\arraystretch}{0.70}

\panelheading{Panel A. Retrieval and Ridge Portfolio Performance}
\begin{tabular*}{\linewidth}{@{\extracolsep{\fill}}l cc cc cc@{}}
\toprule
            & \multicolumn{2}{c}{EW}
            & \multicolumn{2}{c}{VW}
            & \multicolumn{2}{c}{CVW} \\
            \cmidrule(lr){2-3}
            \cmidrule(lr){4-5}
            \cmidrule(lr){6-7}

            Signal
            & Ret & SR
            & Ret & SR
            & Ret & SR \\
            \midrule

            DINO-Retrieval
            & 0.26 & 2.81
            & 0.07 & 0.60
            & 0.07 & 0.87 \\

            DINO-Ridge
            & 0.34 & 2.77
            & 0.10 & 0.71
            & 0.08 & 0.81 \\

            \addlinespace[2pt]

            FT-DINO-Retrieval
            & 0.85 & 6.37
            & 0.24 & 1.45
            & 0.28 & 2.38 \\

            FT-DINO-Ridge
            & 0.82 & 5.99
            & 0.26 & 1.70
            & 0.28 & 2.43 \\

            \bottomrule
\end{tabular*}

\vspace{4pt}
\panelheading{Panel B. DINO Retrieval Spanning Regressions}
\begin{tabular*}{\linewidth}{@{\extracolsep{\fill}}l cccc cccc@{}}
\toprule
            & \multicolumn{4}{c}{EW}
            & \multicolumn{4}{c}{CVW} \\
            \cmidrule(lr){2-5}
            \cmidrule(lr){6-9}

            Variable
            & (1) & (2) & (3) & (4)
            & (1) & (2) & (3) & (4) \\
            \midrule

            Alpha
            & 0.1314 & 0.1332 & 0.0677 & -0.0968
            & 0.0428 & 0.0398 & 0.0240 & -0.0340 \\
            &
            (4.59) & (4.60) & (2.00) & (-2.87)
            & (3.04) & (2.49) & (1.40) & (-2.18) \\

            \addlinespace[1pt]

            DINO Ridge
            & 0.3869 & 0.3798 & 0.3213 & 0.2059
            & 0.3656 & 0.3501 & 0.2975 & 0.2201 \\
            &
            (4.65) & (5.34) & (4.84) & (3.18)
            & (3.10) & (3.91) & (3.53) & (3.10) \\

            \addlinespace[1pt]

            CNN
            &  &  & 0.1489 & -0.0472
            &  &  & 0.1058 & -0.0420 \\
            &
            &  & (4.73) & (-1.06)
            &  &  & (2.89) & (-0.83) \\

            \addlinespace[1pt]

            FT-DINO Retrieval
            &  &  &  & 0.3815
            &  &  &  & 0.3601 \\
            &
            &  &  & (5.34)
            &  &  &  & (4.16) \\

            \midrule

            Factor Controls
            & No & Yes & Yes & Yes
            & No & Yes & Yes & Yes \\

            \(R^2\)
            & 0.2554 & 0.2579 & 0.2923 & 0.4263
            & 0.1959 & 0.2077 & 0.2233 & 0.3808 \\

            Observations
            & 1303 & 1303 & 1303 & 1303
            & 1303 & 1303 & 1303 & 1303 \\

            \bottomrule
\end{tabular*}

\vspace{4pt}
\panelheading{Panel C. FT-DINO Retrieval Spanning Regressions}
\begin{tabular*}{\linewidth}{@{\extracolsep{\fill}}l cccc cccc@{}}
\toprule
            & \multicolumn{4}{c}{EW}
            & \multicolumn{4}{c}{CVW} \\
            \cmidrule(lr){2-5}
            \cmidrule(lr){6-9}

            Variable
            & (1) & (2) & (3) & (4)
            & (1) & (2) & (3) & (4) \\
            \midrule

            Alpha
            & 0.2055 & 0.2235 & 0.2049 & 0.2100
            & 0.0833 & 0.0860 & 0.0830 & 0.0864 \\
            &
            (3.88) & (4.76) & (4.67) & (5.73)
            & (3.51) & (3.34) & (3.25) & (4.36) \\

            \addlinespace[1pt]

            FT-DINO Ridge
            & 0.7921 & 0.7552 & 0.6087 & 0.5019
            & 0.7191 & 0.6538 & 0.5409 & 0.4463 \\
            &
            (11.69) & (14.86) & (10.25) & (6.31)
            & (8.97) & (9.64) & (7.38) & (7.20) \\

            \addlinespace[1pt]

            CNN
            &  &  & 0.2369 & 0.2085
            &  &  & 0.1730 & 0.1248 \\
            &
            &  & (7.33) & (6.08)
            &  &  & (4.44) & (3.58) \\

            \addlinespace[1pt]

            DINO Retrieval
            &  &  &  & 0.3778
            &  &  &  & 0.4851 \\
            &
            &  &  & (2.89)
            &  &  &  & (4.67) \\

            \midrule

            Factor Controls
            & No & Yes & Yes & Yes
            & No & Yes & Yes & Yes \\

            \(R^2\)
            & 0.6522 & 0.6658 & 0.6958 & 0.7472
            & 0.4761 & 0.5026 & 0.5180 & 0.6135 \\

            Observations
            & 1303 & 1303 & 1303 & 1303
            & 1303 & 1303 & 1303 & 1303 \\

            \bottomrule
\end{tabular*}

\end{minipage}
\end{table}
\clearpage

\begin{table}[H]\caption{Temporal Structure Perturbations}
\label{tab:app_temporal}
\tablenote{This table reports high-minus-low Sharpe ratios for the CNN, benchmark DINO, and FT-DINO when the query charts are perturbed. B1--B4 denote the four consecutive five-day blocks of the 20-day chart window, with B4 containing the most recent days. Rows B1 to B4 randomly permute the daily observations within the indicated block and redraw the chart. B1--B3 Reorder permutes the order of the first three blocks while keeping B4 fixed. The perturbations apply only to the query charts: no model is re-trained, and the historical retrieval library is unchanged. DINO and FT-DINO use $K=1{,}000$. $\Delta$SR is the percentage change in the Sharpe ratio relative to the unperturbed model. Panel A reports equal-weighted portfolios, and Panel B reports capped value-weighted (CVW) portfolios, which winsorize market capitalization at the contemporaneous NYSE 80th percentile following \citet{jensen2023}.}

\apptablesize
\centering
\begin{threeparttable}
\setlength{\tabcolsep}{2.8pt}
\panelheading{Panel A. EW}
\begin{tabular*}{\textwidth}{@{\extracolsep{\fill}}l cc cc cc@{}}
\toprule
& \multicolumn{2}{c}{CNN} & \multicolumn{2}{c}{DINO} & \multicolumn{2}{c}{FT-DINO} \\
\cmidrule(lr){2-3}\cmidrule(lr){4-5}\cmidrule(lr){6-7}
Test & SR & $\Delta$SR & SR & $\Delta$SR & SR & $\Delta$SR \\
\midrule
Benchmark & 4.46 & -- & 2.81 & -- & 6.37 & -- \\
B1 & 4.55 & $+$1.92\% & 2.90 & $+$3.25\% & 6.45 & $+$1.23\% \\
B2 & 4.71 & $+$5.44\% & 2.76 & $-$1.92\% & 6.29 & $-$1.33\% \\
B3 & 4.59 & $+$2.82\% & 2.77 & $-$1.62\% & 6.36 & $-$0.20\% \\
B4 & 2.32 & $-$48.08\% & 2.65 & $-$5.82\% & 4.37 & $-$31.43\% \\
B1--B3 Reorder & 4.35 & $-$2.43\% & 1.96 & $-$30.30\% & 5.97 & $-$6.27\% \\
\bottomrule
\end{tabular*}

\vspace{8pt}
\panelheading{Panel B. CVW}
\begin{tabular*}{\textwidth}{@{\extracolsep{\fill}}l cc cc cc@{}}
\toprule
& \multicolumn{2}{c}{CNN} & \multicolumn{2}{c}{DINO} & \multicolumn{2}{c}{FT-DINO} \\
\cmidrule(lr){2-3}\cmidrule(lr){4-5}\cmidrule(lr){6-7}
Test & SR & $\Delta$SR & SR & $\Delta$SR & SR & $\Delta$SR \\
\midrule
Benchmark & 1.82 & -- & 0.87 & -- & 2.38 & -- \\
B1 & 1.86 & $+$1.89\% & 1.01 & $+$15.64\% & 2.33 & $-$1.88\% \\
B2 & 1.95 & $+$7.15\% & 0.90 & $+$2.35\% & 2.23 & $-$6.38\% \\
B3 & 1.90 & $+$4.34\% & 0.88 & $+$0.17\% & 2.19 & $-$7.99\% \\
B4 & 0.88 & $-$51.86\% & 0.78 & $-$10.49\% & 1.74 & $-$26.80\% \\
B1--B3 Reorder & 1.76 & $-$3.51\% & 0.53 & $-$39.96\% & 2.10 & $-$11.77\% \\
\bottomrule
\end{tabular*}
\end{threeparttable}
\end{table}
\clearpage
\begin{table}[H]\caption{Transformed Queries}
\label{tab:app_transform}
\tablenote{This table reports retrieval results when the query chart is flipped horizontally or rotated by 180 degrees over 2001--2025, with $K=1{,}000$. Only the query is transformed; the historical library and the retrieval rule are unchanged. Panel A is based on a single randomly selected query chart and reports the hypothetical rank of the transformed copy: the rank that its cosine similarity to the original query would attain among the original query's similarities to the eligible historical charts. Panel B reports the annualized mean return (Ret) and Sharpe ratio (SR) of the high-minus-low portfolio under equal weighting (EW) and capped value weighting (CVW).}
\apptablesize
\centering
\panelheading{Panel A. Hypothetical Similarity Rank of the Transformed Query}
\begin{tabular*}{\textwidth}{@{\extracolsep{\fill}}l rr@{}}
\toprule
Model & Horizontal Flip & Rotation $180^\circ$ \\
\midrule
DINO & 524{,}306 & 1{,}198{,}188 \\
FT-DINO & 586{,}020 & 16{,}901{,}476 \\
\bottomrule
\end{tabular*}

\vspace{6pt}
\panelheading{Panel B. Portfolio Performance}
\begin{tabular*}{\textwidth}{@{\extracolsep{\fill}}ll cc cc@{}}
\toprule
& & \multicolumn{2}{c}{EW} & \multicolumn{2}{c}{CVW} \\
\cmidrule(lr){3-4}\cmidrule(lr){5-6}
Model & Query & Ret & SR & Ret & SR \\
\midrule
DINO & Original & 0.26 & 2.81 & 0.07 & 0.87 \\
     & Horizontal Flip & 0.11 & 1.53 & 0.02 & 0.32 \\
     & Rotation $180^\circ$ & 0.11 & 1.38 & 0.03 & 0.46 \\
\addlinespace[2pt]
FT-DINO & Original & 0.85 & 6.37 & 0.28 & 2.38 \\
        & Horizontal Flip & -0.04 & -0.45 & 0.00 & 0.04 \\
        & Rotation $180^\circ$ & -0.01 & -0.07 & 0.00 & 0.04 \\
\bottomrule
\end{tabular*}
\end{table}
\clearpage

\begin{table}[H]\caption{Subsample Portfolio Performance}
\label{app:subsample}
\tablenote{This table reports high-minus-low portfolio performance for FT-DINO, benchmark DINO, and CNN in three nonoverlapping subperiods of the out-of-sample period: 2001--2009, 2010--2017, and 2018--2025. FT-DINO and DINO use $K=1{,}000$. All three models use the same weekly decile-formation procedure and subsequent five-trading-day holding period. Panel A reports equal-weighted portfolios, and Panel B reports capped value-weighted (CVW) portfolios, which winsorize market capitalization at the contemporaneous NYSE 80th percentile following \citet{jensen2023}. Ret is 52 times the mean weekly H--L return, Alpha is 52 times the intercept from a regression of weekly H--L returns on the Fama--French five factors, momentum, and short-term reversal, and SR is the mean weekly H--L return divided by its standard deviation and multiplied by $\sqrt{52}$.}

\apptablesize
\centering
\begin{threeparttable}
\setlength{\tabcolsep}{2.8pt}
\panelheading{Panel A. EW}
\begin{tabular*}{\textwidth}{@{\extracolsep{\fill}}l ccc ccc ccc@{}}
\toprule
& \multicolumn{3}{c}{FT-DINO} & \multicolumn{3}{c}{DINO} & \multicolumn{3}{c}{CNN} \\
\cmidrule(lr){2-4}\cmidrule(lr){5-7}\cmidrule(lr){8-10}
Period & Ret & Alpha & SR & Ret & Alpha & SR & Ret & Alpha & SR \\
\midrule
2001--2009 & 1.26 & 1.17 & 8.49 & 0.40 & 0.36 & 4.33 & 0.89 & 0.81 & 6.21 \\
2010--2017 & 0.63 & 0.60 & 7.91 & 0.18 & 0.17 & 2.99 & 0.47 & 0.43 & 5.31 \\
2018--2025 & 0.61 & 0.58 & 4.41 & 0.18 & 0.18 & 1.59 & 0.44 & 0.39 & 2.78 \\
\bottomrule
\end{tabular*}

\vspace{8pt}
\panelheading{Panel B. CVW}
\begin{tabular*}{\textwidth}{@{\extracolsep{\fill}}l ccc ccc ccc@{}}
\toprule
& \multicolumn{3}{c}{FT-DINO} & \multicolumn{3}{c}{DINO} & \multicolumn{3}{c}{CNN} \\
\cmidrule(lr){2-4}\cmidrule(lr){5-7}\cmidrule(lr){8-10}
Period & Ret & Alpha & SR & Ret & Alpha & SR & Ret & Alpha & SR \\
\midrule
2001--2009 & 0.56 & 0.47 & 4.12 & 0.15 & 0.10 & 1.59 & 0.40 & 0.35 & 3.21 \\
2010--2017 & 0.17 & 0.15 & 2.61 & 0.03 & 0.02 & 0.73 & 0.08 & 0.06 & 1.11 \\
2018--2025 & 0.09 & 0.06 & 0.66 & 0.02 & 0.02 & 0.21 & 0.14 & 0.09 & 1.01 \\
\bottomrule
\end{tabular*}
\end{threeparttable}
\end{table}
\clearpage

\begin{table}[H]\caption{Alternative Training Windows and Out-of-Sample Performance}
\label{tab:app_oos_periods}
\tablenote{This table compares high-minus-low portfolio performance for FT-DINO-NEW, FT-DINO, DINO, and CNN over 2010--2025 and the 2010--2017 and 2018--2025 subperiods. FT-DINO-NEW is fine-tuned on charts from 1993--2007 and validated on charts from 2008--2009. FT-DINO is the baseline model fine-tuned on charts from 1993--1999 and validated on charts from 2000. The common evaluation begins in 2010 so that both fine-tuned encoders are fixed before the reported out-of-sample period. Each week, stocks are sorted into deciles by the signal and held over the subsequent five trading days; H--L is the return on decile 10 minus decile 1. EW denotes equal weighting, and CVW denotes capped value weighting, which winsorizes market capitalization at the contemporaneous NYSE 80th percentile following \citet{jensen2023}. Ret is the annualized mean H--L return, and SR is the annualized Sharpe ratio.}

\apptablesize
\centering
\begin{threeparttable}

\setlength{\tabcolsep}{1.8pt}
\begin{tabular*}{\textwidth}{@{\extracolsep{\fill}}l cc cc cc cc cc cc cc cc@{}}
\toprule
& \multicolumn{8}{c}{EW} & \multicolumn{8}{c}{CVW} \\
\cmidrule(lr){2-9}\cmidrule(lr){10-17}
& \multicolumn{2}{c}{FT-DINO-NEW} & \multicolumn{2}{c}{FT-DINO} & \multicolumn{2}{c}{DINO} & \multicolumn{2}{c}{CNN}
& \multicolumn{2}{c}{FT-DINO-NEW} & \multicolumn{2}{c}{FT-DINO} & \multicolumn{2}{c}{DINO} & \multicolumn{2}{c}{CNN} \\
Period & Ret & SR & Ret & SR & Ret & SR & Ret & SR & Ret & SR & Ret & SR & Ret & SR & Ret & SR \\
\midrule
2010--2025 & 0.73 & 6.79 & 0.62 & 5.51 & 0.18 & 1.99 & 0.46 & 3.54 & 0.19 & 1.84 & 0.13 & 1.24 & 0.03 & 0.37 & 0.11 & 0.99 \\
2010--2017 & 0.75 & 9.46 & 0.63 & 7.91 & 0.18 & 2.99 & 0.47 & 5.31 & 0.25 & 3.51 & 0.17 & 2.61 & 0.03 & 0.73 & 0.08 & 1.11 \\
2018--2025 & 0.71 & 5.48 & 0.61 & 4.41 & 0.18 & 1.59 & 0.44 & 2.78 & 0.13 & 1.04 & 0.09 & 0.66 & 0.02 & 0.21 & 0.14 & 1.01 \\
\bottomrule
\end{tabular*}
\end{threeparttable}
\end{table}
\clearpage

\begin{table}[H]\caption{Alternative Fine-Tuning Targets}
\label{tab:app_adjusted_ft}
\tablenote{This table reports equal-weighted high-minus-low portfolio performance over 2001--2025 for FT-DINO models that differ only in the future-return target used for return-rank alignment. Raw Return is the baseline target, the cross-sectional rank of the raw subsequent five-day return. FF6-Residual uses returns residualized on the Fama--French five factors and momentum \citep{fama2015,carhart1997}, FF12-Adj uses Fama--French 12 industry-adjusted returns, and DGTW-Adj uses characteristic-adjusted returns following \citet{daniel1997}. The pretrained encoder, training and validation periods, retrieval rule, and portfolio construction are otherwise unchanged, and the retrieval signal still averages neighbors' raw subsequent five-day returns. $K$ is the number of retrieved neighbors. Ret is the annualized mean H--L return, and SR is the annualized Sharpe ratio.}

\apptablesize
\centering
\begin{threeparttable}

\setlength{\tabcolsep}{2.0pt}
\begin{tabular*}{\textwidth}{@{\extracolsep{\fill}}c cc cc cc cc@{}}
\toprule
& \multicolumn{2}{c}{Raw Return} & \multicolumn{2}{c}{FF6-Residual} & \multicolumn{2}{c}{FF12-Adj} & \multicolumn{2}{c}{DGTW-Adj} \\
\cmidrule(lr){2-3}\cmidrule(lr){4-5}\cmidrule(lr){6-7}\cmidrule(lr){8-9}
$K$ & Ret & SR & Ret & SR & Ret & SR & Ret & SR \\
\midrule
5    & 0.21 & 2.87 & 0.16 & 2.58 & 0.18 & 2.52 & 0.20 & 2.90 \\
10   & 0.28 & 3.59 & 0.23 & 3.20 & 0.28 & 3.72 & 0.26 & 3.57 \\
30   & 0.41 & 4.59 & 0.35 & 4.34 & 0.43 & 5.01 & 0.41 & 4.68 \\
100  & 0.60 & 5.47 & 0.53 & 5.16 & 0.62 & 5.80 & 0.60 & 5.45 \\
500  & 0.81 & 6.27 & 0.73 & 5.77 & 0.81 & 6.33 & 0.80 & 6.25 \\
1{,}000 & 0.85 & 6.37 & 0.78 & 5.82 & 0.87 & 6.55 & 0.83 & 6.27 \\
3{,}000 & 0.88 & 6.31 & 0.84 & 6.15 & 0.90 & 6.57 & 0.87 & 6.37 \\
\bottomrule
\end{tabular*}
\end{threeparttable}
\end{table}
\clearpage
\begin{table}[H]\caption{Factor-Adjusted Neighbor Returns}
\label{tab:app_alpha}
\tablenote{This table reports high-minus-low portfolio performance over 2001--2025 when the retrieval signal averages raw or factor-adjusted subsequent returns of the retrieved neighbors, with $K=1{,}000$. The retrieved neighbors are held fixed. Raw uses each neighbor's realized subsequent five-day return, CAPM Alpha its return net of the market exposure, and FF6 Alpha its return net of exposures to the Fama--French five factors and momentum. Ret is the annualized mean H--L return, and SR is the annualized Sharpe ratio, under equal weighting (EW) and capped value weighting (CVW).}
\apptablesize
\centering
\begin{tabular*}{\textwidth}{@{\extracolsep{\fill}}ll cc cc@{}}
\toprule
& & \multicolumn{2}{c}{EW} & \multicolumn{2}{c}{CVW} \\
\cmidrule(lr){3-4}\cmidrule(lr){5-6}
Model & Neighbor Return & Ret & SR & Ret & SR \\
\midrule
DINO & Raw & 0.26 & 2.81 & 0.07 & 0.87 \\
     & CAPM Alpha & 0.26 & 3.31 & 0.07 & 1.03 \\
     & FF6 Alpha & 0.30 & 3.34 & 0.10 & 1.26 \\
\addlinespace[2pt]
FT-DINO & Raw & 0.85 & 6.37 & 0.28 & 2.38 \\
        & CAPM Alpha & 0.85 & 6.84 & 0.29 & 2.55 \\
        & FF6 Alpha & 0.87 & 6.50 & 0.31 & 2.47 \\
\bottomrule
\end{tabular*}
\end{table}
\clearpage
\begin{table}[H]\caption{FT-DINO and Time-Series Momentum States}
\label{tab:app_momentum}
\tablenote{This table reports equal-weighted sequential sorts on time-series momentum states and the FT-DINO retrieval signal ($K=1{,}000$) over 2001--2025. Each week, stocks are first grouped by the sign of their own past returns and then, within each group, sorted into conditional quintiles of the FT-DINO signal. Positive MOM and Negative MOM split stocks by the sign of the return over the 250 trading days ending on the formation date. The remaining columns interact this sign with the sign of the return over the 20 trading days ending on the formation date, ordered (250-day, 20-day): Bull $(+,+)$, Correction $(+,-)$, Bear $(-,-)$, and Rebound $(-,+)$. FT-DINO 5--1 is the return on the top minus the bottom conditional FT-DINO quintile within each group. Returns are annualized, and $t$-statistics are reported in parentheses.}
\apptablesize
\centering
\setlength{\tabcolsep}{2.0pt}
\begin{tabular*}{\textwidth}{@{\extracolsep{\fill}}l rrrrrr@{}}
\toprule
& Positive MOM & Negative MOM & Bull $(+,+)$ & Correction $(+,-)$ & Bear $(-,-)$ & Rebound $(-,+)$ \\
\midrule
FT-DINO Q1 & -0.09 & -0.24 & -0.14 & -0.06 & -0.19 & -0.38 \\
           & (-1.84) & (-3.39) & (-2.83) & (-1.15) & (-2.53) & (-4.95) \\
FT-DINO Q2 & 0.06 & 0.01 & 0.03 & 0.09 & 0.04 & -0.03 \\
           & (1.55) & (0.09) & (0.81) & (1.96) & (0.68) & (-0.48) \\
FT-DINO Q3 & 0.11 & 0.11 & 0.08 & 0.16 & 0.15 & 0.04 \\
           & (2.79) & (1.88) & (1.93) & (3.62) & (2.40) & (0.74) \\
FT-DINO Q4 & 0.16 & 0.16 & 0.13 & 0.22 & 0.21 & 0.11 \\
           & (3.98) & (2.55) & (3.10) & (4.68) & (3.14) & (1.90) \\
FT-DINO Q5 & 0.34 & 0.50 & 0.29 & 0.40 & 0.62 & 0.35 \\
           & (7.14) & (6.74) & (5.95) & (7.77) & (7.43) & (4.99) \\
\midrule
FT-DINO 5--1 & 0.43 & 0.74 & 0.43 & 0.46 & 0.81 & 0.72 \\
             & (21.13) & (21.08) & (17.96) & (15.76) & (18.71) & (15.76) \\
\bottomrule
\end{tabular*}
\end{table}
\clearpage

\begin{table}[H]\caption{FT-DINO Ablations: Retrieval Library and Chart Design}
\label{tab:app_ft_full_ablation}
\tablenote{This table reports FT-DINO equal-weighted high-minus-low portfolio performance over 2001--2025 across the full grid of neighborhood sizes $K$ for the ablations summarized in Table~\ref{tab:ft_ablation}. Panel A varies the historical retrieval library (Rolling 8Y, Same-Ind., Cross-Ind., and Same-Stock), Panel B varies the color treatment (No Color, Reverse Color, and Random Color), and Panel C varies chart elements (No MA, No Volume, No MA + No Vol., and MA in Price Scale). Baseline uses the unrestricted historical library and the original colored chart with the moving average and volume bars. Each alternative changes only the indicated component while preserving the remaining retrieval and portfolio construction. Definitions follow Table~\ref{tab:ft_ablation}. Ret is the annualized mean H--L return, and SR is the annualized Sharpe ratio.}

\apptablesize
\centering
\begin{minipage}{\textwidth}
\centering

\renewcommand{\arraystretch}{0.90}
\setlength{\tabcolsep}{1.6pt}

\panelheading{Panel A. Historical Retrieval Library}
\begin{tabular*}{\linewidth}{@{\extracolsep{\fill}}c cc cc cc cc cc@{}}
\toprule
& \multicolumn{2}{c}{Baseline} & \multicolumn{2}{c}{Rolling 8Y} & \multicolumn{2}{c}{Same-Ind.} & \multicolumn{2}{c}{Cross-Ind.} & \multicolumn{2}{c}{Same-Stock} \\
\cmidrule(lr){2-3}\cmidrule(lr){4-5}\cmidrule(lr){6-7}\cmidrule(lr){8-9}\cmidrule(lr){10-11}
$K$ & Ret & SR & Ret & SR & Ret & SR & Ret & SR & Ret & SR \\
\midrule
5    & 0.21 & 2.87 & 0.18 & 2.51 & 0.20 & 2.80 & 0.20 & 2.78 & 0.09 & 1.00 \\
10   & 0.28 & 3.59 & 0.24 & 3.20 & 0.26 & 3.31 & 0.26 & 3.45 & 0.10 & 1.13 \\
30   & 0.41 & 4.59 & 0.38 & 4.38 & 0.39 & 4.44 & 0.41 & 4.74 & 0.13 & 1.30 \\
100  & 0.60 & 5.47 & 0.56 & 5.13 & 0.55 & 5.42 & 0.61 & 5.76 & 0.18 & 1.61 \\
500  & 0.81 & 6.27 & 0.75 & 5.72 & 0.75 & 6.01 & 0.80 & 6.19 & 0.17 & 1.36 \\
1{,}000 & 0.85 & 6.37 & 0.78 & 5.77 & 0.79 & 6.06 & 0.85 & 6.24 & 0.17 & 1.38 \\
3{,}000 & 0.88 & 6.31 & 0.83 & 5.69 & 0.82 & 6.07 & 0.87 & 6.24 & 0.06 & 0.46 \\
\bottomrule
\end{tabular*}

\panelheading{Panel B. Color Ablations}
\begin{tabular*}{\linewidth}{@{\extracolsep{\fill}}c cc cc cc cc@{}}
\toprule
& \multicolumn{2}{c}{Baseline} & \multicolumn{2}{c}{No Color} & \multicolumn{2}{c}{Reverse Color} & \multicolumn{2}{c}{Random Color} \\
\cmidrule(lr){2-3}\cmidrule(lr){4-5}\cmidrule(lr){6-7}\cmidrule(lr){8-9}
$K$ & Ret & SR & Ret & SR & Ret & SR & Ret & SR \\
\midrule
5    & 0.21 & 2.87 & 0.13 & 2.11 & 0.22 & 3.16 & 0.21 & 3.04 \\
10   & 0.28 & 3.59 & 0.19 & 2.90 & 0.27 & 3.92 & 0.25 & 3.42 \\
30   & 0.41 & 4.59 & 0.28 & 3.79 & 0.40 & 4.68 & 0.36 & 4.31 \\
100  & 0.60 & 5.47 & 0.41 & 4.89 & 0.60 & 5.57 & 0.55 & 5.33 \\
500  & 0.81 & 6.27 & 0.58 & 5.59 & 0.80 & 6.19 & 0.72 & 5.93 \\
1{,}000 & 0.85 & 6.37 & 0.64 & 5.71 & 0.85 & 6.29 & 0.78 & 5.92 \\
3{,}000 & 0.88 & 6.31 & 0.67 & 5.70 & 0.88 & 6.24 & 0.81 & 5.89 \\
\bottomrule
\end{tabular*}

\panelheading{Panel C. Chart-Element Ablations}
\begin{tabular*}{\linewidth}{@{\extracolsep{\fill}}c cc cc cc cc cc@{}}
\toprule
& \multicolumn{2}{c}{Baseline} & \multicolumn{2}{c}{No MA} & \multicolumn{2}{c}{No Volume} & \multicolumn{2}{c}{No MA + No Vol.} & \multicolumn{2}{c}{MA in Price Scale} \\
\cmidrule(lr){2-3}\cmidrule(lr){4-5}\cmidrule(lr){6-7}\cmidrule(lr){8-9}\cmidrule(lr){10-11}
$K$ & Ret & SR & Ret & SR & Ret & SR & Ret & SR & Ret & SR \\
\midrule
5    & 0.21 & 2.87 & 0.19 & 3.04 & 0.13 & 2.05 & 0.12 & 1.96 & 0.20 & 2.93 \\
10   & 0.28 & 3.59 & 0.26 & 3.71 & 0.17 & 2.53 & 0.17 & 2.53 & 0.28 & 3.91 \\
30   & 0.41 & 4.59 & 0.39 & 4.62 & 0.27 & 3.67 & 0.25 & 3.45 & 0.42 & 4.84 \\
100  & 0.60 & 5.47 & 0.57 & 5.72 & 0.41 & 4.49 & 0.39 & 4.37 & 0.60 & 5.78 \\
500  & 0.81 & 6.27 & 0.78 & 6.33 & 0.59 & 5.09 & 0.56 & 4.79 & 0.81 & 6.55 \\
1{,}000 & 0.85 & 6.37 & 0.83 & 6.36 & 0.62 & 4.98 & 0.61 & 4.83 & 0.85 & 6.43 \\
3{,}000 & 0.88 & 6.31 & 0.86 & 6.36 & 0.67 & 4.98 & 0.68 & 5.00 & 0.89 & 6.47 \\
\bottomrule
\end{tabular*}

\end{minipage}
\end{table}
\clearpage
\begin{table}[H]\caption{Double Sorts on Chart Sparsity}
\label{tab:app_sparsity}
\tablenote{This table reports equal-weighted sequential $5\times5$ double sorts on chart sparsity and the retrieval signal ($K=1{,}000$) over 2001--2025. Each week, stocks are first sorted into quintiles (columns) by the white-pixel ratio (WPR), the fraction of blank pixels in their 20-day price charts, with Q5 denoting the sparsest charts. Within each WPR quintile, stocks are then sorted into conditional quintiles (rows) of the benchmark DINO signal (Panel A) or the FT-DINO signal (Panel B). The 5--1 row is the return on the top minus the bottom conditional signal quintile within each WPR quintile. Returns are annualized, and $t$-statistics are reported in parentheses.}
\apptablesize
\centering
\renewcommand{\arraystretch}{0.9}
\setlength{\tabcolsep}{2.0pt}
\panelheading{Panel A. DINO}
\begin{tabular*}{\textwidth}{@{\extracolsep{\fill}}l rrrrr@{}}
\toprule
& WPR Q1 & WPR Q2 & WPR Q3 & WPR Q4 & WPR Q5 \\
\midrule
DINO Q1 & 0.10 & 0.10 & 0.06 & 0.04 & -0.07 \\
   & (1.87) & (1.85) & (1.07) & (0.77) & (-1.33) \\
DINO Q2 & 0.11 & 0.10 & 0.08 & 0.09 & 0.05 \\
   & (2.23) & (1.94) & (1.56) & (1.72) & (0.99) \\
DINO Q3 & 0.12 & 0.13 & 0.10 & 0.12 & 0.13 \\
   & (2.43) & (2.66) & (1.89) & (2.27) & (2.54) \\
DINO Q4 & 0.15 & 0.12 & 0.14 & 0.15 & 0.19 \\
   & (2.95) & (2.36) & (2.83) & (2.86) & (3.66) \\
DINO Q5 & 0.16 & 0.17 & 0.20 & 0.21 & 0.33 \\
   & (3.07) & (3.13) & (3.72) & (3.63) & (5.13) \\
\midrule
DINO 5--1 & 0.05 & 0.06 & 0.14 & 0.17 & 0.41 \\
   & (3.11) & (3.27) & (8.00) & (7.73) & (12.74) \\
\bottomrule
\end{tabular*}

\vspace{6pt}
\panelheading{Panel B. FT-DINO}
\begin{tabular*}{\textwidth}{@{\extracolsep{\fill}}l rrrrr@{}}
\toprule
& WPR Q1 & WPR Q2 & WPR Q3 & WPR Q4 & WPR Q5 \\
\midrule
FT-DINO Q1 & -0.09 & -0.09 & -0.11 & -0.13 & -0.29 \\
   & (-1.52) & (-1.51) & (-1.87) & (-2.15) & (-4.78) \\
FT-DINO Q2 & 0.08 & 0.08 & 0.05 & 0.05 & 0.03 \\
   & (1.66) & (1.57) & (1.01) & (0.96) & (0.53) \\
FT-DINO Q3 & 0.13 & 0.12 & 0.11 & 0.12 & 0.11 \\
   & (2.74) & (2.59) & (2.28) & (2.43) & (2.26) \\
FT-DINO Q4 & 0.18 & 0.16 & 0.16 & 0.17 & 0.21 \\
   & (3.74) & (3.18) & (3.11) & (3.28) & (4.04) \\
FT-DINO Q5 & 0.34 & 0.35 & 0.37 & 0.40 & 0.58 \\
   & (5.99) & (6.23) & (6.49) & (6.44) & (8.29) \\
\midrule
FT-DINO 5--1 & 0.43 & 0.45 & 0.48 & 0.53 & 0.86 \\
   & (16.78) & (17.33) & (19.11) & (16.90) & (20.28) \\
\bottomrule
\end{tabular*}
\end{table}
\clearpage

\begin{table}[H]\caption{Alternative Image Windows: DINO versus FT-DINO}
\label{tab:app_window_compare}
\tablenote{This table compares benchmark DINO and FT-DINO equal-weighted high-minus-low portfolio performance over 2001--2025 using D20, D40, and D60 price images across the full grid of neighborhood sizes $K$. The three specifications display OHLC bars, volume, and MA20, MA40, and MA60 over the prior 20, 40, and 60 trading days, respectively (Figure~\ref{fig:app_windows}). Signals are formed weekly from neighbors' realized subsequent five-trading-day returns, and portfolios are held for five trading days. The remaining chart-construction, historical-library, retrieval, and portfolio rules are held fixed within each model. Ret is the annualized mean H--L return, and SR is the annualized Sharpe ratio.}

\apptablesize
\centering
\setlength{\tabcolsep}{1.5pt}
\begin{tabular*}{\textwidth}{@{\extracolsep{\fill}}c cc cc cc cc cc cc@{}}
            \toprule

            & \multicolumn{4}{c}{D20}
            & \multicolumn{4}{c}{D40}
            & \multicolumn{4}{c}{D60} \\

            \cmidrule(lr){2-5}
            \cmidrule(lr){6-9}
            \cmidrule(lr){10-13}

            & \multicolumn{2}{c}{DINO}
            & \multicolumn{2}{c}{FT-DINO}
            & \multicolumn{2}{c}{DINO}
            & \multicolumn{2}{c}{FT-DINO}
            & \multicolumn{2}{c}{DINO}
            & \multicolumn{2}{c}{FT-DINO} \\

            \cmidrule(lr){2-3}
            \cmidrule(lr){4-5}
            \cmidrule(lr){6-7}
            \cmidrule(lr){8-9}
            \cmidrule(lr){10-11}
            \cmidrule(lr){12-13}

            $K$
            & Ret & SR
            & Ret & SR
            & Ret & SR
            & Ret & SR
            & Ret & SR
            & Ret & SR \\

            \midrule

            5
            & 0.05 & 0.83
            & 0.21 & 2.87
            & 0.02 & 0.34
            & 0.22 & 2.80
            & -0.04 & -0.39
            & 0.22 & 2.61 \\

            10
            & 0.07 & 1.12
            & 0.28 & 3.59
            & 0.04 & 0.66
            & 0.29 & 3.53
            & -0.01 & -0.12
            & 0.31 & 3.54 \\

            30
            & 0.11 & 1.83
            & 0.41 & 4.59
            & 0.09 & 1.25
            & 0.46 & 4.72
            & 0.05 & 0.60
            & 0.45 & 4.62 \\

            100
            & 0.16 & 2.15
            & 0.60 & 5.47
            & 0.17 & 2.23
            & 0.66 & 5.61
            & 0.12 & 1.47
            & 0.64 & 5.19 \\

            500
            & 0.23 & 2.74
            & 0.81 & 6.27
            & 0.27 & 2.83
            & 0.89 & 6.08
            & 0.21 & 2.25
            & 0.85 & 5.68 \\

            1{,}000
            & 0.26 & 2.81
            & 0.85 & 6.37
            & 0.31 & 2.97
            & 0.93 & 6.25
            & 0.26 & 2.63
            & 0.89 & 5.86 \\

            3{,}000
            & 0.29 & 2.90
            & 0.88 & 6.31
            & 0.35 & 3.07
            & 0.96 & 6.24
            & 0.33 & 2.82
            & 0.92 & 5.90 \\

            \bottomrule
\end{tabular*}
\end{table}
\clearpage

\begin{table}[H]\caption{20-Day Return Horizon: DINO versus FT-DINO}
\label{tab:app_r20}
\tablenote{This table reports benchmark DINO and FT-DINO equal-weighted high-minus-low portfolio performance for the 20-trading-day return horizon across D20, D40, and D60 price images and the full grid of neighborhood sizes $K$. D20, D40, and D60 use the prior 20, 40, and 60 trading days, respectively, with the corresponding moving-average window. Queries are formed monthly on the last trading day of each month, the neighbor outcome is the realized subsequent 20-trading-day return, and portfolios are held for 20 trading days. A historical observation is eligible only after its subsequent 20-day return is fully realized. FT-DINO is re-trained using cross-sectional ranks of subsequent 20-trading-day returns as the fine-tuning target. All other historical-retrieval settings follow the main analysis. Ret is the annualized mean H--L return, and SR is the annualized Sharpe ratio.}

\apptablesize
\centering
\setlength{\tabcolsep}{1.5pt}
\begin{tabular*}{\textwidth}{@{\extracolsep{\fill}}c cc cc cc cc cc cc@{}}
            \toprule

            & \multicolumn{4}{c}{D20}
            & \multicolumn{4}{c}{D40}
            & \multicolumn{4}{c}{D60} \\

            \cmidrule(lr){2-5}
            \cmidrule(lr){6-9}
            \cmidrule(lr){10-13}

            & \multicolumn{2}{c}{DINO}
            & \multicolumn{2}{c}{FT-DINO}
            & \multicolumn{2}{c}{DINO}
            & \multicolumn{2}{c}{FT-DINO}
            & \multicolumn{2}{c}{DINO}
            & \multicolumn{2}{c}{FT-DINO} \\

            \cmidrule(lr){2-3}
            \cmidrule(lr){4-5}
            \cmidrule(lr){6-7}
            \cmidrule(lr){8-9}
            \cmidrule(lr){10-11}
            \cmidrule(lr){12-13}

            $K$
            & Ret & SR
            & Ret & SR
            & Ret & SR
            & Ret & SR
            & Ret & SR
            & Ret & SR \\

            \midrule

            5
            & 0.02 & 0.32
            & 0.06 & 0.90
            & 0.04 & 0.64
            & 0.09 & 1.16
            & 0.05 & 0.70
            & 0.10 & 1.19 \\

            10
            & 0.04 & 0.74
            & 0.09 & 1.38
            & 0.05 & 0.76
            & 0.12 & 1.43
            & 0.06 & 0.89
            & 0.12 & 1.50 \\

            30
            & 0.04 & 0.72
            & 0.13 & 1.74
            & 0.08 & 1.23
            & 0.18 & 1.77
            & 0.08 & 1.10
            & 0.19 & 1.82 \\

            100
            & 0.08 & 1.21
            & 0.20 & 2.02
            & 0.12 & 1.51
            & 0.25 & 1.94
            & 0.11 & 1.34
            & 0.26 & 1.96 \\

            500
            & 0.11 & 1.55
            & 0.26 & 1.96
            & 0.18 & 2.12
            & 0.34 & 2.08
            & 0.16 & 1.70
            & 0.36 & 2.15 \\

            1{,}000
            & 0.13 & 1.54
            & 0.30 & 2.13
            & 0.20 & 1.99
            & 0.36 & 2.11
            & 0.19 & 1.86
            & 0.38 & 2.19 \\

            3{,}000
            & 0.14 & 1.52
            & 0.33 & 2.08
            & 0.20 & 1.87
            & 0.40 & 2.09
            & 0.21 & 1.76
            & 0.40 & 2.17 \\

            \bottomrule
\end{tabular*}
\end{table}
\clearpage

\begin{table}[H]\caption{Holding Horizons and Transaction Costs}
\label{tab:app_holding_costs}
\tablenote{This table reports benchmark DINO and FT-DINO high-minus-low portfolio performance across rebalancing intervals over 2001--2025. All specifications use D20 price charts and the same D20R5 retrieval signal. H5, H10, H15, and H20 rebalance every one, two, three, and four weeks, respectively. Panel A reports equal-weighted portfolios, and Panel B reports CVW portfolios. Turn. is four-week-equivalent one-way turnover, computed as average rebalancing turnover multiplied by $4/s$, where $s$ is the rebalancing interval in weeks. For each leg, turnover is one-half of the absolute changes from return-drifted prior weights to current target weights, and H--L turnover sums the two legs. Transaction costs are charged on the full absolute weight changes at 10 basis points for stocks above the contemporaneous NYSE 80th market-capitalization percentile and 20 basis points otherwise. Net H--L returns deduct costs on both legs; initial establishment costs are included in net returns but excluded from reported turnover. Ret, Vol, and SR are annualized gross mean return, volatility, and Sharpe ratio, using $52/s$ periods per year; Net Ret and Net SR are the corresponding net-of-cost values.}

\apptablesize
\centering
\begin{minipage}{\textwidth}
\centering

\setlength{\tabcolsep}{1.4pt}
\renewcommand{\arraystretch}{0.94}

\panelheading{Panel A. EW}
\begin{tabular*}{\linewidth}{@{\extracolsep{\fill}}c cccccc cccccc@{}}
\toprule
& \multicolumn{6}{c}{DINO} & \multicolumn{6}{c}{FT-DINO} \\
\cmidrule(lr){2-7}\cmidrule(lr){8-13}
Holding & Ret & Vol & SR & Turn. & Net Ret & Net SR & Ret & Vol & SR & Turn. & Net Ret & Net SR \\
\midrule
H5  & 0.26 & 0.09 & 2.81 & 645\% & -0.06 & -0.63 & 0.85 & 0.13 & 6.37 & 600\% & 0.55 & 4.12 \\
H10 & 0.15 & 0.20 & 0.78 & 350\% & -0.02 & -0.10 & 0.50 & 0.22 & 2.32 & 329\% & 0.34 & 1.55 \\
H15 & 0.16 & 0.10 & 1.60 & 239\% & 0.05 & 0.45 & 0.40 & 0.14 & 2.89 & 228\% & 0.28 & 2.06 \\
H20 & 0.13 & 0.09 & 1.36 & 181\% & 0.04 & 0.41 & 0.32 & 0.13 & 2.43 & 173\% & 0.24 & 1.78 \\
\bottomrule
\end{tabular*}

\panelheading{Panel B. CVW}
\begin{tabular*}{\linewidth}{@{\extracolsep{\fill}}c cccccc cccccc@{}}
\toprule
& \multicolumn{6}{c}{DINO} & \multicolumn{6}{c}{FT-DINO} \\
\cmidrule(lr){2-7}\cmidrule(lr){8-13}
Holding & Ret & Vol & SR & Turn. & Net Ret & Net SR & Ret & Vol & SR & Turn. & Net Ret & Net SR \\
\midrule
H5  & 0.07 & 0.08 & 0.87 & 675\% & -0.20 & -2.39 & 0.28 & 0.12 & 2.38 & 677\% & 0.00 & 0.01 \\
H10 & 0.04 & 0.08 & 0.50 & 362\% & -0.11 & -1.41 & 0.18 & 0.12 & 1.53 & 362\% & 0.03 & 0.22 \\
H15 & 0.03 & 0.08 & 0.42 & 244\% & -0.07 & -0.88 & 0.14 & 0.11 & 1.29 & 247\% & 0.04 & 0.36 \\
H20 & 0.01 & 0.07 & 0.15 & 184\% & -0.06 & -0.85 & 0.09 & 0.11 & 0.87 & 187\% & 0.02 & 0.16 \\
\bottomrule
\end{tabular*}

\end{minipage}
\end{table}
\clearpage

\begin{table}[H]\caption{Task-Vector Steering}
\label{tab:app_task_vector}
\tablenote{This table reports equal-weighted high-minus-low portfolio performance over 2001--2025 for encoders obtained by task-vector steering between DINO and FT-DINO. The task vector is $\Delta\theta=\theta_{\mathrm{FT\text{-}DINO}}-\theta_{\mathrm{DINO}}$, and the steered parameters are $\theta(\lambda)=\theta_{\mathrm{DINO}}+\lambda\Delta\theta$. Thus, $\lambda=0$ reproduces DINO, $\lambda=1$ reproduces FT-DINO, $\lambda=0.5$ interpolates between them, $\lambda=1.5$ extrapolates along the fine-tuning direction, and $\lambda=-0.5$ moves in the opposite direction. Each steered encoder re-encodes both queries and the historical library, and the historical-retrieval and portfolio-construction rules are otherwise unchanged. $K$ is the number of retrieved neighbors. Ret is the annualized mean H--L return, and SR is the annualized Sharpe ratio.}

\apptablesize
\centering
\setlength{\tabcolsep}{2.0pt}
\begin{tabular*}{\textwidth}{@{\extracolsep{\fill}}c cc cc cc cc cc@{}}
            \toprule

            & \multicolumn{2}{c}{$\lambda=-0.5$}
            & \multicolumn{2}{c}{$\lambda=0$ (DINO)}
            & \multicolumn{2}{c}{$\lambda=0.5$}
            & \multicolumn{2}{c}{$\lambda=1$ (FT-DINO)}
            & \multicolumn{2}{c}{$\lambda=1.5$} \\

            \cmidrule(lr){2-3}
            \cmidrule(lr){4-5}
            \cmidrule(lr){6-7}
            \cmidrule(lr){8-9}
            \cmidrule(lr){10-11}

            $K$
            & Ret & SR
            & Ret & SR
            & Ret & SR
            & Ret & SR
            & Ret & SR \\

            \midrule

            5
            & 0.05 & 0.79
            & 0.05 & 0.83
            & 0.15 & 2.07
            & 0.21 & 2.87
            & 0.19 & 2.60 \\

            10
            & 0.05 & 0.78
            & 0.07 & 1.12
            & 0.20 & 2.92
            & 0.28 & 3.59
            & 0.25 & 3.28 \\

            30
            & 0.09 & 1.22
            & 0.11 & 1.83
            & 0.27 & 3.33
            & 0.41 & 4.59
            & 0.38 & 4.64 \\

            100
            & 0.13 & 1.93
            & 0.16 & 2.15
            & 0.40 & 4.15
            & 0.60 & 5.47
            & 0.55 & 5.37 \\

            500
            & 0.20 & 2.25
            & 0.23 & 2.74
            & 0.57 & 4.81
            & 0.81 & 6.27
            & 0.77 & 6.21 \\

            1{,}000
            & 0.22 & 2.30
            & 0.26 & 2.81
            & 0.63 & 5.15
            & 0.85 & 6.37
            & 0.81 & 6.15 \\

            3{,}000
            & 0.25 & 2.28
            & 0.29 & 2.90
            & 0.69 & 5.27
            & 0.88 & 6.31
            & 0.86 & 6.28 \\

            \bottomrule
\end{tabular*}
\end{table}

\clearpage
\setapptab{D}
\subsection*{Appendix D: Economic Mechanism}
\begin{table}[H]\caption{DINO and Limits to Arbitrage}
\label{tab:app_dino_limits}
\tablenote{This table reports equal-weighted sequential $5\times5$ double sorts on individual limits-to-arbitrage proxies and the benchmark DINO retrieval signal over 2001--2025. Each week, stocks are first sorted into quintiles of the indicated limits-to-arbitrage proxy and then, within each proxy quintile, into conditional quintiles of the DINO retrieval signal ($K=1{,}000$). Columns correspond to proxy quintiles and rows to conditional DINO quintiles; DINO 5--1 is the return on the top minus the bottom conditional signal quintile within each proxy quintile, and the final column reports the difference between proxy quintiles 5 and 1. Proxy quintiles retain the natural ordering of each raw variable, so greater limits to arbitrage correspond to lower Size, Dollar Volume, and Turnover and to higher Amihud Illiquidity, Securities Lending Fee, Analyst Forecast Dispersion, and Days-to-Cover. Portfolios are equal weighted, returns are annualized, and $t$-statistics are reported in parentheses. Panel A conditions on Size and Panel B on Amihud Illiquidity; the continuation tables report Dollar Volume (Panel C), Securities Lending Fee (Panel D, available from 2008), Analyst Forecast Dispersion (Panel E), Turnover (Panel F), and Days-to-Cover (Panel G).}

\apptablesize
\centering
\begin{minipage}{\textwidth}
\centering

\renewcommand{\arraystretch}{0.86}
\setlength{\tabcolsep}{2.0pt}
\panelheading{Panel A. Size}
\begin{tabular*}{\linewidth}{@{\extracolsep{\fill}}lrrrrrr@{}}
\toprule
& Size Q1 & Size Q2 & Size Q3 & Size Q4 & Size Q5 & Size 5--1 \\
\midrule
\multirow[c]{2}{*}{DINO Q1}
    & -0.05 & 0.05 & 0.09 & 0.10 & 0.10 & 0.15 \\
    & (-0.70) & (0.81) & (1.54) & (2.01) & (2.23) & (3.15) \\[0.15em]

    \multirow[c]{2}{*}{DINO Q2}
    & 0.09 & 0.11 & 0.10 & 0.10 & 0.10 & 0.01 \\
    & (1.41) & (1.89) & (1.93) & (2.05) & (2.36) & (0.15) \\[0.15em]

    \multirow[c]{2}{*}{DINO Q3}
    & 0.16 & 0.12 & 0.10 & 0.11 & 0.11 & -0.06 \\
    & (2.53) & (2.24) & (1.98) & (2.39) & (2.58) & (-1.29) \\[0.15em]

    \multirow[c]{2}{*}{DINO Q4}
    & 0.21 & 0.12 & 0.13 & 0.14 & 0.12 & -0.09 \\
    & (3.37) & (2.13) & (2.51) & (2.89) & (2.79) & (-2.12) \\[0.15em]

    \multirow[c]{2}{*}{DINO Q5}
    & 0.37 & 0.22 & 0.18 & 0.14 & 0.15 & -0.23 \\
    & (5.38) & (3.64) & (3.04) & (2.64) & (3.10) & (-4.60) \\[0.15em]

    \midrule

    \multirow[c]{2}{*}{DINO 5--1}
    & 0.42 & 0.17 & 0.09 & 0.04 & 0.05 &  \\
    & (12.45) & (7.05) & (5.34) & (2.36) & (3.08) &  \\

\bottomrule
\end{tabular*}
\panelheading{Panel B. Amihud Illiquidity}
\begin{tabular*}{\linewidth}{@{\extracolsep{\fill}}lrrrrrr@{}}
\toprule
& Amihud Q1 & Amihud Q2 & Amihud Q3 & Amihud Q4 & Amihud Q5 & Amihud 5--1 \\
\midrule
\multirow[c]{2}{*}{DINO Q1}
    & 0.09 & 0.10 & 0.08 & 0.02 & -0.04 & -0.13 \\
    & (2.17) & (1.98) & (1.35) & (0.30) & (-0.56) & (-2.95) \\[0.15em]

    \multirow[c]{2}{*}{DINO Q2}
    & 0.08 & 0.10 & 0.11 & 0.06 & 0.10 & 0.02 \\
    & (2.12) & (2.03) & (1.99) & (1.13) & (1.55) & (0.42) \\[0.15em]

    \multirow[c]{2}{*}{DINO Q3}
    & 0.10 & 0.11 & 0.11 & 0.10 & 0.18 & 0.08 \\
    & (2.63) & (2.32) & (1.99) & (1.69) & (2.82) & (1.83) \\[0.15em]

    \multirow[c]{2}{*}{DINO Q4}
    & 0.11 & 0.13 & 0.12 & 0.15 & 0.21 & 0.10 \\
    & (2.74) & (2.72) & (2.31) & (2.56) & (3.26) & (2.27) \\[0.15em]

    \multirow[c]{2}{*}{DINO Q5}
    & 0.14 & 0.15 & 0.16 & 0.20 & 0.42 & 0.29 \\
    & (3.04) & (2.86) & (2.73) & (3.23) & (5.94) & (5.84) \\[0.15em]

    \midrule

    \multirow[c]{2}{*}{DINO 5--1}
    & 0.04 & 0.05 & 0.08 & 0.18 & 0.46 &  \\
    & (3.09) & (3.27) & (4.62) & (8.22) & (13.27) &  \\

\bottomrule
\end{tabular*}
\end{minipage}
\end{table}
\clearpage
\begin{table}[H]
\apptablesize
\centering
\ContinuedFloat
\caption[]{DINO and Limits to Arbitrage (continued)}
\tablenote{This table continues Table~\ref{tab:app_dino_limits}. Panel C conditions on Dollar Volume, and Panel D conditions on Securities Lending Fee, which is available from 2008. See the notes to Table~\ref{tab:app_dino_limits} for details.}
\begin{minipage}{\textwidth}
\centering

\renewcommand{\arraystretch}{0.86}
\setlength{\tabcolsep}{2.0pt}
\panelheading{Panel C. Dollar Volume}
\begin{tabular*}{\linewidth}{@{\extracolsep{\fill}}lrrrrrr@{}}
\toprule
& DVol Q1 & DVol Q2 & DVol Q3 & DVol Q4 & DVol Q5 & DVol 5--1 \\
\midrule
\multirow[c]{2}{*}{DINO Q1}
    & 0.01 & 0.02 & 0.05 & 0.08 & 0.08 & 0.06 \\
    & (0.23) & (0.37) & (0.80) & (1.57) & (1.65) & (1.75) \\[0.15em]

    \multirow[c]{2}{*}{DINO Q2}
    & 0.13 & 0.08 & 0.08 & 0.09 & 0.07 & -0.05 \\
    & (2.08) & (1.54) & (1.46) & (1.94) & (1.70) & (-1.46) \\[0.15em]

    \multirow[c]{2}{*}{DINO Q3}
    & 0.19 & 0.11 & 0.09 & 0.11 & 0.09 & -0.10 \\
    & (3.25) & (1.99) & (1.81) & (2.31) & (2.17) & (-2.77) \\[0.15em]

    \multirow[c]{2}{*}{DINO Q4}
    & 0.22 & 0.14 & 0.13 & 0.11 & 0.11 & -0.11 \\
    & (3.68) & (2.56) & (2.57) & (2.28) & (2.59) & (-2.93) \\[0.15em]

    \multirow[c]{2}{*}{DINO Q5}
    & 0.42 & 0.23 & 0.15 & 0.14 & 0.13 & -0.29 \\
    & (6.36) & (3.72) & (2.58) & (2.56) & (2.70) & (-6.88) \\[0.15em]

    \midrule

    \multirow[c]{2}{*}{DINO 5--1}
    & 0.40 & 0.21 & 0.10 & 0.06 & 0.06 &  \\
    & (12.87) & (8.32) & (5.23) & (3.49) & (3.55) &  \\

\bottomrule
\end{tabular*}
\panelheading{Panel D. Securities Lending Fee}
\begin{tabular*}{\linewidth}{@{\extracolsep{\fill}}lrrrrrr@{}}
\toprule
& SAF Q1 & SAF Q2 & SAF Q3 & SAF Q4 & SAF Q5 & SAF 5--1 \\
\midrule
\multirow[c]{2}{*}{DINO Q1}
    & 0.13 & 0.13 & 0.12 & 0.11 & -0.10 & -0.23 \\
    & (2.41) & (2.29) & (2.11) & (1.77) & (-1.11) & (-3.89) \\[0.15em]

    \multirow[c]{2}{*}{DINO Q2}
    & 0.12 & 0.12 & 0.13 & 0.11 & 0.02 & -0.11 \\
    & (2.27) & (2.20) & (2.35) & (1.77) & (0.19) & (-1.99) \\[0.15em]

    \multirow[c]{2}{*}{DINO Q3}
    & 0.12 & 0.13 & 0.14 & 0.14 & 0.06 & -0.06 \\
    & (2.32) & (2.41) & (2.42) & (2.14) & (0.74) & (-1.08) \\[0.15em]

    \multirow[c]{2}{*}{DINO Q4}
    & 0.13 & 0.14 & 0.15 & 0.15 & 0.07 & -0.06 \\
    & (2.36) & (2.45) & (2.51) & (2.37) & (0.84) & (-1.09) \\[0.15em]

    \multirow[c]{2}{*}{DINO Q5}
    & 0.15 & 0.16 & 0.18 & 0.21 & 0.16 & 0.01 \\
    & (2.66) & (2.70) & (2.78) & (3.04) & (1.77) & (0.12) \\[0.15em]

    \midrule

    \multirow[c]{2}{*}{DINO 5--1}
    & 0.02 & 0.03 & 0.06 & 0.10 & 0.26 &  \\
    & (1.48) & (1.61) & (3.27) & (3.74) & (5.94) &  \\

\bottomrule
\end{tabular*}
\end{minipage}
\end{table}
\clearpage
\begin{table}[H]
\apptablesize
\centering
\ContinuedFloat
\caption[]{DINO and Limits to Arbitrage (continued)}
\tablenote{This table continues Table~\ref{tab:app_dino_limits}. Panel E conditions on Analyst Forecast Dispersion, and Panel F conditions on Turnover. See the notes to Table~\ref{tab:app_dino_limits} for details.}
\begin{minipage}{\textwidth}
\centering

\renewcommand{\arraystretch}{0.86}
\setlength{\tabcolsep}{2.0pt}
\panelheading{Panel E. Analyst Forecast Dispersion}
\begin{tabular*}{\linewidth}{@{\extracolsep{\fill}}lrrrrrr@{}}
\toprule
& FDisp Q1 & FDisp Q2 & FDisp Q3 & FDisp Q4 & FDisp Q5 & FDisp 5--1 \\
\midrule
\multirow[c]{2}{*}{DINO Q1}
    & 0.08 & 0.09 & 0.10 & 0.08 & 0.01 & -0.07 \\
    & (1.99) & (1.95) & (1.98) & (1.30) & (0.17) & (-1.48) \\[0.15em]

    \multirow[c]{2}{*}{DINO Q2}
    & 0.09 & 0.09 & 0.10 & 0.10 & 0.10 & 0.02 \\
    & (2.16) & (2.12) & (2.02) & (1.75) & (1.41) & (0.36) \\[0.15em]

    \multirow[c]{2}{*}{DINO Q3}
    & 0.11 & 0.11 & 0.12 & 0.12 & 0.15 & 0.04 \\
    & (2.76) & (2.44) & (2.38) & (2.12) & (2.05) & (0.87) \\[0.15em]

    \multirow[c]{2}{*}{DINO Q4}
    & 0.12 & 0.12 & 0.13 & 0.15 & 0.19 & 0.07 \\
    & (3.00) & (2.86) & (2.52) & (2.64) & (2.63) & (1.57) \\[0.15em]

    \multirow[c]{2}{*}{DINO Q5}
    & 0.17 & 0.14 & 0.15 & 0.19 & 0.29 & 0.12 \\
    & (3.97) & (3.03) & (2.94) & (3.07) & (3.55) & (2.20) \\[0.15em]

    \midrule

    \multirow[c]{2}{*}{DINO 5--1}
    & 0.08 & 0.05 & 0.05 & 0.11 & 0.27 &  \\
    & (5.79) & (3.65) & (3.15) & (5.23) & (8.12) &  \\

\bottomrule
\end{tabular*}
\panelheading{Panel F. Turnover}
\begin{tabular*}{\linewidth}{@{\extracolsep{\fill}}lrrrrrr@{}}
\toprule
& Turnover Q1 & Turnover Q2 & Turnover Q3 & Turnover Q4 & Turnover Q5 & Turnover 5--1 \\
\midrule
\multirow[c]{2}{*}{DINO Q1}
    & 0.03 & 0.09 & 0.08 & 0.07 & -0.03 & -0.05 \\
    & (0.57) & (1.74) & (1.60) & (1.17) & (-0.35) & (-1.26) \\[0.15em]

    \multirow[c]{2}{*}{DINO Q2}
    & 0.08 & 0.12 & 0.12 & 0.10 & 0.03 & -0.04 \\
    & (1.72) & (2.51) & (2.51) & (1.86) & (0.52) & (-1.19) \\[0.15em]

    \multirow[c]{2}{*}{DINO Q3}
    & 0.11 & 0.12 & 0.13 & 0.13 & 0.07 & -0.04 \\
    & (2.42) & (2.52) & (2.81) & (2.59) & (1.06) & (-1.01) \\[0.15em]

    \multirow[c]{2}{*}{DINO Q4}
    & 0.13 & 0.15 & 0.14 & 0.14 & 0.15 & 0.02 \\
    & (3.01) & (3.39) & (3.06) & (2.75) & (2.18) & (0.40) \\[0.15em]

    \multirow[c]{2}{*}{DINO Q5}
    & 0.28 & 0.24 & 0.23 & 0.21 & 0.16 & -0.11 \\
    & (5.81) & (4.70) & (4.21) & (3.65) & (2.16) & (-2.54) \\[0.15em]

    \midrule

    \multirow[c]{2}{*}{DINO 5--1}
    & 0.25 & 0.15 & 0.15 & 0.15 & 0.19 &  \\
    & (11.52) & (7.58) & (7.79) & (7.35) & (6.34) &  \\

\bottomrule
\end{tabular*}
\end{minipage}
\end{table}
\clearpage
\begin{table}[H]
\apptablesize
\centering
\ContinuedFloat
\caption[]{DINO and Limits to Arbitrage (continued)}
\tablenote{This table continues Table~\ref{tab:app_dino_limits}. Panel G conditions on Days-to-Cover. See the notes to Table~\ref{tab:app_dino_limits} for details.}
\begin{minipage}{\textwidth}
\centering

\renewcommand{\arraystretch}{0.86}
\setlength{\tabcolsep}{2.0pt}
\panelheading{Panel G. Days-to-Cover}
\begin{tabular*}{\linewidth}{@{\extracolsep{\fill}}lrrrrrr@{}}
\toprule
& DTC Q1 & DTC Q2 & DTC Q3 & DTC Q4 & DTC Q5 & DTC 5--1 \\
\midrule
\multirow[c]{2}{*}{DINO Q1}
    & -0.07 & 0.09 & 0.11 & 0.10 & 0.09 & 0.16 \\
    & (-1.04) & (1.85) & (2.16) & (1.89) & (1.69) & (3.90) \\[0.15em]

    \multirow[c]{2}{*}{DINO Q2}
    & 0.02 & 0.12 & 0.10 & 0.11 & 0.11 & 0.08 \\
    & (0.37) & (2.52) & (2.20) & (2.38) & (2.21) & (2.43) \\[0.15em]

    \multirow[c]{2}{*}{DINO Q3}
    & 0.07 & 0.13 & 0.12 & 0.12 & 0.11 & 0.04 \\
    & (1.14) & (2.90) & (2.65) & (2.40) & (2.32) & (1.07) \\[0.15em]

    \multirow[c]{2}{*}{DINO Q4}
    & 0.13 & 0.16 & 0.14 & 0.15 & 0.12 & -0.02 \\
    & (2.17) & (3.34) & (2.87) & (3.01) & (2.37) & (-0.44) \\[0.15em]

    \multirow[c]{2}{*}{DINO Q5}
    & 0.26 & 0.21 & 0.20 & 0.20 & 0.17 & -0.10 \\
    & (3.76) & (4.02) & (3.81) & (3.62) & (3.27) & (-2.26) \\[0.15em]

    \midrule

    \multirow[c]{2}{*}{DINO 5--1}
    & 0.33 & 0.11 & 0.09 & 0.10 & 0.08 &  \\
    & (11.33) & (6.58) & (5.68) & (5.98) & (4.70) &  \\

\bottomrule
\end{tabular*}
\end{minipage}
\end{table}
\clearpage
\begin{table}[H]\caption{FT-DINO and Limits to Arbitrage}
\label{tab:app_ft_limits}
\tablenote{This table reports equal-weighted sequential $5\times5$ double sorts on individual limits-to-arbitrage proxies and the FT-DINO retrieval signal over 2001--2025. Each week, stocks are first sorted into quintiles of the indicated limits-to-arbitrage proxy and then, within each proxy quintile, into conditional quintiles of the FT-DINO retrieval signal ($K=1{,}000$). Columns correspond to proxy quintiles and rows to conditional FT-DINO quintiles; FT-DINO 5--1 is the return on the top minus the bottom conditional signal quintile within each proxy quintile, and the final column reports the difference between proxy quintiles 5 and 1. Proxy quintiles retain the natural ordering of each raw variable, so greater limits to arbitrage correspond to lower Size, Dollar Volume, and Turnover and to higher Amihud Illiquidity, Securities Lending Fee, Analyst Forecast Dispersion, and Days-to-Cover. Portfolios are equal weighted, returns are annualized, and $t$-statistics are reported in parentheses. Panel A conditions on Size and Panel B on Amihud Illiquidity; the continuation tables report Dollar Volume (Panel C), Securities Lending Fee (Panel D, available from 2008), Analyst Forecast Dispersion (Panel E), Turnover (Panel F), and Days-to-Cover (Panel G).}

\apptablesize
\centering
\begin{minipage}{\textwidth}
\centering

\renewcommand{\arraystretch}{0.86}
\setlength{\tabcolsep}{2.0pt}
\panelheading{Panel A. Size}
\begin{tabular*}{\linewidth}{@{\extracolsep{\fill}}lrrrrrr@{}}
\toprule
& Size Q1 & Size Q2 & Size Q3 & Size Q4 & Size Q5 & Size 5--1 \\
\midrule
\multirow[c]{2}{*}{FT-DINO Q1}
    & -0.44 & -0.14 & 0.00 & 0.05 & 0.05 & 0.50 \\
    & (-6.03) & (-2.37) & (0.05) & (0.88) & (1.19) & (9.67) \\[0.15em]

    \multirow[c]{2}{*}{FT-DINO Q2}
    & -0.06 & 0.04 & 0.08 & 0.09 & 0.09 & 0.15 \\
    & (-0.94) & (0.77) & (1.46) & (2.05) & (2.26) & (3.59) \\[0.15em]

    \multirow[c]{2}{*}{FT-DINO Q3}
    & 0.15 & 0.11 & 0.11 & 0.12 & 0.11 & -0.04 \\
    & (2.53) & (2.21) & (2.11) & (2.52) & (2.66) & (-1.06) \\[0.15em]

    \multirow[c]{2}{*}{FT-DINO Q4}
    & 0.32 & 0.19 & 0.15 & 0.12 & 0.13 & -0.18 \\
    & (5.09) & (3.49) & (2.83) & (2.38) & (3.15) & (-4.34) \\[0.15em]

    \multirow[c]{2}{*}{FT-DINO Q5}
    & 0.82 & 0.41 & 0.27 & 0.22 & 0.18 & -0.64 \\
    & (11.10) & (6.28) & (4.37) & (3.95) & (3.71) & (-11.77) \\[0.15em]

    \midrule

    \multirow[c]{2}{*}{FT-DINO 5--1}
    & 1.27 & 0.55 & 0.27 & 0.17 & 0.13 &  \\
    & (25.73) & (16.45) & (10.14) & (8.04) & (6.40) &  \\

\bottomrule
\end{tabular*}
\panelheading{Panel B. Amihud Illiquidity}
\begin{tabular*}{\linewidth}{@{\extracolsep{\fill}}lrrrrrr@{}}
\toprule
& Amihud Q1 & Amihud Q2 & Amihud Q3 & Amihud Q4 & Amihud Q5 & Amihud 5--1 \\
\midrule
\multirow[c]{2}{*}{FT-DINO Q1}
    & 0.05 & 0.05 & -0.01 & -0.18 & -0.46 & -0.51 \\
    & (1.14) & (0.90) & (-0.15) & (-2.70) & (-6.33) & (-10.47) \\[0.15em]

    \multirow[c]{2}{*}{FT-DINO Q2}
    & 0.09 & 0.10 & 0.07 & 0.04 & -0.04 & -0.13 \\
    & (2.20) & (2.07) & (1.35) & (0.65) & (-0.68) & (-3.08) \\[0.15em]

    \multirow[c]{2}{*}{FT-DINO Q3}
    & 0.10 & 0.11 & 0.11 & 0.11 & 0.14 & 0.04 \\
    & (2.66) & (2.39) & (2.17) & (2.04) & (2.33) & (1.00) \\[0.15em]

    \multirow[c]{2}{*}{FT-DINO Q4}
    & 0.12 & 0.13 & 0.14 & 0.18 & 0.35 & 0.23 \\
    & (3.06) & (2.70) & (2.56) & (3.19) & (5.55) & (5.44) \\[0.15em]

    \multirow[c]{2}{*}{FT-DINO Q5}
    & 0.16 & 0.20 & 0.26 & 0.38 & 0.89 & 0.73 \\
    & (3.58) & (3.78) & (4.26) & (5.69) & (11.40) & (12.45) \\[0.15em]

    \midrule

    \multirow[c]{2}{*}{FT-DINO 5--1}
    & 0.11 & 0.16 & 0.27 & 0.56 & 1.35 &  \\
    & (6.90) & (7.83) & (10.58) & (17.86) & (24.35) &  \\

\bottomrule
\end{tabular*}
\end{minipage}
\end{table}
\clearpage
\begin{table}[H]
\apptablesize
\centering
\ContinuedFloat
\caption[]{FT-DINO and Limits to Arbitrage (continued)}
\tablenote{This table continues Table~\ref{tab:app_ft_limits}. Panel C conditions on Dollar Volume, and Panel D conditions on Securities Lending Fee, which is available from 2008. See the notes to Table~\ref{tab:app_ft_limits} for details.}
\begin{minipage}{\textwidth}
\centering

\renewcommand{\arraystretch}{0.86}
\setlength{\tabcolsep}{2.0pt}
\panelheading{Panel C. Dollar Volume}
\begin{tabular*}{\linewidth}{@{\extracolsep{\fill}}lrrrrrr@{}}
\toprule
& DVol Q1 & DVol Q2 & DVol Q3 & DVol Q4 & DVol Q5 & DVol 5--1 \\
\midrule
\multirow[c]{2}{*}{FT-DINO Q1}
    & -0.42 & -0.16 & -0.06 & 0.01 & 0.02 & 0.45 \\
    & (-6.35) & (-2.42) & (-0.99) & (0.20) & (0.48) & (11.26) \\[0.15em]

    \multirow[c]{2}{*}{FT-DINO Q2}
    & -0.01 & 0.04 & 0.06 & 0.10 & 0.08 & 0.09 \\
    & (-0.18) & (0.79) & (1.18) & (2.02) & (1.86) & (2.50) \\[0.15em]

    \multirow[c]{2}{*}{FT-DINO Q3}
    & 0.17 & 0.12 & 0.11 & 0.10 & 0.10 & -0.07 \\
    & (3.16) & (2.21) & (2.12) & (2.17) & (2.35) & (-2.38) \\[0.15em]

    \multirow[c]{2}{*}{FT-DINO Q4}
    & 0.35 & 0.18 & 0.14 & 0.12 & 0.12 & -0.23 \\
    & (6.02) & (3.35) & (2.66) & (2.53) & (2.76) & (-6.37) \\[0.15em]

    \multirow[c]{2}{*}{FT-DINO Q5}
    & 0.88 & 0.41 & 0.25 & 0.20 & 0.17 & -0.71 \\
    & (11.89) & (6.20) & (4.15) & (3.71) & (3.32) & (-13.75) \\[0.15em]

    \midrule

    \multirow[c]{2}{*}{FT-DINO 5--1}
    & 1.30 & 0.57 & 0.31 & 0.19 & 0.14 &  \\
    & (24.86) & (17.16) & (11.49) & (9.33) & (7.58) &  \\

\bottomrule
\end{tabular*}
\panelheading{Panel D. Securities Lending Fee}
\begin{tabular*}{\linewidth}{@{\extracolsep{\fill}}lrrrrrr@{}}
\toprule
& SAF Q1 & SAF Q2 & SAF Q3 & SAF Q4 & SAF Q5 & SAF 5--1 \\
\midrule
\multirow[c]{2}{*}{FT-DINO Q1}
    & 0.07 & 0.07 & 0.06 & -0.02 & -0.32 & -0.39 \\
    & (1.25) & (1.25) & (0.93) & (-0.23) & (-3.38) & (-5.96) \\[0.15em]

    \multirow[c]{2}{*}{FT-DINO Q2}
    & 0.12 & 0.10 & 0.11 & 0.09 & -0.06 & -0.17 \\
    & (2.27) & (1.85) & (1.93) & (1.46) & (-0.69) & (-3.38) \\[0.15em]

    \multirow[c]{2}{*}{FT-DINO Q3}
    & 0.12 & 0.13 & 0.14 & 0.13 & 0.03 & -0.09 \\
    & (2.37) & (2.55) & (2.56) & (2.06) & (0.39) & (-1.93) \\[0.15em]

    \multirow[c]{2}{*}{FT-DINO Q4}
    & 0.14 & 0.14 & 0.16 & 0.16 & 0.12 & -0.02 \\
    & (2.51) & (2.47) & (2.76) & (2.55) & (1.50) & (-0.29) \\[0.15em]

    \multirow[c]{2}{*}{FT-DINO Q5}
    & 0.21 & 0.24 & 0.26 & 0.36 & 0.43 & 0.22 \\
    & (3.51) & (3.55) & (3.80) & (5.06) & (4.39) & (3.32) \\[0.15em]

    \midrule

    \multirow[c]{2}{*}{FT-DINO 5--1}
    & 0.14 & 0.17 & 0.20 & 0.38 & 0.75 &  \\
    & (5.65) & (5.06) & (8.76) & (12.85) & (14.90) &  \\

\bottomrule
\end{tabular*}
\end{minipage}
\end{table}
\clearpage
\begin{table}[H]
\apptablesize
\centering
\ContinuedFloat
\caption[]{FT-DINO and Limits to Arbitrage (continued)}
\tablenote{This table continues Table~\ref{tab:app_ft_limits}. Panel E conditions on Analyst Forecast Dispersion, and Panel F conditions on Turnover. See the notes to Table~\ref{tab:app_ft_limits} for details.}
\begin{minipage}{\textwidth}
\centering

\renewcommand{\arraystretch}{0.86}
\setlength{\tabcolsep}{2.0pt}
\panelheading{Panel E. Analyst Forecast Dispersion}
\begin{tabular*}{\linewidth}{@{\extracolsep{\fill}}lrrrrrr@{}}
\toprule
& FDisp Q1 & FDisp Q2 & FDisp Q3 & FDisp Q4 & FDisp Q5 & FDisp 5--1 \\
\midrule
\multirow[c]{2}{*}{FT-DINO Q1}
    & 0.02 & 0.03 & 0.01 & -0.06 & -0.23 & -0.25 \\
    & (0.49) & (0.69) & (0.20) & (-1.06) & (-2.82) & (-4.73) \\[0.15em]

    \multirow[c]{2}{*}{FT-DINO Q2}
    & 0.08 & 0.07 & 0.08 & 0.09 & 0.06 & -0.02 \\
    & (1.94) & (1.65) & (1.60) & (1.64) & (0.82) & (-0.45) \\[0.15em]

    \multirow[c]{2}{*}{FT-DINO Q3}
    & 0.10 & 0.10 & 0.12 & 0.12 & 0.17 & 0.06 \\
    & (2.74) & (2.44) & (2.50) & (2.17) & (2.49) & (1.54) \\[0.15em]

    \multirow[c]{2}{*}{FT-DINO Q4}
    & 0.12 & 0.13 & 0.14 & 0.17 & 0.21 & 0.09 \\
    & (3.16) & (3.00) & (2.83) & (3.05) & (2.95) & (1.96) \\[0.15em]

    \multirow[c]{2}{*}{FT-DINO Q5}
    & 0.24 & 0.22 & 0.25 & 0.31 & 0.53 & 0.29 \\
    & (5.46) & (4.40) & (4.56) & (4.86) & (6.01) & (4.91) \\[0.15em]

    \midrule

    \multirow[c]{2}{*}{FT-DINO 5--1}
    & 0.21 & 0.19 & 0.24 & 0.38 & 0.76 &  \\
    & (11.62) & (9.48) & (12.28) & (13.55) & (15.34) &  \\

\bottomrule
\end{tabular*}
\panelheading{Panel F. Turnover}
\begin{tabular*}{\linewidth}{@{\extracolsep{\fill}}lrrrrrr@{}}
\toprule
& Turnover Q1 & Turnover Q2 & Turnover Q3 & Turnover Q4 & Turnover Q5 & Turnover 5--1 \\
\midrule
\multirow[c]{2}{*}{FT-DINO Q1}
    & -0.32 & -0.11 & -0.04 & -0.03 & -0.18 & 0.14 \\
    & (-6.12) & (-2.03) & (-0.80) & (-0.54) & (-2.34) & (2.82) \\[0.15em]

    \multirow[c]{2}{*}{FT-DINO Q2}
    & -0.00 & 0.08 & 0.08 & 0.09 & 0.04 & 0.04 \\
    & (-0.00) & (1.75) & (1.68) & (1.84) & (0.65) & (1.18) \\[0.15em]

    \multirow[c]{2}{*}{FT-DINO Q3}
    & 0.11 & 0.12 & 0.14 & 0.12 & 0.09 & -0.02 \\
    & (2.59) & (2.84) & (3.18) & (2.32) & (1.41) & (-0.67) \\[0.15em]

    \multirow[c]{2}{*}{FT-DINO Q4}
    & 0.23 & 0.18 & 0.17 & 0.17 & 0.13 & -0.10 \\
    & (5.37) & (4.01) & (3.59) & (3.26) & (1.92) & (-2.81) \\[0.15em]

    \multirow[c]{2}{*}{FT-DINO Q5}
    & 0.60 & 0.44 & 0.36 & 0.31 & 0.32 & -0.28 \\
    & (11.25) & (7.67) & (6.13) & (5.15) & (4.02) & (-6.09) \\[0.15em]

    \midrule

    \multirow[c]{2}{*}{FT-DINO 5--1}
    & 0.92 & 0.55 & 0.40 & 0.34 & 0.50 &  \\
    & (23.91) & (16.25) & (14.58) & (13.43) & (14.63) &  \\

\bottomrule
\end{tabular*}
\end{minipage}
\end{table}
\clearpage
\begin{table}[H]
\apptablesize
\centering
\ContinuedFloat
\caption[]{FT-DINO and Limits to Arbitrage (continued)}
\tablenote{This table continues Table~\ref{tab:app_ft_limits}. Panel G conditions on Days-to-Cover. See the notes to Table~\ref{tab:app_ft_limits} for details.}
\begin{minipage}{\textwidth}
\centering

\renewcommand{\arraystretch}{0.86}
\setlength{\tabcolsep}{2.0pt}
\panelheading{Panel G. Days-to-Cover}
\begin{tabular*}{\linewidth}{@{\extracolsep{\fill}}lrrrrrr@{}}
\toprule
& DTC Q1 & DTC Q2 & DTC Q3 & DTC Q4 & DTC Q5 & DTC 5--1 \\
\midrule
\multirow[c]{2}{*}{FT-DINO Q1}
    & -0.41 & -0.05 & -0.01 & -0.00 & -0.04 & 0.37 \\
    & (-5.80) & (-0.88) & (-0.27) & (-0.03) & (-0.81) & (8.60) \\[0.15em]

    \multirow[c]{2}{*}{FT-DINO Q2}
    & -0.07 & 0.09 & 0.10 & 0.10 & 0.07 & 0.15 \\
    & (-1.22) & (1.96) & (2.20) & (2.13) & (1.58) & (4.38) \\[0.15em]

    \multirow[c]{2}{*}{FT-DINO Q3}
    & 0.07 & 0.12 & 0.12 & 0.11 & 0.12 & 0.05 \\
    & (1.20) & (2.62) & (2.54) & (2.40) & (2.47) & (1.44) \\[0.15em]

    \multirow[c]{2}{*}{FT-DINO Q4}
    & 0.20 & 0.17 & 0.16 & 0.16 & 0.16 & -0.04 \\
    & (3.32) & (3.53) & (3.32) & (3.16) & (3.29) & (-1.21) \\[0.15em]

    \multirow[c]{2}{*}{FT-DINO Q5}
    & 0.63 & 0.39 & 0.31 & 0.30 & 0.28 & -0.35 \\
    & (8.39) & (7.11) & (5.66) & (5.42) & (5.22) & (-7.84) \\[0.15em]

    \midrule

    \multirow[c]{2}{*}{FT-DINO 5--1}
    & 1.04 & 0.43 & 0.33 & 0.30 & 0.32 &  \\
    & (25.63) & (16.11) & (14.12) & (15.35) & (13.70) &  \\

\bottomrule
\end{tabular*}
\end{minipage}
\end{table}
\clearpage
\begin{table}[H]\caption{DINO and Retail Trading}
\label{tab:app_dino_retail}
\tablenote{This table reports equal-weighted sequential $5\times5$ double sorts on retail-trading proxies and the benchmark DINO retrieval signal. Each week, stocks are first sorted into quintiles of the indicated retail-trading proxy and then, within each proxy quintile, into conditional quintiles of the DINO retrieval signal ($K=1{,}000$). Columns correspond to proxy quintiles, with Q5 denoting the most intensive retail activity, and rows correspond to conditional DINO quintiles; DINO 5--1 is the return on the top minus the bottom conditional signal quintile within each proxy quintile. Panel A uses RSVOL, the retail share of trading volume of \citet{laarits2025}, over 2006--2024, and Panel B uses the Robinhood demand shock over July 2018 to September 2020. Portfolios are equal weighted, returns are annualized, and $t$-statistics are reported in parentheses.}

\apptablesize
\centering
\begin{minipage}{\textwidth}
\centering

\renewcommand{\arraystretch}{0.86}
\setlength{\tabcolsep}{2.0pt}
\panelheading{Panel A. RSVOL}
\begin{tabular*}{\linewidth}{@{\extracolsep{\fill}}lrrrrrr@{}}
\toprule
& RSVOL Q1 & RSVOL Q2 & RSVOL Q3 & RSVOL Q4 & RSVOL Q5 & RSVOL 5--1 \\
\midrule
\multirow[c]{2}{*}{DINO Q1}
    & 0.10 & 0.10 & 0.10 & 0.07 & -0.11 & -0.21 \\
    & (2.02) & (1.98) & (1.84) & (1.04) & (-1.43) & (-4.28) \\[0.15em]

    \multirow[c]{2}{*}{DINO Q2}
    & 0.10 & 0.10 & 0.09 & 0.11 & 0.02 & -0.08 \\
    & (1.97) & (1.99) & (1.65) & (1.72) & (0.21) & (-1.63) \\[0.15em]

    \multirow[c]{2}{*}{DINO Q3}
    & 0.10 & 0.11 & 0.11 & 0.09 & 0.08 & -0.02 \\
    & (1.98) & (2.10) & (2.08) & (1.41) & (1.09) & (-0.31) \\[0.15em]

    \multirow[c]{2}{*}{DINO Q4}
    & 0.10 & 0.10 & 0.13 & 0.13 & 0.13 & 0.03 \\
    & (2.05) & (1.89) & (2.30) & (2.05) & (1.72) & (0.52) \\[0.15em]

    \multirow[c]{2}{*}{DINO Q5}
    & 0.11 & 0.14 & 0.14 & 0.17 & 0.23 & 0.12 \\
    & (2.08) & (2.58) & (2.31) & (2.44) & (2.82) & (2.08) \\[0.15em]

    \midrule

    \multirow[c]{2}{*}{DINO 5--1}
    & 0.01 & 0.04 & 0.03 & 0.10 & 0.34 &  \\
    & (0.69) & (2.19) & (1.99) & (4.33) & (10.84) &  \\

\bottomrule
\end{tabular*}
\panelheading{Panel B. Robinhood Demand Shock}
\begin{tabular*}{\linewidth}{@{\extracolsep{\fill}}lrrrrrr@{}}
\toprule
& RH Shock Q1 & RH Shock Q2 & RH Shock Q3 & RH Shock Q4 & RH Shock Q5 & RH Shock 5--1 \\
\midrule
\multirow[c]{2}{*}{DINO Q1}
    & -0.02 & 0.11 & 0.09 & 0.11 & 0.16 & 0.18 \\
    & (-0.08) & (0.49) & (0.44) & (0.45) & (0.65) & (1.79) \\[0.15em]

    \multirow[c]{2}{*}{DINO Q2}
    & 0.06 & 0.09 & 0.02 & 0.09 & 0.11 & 0.05 \\
    & (0.27) & (0.39) & (0.10) & (0.40) & (0.51) & (0.48) \\[0.15em]

    \multirow[c]{2}{*}{DINO Q3}
    & 0.01 & 0.03 & 0.04 & 0.11 & 0.16 & 0.15 \\
    & (0.03) & (0.12) & (0.19) & (0.50) & (0.66) & (1.42) \\[0.15em]

    \multirow[c]{2}{*}{DINO Q4}
    & 0.02 & 0.09 & 0.09 & 0.12 & 0.29 & 0.27 \\
    & (0.09) & (0.42) & (0.41) & (0.48) & (1.21) & (2.12) \\[0.15em]

    \multirow[c]{2}{*}{DINO Q5}
    & 0.10 & 0.05 & 0.16 & 0.19 & 0.39 & 0.28 \\
    & (0.42) & (0.19) & (0.63) & (0.74) & (1.30) & (1.67) \\[0.15em]

    \midrule

    \multirow[c]{2}{*}{DINO 5--1}
    & 0.12 & -0.06 & 0.07 & 0.09 & 0.23 &  \\
    & (1.80) & (-0.75) & (0.84) & (1.04) & (2.30) &  \\

\bottomrule
\end{tabular*}
\end{minipage}
\end{table}
\clearpage
\begin{table}[H]\caption{FT-DINO and Retail Trading}
\label{tab:app_ft_retail}
\tablenote{This table reports equal-weighted sequential $5\times5$ double sorts on retail-trading proxies and the FT-DINO retrieval signal. Each week, stocks are first sorted into quintiles of the indicated retail-trading proxy and then, within each proxy quintile, into conditional quintiles of the FT-DINO retrieval signal ($K=1{,}000$). Columns correspond to proxy quintiles, with Q5 denoting the most intensive retail activity, and rows correspond to conditional FT-DINO quintiles; FT-DINO 5--1 is the return on the top minus the bottom conditional signal quintile within each proxy quintile. Panel A uses RSVOL, the retail share of trading volume of \citet{laarits2025}, over 2006--2024, and Panel B uses the Robinhood demand shock over July 2018 to September 2020. Portfolios are equal weighted, returns are annualized, and $t$-statistics are reported in parentheses.}

\apptablesize
\centering
\begin{minipage}{\textwidth}
\centering

\renewcommand{\arraystretch}{0.86}
\setlength{\tabcolsep}{2.0pt}
\panelheading{Panel A. RSVOL}
\begin{tabular*}{\linewidth}{@{\extracolsep{\fill}}lrrrrrr@{}}
\toprule
& RSVOL Q1 & RSVOL Q2 & RSVOL Q3 & RSVOL Q4 & RSVOL Q5 & RSVOL 5--1 \\
\midrule
\multirow[c]{2}{*}{FT-DINO Q1}
    & 0.06 & 0.04 & 0.02 & -0.07 & -0.38 & -0.43 \\
    & (1.18) & (0.87) & (0.45) & (-1.08) & (-4.58) & (-7.87) \\[0.15em]

    \multirow[c]{2}{*}{FT-DINO Q2}
    & 0.08 & 0.08 & 0.09 & 0.05 & -0.07 & -0.15 \\
    & (1.70) & (1.64) & (1.67) & (0.76) & (-0.90) & (-3.06) \\[0.15em]

    \multirow[c]{2}{*}{FT-DINO Q3}
    & 0.10 & 0.10 & 0.11 & 0.12 & 0.09 & -0.00 \\
    & (2.03) & (1.98) & (1.91) & (1.91) & (1.29) & (-0.10) \\[0.15em]

    \multirow[c]{2}{*}{FT-DINO Q4}
    & 0.11 & 0.14 & 0.13 & 0.16 & 0.19 & 0.08 \\
    & (2.05) & (2.66) & (2.35) & (2.47) & (2.50) & (1.63) \\[0.15em]

    \multirow[c]{2}{*}{FT-DINO Q5}
    & 0.16 & 0.19 & 0.22 & 0.31 & 0.50 & 0.34 \\
    & (3.04) & (3.24) & (3.59) & (4.47) & (6.11) & (5.99) \\[0.15em]

    \midrule

    \multirow[c]{2}{*}{FT-DINO 5--1}
    & 0.10 & 0.14 & 0.19 & 0.39 & 0.88 &  \\
    & (6.87) & (7.41) & (7.82) & (11.59) & (21.13) &  \\

\bottomrule
\end{tabular*}
\panelheading{Panel B. Robinhood Demand Shock}
\begin{tabular*}{\linewidth}{@{\extracolsep{\fill}}lrrrrrr@{}}
\toprule
& RH Shock Q1 & RH Shock Q2 & RH Shock Q3 & RH Shock Q4 & RH Shock Q5 & RH Shock 5--1 \\
\midrule
\multirow[c]{2}{*}{FT-DINO Q1}
    & -0.16 & -0.00 & 0.02 & 0.00 & 0.21 & 0.37 \\
    & (-0.69) & (-0.01) & (0.07) & (0.00) & (0.88) & (2.64) \\[0.15em]

    \multirow[c]{2}{*}{FT-DINO Q2}
    & 0.05 & 0.04 & 0.03 & 0.06 & 0.09 & 0.04 \\
    & (0.23) & (0.21) & (0.13) & (0.29) & (0.38) & (0.43) \\[0.15em]

    \multirow[c]{2}{*}{FT-DINO Q3}
    & -0.01 & 0.02 & 0.02 & 0.06 & 0.18 & 0.18 \\
    & (-0.02) & (0.11) & (0.10) & (0.27) & (0.76) & (1.57) \\[0.15em]

    \multirow[c]{2}{*}{FT-DINO Q4}
    & 0.03 & 0.02 & 0.07 & 0.11 & 0.13 & 0.10 \\
    & (0.14) & (0.10) & (0.34) & (0.47) & (0.56) & (0.99) \\[0.15em]

    \multirow[c]{2}{*}{FT-DINO Q5}
    & 0.26 & 0.28 & 0.27 & 0.38 & 0.50 & 0.24 \\
    & (1.01) & (1.08) & (1.04) & (1.36) & (1.60) & (1.42) \\[0.15em]

    \midrule

    \multirow[c]{2}{*}{FT-DINO 5--1}
    & 0.41 & 0.28 & 0.26 & 0.38 & 0.29 &  \\
    & (5.65) & (3.48) & (3.28) & (4.36) & (1.95) &  \\

\bottomrule
\end{tabular*}
\end{minipage}
\end{table}
\clearpage
\begin{table}[H]\caption{FT-DINO Performance by Neighbor Similarity Rank}
\label{tab:app_similarity}
\tablenote{This table reports high-minus-low portfolio performance for stand-alone FT-DINO signals formed from bands of 200 neighbors by similarity rank over 2001--2025. For each query, the 1,000 most similar historical charts are split into five bands of 200 by cosine similarity, from ranks 1--200 (most similar) to ranks 801--1,000 (least similar), and each band forms a separate retrieval signal. The benchmark uses all 1,000 neighbors. Each week, stocks are sorted into deciles by the signal, and H--L is the return on decile 10 minus decile 1 under equal weighting (EW) or capped value weighting (CVW). Returns are annualized.}
\apptablesize
\centering
\begin{tabular*}{\textwidth}{@{\extracolsep{\fill}}lc cc cc@{}}
\toprule
& & \multicolumn{2}{c}{EW} & \multicolumn{2}{c}{CVW} \\
\cmidrule(lr){3-4}\cmidrule(lr){5-6}
Neighbors & Similarity rank & Return & Sharpe & Return & Sharpe \\
\midrule
Benchmark (all 1,000) & 1--1,000 & 0.85 & 6.37 & 0.28 & 2.38 \\
\addlinespace[2pt]
Band 1 (most similar) & 1--200 & 0.71 & 6.00 & 0.20 & 2.11 \\
Band 2 & 201--400 & 0.70 & 5.89 & 0.21 & 2.21 \\
Band 3 & 401--600 & 0.68 & 5.77 & 0.21 & 2.10 \\
Band 4 & 601--800 & 0.69 & 5.75 & 0.21 & 2.12 \\
Band 5 (least similar) & 801--1,000 & 0.67 & 5.63 & 0.19 & 1.95 \\
\bottomrule
\end{tabular*}
\end{table}
\clearpage
\begin{table}[H]\caption{FT-DINO Performance under Alternative Neighbor Weights}
\label{tab:app_memory_weights}
\tablenote{This table reports high-minus-low portfolio performance for the FT-DINO signal ($K=1{,}000$) when the retrieved neighbors are weighted by characteristics associated with memorable past episodes, over 2001--2025. Equal weighting is the baseline. Composite salience is the equally weighted average of the percentile ranks of each neighbor's absolute subsequent five-day return, abnormal trading volume, abnormal turnover, Google search attention, same-FF12-industry status, and recency. Absolute return weights by the neighbor's absolute subsequent five-day return, and abnormal volume and abnormal turnover by its trading activity relative to normal levels. Industry assigns a weight of one to neighbors in the query stock's FF12 industry and zero otherwise. Time decay weights by the percentile rank of the neighbor's date among the 1,000 neighbors, so more recent neighbors receive higher weights. Google search weights use the monthly cross-sectional rank of abnormal search volume for the neighbor's firm in the last completed calendar month before the neighbor's formation date (historical) or before the query date (current). Each week, stocks are sorted into deciles by the signal, and H--L is the return on decile 10 minus decile 1 under equal weighting (EW) or capped value weighting (CVW). Returns are annualized.}
\apptablesize
\centering
\begin{tabular*}{\textwidth}{@{\extracolsep{\fill}}l cc cc@{}}
\toprule
& \multicolumn{2}{c}{EW} & \multicolumn{2}{c}{CVW} \\
\cmidrule(lr){2-3}\cmidrule(lr){4-5}
Neighbor weighting & Return & Sharpe & Return & Sharpe \\
\midrule
Equal (baseline) & 0.85 & 6.37 & 0.28 & 2.38 \\
\addlinespace[2pt]
Composite salience & 0.74 & 5.50 & 0.24 & 2.08 \\
Absolute return & 0.51 & 2.94 & 0.16 & 1.23 \\
Abnormal volume & 0.20 & 1.30 & 0.10 & 0.73 \\
Abnormal turnover & 0.21 & 1.33 & 0.10 & 0.73 \\
Industry & 0.55 & 5.37 & 0.13 & 1.56 \\
Time decay & 0.80 & 5.98 & 0.26 & 2.29 \\
Google search (historical) & 0.22 & 2.68 & 0.06 & 0.69 \\
Google search (current) & 0.41 & 3.72 & 0.14 & 1.34 \\
\bottomrule
\end{tabular*}
\end{table}
\clearpage
\setapptab{E}
\subsection*{Appendix E: Sparse-Feature Interpretation}
\begin{table}[H]\caption{SAE Visual States and Constituent Features}
\label{app:saestates}
\tablenote{This table maps the 30 visually consistent SAE features into 11 visual states. The SAE is a JumpReLU sparse autoencoder with 16,384 latent features trained on FT-DINO representations from 1993--2000. To rank features by return-related importance, for each test year we standardize the SAE activations and reduce them to 500 truncated-SVD components. An $L_2$-penalized logistic regression then predicts the sign of the subsequent five-day return using these components. The estimated coefficients are projected back to the original 16,384-dimensional SAE feature space, and each feature's year-specific importance is measured as its share of the absolute back-projected coefficients. These importance shares are aggregated ex post across test years. Among the 100 highest-ranked features, a feature is retained when its representative charts from the top 5\% of positive activations share the same visual characteristic. Retained features are grouped into visual states based on descriptions of their representative charts. SAE Feature IDs are listed in the order displayed in Figures~\ref{fig:sae_states_c01_c03}--\ref{fig:sae_states_c09_c11}, and $N$ is the number of constituent features.}

\apptablesize
\centering

\begin{threeparttable}

\setlength{\tabcolsep}{4.5pt}
\renewcommand{\arraystretch}{1.14}

\begin{tabularx}{\textwidth}{
    @{}
    >{\centering\arraybackslash}p{1.15cm}
    >{\raggedright\arraybackslash}p{3.60cm}
    >{\raggedright\arraybackslash}X
    >{\centering\arraybackslash}p{0.50cm}
    @{}
}
\toprule
Cluster
& Visual State
& SAE Feature IDs
& $N$ \\
\midrule

C01
& Sustained Uptrend
& 16299
& 1 \\

C02
& Peak Pullback
& 1302, 13116, 712
& 3 \\

C03
& Downtrend Recovery
& 9216, 2838
& 2 \\

C04
& Late Upside Breakout
& 10529, 6232, 1580
& 3 \\

C05
& Stable Downtrend
& 5178, 7435, 9690
& 3 \\

C06
& Stepwise Downtrend
& 8031, 3657, 2699
& 3 \\

C07
& Accelerating Downtrend
& 7123
& 1 \\

C08
& Late Breakdown
& 15536, 12488, 35, 2751, 3560, 7040
& 6 \\

C09
& High-Volatility Range
& 4500, 1984, 12418
& 3 \\

C10
& Isolated Upward Spike
& 12695, 8453
& 2 \\

C11
& High-Price Plateau
& 9418, 9776, 3635
& 3 \\

\midrule
\multicolumn{3}{r}{Total}
& 30 \\
\bottomrule
\end{tabularx}
\end{threeparttable}
\end{table}
\clearpage

\begin{table}[H]
\centering
\caption{Literature-Based Visual Characteristics}
\label{tab:literature_patterns}
\tablenote{This table lists the visual characteristics motivated by the cited finance literature and located in the visual atlas. They comprise eight classical formations from \citet{lo2000}, continuous and discrete winner and loser paths from \citet{da2014}, MAX(1) from \citet{bali2011}, negative overnight jumps from \citet{jiangzhu2017}, drop-and-rebound patterns from \citet{bremer1991}, combinations of price direction and price-path convexity from \citet{gulenwoeppel2026}, daytime reversals from \citet{akbas2022}, and price-volume lead-lag dynamics from \citet{liao2022}. For each characteristic, we select an illustrative real-market 20-day chart that captures a visual configuration motivated by the cited study and map it to the nearest atlas prototype. Figure~\ref{fig:landmarks_main} displays the corresponding locations.}

\apptablesize

\begin{threeparttable}
\renewcommand{\arraystretch}{1.18}
\begin{tabularx}{\textwidth}{@{}Y>{\raggedright\arraybackslash}p{4.2cm}@{}}
\toprule
Visual Characteristics & Reference \\
\midrule

\textbf{No.01} H\&S;
\textbf{No.02} Inv. H\&S;
\textbf{No.03} Broadening Top;
\textbf{No.04} Broadening Bottom;
\textbf{No.05} Triangle Top;
\textbf{No.06} Triangle Bottom;
\textbf{No.07} Double Top;
\textbf{No.08} Double Bottom
& \citet{lo2000}, JF \\

\textbf{No.09} Continuous Winner;
\textbf{No.10} Discrete Winner;
\textbf{No.11} Continuous Loser;
\textbf{No.12} Discrete Loser
& \citet{da2014}, RFS \\

\textbf{No.13} MAX(1)
& \citet{bali2011}, JFE \\

\textbf{No.14} Negative Overnight Jump
& \citet{jiangzhu2017}, JFE \\

\textbf{No.15} Drop + Rebound
& \citet{bremer1991}, JF \\

\textbf{No.16} Up + Positive Convexity;
\textbf{No.17} Up + Negative Convexity;
\textbf{No.18} Down + Positive Convexity;
\textbf{No.19} Down + Negative Convexity
& \citet{gulenwoeppel2026}, JFQA \\

\textbf{No.20} Daytime Reversal
& \citet{akbas2022}, JFE \\

\textbf{No.21} Price--Volume Lead--Lag
& \citet{liao2022}, RFS \\

\bottomrule
\end{tabularx}
\end{threeparttable}
\end{table}
\clearpage